\documentclass{aa} 

\usepackage{graphicx}
\usepackage{txfonts}
\usepackage{lipsum}
\usepackage{subcaption}
\usepackage{lscape}
\usepackage{placeins}
\usepackage{longtable}
\usepackage{tablefootnote}
\usepackage{upgreek}
\usepackage{amsmath}
\usepackage{hyperref}
\usepackage{gensymb}
\usepackage{color}
\usepackage{ulem}
\newcommand{\msun}{$\mathrm{M_{\odot}}$}

\newcommand{\Ni}{$^{56}$Ni\ }

\newcommand{\lam}{$\uplambda$}
\newcommand{\Ha}{H$\upalpha$}
\newcommand{\Hb}{H$\upbeta$}

\begin{document}
	\title{Spectral Dataset of Stripped-Envelope Supernovae from the Tsinghua Supernova Group}
	
	\author{Danfeng Xiang\inst{1,2}\thanks{xiangdanfeng@bjp.org.cn}
	\and Xiaofeng Wang\inst{2}\fnmsep\thanks{wang\_xf@mail.tsinghua.edu.cn} 
	\and Jujia Zhang\inst{3,4,5}\thanks{jujia@ynao.ac.cn}
	\and Shengyu Yan \inst{2}
	\and Han Lin\inst{3}
	\and Liming Rui\inst{2}
	\and Jun Mo\inst{2}
	\and Xinghan Zhang\inst{2,6}
	\and Hanna Sai\inst{2}
	\and Cheng Miao\inst{2}
	\and Gaobo Xi\inst{2}
	\and Zhihao Chen\inst{2}
	\and Fangzhou Guo\inst{2}
	\and Xiaoran Ma\inst{2}
	\and Gaici Li\inst{2}
	\and Tianmeng Zhang\inst{7,8,9}
	\and Liyang Chen\inst{2}
	\and Jialian Liu\inst{2}
	\and Wenxiong Li\inst{2,10}
	\and Xulin Zhao\inst{11}
	\and Fang Huang\inst{12}
	\and Yongzhi Cai\inst{3,4,5} 
	\and Weili Lin\inst{2}
	\and Jie Lin\inst{13}
	\and Chengyuan Wu\inst{3,4}
	\and Maokai Hu\inst{2}
	\and Cuiying Song\inst{2}
	\and Jicheng Zhang\inst{14}
	\and Qiqi Xia\inst{2}
	\and Zhitong Li\inst{9,15}
	\and Linyi Li\inst{2}
	\and Kaicheng Zhang\inst{2}
	\and Qian Zhai\inst{3}
	\and Juncheng Chen\inst{16}
	\and Zhou Fan\inst{10,9}
	\and Jianning Fu\inst{17,14}
	\and Shengbang Qian\inst{18}
	\and Hong Wu\inst{10}
	\and Xue-Bing Wu\inst{19}
	\and Huawei Zhang\inst{19}
	\and Junbo Zhang\inst{10}
	\and Liyun Zhang\inst{20}
	\and Jie Zheng\inst{15}
}

\institute{Beijing Planetarium, Beijing Academy of Sciences and Technology, Beijing, 100044, China
	\and Department of Physics, Tsinghua University, Haidian District, Beijing 100084, China
	\and Yunnan Observatories, Chinese Academy of Sciences, Kunming 650216, China
	\and International Centre of Supernovae, Yunnan Key Laboratory, Kunming 650216, China
	\and Key Laboratory for the Structure and Evolution of Celestial Objects, Chinese Academy of Sciences, Kunming 650011, China
	\and School of Physics and Information Engineering, Jiangsu Second Normal University, Nanjing, Jiangsu 211200, China
	\and Institute for Frontiers in Astronomy and Astrophysics, Beijing Normal University, Beijing 102206, China
	\and Key Laboratory of Space Astronomy and Technology, National Astronomical Observatories, Chinese Academy of Sciences, 20A Datun Road, Beijing 100101, China
	\and School of Astronomy and Space Science, University of Chinese Academy of Sciences, Beijing 100049, China
	\and National Astronomical Observatories, Chinese Academy of Sciences, Beijing 100101, China
	\and School of Science, Tianjin University of Technology, Tianjin 300384, China
	\and Department of Astronomy, Shanghai Jiao Tong University, Shanghai 200240, China
	\and Department of Astronomy, University of Science and Technology, Hefei 230026, China
	\and School of Physics and Astronomy, Beijing Normal University, Beijing 100875, China
	\and Key Laboratory of Optical Astronomy, National Astronomical Observatories, Chinese Academy of Sciences, Beijing 100101, China
	\and Schools of Electronics and Information Engineering, Wuzhou University, Wuzhou 543002, China
	\and Institute for Frontiers in Astronomy and Astrophysics, Beijing Normal University, Beijing 102206, China
	\and Department of Astronomy, School of Physics and Astronomy, Yunnan University, Kunming 650091, China
	\and Department of Astronomy, School of Physics, Peking University, Beijing 100871, China
	\and College of Physics, Guizhou University, Guiyang 550025, China
}

\date{Received November 1, 2025}

\abstract
{The extent of envelope stripping in the progenitor stars is directly reflected in the diversity of spectral features observed in stripped-envelope supernovae (SESNe).}
{Through extensive spectral observation and analysis, we aim to clarify the statistical differences between the subclasses of SESNe.
} 
{The Tsinghua Supernova group obtained 249 optical spectra of 62 SESNe during the years from 2010 to 2020, covering phases from $-$16 to over 190 days relative to maximum light. 
Most spectra were obtained during the photospheric phases after the supernova explosion. For each spectrum, the pseudo-equivalent widths (pEWs) and blueshift velocities of principal lines were measured.
We further investigated the common spectral features by analysing their velocity and strength correlations across all subtypes.}
{We identify the feature near 6200~\AA\ in SNe Ib as H$\mathrm{\alpha}$ through comparison with SNe IIb and Ic, which resolves inconsistent literature interpretations. Our finding reveals prevalent residual hydrogen in SNe Ib, further supporting a continuous stripping sequence from SNe IIb to Ib. 
We observe a trend in increasing velocity among different subtypes of stripped-envelope SNe, with SNe IIb exhibiting the lowest line velocities, followed by Ib, Ic, and Ic-BL. Typically, the O~I lines in SNe Ic/Ic-BL are stronger than those seen in SNe IIb/Ib. In nebular phases, the [Ca II] emission dominates over [O I] in SNe IIb/Ib while [O I] is stronger in SNe Ic, including the He-rich SN 2016coi. This spectral dichotomy implies that progenitors of SNe Ic (BL) have more massive CO cores and hence higher initial masses.}
{}


\maketitle

\nolinenumbers
\section{Introduction}
Core-collapse supernovae (CCSNe) are the violent deaths of massive stars that have initial masses $\gtrsim$~8~\msun. During the evolution of massive stars, their envelope materials may be stripped by stellar winds, binary interaction or eruptive mass ejection. Therefore, at the moment of explosion, the stars become H-poor or even He-poor, thus deficient of corresponding spectral features in SN spectra \citep{1997ARA&A..35..309F,2017hsn..book..195G}. These SNe are classified as stripped-envelope supernovae (SESNe). The spectra of SNe IIP/IIL or IIn exhibit persistent H lines, whereas in SNe IIb, the H lines weaken and eventually fade, with He~I lines progressively becoming the dominant feature. Weaker H lines indicate that the H envelope in SN IIb progenitors has been partially stripped.
SNe Ib lack H lines but show strong He lines, indicating that their progenitors have lost their H envelopes but still retain their He envelopes. Spectra of SNe Ic lack both H and He lines, implying that both H and He envelopes of their progenitors have been lost before the explosion. 

During the last decade, a new class of peculiar SN, spectroscopically similar to SESN, has been discovered. Their observed properties are best explained by the ultra-stripped supernovae model \citep{2015MNRAS.451.2123T,2020ApJ...900...46Y,2023ApJ...959L..32Y,2024ApJ...969L..11D}.
These events are characterized by fast-evolving light curves, which imply unusually low ejecta masses, and extensive stripping of the progenitor material. Observations indicate that several such SNe have likely lost their CNO layers  \citep{2022ApJ...941L..32K,2022Natur.601..201G,2022ApJ...938...73P,2023A&A...673A..27N,2024ApJ...967L..45W}, even the inner O/Si/S layers \citep{2025Natur.644..634S}.

In classical stellar evolutionary theories where stars evolve alone, stars with $M_{\mathrm{ZAMS}}\gtrsim$~25~\msun\ can evolve to H-poor Wolf-Rayet stars and explode as SNe Ib or Ic. However, it is not settled whether SESNe are produced by single stars or binaries, nor the mass range of the progenitors. However, mounting evidence now favours lower-mass binary systems over very massive single stars as the dominant formation channel for SESNe  \citep{2012Sci...337..444S,2015PASA...32...15Y,2019MNRAS.485.1559P,2025MNRAS.tmp.2097Z}. The few studies of direct observations of SN IIb or Ib progenitors to date have favoured progenitors of lower initial masses in binary systems. For instance, strong evidence points to a companion for the SN Ib iPTF13bvn progenitor \citep{2013ApJ...775L...7C,2014AJ....148...68B,2015MNRAS.446.2689E,2015A&A...579A..95K,2016MNRAS.461L.117E}. Similarly, a yellow hypergiant companion has been suggested for SN 2019yvr \citep{2021MNRAS.504.2073K,2022MNRAS.510.3701S}. No direct detection of SNe Ic progenitors has been reported to date, despite indirect evidence for the existence of binary companions has been found \citep{2019ApJ...871..176X,2022ApJ...929L..15F,2025ApJ...980L...6Z}.  The most massive stars with $M_{\mathrm{ZAMS}}\gtrsim$~25~\msun\ may instead result in super luminous SNe or failed SNe. Moreover, binary progenitors have also been proposed to contribute to the diversity of H-rich SNe II \citep{1992PASP..104..717P,2019A&A...631A...5Z}, suggesting that this channel may be common for all types of core-collapse supernovae.

One way to study the properties of SESN progenitors is through their spectra. Spectral line intensities during the photospheric phase reflect the envelope composition, providing insight into the extent of envelope stripping prior to the explosion. Additionaly, in the nebular phase, luminosity ratios of [O~I] and [Ca~II] lines  are strongly correlated with progenitor CO core masses \citep{2014MNRAS.439.3694J,2015A&A...573A..12J,2021A&A...656A..61D,2019NatAs...3..434F,2023ApJ...949...93F}. And nebular spetra models show that higher $L_{\mathrm{[OI]}}$ and $L_{\mathrm{[NII]}}$ correspond to higher initial masses. Through spectroscopic observations and large-sample statistical analysis, it is possible to clarify differences in envelope stripping among progenitor stars of different types of SESNe. This approach can address questions such as whether the degree of envelope stripping in progenitor stars of SNe with different spectral types is continuous or distinct across spectral types, how the composition of the remaining envelope influences spectral characteristics, etc.

Substantial optical spectroscopic databases for SESNe, now comprising over 2,500 spectra for objects discovered between 1994 and 2016 \citep{2014AJ....147...99M,2018A&A...618A..37F,2019MNRAS.482.1545S,2023A&A...675A..82S}.
These data facilitate more accurate classification, reveal ejecta composition across subtypes, and probe progenitor properties \cite{2016ApJ...827...90L,2019MNRAS.482.1545S,2018A&A...618A..37F}.
Though the classifications of SESNe are based on the presence or absence of several specific lines, some SNe show spectral evidence of residual hydrogen in SNe Ib, \citep[e.g.,][]{2014AJ....147...99M}, or helium in SNe Ic \citep[e.g.,][]{2016ApJ...821...57D,2018MNRAS.478.4162P}. 
In fact, hydrogen features are now considered common in SNe Ib. Comparison of spectra near maximum light of SNe IIP, IIb and Ib shows a sequence of increasing prominence of He~I lines and decreasing H~I line strength \citep{2017hsn..book..195G}. 
Moreover, measurements of spectral lines reveal a continuous variation in \Ha\ line strength from SNe IIb to Ib, suggesting that envelope stripping in their progenitors may be a gradual process. 
In contrast, the photospheric spectra of SNe Ic are distinct from those of SNe IIb and Ib, favouring models with negligible helium over those with gradually stripped He envelopes \citep{2016ApJ...827...90L,2018A&A...618A..37F}. 
Furthermore, nebular-phase spectroscopic analysis reveals that while SNe IIb and Ib share similar CO core mass distributions, SNe Ic show significantly higher CO core masses, implying more massive progenitor stars \cite{Fang2019NatAs}. 
These observational studies appear to confirm that progenitor stars of SNe Ic have completely lost their He envelopes. Although some theoretical studies rule out substantial residual He in SNe Ic \citep{2013ApJ...773L...7F,2021ApJ...908..150W}, others propose that mixing of He could influence the spectral types of SNe Ib/c thus He might be hidden in SNe Ic \citep{2012MNRAS.424.2139D}.
Determining the true stripping history of SESNe progenitors thus requires further observations and theoretical modelling.

In this work, we present the spectra of SESNe obtained by the supernova group at Tsinghua University (THU). A total of 249 low-resolution spectra were obtained for 62 SNe during the phase from late 2010 to 2020. Our database extends the temporal coverage of previous homogeneous samples by including SNe up to 2020, compared to the 2016 cutoff in earlier works. Note that a portion of spectra from our group have been published in \cite{2024MNRAS.528.3092L} for SNe II, and \cite{2021ApJ...906...99L} for some SNe Ia.
All spectral data used in this work will be made available via the Weizmann Interactive Supernova Data Repository \cite[WiseREP\footnote{\url{https://www.wiserep.org/}};][]{2012PASP..124..668Y}.

\section{Observations}\label{sec:data-observation}
Since 2010, the supernova group at THU has been conducting continuous observations of SNe, including multi-band photometry and spectroscopy in optical wavelengths.
Optical spectra of SNe were obtained using the Beijing Faint Object Spectrograph and Camera (BFOSC) or the spectrograph made by Optomechanics Research Inc. (OMR), mounted on the 2.16-m telescope at Beijing Xinglong Observatory (BAO) \cite[hereafter XLT;][]{2016PASP..128k5005F}, and the Yunnan Faint Object Spectrograph and Camera (YFOSC) on the 2.4-m telescope at Lijiang Observatory \cite[hereafter LJT;][]{2015RAA....15..918F}. We selected recently discovered supernova candidates from public databases such as the Latest Supernovae website\footnote{\url{https://www.rochesterastronomy.org/snimages/}} and the Transient Name Server (TNS\footnote{\url{https://www.wis-tns.org/}}). To ensure data quality, we only observed SNe or SNe candidates brighter than $\sim$18~mag. Both telescopes monitored low-resolution (R$\sim$500--2000) spectroscopy for SNe.
All the spectra were reduced using standard IRAF pipelines \citep{1986SPIE..627..733T,1993ASPC...52..173T}, including bias and flat-field corrections. Wavelength calibration solution was determined using the arc lamps of He/Ar or Fe/Ar and applied to the one-dimensional spectra. Flux calibrations were obtained using the spectra of nearby standard stars taken on the same night.
Finally, telluric absorptions were removed from the spectra using the atom lines produced by the reduction process. 

For some SNe, photometry was conducted in Johnson-Cousins $BVRI$ or $BV$ plus sloan $gri$ bands with the 80-cm THU-NAOC Telescope (TNT; \citealt{2008ApJ...675..626W,2012RAA....12.1585H}) at BAO. 
The TNT light curves are used to derive the date of maximum light when available.  

Through a decade-long observational campaign conducted from 2010 to 2020, we ultimately obtained a total of 249 spectra of 62 SESNe. Details of these SNe and the corresponding observations are listed in Tab.~\ref{tab:sne-list} and Tab.~\ref{tab:spec-list}, respectively. As a summary, 106 spectra are from XLT and 139 are from LJT. Spectroscopic data for eight SNe have been previously published in separate papers (see details in Tab.~\ref{tab:spec-list}). These published data include four extra spectra for SN~2017ein from telescopes outside China, including the 2-m Faulkes Telescope North (FTN) of the Las Cumbres Observatory network, and the 9.2 m Hobby-Eberly Telescope (HET), the ARC 3.5-m telescope at Apache Point Observatory (hereafter APO-3.5~m). These spectra are also included in this work. Specifically, spectra for seven SNe (CSS141005, PSN J0123, SNe 2010ln, 2011bl, 2012C, 2012ej, and 2017iro) are presented here for the first time. Furthermore, spectral data for eight SNe (2016bll, 2016cce, 2016G, 2016M, 2017cxz, 2017jdn, 2018ie, and 2018if) have only their classification spectra publicly available on TNS, all of which originate from our research group.

\begin{figure}
	\centering
	\includegraphics[width=0.48\columnwidth]{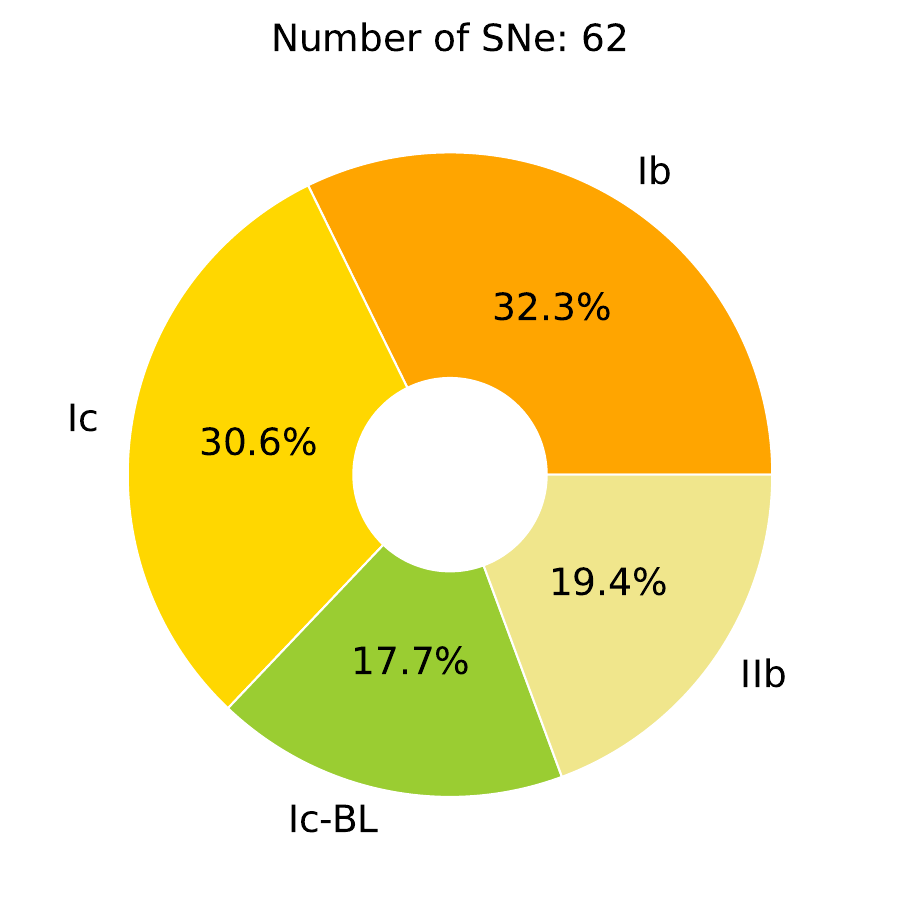}
	\includegraphics[width=0.48\columnwidth]{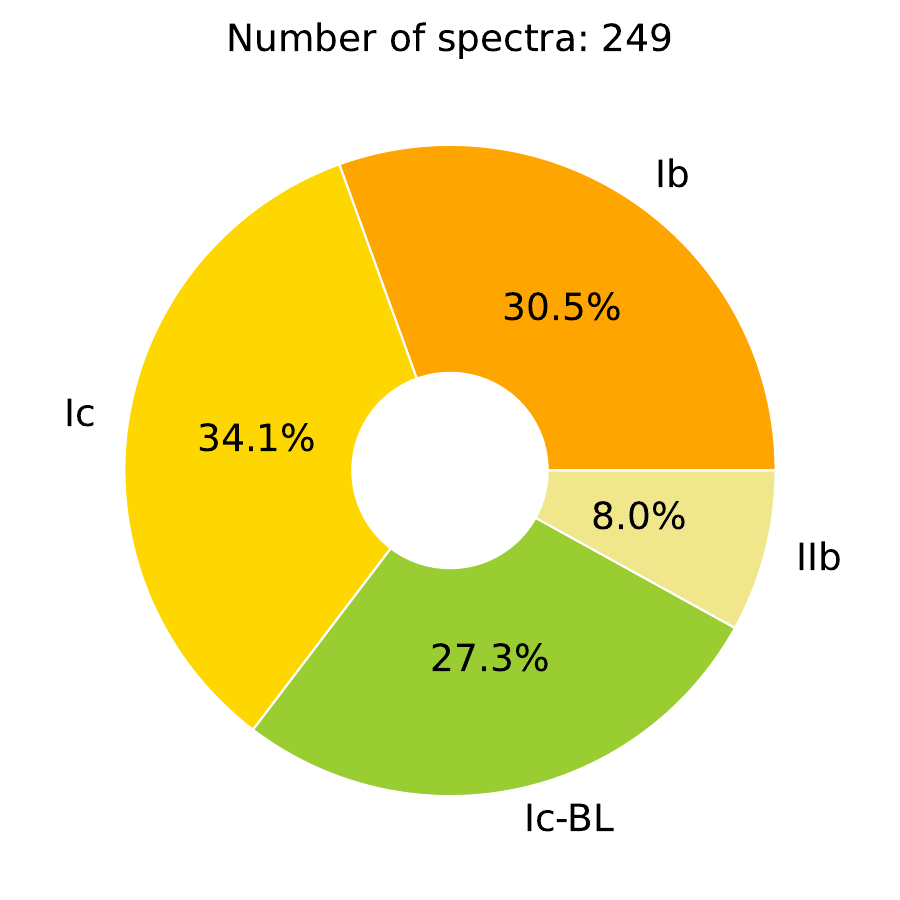}
	\caption{Pie charts of numbers of SNe (left) and numbers of spectra (right) by types of our sample. }
	\label{fig:static-pie}
\end{figure}

Fig.~\ref{fig:static-pie} show the pie chart of the observed SNe and the spectra by type. The number of SNe Ib is comparable to that of SNe Ic, and approximately twice that of SNe Ic-BL. The local rate of different subtypes of SESNe are recently presented by \cite{2025A&A...698A.305M,2025A&A...698A.306M}. We notice that the type fractions of our spectral sample differ significantly from those in \cite{2025A&A...698A.305M}, who reported 35\%, 17.5\%, 32.5\%, and 15\% for IIb, Ib, Ic, and Ic-BL, respectively. While the fractions of Ic and Ic-BL in our sample are consistent with those in the literature, we find notable discrepancies for types IIb and Ib. This may be attributed to several factors. First, it is important to emphasize that our sample does not include a generic ``Ib/c'' subtype. First, objects classified as ``Ib/c'' on public platforms, e.g. TNS or the Weizmann Interactive Supernova Data Repository \cite[WiseREP;][]{2012PASP..124..668Y}, were reclassified as Ib or Ic based on detailed spectroscopic analysis (see the footnote of Tab.~\ref{tab:sne-list}). Second, \cite{2025A&A...698A.305M} imposes a strict cutoff distance of 40~Mpc ($z\approx0.01$) to ensure sample completeness. However, 34 supernovae ($\sim$55\%) in our sample have redshifts exceeding 0.01, suggesting a potential bias in the sample selection.
Additionally, SNe IIb have the lowest mean peak luminosity among the four subclasses \citep{2014AJ....147..118R}. Consequently, a brightness-limited spectroscopic sample may miss a substantial fraction of these dimmer events, leading to an underrepresentation of SNe IIb. Furthermore, as we will discuss later, some SNe Ib also show hydrogen lines in their spectra and SNe IIb may be misclassified as type Ib, which could also affect the observed fractions. 

The right panel of Fig.~\ref{fig:static-pie} shows the proportions of spectra across different subtypes. The fractions of type Ib and Ic spectra are similar to the actual fractions of SN types estimated for the local universe. The significant increase in the number of type Ic-BL spectra is likely due to the selection bias, with the tendency of obtaining more spectra for rare SN types. The relatively small number of type IIb spectra is likely due to their intrisic lower luminosity, which makes obtaining high-quality spectra challenging. As a result, type IIb SNe are often deprioritized as observational targets, or the obtained spectra exhibit excessively low signal-to-noise ratios.

\begin{figure}
	\centering
	\includegraphics[width=0.48\columnwidth]{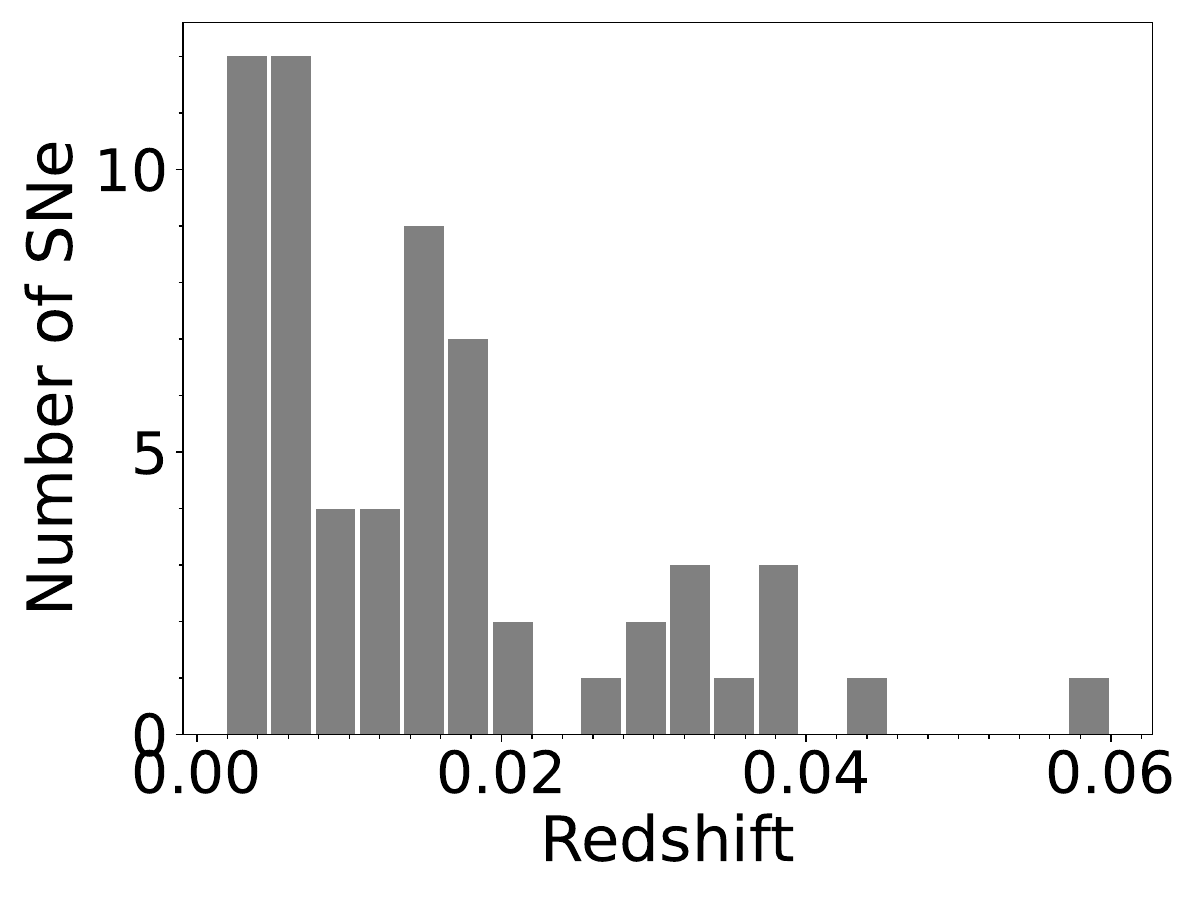}
	\includegraphics[width=0.48\columnwidth]{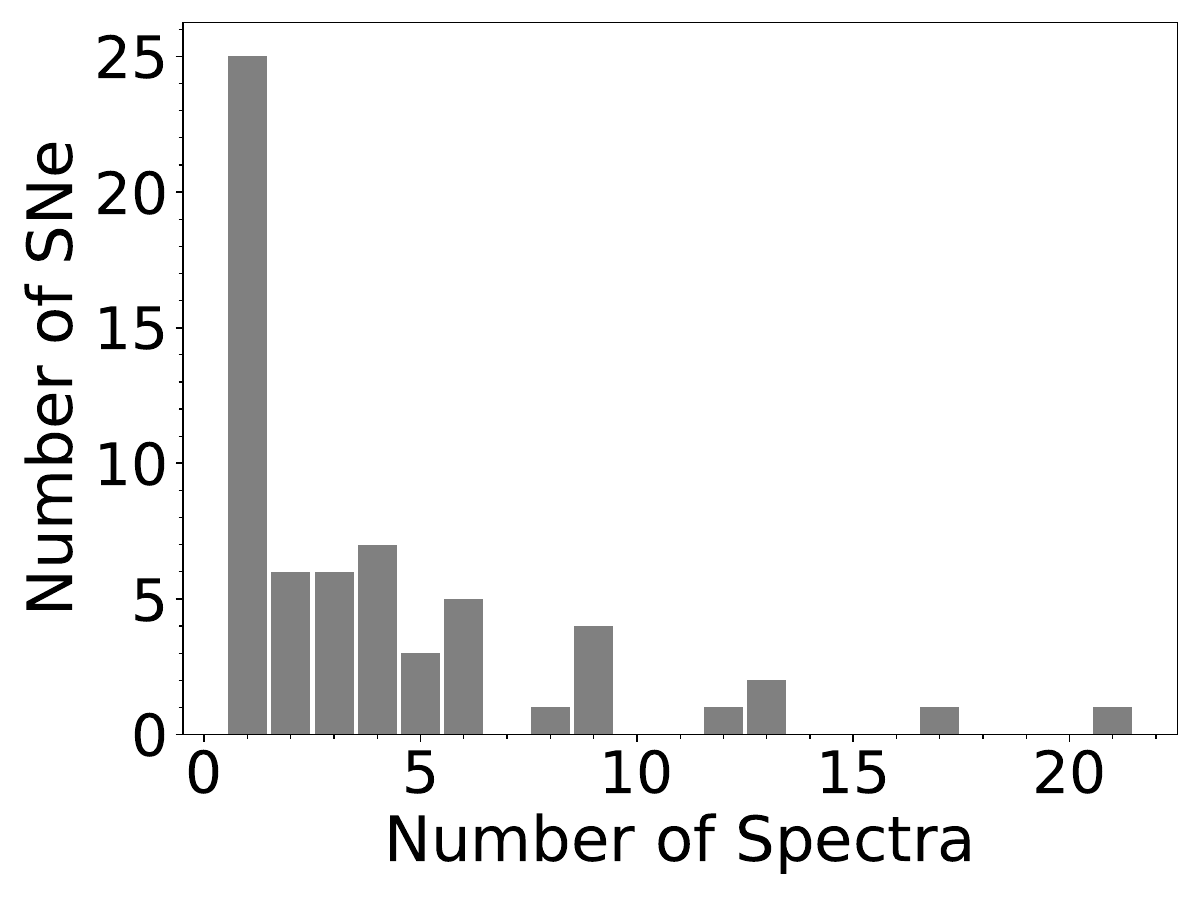}
	\includegraphics[width=0.48\columnwidth]{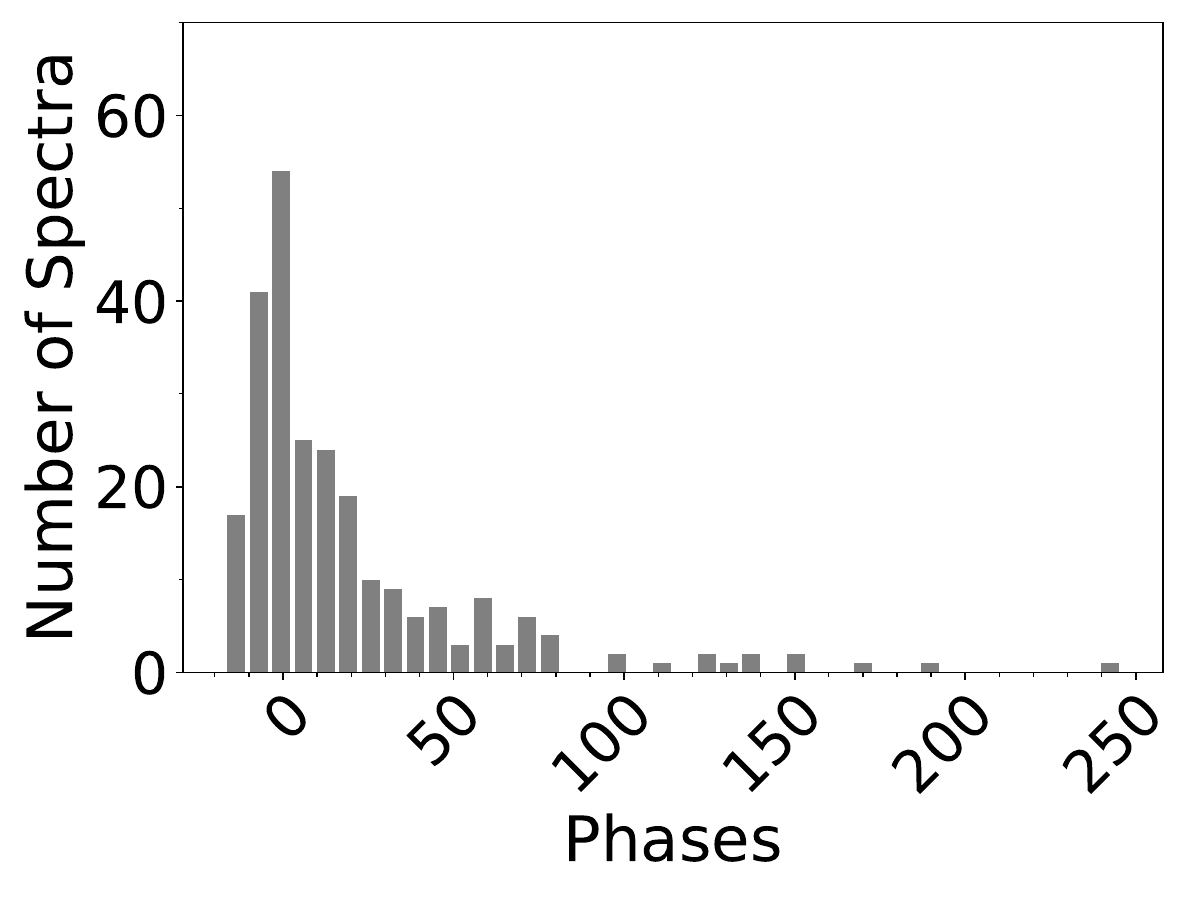}
	\includegraphics[width=0.48\columnwidth]{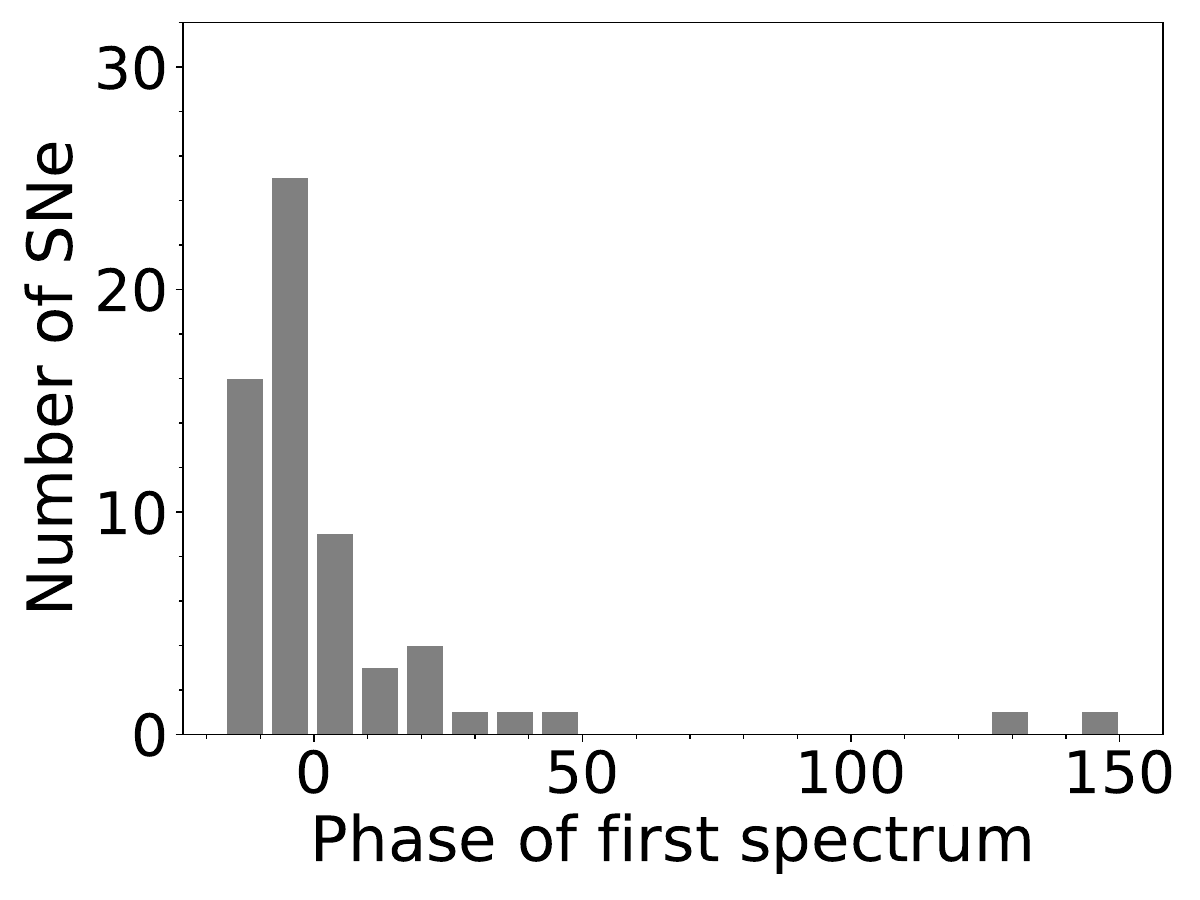}
	\caption{Distribution of the SN redshift (upper-left), number of spectra for a single SN (upper-right), phases of spectra (lower-left), and the phase of the first spectrum of each SN (lower-right).}
	\label{fig:static-hist}
\end{figure}

In Fig.~\ref{fig:static-hist} we show the distribution of SN redshift, number of spectra for each one single SN, phases of spectra and phases of the first spectrum for an SN.
The SNe of our spectral sample have an average redshift of 0.015, with the maximum redshift being as 0.06 (SN~2018beh). Most SNe of our sample were glanced by only once, and an average of 4 spectra were obtained for each SN. 

\section{Spectral analysis methods}\label{sec:methods}
Before this work, several groups have published spectroscopic samples of SESNe, including \cite{2014AJ....147...99M} from the Harvard-Smithsonian Center for Astrophysics, \cite{2018A&A...618A..37F} from the Palomar Transient Factory, \cite{2019MNRAS.482.1545S} from the Lick Observatory Supernova Search collaboration of Berkeley, and \cite{2023A&A...675A..82S} from the Carnegie Supernova Project. The relevant analysis papers include \citet{2016ApJ...832..108M}, \citet{2016ApJ...827...90L}, \citet{2019ApJ...880L..22W}, and \citet{2023A&A...675A..83H}. Subsequent analyses of such samples commonly involve measuring line velocities and intensities, comparing spectral properties across subtypes, and applying principle component analysis (PCA) for classifications. Our analysis references and builds upon this body of work.

\begin{figure}
	\centering
	\includegraphics[width=1.\columnwidth]{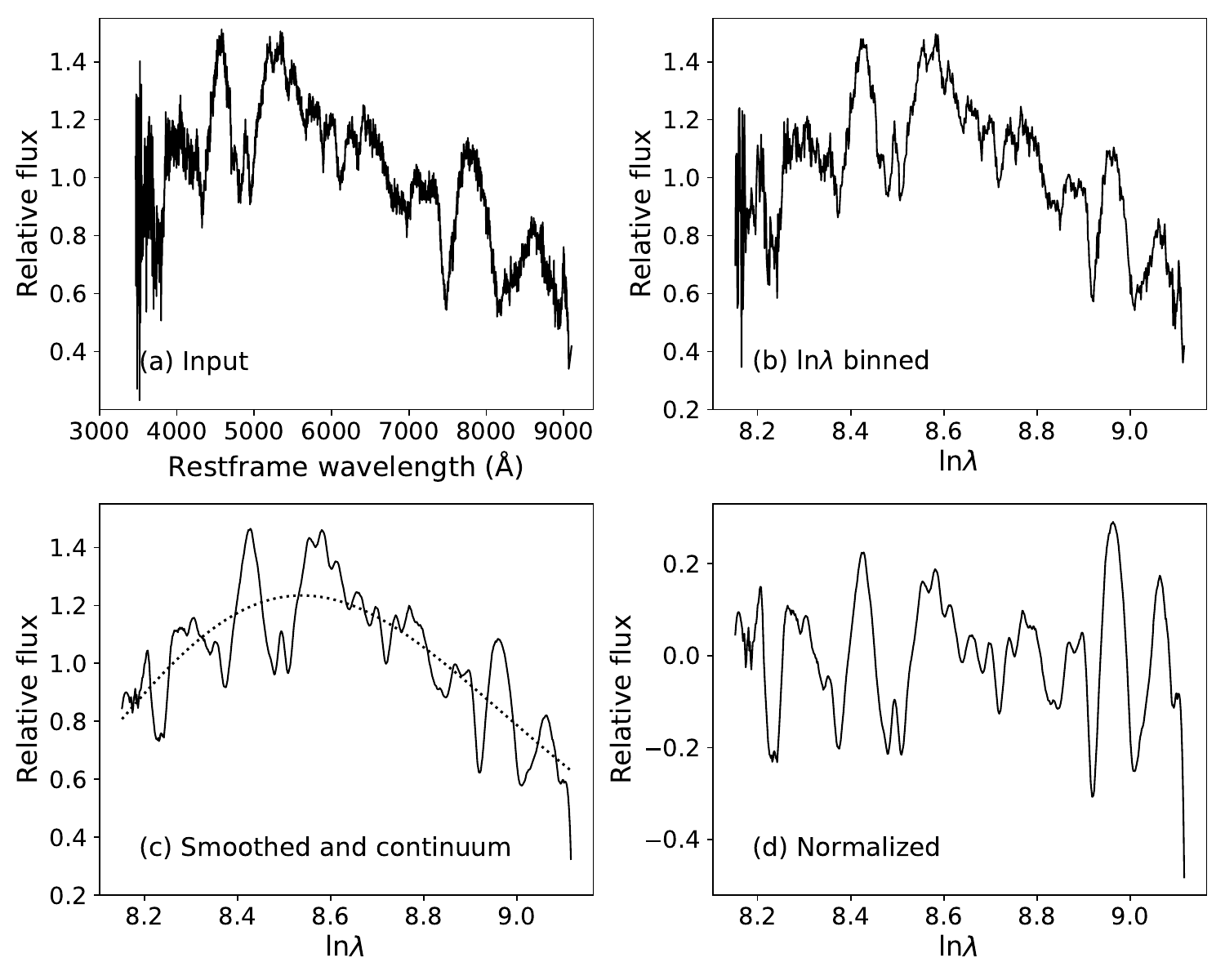}
	\caption{Pre-processing method for our spectral data. (a) The input spectrum after corrections of redshift; (b) Spectrum after binning into logarithmic scale with d$\lambda_{\mathrm{ln}} = 0.0015$; (c) Smoothed spectrum with window length of 21, the pseudo continuum is plotted as a dotted line; (d) The normalized spectrum by dividing the continuum in (c). }
	\label{fig:pre-process}
\end{figure}

\subsection{Pre-processing of spectra}\label{sec:pre-process}
The original SN spectra, obtained from different instruments under varying conditions, have different wavelength resolution and signal-to-noise ratios. To enable a consistent comparison, we homogenized the dataset by convolving all spectra to a common resolution and applying a smoothing kernel.
Fig.~\ref{fig:pre-process} shows the procedure of spectra pre-processing.
First, observed wavelengths are converted to the rest frame by dividing $(1+z)$, where $z$ is the SN redshift. Following \cite{2007ApJ...666.1024B}, we then bin the wavelength axis logarithmically with d$\lambda_{\mathrm{ln}} = 0.0015$, corresponding to d$\lambda = 10$~\AA (d$v=450$~km~s$^{-1}$) at \Ha. 
For spectra containing strong host-galaxy emission lines, emission lines were masked out by applying a third-order polynomial fit to the adjacent continuum. Then the binned spectra were smoothed with a Savitzky-Golay filter using the \texttt{Python} function \texttt{scipy.signal.savgol\_filter}. 
Finally, spectra in photospheric phases are normalized by $f_{\mathrm{flat}} = f_{\mathrm{obs}}/{f_{\mathrm{con}}}-1$, where $f_{\mathrm{flat}}$ is the normalized flux and $f_{\mathrm{con}}$ is the pseudo continuum.
To calculate the pseudo continuum, we first chose a certain number of knots and calculate the mean flux between each pair of knots. Then fitted the knots with a smoothing spline fit using the \texttt{Python} function \texttt{scipy.interpolate.UnivariateSpline}.

\subsection{Determining dates of maximum light}\label{sec:peaktime}
To determine the phase of each spectrum, time of maximum light of each SN is needed. 
We first checked the light curves by TNT and found 14 SNe in our sample have abundant photometric observations around peak in $B$- or $V$-band. 
For the remaining SNe, we collected light curve data from literature and open data archive such as the Zwicky Transient Facility \citep[ZTF;][]{2019PASP..131a8003M,2019PASP..131a8002B}. When only ZTF $gr$-band photometry was available, we converted the Sloan magnitudes to standard Johnson $BV$-band magnitudes by applying the relation presented by \cite{2006A&A...460..339J}. Dates of maximum light were then estimated using a third-order polynomial fit to the light curve data around peak. Using this approach, we determined the time of maximum for 41 SNe in total.

\begin{figure}
	\centering
	\includegraphics[width=0.85\columnwidth]{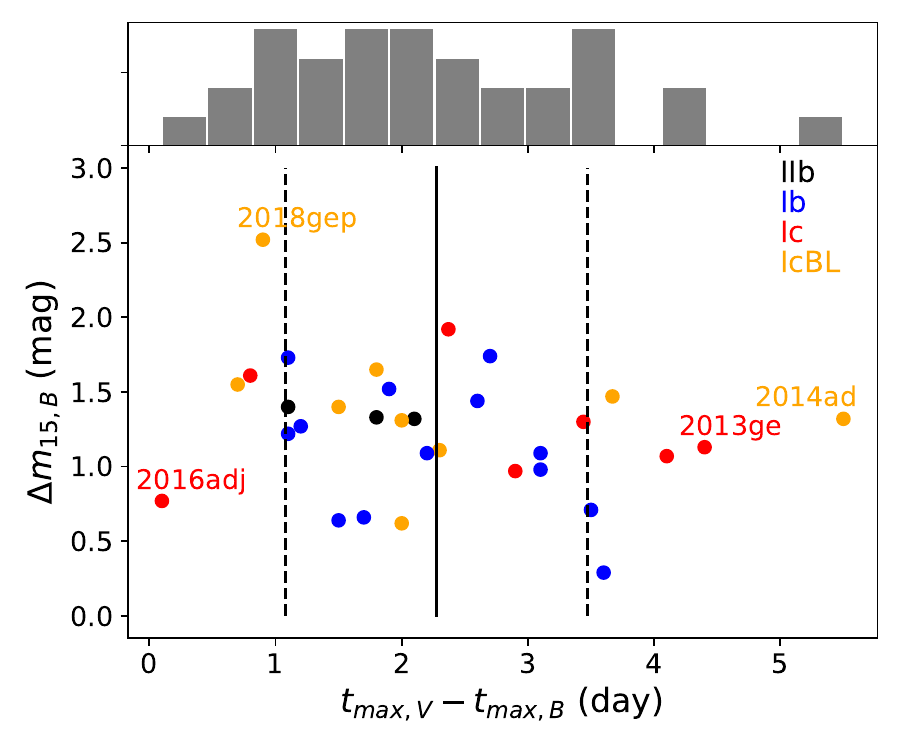}
	\caption{The distribution of $t_{\mathrm{max},V}-t_{\mathrm{max},B}$ versus $\Delta m_{15,B}$ of our SESN sample. The vertical lines mark the average delay (solid line) and 1-$\sigma$ range (dashed lines). Different types are plotted in different colours which are denoted at the upper right corner.}
	\label{fig:tV-tB}
\end{figure}

For those without sufficient light curve data, we used SNID \citep{2007ApJ...666.1024B} to determine the phase via spectral cross-correlation, thereby estimating the date of $V$-band maximum. This method yielded peak dates for 21 SNe in our sample. All derived times of maximum light are listed in Tab.~\ref{tab:sne-list}.
As shown in Fig.~\ref{fig:tV-tB}, most SESNe reach to $V$-band peak 2.2$\pm$1.2 days after the $B$-band peak, consistent with the finding of \cite{2014Ap&SS.354...89B}. Therefore, for SNe with valid estimate of peak dates only in $B$-band, peak dates in $V$-band were estimated by adding an average offset of 2.2~days. This approach was used for 3 SNe in our sample. Throughout this work, phases are the rest frame days since $V$-band maximum, matching the convention used by SNID.

Several SNe are outliers in the $t_{\mathrm{max},V}-t_{\mathrm{max},B}$ distribution (Fig.~\ref{fig:tV-tB}), lying outside the 1-$\sigma$ range of the average. We find a weak negative correlation between this peak-time difference and the decline rates $\Delta m_{15}$ in $B$-band. Most outliers still follow the relation, except for SN~2014ad and SN~2016adj. 
These outliers may originate from energy sources that depart from the typical case. 
We note that SN~2013ge is a type Ic with double-peaked light curves \citep{2016ApJ...821...57D}. Two scenarios are proposed for the extra energy input of the first peak: an extended envelope or outward mixing of \Ni. Both scenarios would result in faster rise in bluer bands, therefore larger separation between the $B$ and $V$ band peaks. SN 2014ad is a broad-lined type Ic that was exceptionally energetic compared to other SNe Ic-BL \citep{2018MNRAS.475.2591S}. SN~2016adj is a carbon-rich SN Ic that exhibits CO emission two months after explosion, and hydrogen emission in NIR about 3 months after explosion \citep{2024A&A...686A..79S,2022ApJ...939L...8S,2018MNRAS.481..806B}.
SN~2016iae is a normal SN Ic with relatively flat velocity evolution \citep{2019MNRAS.485.1559P}. The energy sources of SNe~2016adj and 2014ad may differ from those of other SESNe. SN~2018gep is found to have faster temporal evolution in light curves. Though classified as broad-lined, SN~2018gep is an outlier when compared with other SNe Ic-BL except for iPTF16asu, transitioning from a fast blue transient to an SN Ic-BL \citep{2021ApJ...915..121P}. Its early light curve has been attributed to interaction between the ejecta and circumstellar material lost through eruptive mass loss shortly before explosion \citep{2019ApJ...887..169H}. 

\subsection{Line identification and measurements}\label{sec:line}
Accurate line identification is required prior to measuring line intensities and velocities. For this task, we consulted established line lists from previous studies  \citep{2016ApJ...827...90L,2023A&A...675A..83H}and utilized the spectral synthesis code SYNOW/SYNAPPS \citep{2011PASP..123..237T} for verification.

\cite{2023A&A...675A..83H} identified the main features by fitting the mean spectra of each subtype at multiple epochs with 11 ionic species.  
For SNe IIb, the primary spectral features include H~I, He~I, O~I, Na~I, Fe~II, Ti~II, Ca~II and possible high-velocity Si~II. For SNe~Ib, the main elements are the same as for SNe~IIb, with ambiguous detections of H~I. For SNe~Ic, the main features are identified as O~I, C~II, Na~I, Si~II, Fe~II, Ti~II, Co~II, and Ca~II. The C~II lines are visible only in early phases and faded after maximum light.
In our preliminary spectral fitting, we utilized identical ion species. Thus the main features in our SN spectra were identified according to previous studies and our fitting results. Consequently, for all SNe, lines of Fe~II, Ca~II~NIR triplet, and OI~\lam7774 were identified and measured. Specifically, lines of \Ha, \Hb, He~I\lam5876, He~I\lam6678 and He~I\lam7065 were identified for SNe IIb; and lines of He~I\lam5876, He~I\lam6678, He~I\lam7065 and He~I\lam7281 were identified for SNe Ib; lines of Si~II\lam6355, C~II\lam6580 and C~II\lam7234 can be identified for SNe Ic/Ic-BL at phases earlier than +10~d. For SNe Ic, which we treated as helium-free, the 5700\AA\ absorption was interpreted as Na~I~D, neglecting potential contamination from He~I\lam5876. And the 6200\AA\ absorption was identified as Si~II\lam6355. 
We note that for SNe Ib, the feature at similar wavelength can be residual H or Si~II, which will be discussed further in Sec.~\ref{sec:results}. 

\begin{figure}
	\includegraphics[width=1.0\linewidth]{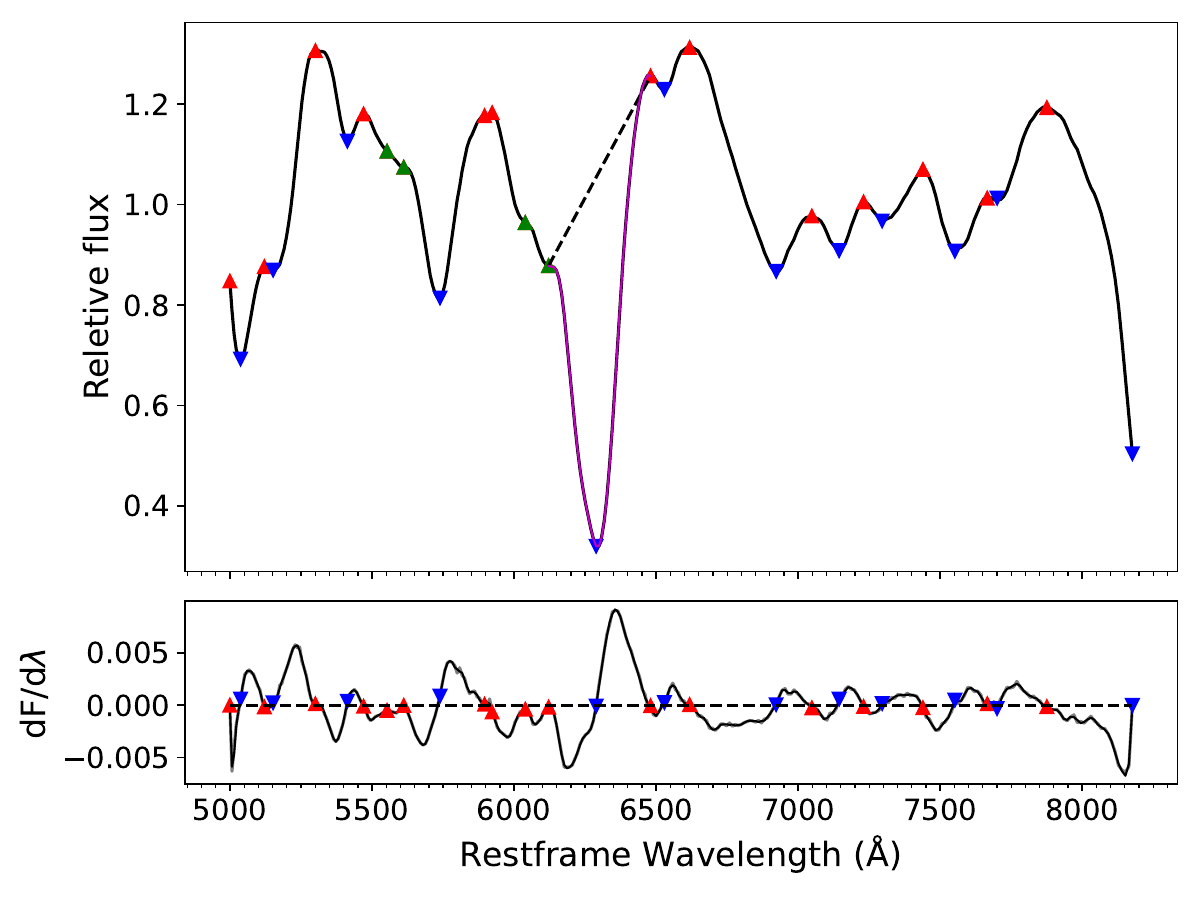}
	\caption{The schematic diagram illustrating the method of defining the spectral line range. For the smoothed spectrum (upper panel), the first derivative of flux with respect to wavelength is computed (lower panel). Local maxima and minima are identified where the sign of the derivative changes, marked with red upward-pointing and blue downward-pointing triangles, respectively. At certain wavelengths, distinct inflection points are observed where the derivative approaches zero without a sign change; these are indicated by green upward-pointing triangles. The wavelength range of the \Ha\ line is thus delineated by purple colour, while the local pseudo-continuum is represented by the black dashed line.}
	\label{fig:wave-range}
\end{figure}

For each spectrum, we measured the pseudo equivalent widths (pEWs) and expansion velocities of each line. 
The procedure was as follows. First, on the smoothed (but not normalized) spectra, we defined the wavelength range for each line by locating the local maxima flanking the absorption trough or distinct inflection points (where $\mathrm{d}f/\mathrm{d}\lambda\approx0$). This range selection is illustrated in Fig.~\ref{fig:wave-range}. We then constructed a local pseudo-continuum by drawing a straight line between the two endpoints of the defined range. Finally, the pEW was calculated by normalizing the flux within the line range to this local pseudo-continuum and integrating across the wavelength interval.

\begin{figure*} 
	\centering
	\includegraphics[width=0.95\columnwidth]{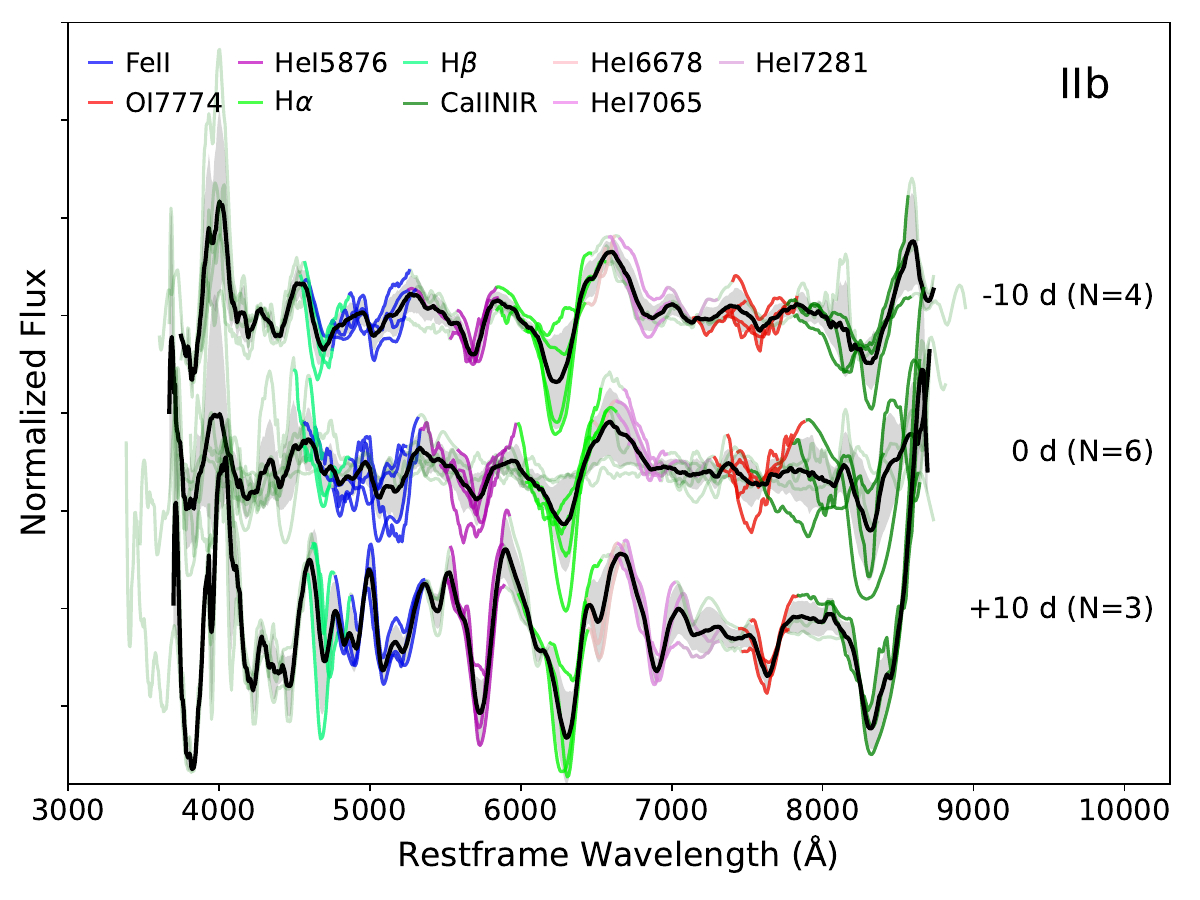}
	\includegraphics[width=0.95\columnwidth]{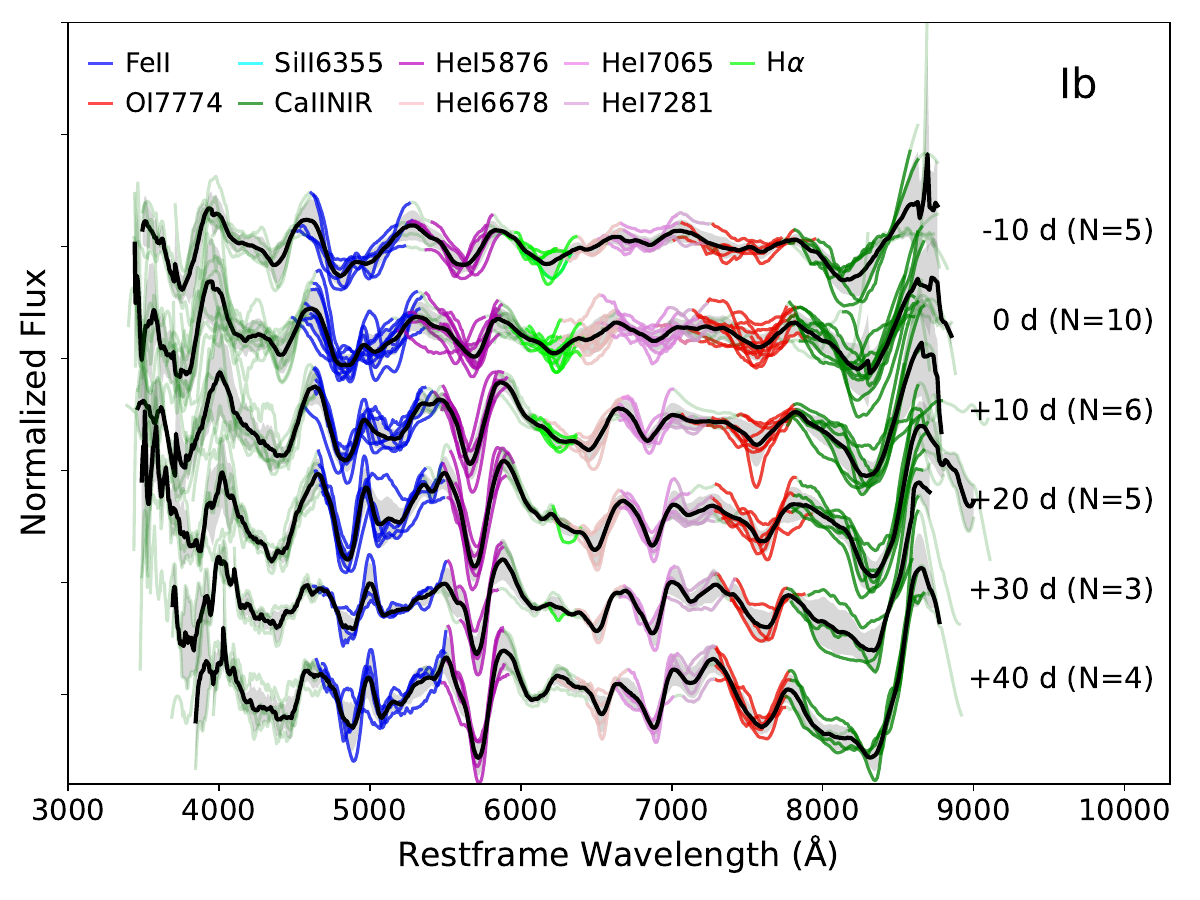}
	\includegraphics[width=0.95\columnwidth]{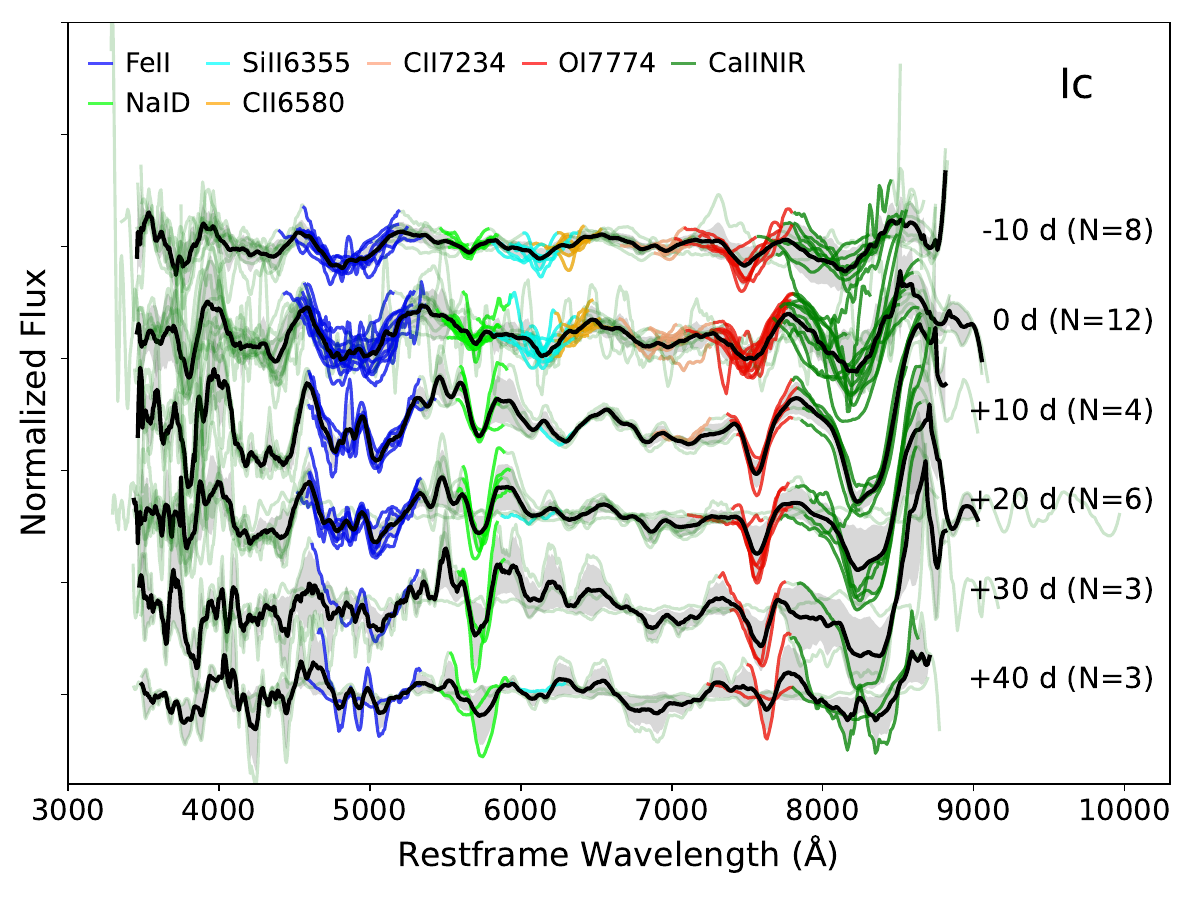}
	\includegraphics[width=0.95\columnwidth]{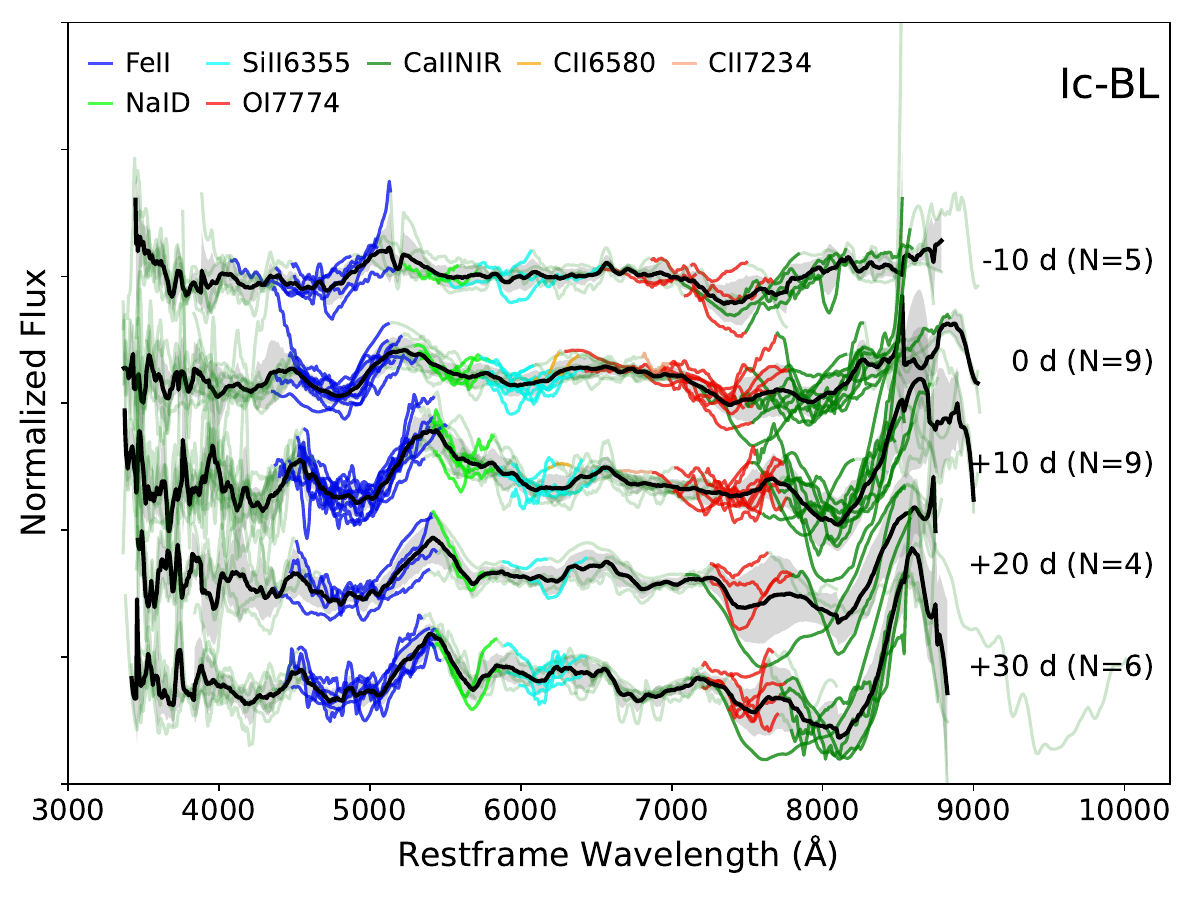}
	\caption{Mean spectra and their corresponding standard deviations of different subtypes of SESNe at four different phases spanning from $-$10 to 40 days since $V$-band maximum. Numbers on the right of each set of spectra are the corresponding phases and number of spectra included to calculate the mean spectra are shown in the brackets. The mean spectra are shown by solid black lines, and 1-$\sigma$ uncertainty is shown by shaded gray regions. Each single spectrum is plotted in light green. For each spectrum, lines are marked by colors denoted on top of each figure. }
	\label{fig:line-identify}
\end{figure*}

In most cases, where a spectral line does not blend with others, its pEW was calculated by directly integrating over the wavelength interval, and the expansion velocity was taken as the blueshift of the absorption minimum. While in cases where spectral line blending occurs, multi-Gaussian fitting was employed to deconvolve the individual lines. The wavelength range of each blended line is defined by the intersection points of the fitted Gaussian functions, while the line velocity is determined by the blueshifted minimum of the corresponding Gaussian component. Common blends treated in this way include the O~I\lam7774+Ca~II~NIR triplet in high velocity SNe Ic and Ic-BL, and the \Ha+He~I\lam6678 complex in SNe Ib.
In particular, we found that Fe~II lines (\lam\lam4924, 5018, 5091, 5169, 5198) in most spectra are severely blended and challenging to separate each component (see Fig.~\ref{fig:line-identify} and Fig.~7 of \citealt{2016ApJ...827...90L}). To ensure uniformity and avoid inconsistencies in defining individual component boundaries, we measured only the total pEW of the blended Fe~II feature. Its velocity was measured from the blueshifted minimum of the Fe~II\lam5169 line on the red side of the blend. In SNe IIb, where the Fe~II lines blend with \Hb, we applied multi-Gaussian fitting to separate the contributions and measure their individual pEWs.

We constructed mean spectra for each SESN subtype at various epochs using the normalized spectra.
The phases were set to be in range of $-$10 d to $+$40 d, with interval of 10 days. For each phase, we selected spectra obtained within a 5-day interval around that age. To avoid over-representing individual SN with multiple spectra, only the spectrum closest to the target phase was used from each object. 
At least three SNe (spectra) were required in deriving the mean spectrum at a given phase. Fig.~\ref{fig:line-identify} shows the mean spectra of each type, with the dominant spectral features marked in different colours. 
As shown in Fig.~\ref{fig:line-identify}, for each subtype, line profiles of different SNe at the same epoch exhibit varying degrees of diversity. In SNe IIb, the \Ha\ region and Fe~II lines show greater scatter at a given phase than the He~I lines. For SNe Ib, the strength of \Ha\ displays considerable variation at early times but converges gradually over time. 
In SNe Ic, significant dispersion is present in the C~II and Si~II regions during early phases; except for Fe~II lines, the post-peak dispersion ($\gtrsim$30~days) appears to mainly arise from differences in the strength of the emission components of the lines, possibly reflecting diversity in ejecta thickness. SNe Ic-BL do not show pronounced changes in diversity across epochs, with the Si~II and O~I lines exhibiting the greatest variation.

Mean spectra of different subtypes at various epochs were also presented by \cite{2016ApJ...827...90L} and \cite{2023A&A...675A..83H}. Those from the former are publicly available online. Our mean spectra show excellent agreement with theirs, particularly for types IIb and Ib, where differences are minimal (Fig.~\ref{fig:mean-spec-comp}). A notable discrepancy is observed for SNe IIb near maximum light: the absorption around 6900~\AA\ (He~I\lam7065) is weaker in our sample. For SNe Ic, the mean spectra also exhibit only minor differences, with the main deviation in the region near 5700~\AA\ (Na~I~D) after maximum light. In SNe Ic-BL, our mean spectra display stronger O~I\lam7774 absorption with larger scatter, along with slight variations in the red wing of the Fe~II complex.

\subsection{Transition to nebular phase}\label{sec:neb}
Measurements were only performed on spectra during photospheric phase. We defined a spectrum as having exited the photospheric phase if, within the line-profile range defined in Fig.~\ref{fig:wave-range}, the flux at the red end exceeded that at the blue end by more than a factor of three. This criterion indicates that the ejecta have become optically thin, and measuring the expansion velocity through the blueshift of the local minima is no longer applicable. For all subtypes of SESNe, the Ca~II~NIR triplet is the first feature to exhibit this nebular characteristic, followed by the O~I\lam7774 line. All SNe ended their photometric phases before 80 days since the $V$-band peak. In total, we have 11 nebular spectra of six SNe. 

\section{Results}\label{sec:results}

\subsection{Individual objects}\label{sec:objects}
Some well observed SNe in our sample are representative for peculiar features of SESNe. For example, SN 2013ge is typical for SNe Ic with significant residual carbon in its spectra. SN 2016coi is broad-lined but with He present in spectra, representing a ``Ib-BL''. SN 2019ehk is a peculiar Ca-rich SN Ib, possibly not originated from core-collapse of a massive star. Below discussions of these three SNe is presented.
\subsubsection{SN~2013ge}
\begin{figure}
	\includegraphics[width=1.0\linewidth]{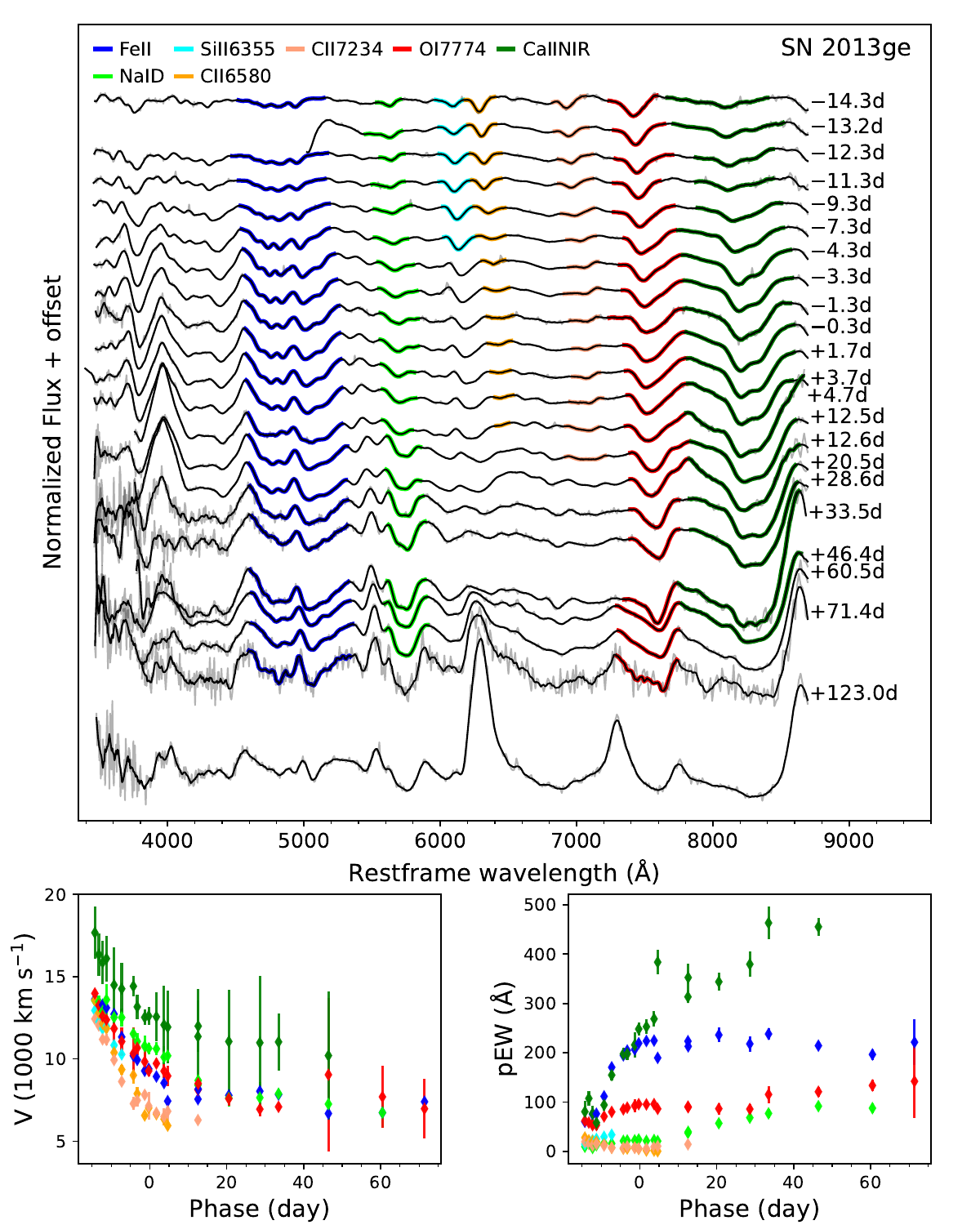}
	\caption{Spectral evolution (\textit{upper}) and line properties (\textit{lower}) of the double-peaked type Ic SN~2013ge. Different spectral lines are denoted with the colours indicated in the upper panel.}
	\label{fig:spec-13ge}
\end{figure}

SN~2013ge is a type Ic supernova with double peaks in its multi-band light curves, especially in UV bands. A full dataset from radio to X-ray (upper limit) was presented by \cite{2016ApJ...821...57D}. 
In this work we present the unpublished optical spectra of SN~2013ge obtained independently by THU supernova group. Shown in Fig.~\ref{fig:spec-13ge}, our data spans from 2 weeks before peak to +123~d after that. C~II lines were prominent in spectra before the maximum light and vanished within one week post the peak. Absorption lines in early phases are of high blueshift velocity but are relatively narrow, implying a restricted line-forming region.
With the aid of NIR spectra, \cite{2016ApJ...821...57D} identified weak He features in SN~2013ge, leading to a classification of SN~2013ge as type Ib/c subclass. However, more recent studies have consistently treated it as a type Ic \citep{2019MNRAS.482.1545S,2022ApJ...925..175S,2022ApJ...928..151F}. While the early-time spectra allow for ambiguity, the putative He lines can be plausibly attributed to other species such as Na~I, C~II, or Si~II at similar wavelengths (Fig.~\ref{fig:spec-13ge}). This, combined with its spectral similarity to normal SNe Ic like SN~2007gr and SN~2017ein, and the definitive evidence from its nebular-phase spectrum (Section~\ref{sec:neb-phase}), strongly favours a type Ic classification for SN~2013ge.

\subsubsection{SN~2016coi}
\begin{figure}
	\centering
	\includegraphics[width=1.0\linewidth]{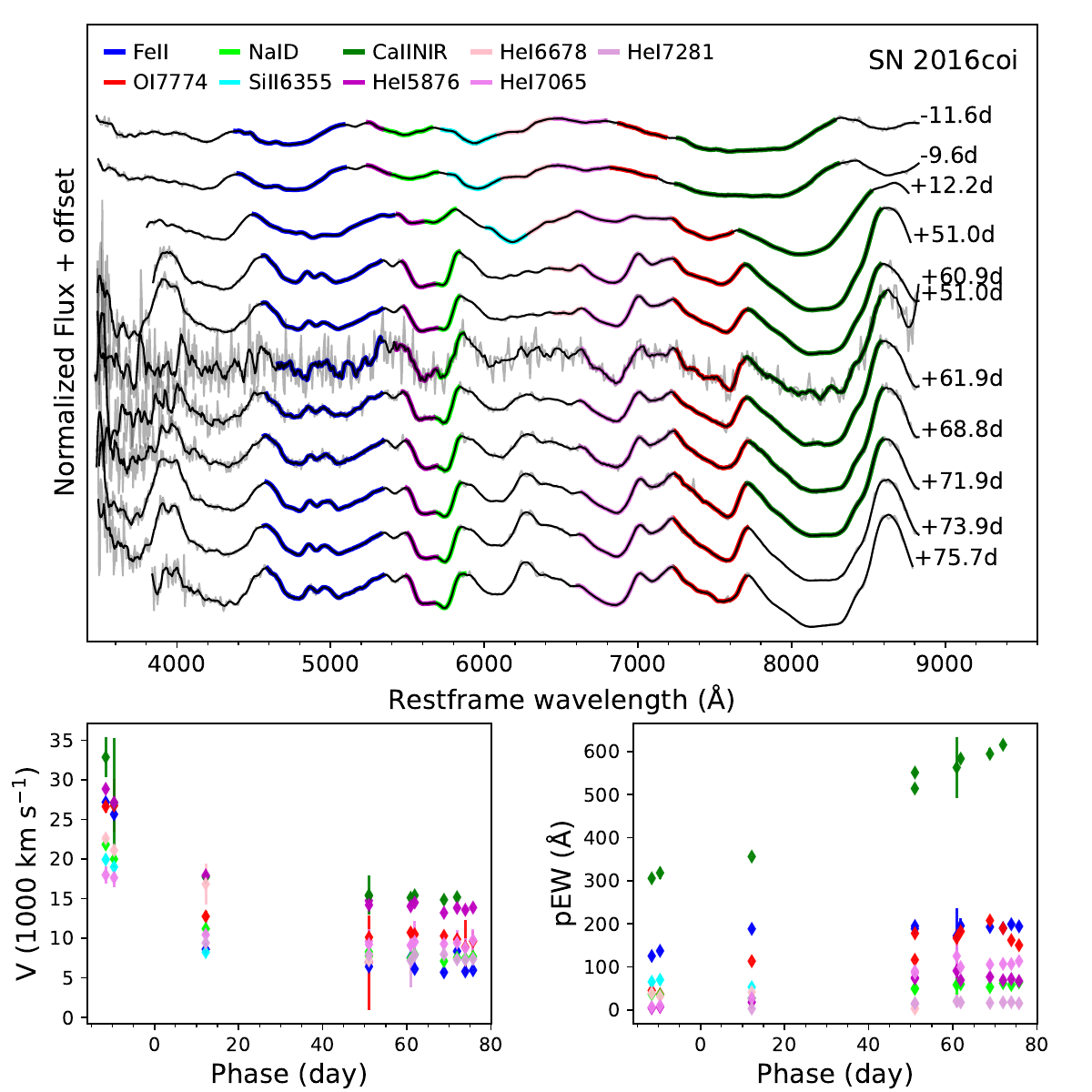}
	\caption{Same as Fig.~\ref{fig:spec-13ge} but for the He-rich broad-lined SN~2016coi.}
	\label{fig:spec-sn2016coi}
\end{figure}

SN~2016coi is characterized by broad absorption lines like SNe Ic-BL, but with He~I features in the spectra \citep{2018MNRAS.478.4162P}. The pEWs of He~I lines seen in SN~2016coi is located at the lower end of SNe~Ib. At approximately one week post-maximum light, the absorption feature near 5200~\AA\ in the spectra of SN~2016coi exhibits a distinct double-Gaussian profile, as shown in Fig.~\ref{fig:spec-sn2016coi}. 
In contrast, typical Type Ib/c spectra show only a single Gaussian feature at this wavelength, which originates from either He~I (Ib) or Na~I (Ic) alone. Therefore, the corresponding line in SN~2016coi contains contributions from both elements. Furthermore, the He~I lines at 6678~\AA\ and 7065~\AA\ can be also identified. It can be seen from Fig.~\ref{fig:FeII5169-SiII6355-Ib-Ic} that the pEWs and velocity evolution of SN~2016coi lie well within the range of SNe Ic-BL. Consequently, SN~2016coi is established as an exemplary transitional object linking SNe Ib to SNe Ic-BL, which are characterized by higher explosion energy while retaining a residual helium envelope \citep{2017ApJ...837....1Y,2018MNRAS.478.4162P,2019ApJ...883..147T}.

\subsubsection{SN 2019ehk}
SN 2019ehk is a calcium-rich supernova with double-peak light curves. The presentation of a full observation dataset from NIR to X-ray band and analysis of its possible progenitor system were given by \cite{2020ApJ...898..166J}. The very early spectrum of SN~2019ehk shows narrow emission lines of H~I and He~II, implying interaction with dense material surrounding the progenitor. The spectra of SN~2019ehk were characterized by strong Ca~II absorption, prominent He~I profiles, and fast emergence of a [Ca II] profile a few days before the maximum light. Although the He lines are present in spectra, SN~2019ehk is classified as Ca-rich supernova. The progenitor system of SN 2019ehk remains debated. \cite{2020ApJ...898..166J} proposed a low mass binary white dwarfs for the origin of SN 2019ehk based on its explosion and environment properties. While other studies proposed alternative scenarios invloving collapse of stripped massive stars for these Ca-rich envents \citep{2021ApJ...907L..18D,2021ApJ...912...30N,2023MNRAS.526..279E}.

Compared with SNe Ib, SN~2019ehk has very weak absorption lines before 10~days since the maximum light. Such a time coincidence with the first light-curve peak can be attributed to interaction with a dense circumstellar matter shell \citep{2020ApJ...898..166J}. High temperature of the ejecta blocks any information from spectral features, producing nearly featureless spectra.

\subsection{Hydrogen lines in SNe Ib and IIb}\label{sec:H-in-Ib}
In both SNe Ib and Ic, the blended line feature in the wavelength range of 6000--6400~\AA\ can be attributed to several different ions.
In early-phase SNe Ib spectra, potential contributors include Fe~II, Ca~I, Si~II, and possibly H~I, with an additional contribution from He I that strengthens over time \citep{2012MNRAS.424.2139D}.
While in SNe Ic, the He~I\lam6678 line is replaced by C~II\lam6580 line which gradually fades.
Previous studies have identified the presence of H features in most SNe Ib, suggesting a continuous spectroscopic sequence between types IIb and Ib \citep{2016ApJ...827...90L,2023A&A...675A..83H}. 
Possible contribution of Si~II to the spectra of SNe Ib was also discussed by \cite{2023A&A...675A..83H}. In this section, we will focus on the evolutionary properties of H features in SN Ib spectra and compare their similarities and differences with those of SNe IIb. 

\begin{figure}
	\centering
	\includegraphics[width=0.8\linewidth]{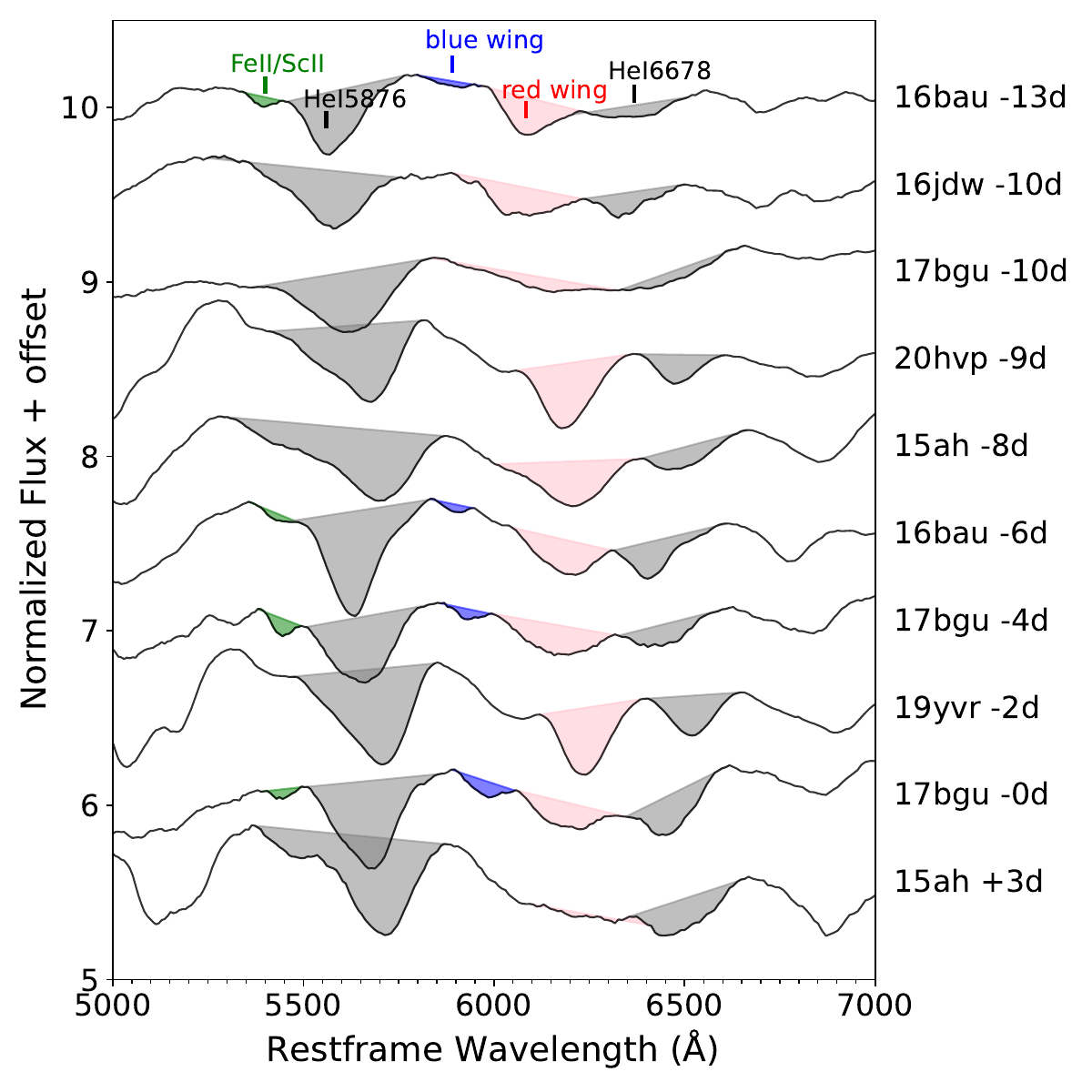}
	\includegraphics[width=0.8\linewidth]{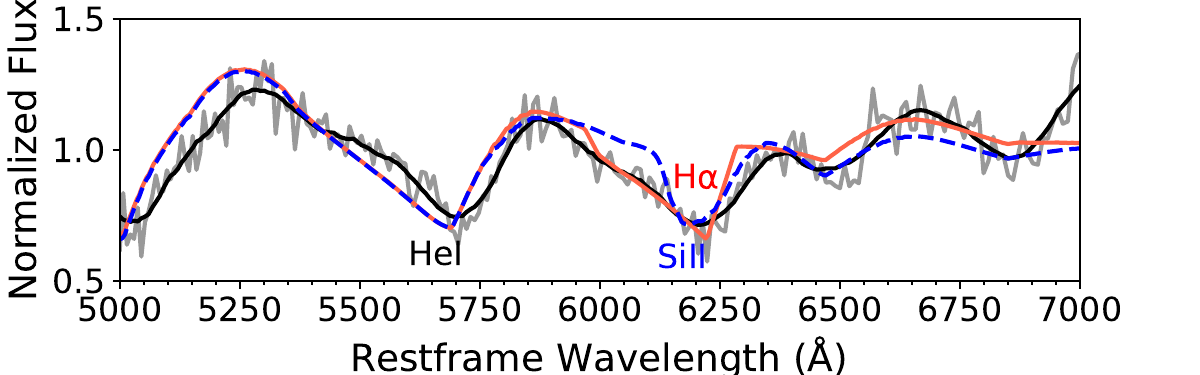}
	\caption{\textit{Upper}: Normalized spectra of SNe Ib in our sample till maximum light in the range of 5000--7000~\AA, showing the features around \Ha. Spectra of different SNe are plotted in distinct colors and shifted vertically for better illustration. The main absorption features are highlighted with distinct coloured shaded areas. Among these, the green bands denote an absorption line of ambiguous origin that appears concurrently with the blue-wing structure. \textit{Lower}: Normalized spectrum of SN~2015ah at $-$8-d showing the range of the doublet feature near 6200~\AA. Synthetic spectra by SYNOW with either HI (red solid line) or Si~II (blue dashed line) are also overplotted.}
	\label{fig:synow-6200}
\end{figure}

\begin{figure*}
	\centering
	\includegraphics[width=0.9\columnwidth]{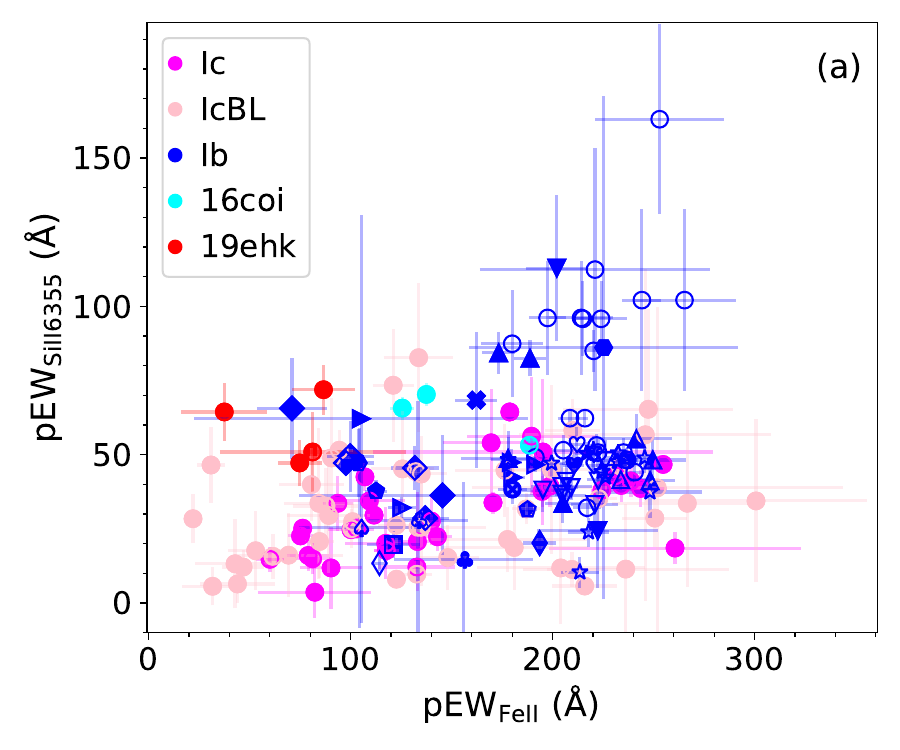}
	\includegraphics[width=0.9\columnwidth]{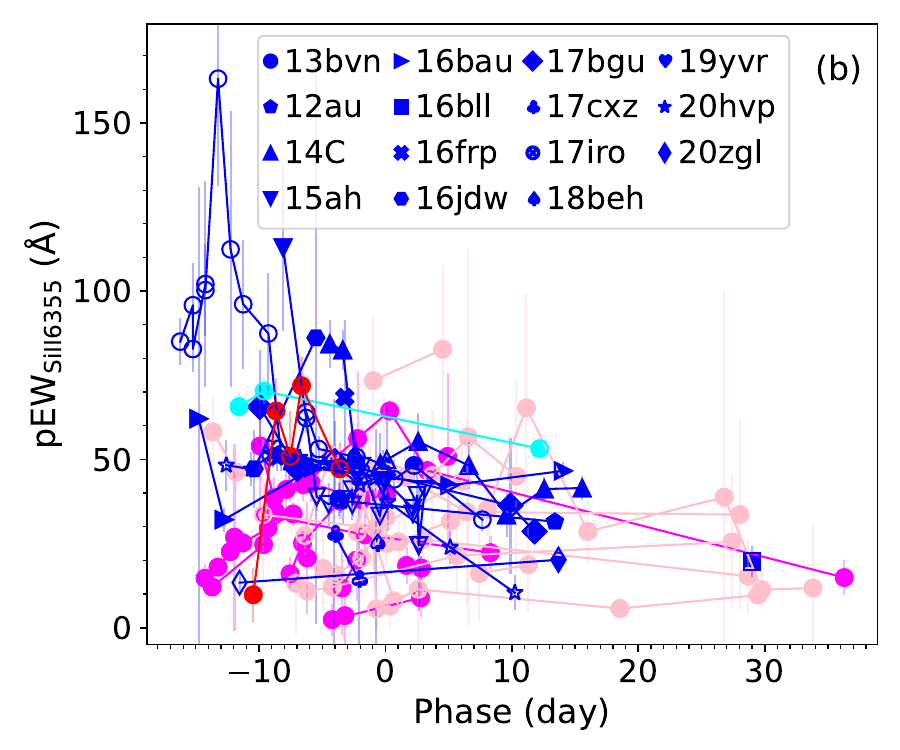}
	\includegraphics[width=0.9\columnwidth]{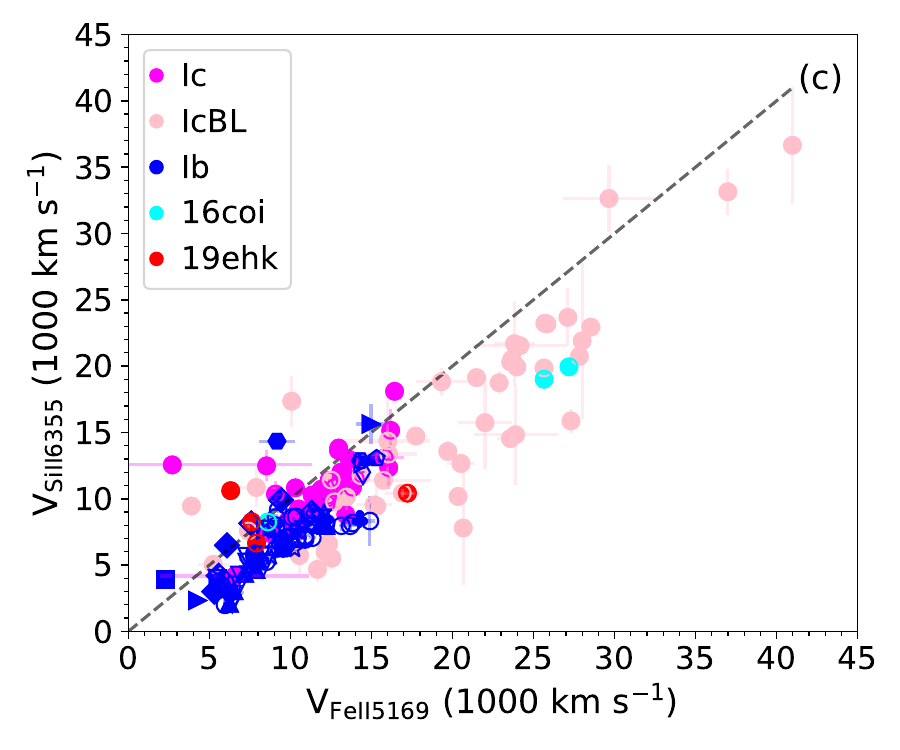}
	\includegraphics[width=0.9\columnwidth]{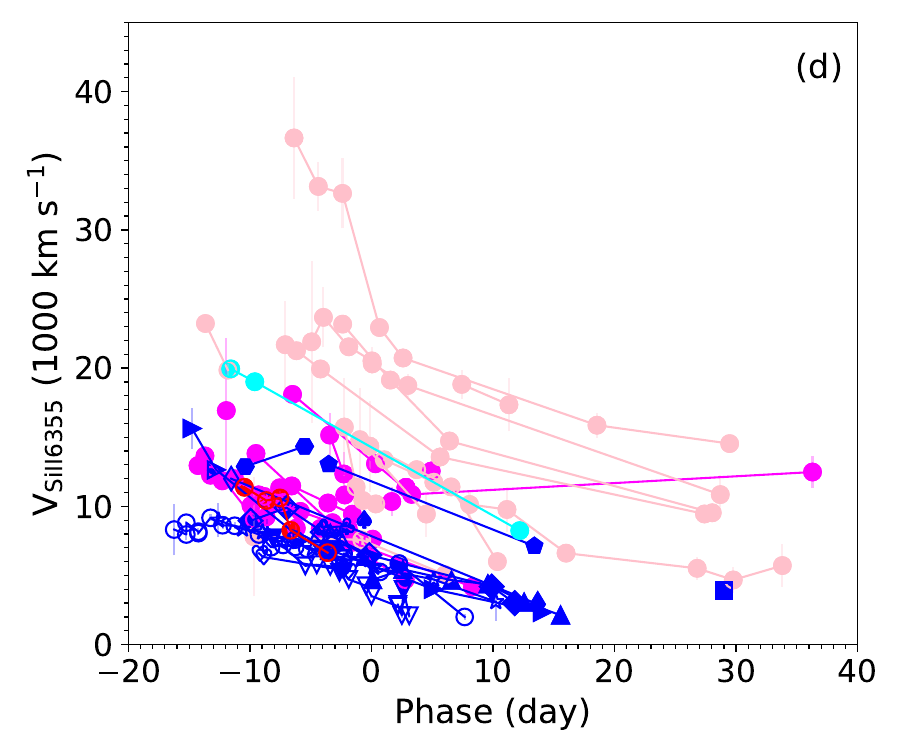}
	\caption{Pseudo-Equivalent Widths (pEWs) and velocities of the red-wing feature (attributed as Si~II\lam6355) and Fe~II lines in spectra of SNe Ib compared with SNe Ic and Ic-BL. Different subtypes are colour-coded, including the peculiar objects SN 2016coi and SN 2019ehk. The data points belonging to the same SN are connected by lines, and those measured from WiseRep spectra are shown with empty symbols. Each SN Ib is represented by a distinct symbol denoted in the legend of (b). }
	\label{fig:FeII5169-SiII6355-Ib-Ic}
\end{figure*}

\begin{figure*}
	\centering
	\includegraphics[width=0.9\columnwidth]{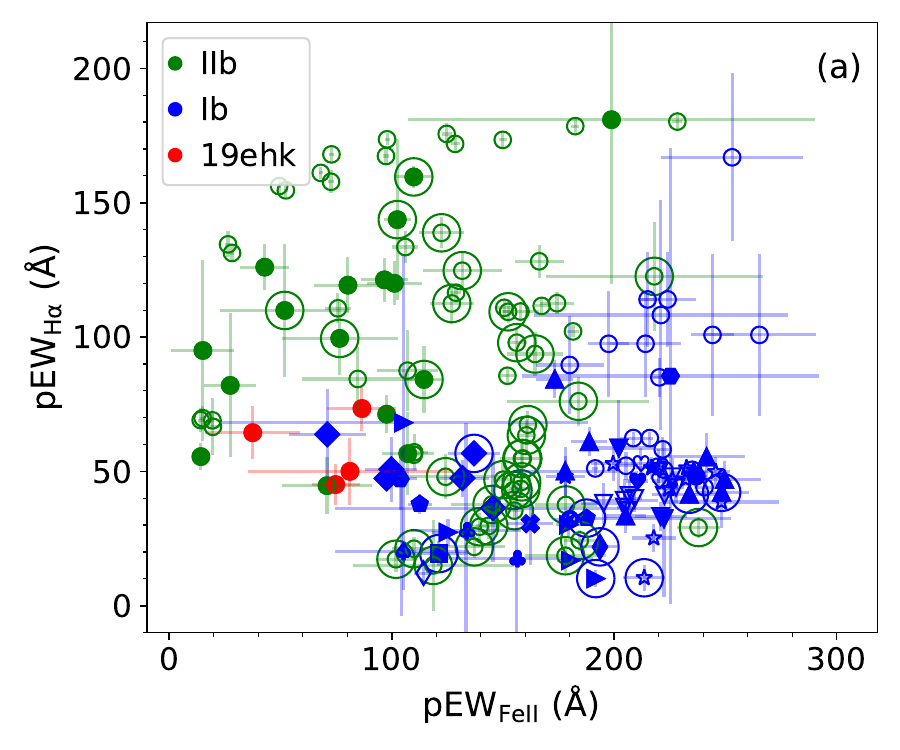}
	\includegraphics[width=0.9\columnwidth]{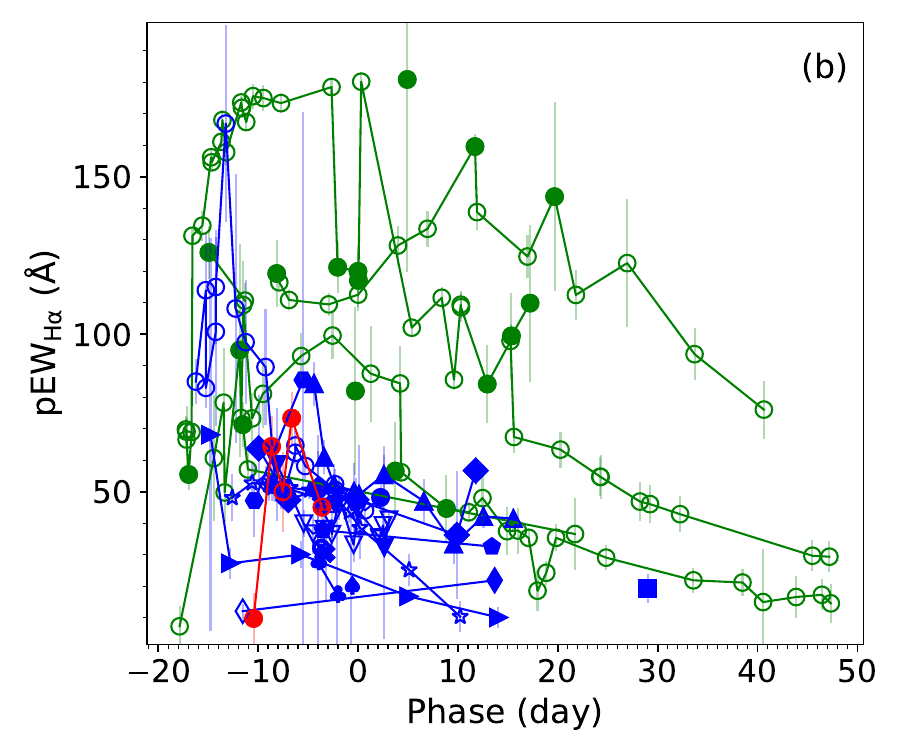}
	\includegraphics[width=0.9\columnwidth]{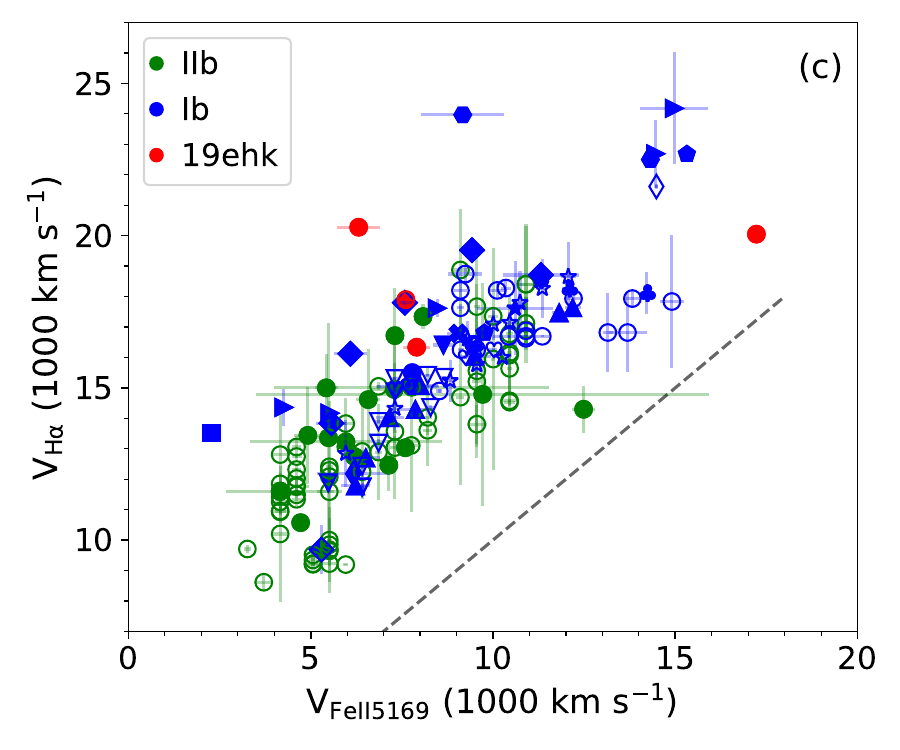}
	\includegraphics[width=0.9\columnwidth]{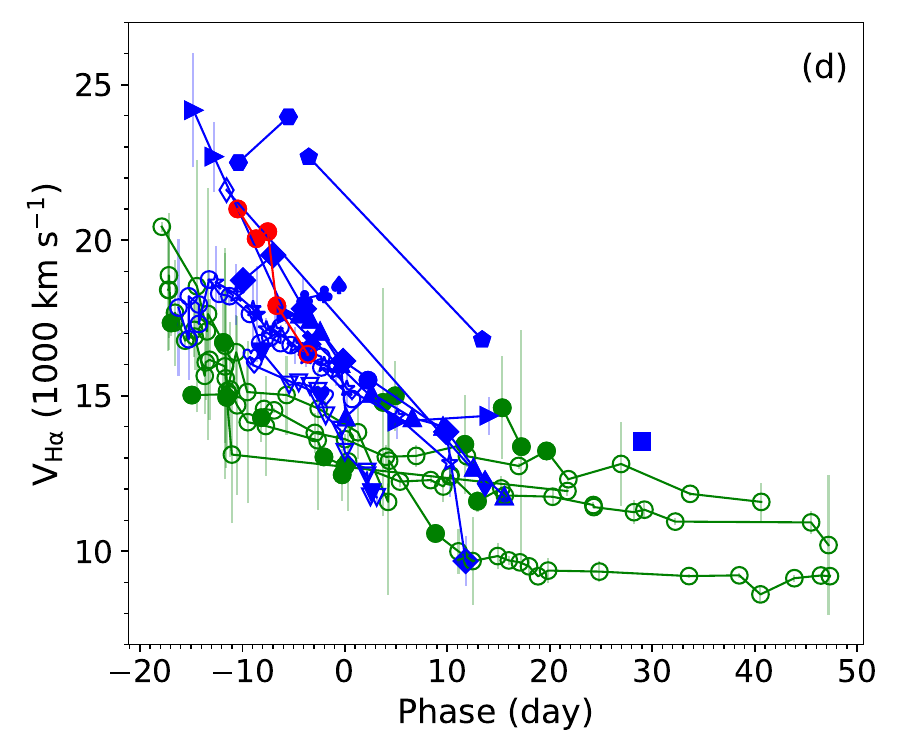}
	\caption{Same as Fig.~\ref{fig:FeII5169-SiII6355-Ib-Ic} but the red-wing feature attributted as \Ha\ in spectra of SNe Ib and compared with SNe IIb. The data points circled in (a) correspond to phases later than +10 days. See the main text for details. }
	\label{fig:FeII5169-Halpha-Ib-Ic}
\end{figure*}

In SN Ib spectra at phases earlier than +10 days, the absorption feature near 6200~\AA can be attributed to \Ha, Si~II\lam6355, or a blend of both.
Prior to maximum light, all SN Ib spectra in our sample exhibit an absorption feature near 6200~\AA. The top panel of Fig.~\ref{fig:synow-6200} shows early time spectra of SNe Ib, covering the wavelength range of 5000--7000~\AA. The spectral profiles between 6000 and 6500~\AA\ are generally consistent across different objects, though variations in absorption depth and velocity are observed.
\cite{2023A&A...675A..83H} attributed the blue part of this feature (the blue-wing) in SNe Ib to high-velocity Si~II\lam6355, and the red part (the red-wing) to \Ha. These two components, combined with the He~I\lam6678 line, form a triplet line structure. 
When inspecting this feature, we notice that the triplet feature is not present in all early-phase (i.e., t$\lesssim$+10 d) spectra (Fig.~\ref{fig:synow-6200}, upper panel). If the blue-wing is attributed to Si~II\lam6355, its measured blueshift velocity would be significantly higher than that of the Fe~II\lam5169 in the same spectrum. However, our measurements of SN Ic spectra show that Si~II velocities are typically slightly lower than those of Fe~II lines (Fig.~\ref{fig:FeII5169-SiII6355-Ib-Ic}(c)). Furthermore, after +10 days, the blue-wing component strengthens and emerges concurrently with an absorption structure near 5350~\AA, which is marked by green shading  in the upper panel of Fig.~\ref{fig:synow-6200}. This feature near 5350~\AA\ could be Sc~II/Fe~II\lam5531, and thus, this blue-wing feature may also originate from Sc~II/Fe~II.Considering this consistent evidence -- the atypical velocity, the late emergence, and the association with the 5350~\AA\ line -- we conclude that the blue-wing absorption is unlikely to be Si~II.

Regarding the absorption feature on the red-wing, it may originate from either H~I or Si~II. \cite{2016ApJ...827...90L} identified it as H$\alpha$ and found that its pEWs and velocity evolution differ significantly from those of SNe IIb, without ruling out a possible Si II contribution. In contrast, \cite{2023A&A...675A..83H} attributed the aforementioned blue-wing to Si~II and the red-wing to \Ha, finding no substantial difference in H$\alpha$ line velocities between SNe IIb and Ib. 
Our SYNOW spectral synthesis (Fig.~\ref{fig:synow-6200}, lower panel) also shows that the asymmetric and broad line profile can be better synthesized with an H~I component. In the model including H~I, the H~I has a velocity higher than that of other ions by about 7000~km~s$^{-1}$.
While definitive confirmation of the carrier (Si~II or H~I) from a single spectrum is challenging, the dominant role of H~I is supported by our fitting. To further investigate this, we next perform a statistical analysis of the properties of this feature under both assumptions.

First, we consider the line as Si~II\lam6355. The line velocities and pEWs of Si~II\lam6355 alongside those of Fe~II is presented in Fig.~\ref{fig:FeII5169-SiII6355-Ib-Ic}. For comparison, measurements for SNe Ic and Ic-BL are included in the same figure. 
To enable more meaningful comparison of the subtypes, we supplemented the spectral dataset from WISeREP or published literatures of several SNe Ib in our sample which exhibit the distinct red-wing feature (as shown in Fig.~\ref{fig:synow-6200}) and also lack early time spectroscopic data. Accordingly, we retrieved and uniformly processed publicly available data for the following SNe: iPTF13bvn \citep{2013ApJ...775L...7C,2014MNRAS.445.1932S,2016PASA...33...55C},  2015ah \citep{2019MNRAS.482.1545S,2019MNRAS.485.1559P}, 2019yvr, 2020hvp \citep{2025A&A...693A.307Y} and 2020zgl. These additional measurements are plotted as open symbols in Fig.~\ref{fig:FeII5169-SiII6355-Ib-Ic}. Notably, the inclusion of these objects—particularly iPTF13bvn—extends the observed parameter space for SNe Ib, which is representative of objects with relatively strong red-wing features.
In terms of velocity, the three subtypes follow a similar correlation, while SNe Ib and Ic are similar and SNe Ic-BL are the fastest, which is consistent with previous studies \citep[e.g.,][]{2016ApJ...827...90L,2023A&A...675A..83H}. 
However, pEWs of this line in SNe Ib are larger than those in SNe Ic and Ic-BL, especially at early phases. While the latter two types are not distinguishable. This implies that the red-wing feature in SNe Ib has a different origin from that in SNe Ic and Ic-BL. Thus, we conclude that the red-wing feature in spectra of SNe Ib is very likely due to H~I rather than Si~II. Given the symmetric Gaussian profile of the red-wing feature, any contribution from Si~II to this feature would be negligible.

On the other hand, if the absorption feature is attributed to \Ha, the comparison between SNe Ib and IIb is presented Fig.~\ref{fig:FeII5169-Halpha-Ib-Ic}. We also incorporated supplementary spectral data from the literature for the SNe IIb in our sample: including 2011dh \citep{2011ApJ...742L..18A,2014A&A...562A..17E,2019MNRAS.482.1545S}, 2011fu \citep{2013MNRAS.431..308K,2015MNRAS.454...95M,2019MNRAS.482.1545S}, 2014ds \citep{2019MNRAS.482.1545S}, 2015bi \citep{2019MNRAS.482.1545S}, and 2017gpn \citep{2019MNRAS.485.1559P}.
It shows that the two subtypes exhibit distinct distributions at early phases (<+10 days): most SNe Ib show weaker H$\alpha$ line but stronger Fe~II lines. In early phases, the two subtypes occupy opposite extremes of this population with little overlap: SNe IIb display lower H$\alpha$ velocities but greater line strengths, consistent with \cite{2016ApJ...827...90L}. Although several SNe Ib (e.g., the most H-rich iPTF13bvn) have strong \Ha\ lines at very early epochs (t$<-$10 d), their pEWs decline rapidly approaching maximum light. In iPTF13bvn, the early strong \Ha\ is accompanied by notably strong Fe~II lines, further distinguishing it from SNe IIb.
After +10 days post-maximum, the \Ha\ line strengths in SNe IIb fade. Consequently, their locus in the \Ha-Fe~II line-strength relation shifts towards the region occupied by SNe Ib, resulting in reduced distinction between the two subtypes. For objects displaying both He~I and prominent \Ha\ lines, the Fe~II line strength can aid in distinguishing between types IIb and Ib. However, accurate classification may not be possible for SNe with very weak \Ha\ lines based on spectral line strengths alone.

In terms of line velocity, the \Ha\ and Fe~II velocity distributions of SNe IIb largely overlap with the slower portion of the SNe Ib distributions. Moreover, the \Ha\ line velocities in SNe Ib decline to a level comparable to those in SNe IIb within 10 days after maximum light (Fig.~\ref{fig:FeII5169-Halpha-Ib-Ic}(d)). This implies that the hydrogen shell present in SNe Ib may be thinner than that in SNe IIb. Our findings are consistent with those of \cite{2016ApJ...827...90L}.

To conclude, the absorption feature near 6200~\AA\ can be conclusively identified as \Ha. 
However, it should be noted that the dataset of SNe IIb is quite small, which will increase the uncertainty of our conclusion.

\subsection{Evolution and correlation of line velocities}\label{sec:line_V}
Fig.~\ref{fig:evol-V-all} shows the temporal evolution of expansion velocities of all lines present in each subclass. In general, expansion velocities of all lines decline over time.
SNe Ic-BL exhibit distinctly higher velocities than the other subclasses. SNe Ic show higher average velocities than SNe Ib, although their distributions overlap considerably. And SNe IIb have the lowest velocities. This trend of increasing velocities among SNe IIb to Ic and Ic-BL is not new but previously reported by \cite{2016ApJ...827...90L,2016ApJ...832..108M}. These differences are caused by either distinction in the distribution of explosion energy or ejecta mass. Given that SNe Ic-BL are found to have the largest ejecta masses among all subtypes, as inferred from light-curve analyses \citep[e.g.,][]{2019MNRAS.485.1559P}, their exceptionally high velocities must originate from intrinsically greater explosion energies. This points to a different explosion mechanism or energy source (e.g., a central engine). For the other three subtypes, the velocity differences are more subtle, and their ejecta mass distributions are similar according to the literature.Therefore, the progressive velocity increase from type IIb to type Ic, which corresponds to more extensive envelope stripping, may arise from relatively smaller differences in explosion energy. This energy difference could be linked to the stripping process itself, for instance, faster rotation of the progenitor or more intense binary interaction may result in higher explosion energy \citep{2021A&A...645A...5S}, or extra energy input similar to that proposed for SNe Ic-BL. Numerous other factors, such as the excitation state of the materials and the degree of elemental mixing, may also contribute to variations in spectral line velocities. These aspects, however, fall beyond the scope of this study.

Ions are distributed across different layers within the SNe ejecta, resulting in differences in expansion velocity, temperature, density, and other properties of their respective spectral line-forming regions. Typically, lighter elements are located in outer layers and thus exhibit higher velocities under the assumption of homologous expansion. The difference in velocity between elements at adjacent layers is generally smaller.

To quantify correlations between different spectral line measurements, we used Spearman's rank correlation coefficient $\rho$. The absolute value of $\rho$ indicates the strength of the correlation, and its sign indicates a positive ($\rho > 0$) or negative ($\rho < 0$) relationship.

\begin{figure*}
	\centering
	\includegraphics[width=1.6\columnwidth]{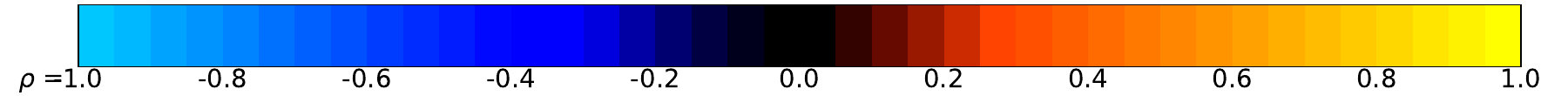}
	\includegraphics[width=0.8\columnwidth]{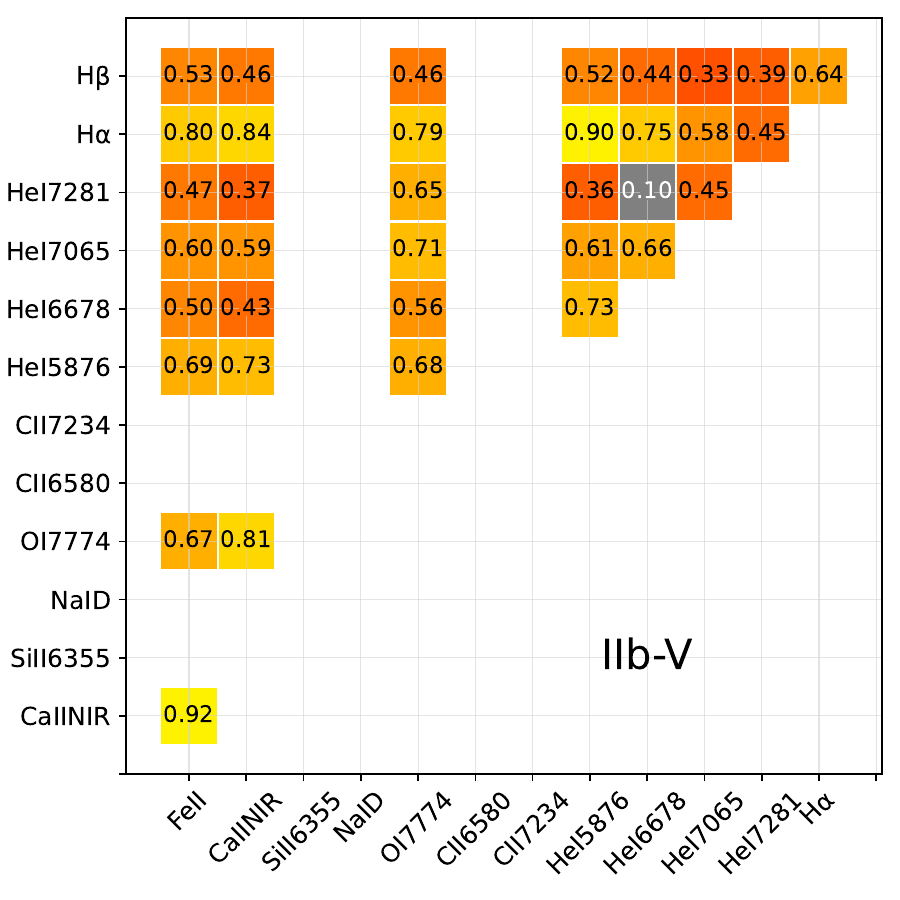}
	\includegraphics[width=0.8\columnwidth]{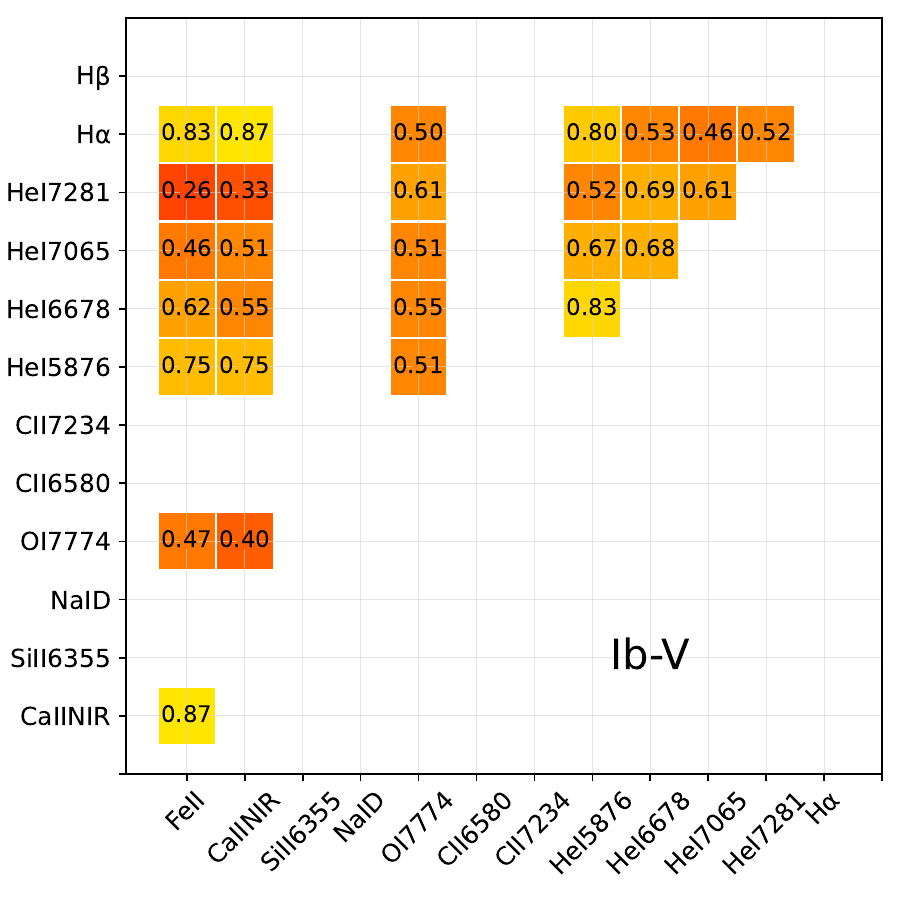}
	
	\includegraphics[width=0.8\columnwidth]{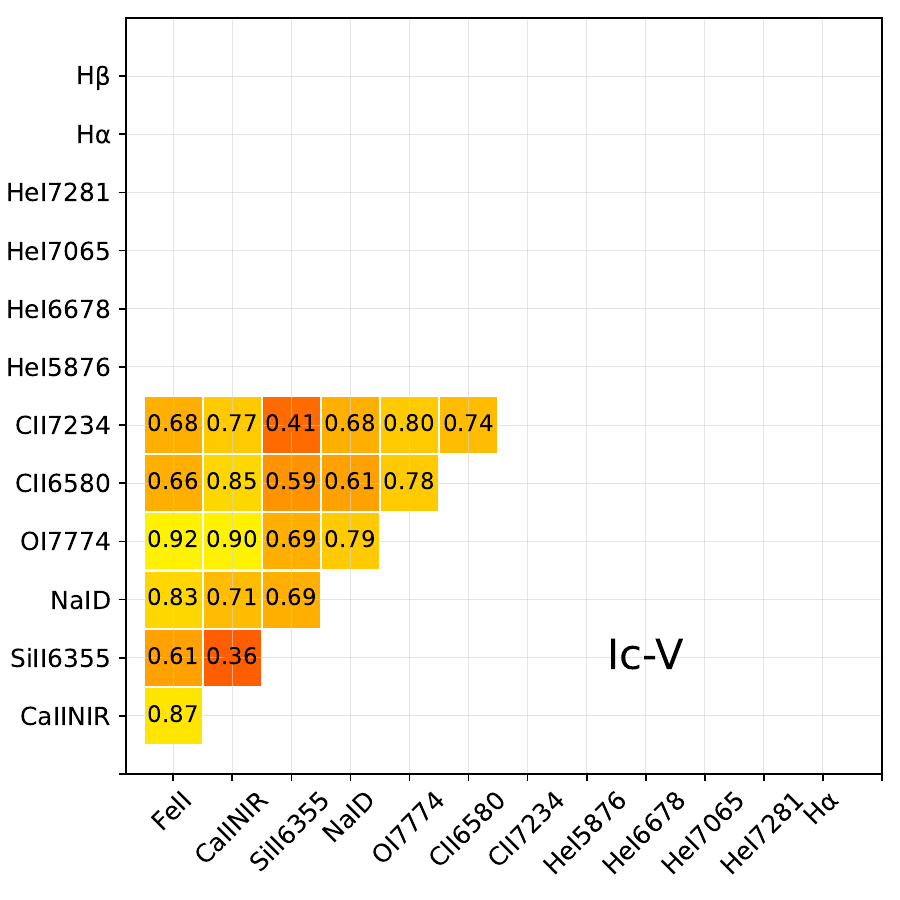}
	\includegraphics[width=0.8\columnwidth]{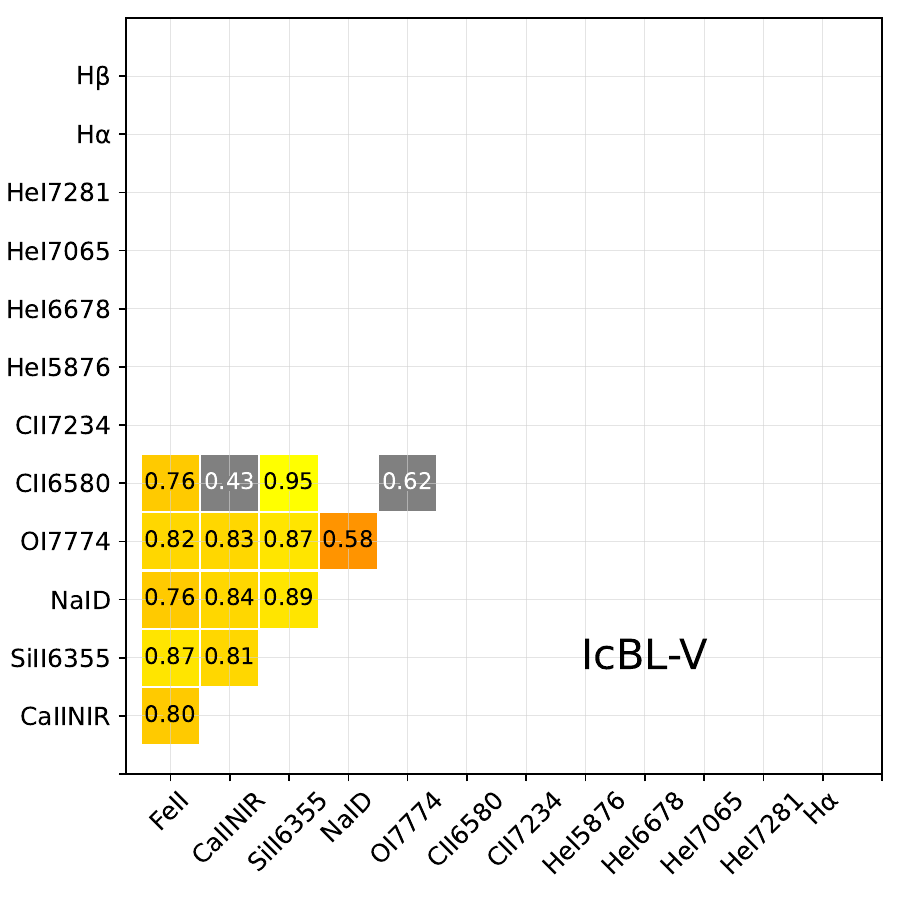}
	
	\caption{Spearman's rank correlation coefficients ($\rho$) between the measured velocities of lines of different types of SESNe. Ions are arranged such that lighter elements are positioned toward the right and top. Numbers in each box show the coefficients of the corresponding line pairs. Boxes with gray color show low statistical significance with p-value larger than 0.05. These cases are all due to very small dataset. Numeric values are indicated by the colorbar on the top of the figure, with lighter colors indicating stronger correlation. The larger is $\rho$, the stronger the correlation exists between the corresponding pair of measurements, i.e., weak (0.4 < |$\rho$| $\leq$ 0.6), moderate (0.6 < |$\rho$| $\leq$ 0.8), and strong (|$\rho$|>0.8) correlation. The exact p-value for each correlation test is available at the CDS. \label{fig:Spearman-V}}
	
\end{figure*}

Here we examine the correlations between the expansion velocities of the various spectral lines of each subtype. The raw data of line velocity correlation are presented in Fig.~\ref{fig:corr-V-all} and the correlation matrices are shown in Fig.~\ref{fig:Spearman-V}. The ions in Fig.~\ref{fig:Spearman-V} are arranged such that lighter elements are positioned toward the right and top to reflect the expected stratification in the ejecta. And all spectral lines are included to facilitate the comparison across subtypes.

Spectral line velocities across all subtypes exhibit positive correlations of varying strength, in agreement with the expectation of homologously expanding ejecta. 
Figure~\ref{fig:Spearman-V} shows that the correlation between Fe~II lines and other lines progressively strengthens from type IIb to type Ic/Ic-BL. In SNe IIb and Ib, He~I line velocities exhibit a weaker correlation with those of heavier elements, whereas \Ha\ velocities show a stronger correlation with metal lines. 
The generally weaker correlations in SNe IIb may be attributable to their smaller sample size.
Velocities of nearly all common spectral lines have overlapping distributions across subtypes and follow similar correlation patterns. A notable exception is the \Ha–O~I\lam7774 pair, for which SNe Ib show systematically higher \Ha\ velocities than SNe IIb.
Although the helium shell is situated in the outer ejecta, our measurements show that He~I line velocities are not consistently higher than those of Fe~II lines. For example, the velocity of He~I\lam7065 shows virtually no correlation with Fe~II\lam5169 (excluding a few high-velocity outliers), and He~I\lam6678 is generally slower than Fe~II\lam5169. Among the four He~I lines studied, He~I\lam7281 exhibits the highest velocities, yet approximately half of its measurements still fall below those of Fe~II\lam5169.

By examining the spectral lines common to all four subtypes (namely Fe~II, O~I, and Ca~II), we found the following. First, the Ca~II lines exhibit the highest expansion velocities. Second, SNe Ib are more prevalent at the lower-velocity end of the distributions. Third, velocity correlations in SNe Ic are stronger than in the other subtypes, indicating a lower dispersion in their velocities.
When comparing between SNe Ic and SNe Ic-BL, in addition to exhibiting generally lower velocities in SNe Ic, they differ in the relationship between the velocities of the Na~I~D and Si~II\lam6355 lines: SNe Ic show lower velocities for Na~I~D line. The distributions for other spectral lines are similar between these two subtypes.

\subsection{Evolution and correlation of line intensities}\label{sec:line_EW}

The temporal evolution of line intensities (pEWs) of the four subtypes SNe is shown in Fig.~\ref{fig:evol-EW-all}. 
The spectral lines of residual elements (i.e. H in SNe Ib and C in SNe Ic/Ic-BL) weaken over time and become undetectable around 10 days after the maximum light. Although the Si lines in SNe Ic/Ic-BL eventually vanish as well, they undergo a short period of slight strengthening  before fading.

In early phases ($t < -$10 d), the \Ha\ line in SNe IIb exhibits pEWs comparable to those in SNe Ib. However, in SNe IIb, the \Ha\ strength continues to increase until after maximum light, after which it weakens, though the onset of weakening varies among objects. In SNe Ib, the intensities of most He~I lines generally decline after maximum. An exception is the He~I\lam7281 line, which shows no clear weakening in our sample. This may be due to contamination from the gradually emerging [Ca~II] emission line near 7300~\AA\ at later phases, which could distort the pEW measurement of the He~I\lam7281 feature.
The intensities of the He~I\lam5876 and He~I\lam6678 lines are significantly stronger in SNe Ib, consistent with results from \cite{2016ApJ...827...90L}. However, the two subtypes exhibit no difference in the intensity evolution of the He~I\lam7281 line, consistent with \cite{2023A&A...675A..83H}. For the He~I\lam7065 line,  our data show that its strength in SNe IIb exceeds that in SNe Ib at later phases. This finding is consistent with \cite{2016ApJ...827...90L} but contrasts with \cite{2018A&A...618A..37F} and \cite{2023A&A...675A..83H}.

Among the subtypes, O~I lines are strongest in SNe Ic/Ic-BL, intermediate in SNe Ib, and weakest in SNe IIb.
The intensity of the O~I\lam7774 line in SNe Ic covers the upper range of the distribution observed in SNe Ib, and there is no significant difference in the intensity distribution of this line between the broad-lined and normal SNe Ic. Ca~II and Fe~II lines are weakest in SNe IIb, while no statistically significant differences are observed among the other three subtypes.

\begin{figure*}
	\centering
	\includegraphics[width=1.6\columnwidth]{figures/spearman_colorbar.pdf}
	\includegraphics[width=0.8\columnwidth]{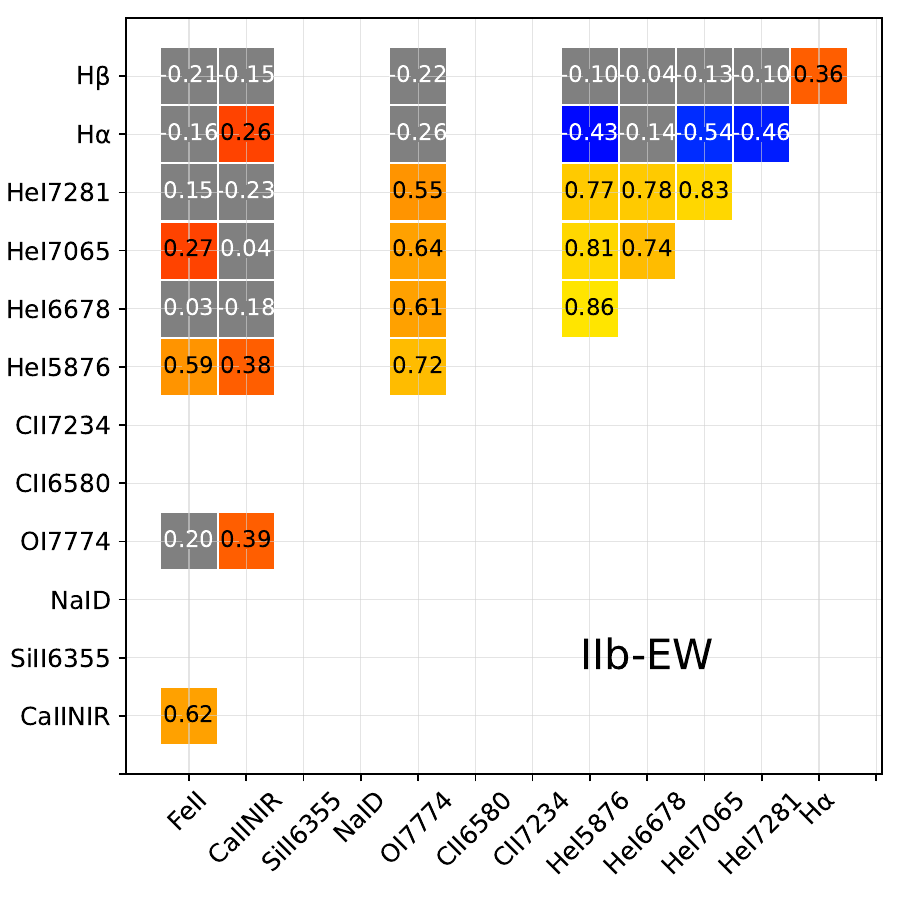}
	\includegraphics[width=0.8\columnwidth]{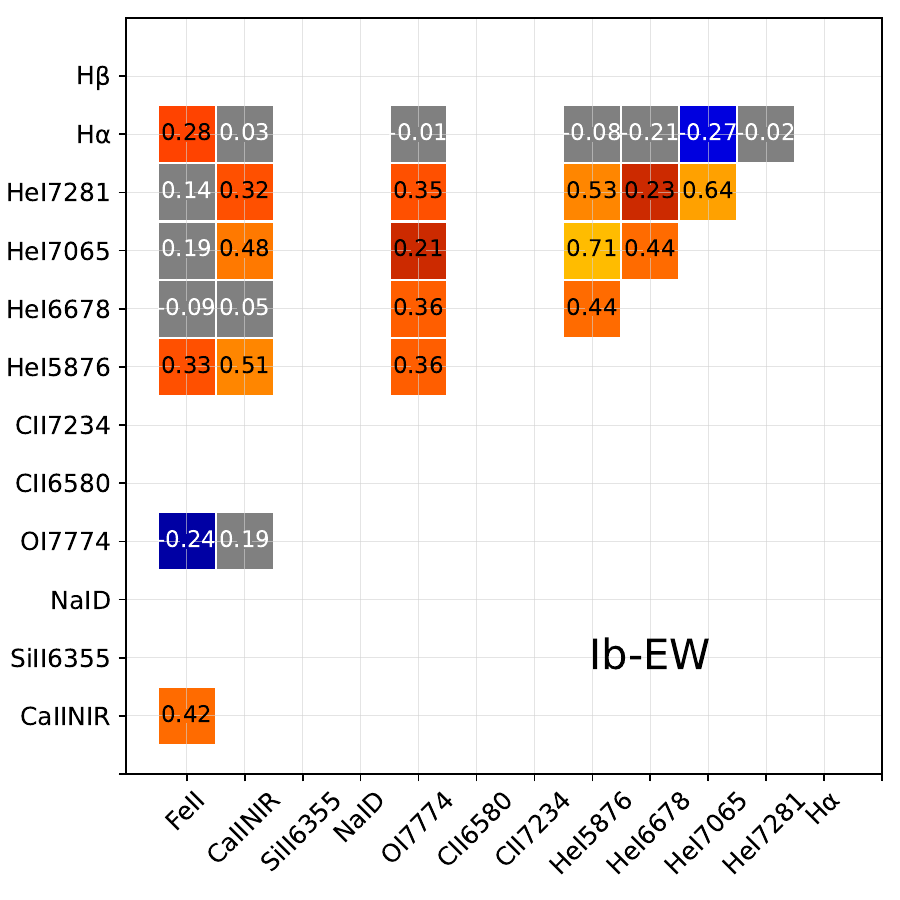}
	
	\includegraphics[width=0.8\columnwidth]{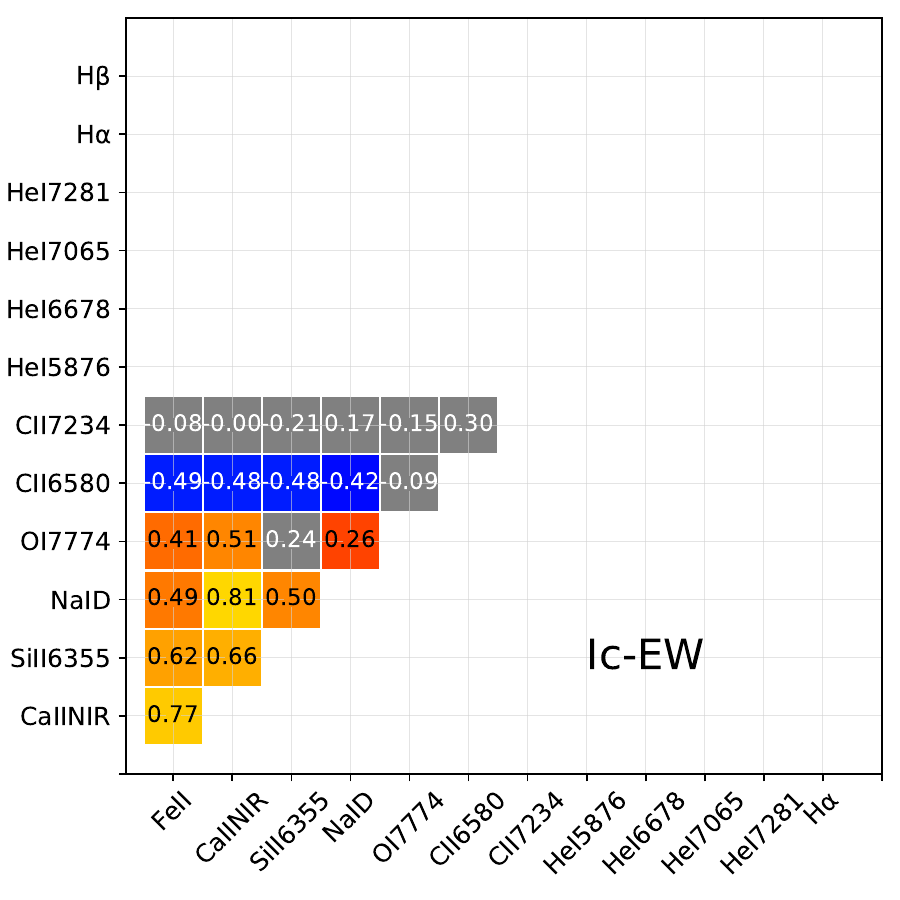}
	\includegraphics[width=0.8\columnwidth]{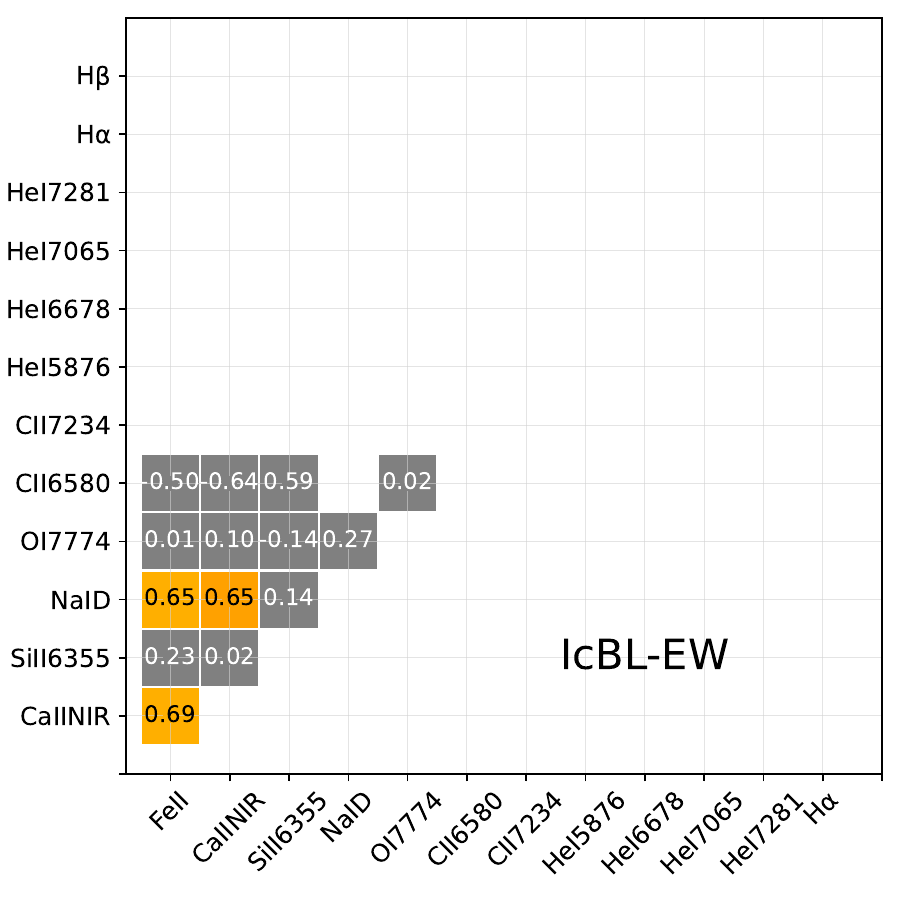}
	\caption{Same as in Fig.~\ref{fig:Spearman-V} but for measurements of pEWs of each line.}
	\label{fig:Spearman-EW}
\end{figure*}

Fig.~\ref{fig:evol-EW-all} shows the correlation of line intensities of the four subclasses of SESNe, and the Spearman's rank correlation coefficients between the pEWs of lines are shown in Fig.~\ref{fig:Spearman-EW}.
Regarding to light elements (H and He), as discussed in Sec.~\ref{sec:H-in-Ib}, SNe IIb and SNe Ib are distinguishable in the \Ha-Fe~II space, where SNe IIb are characterized by stronger \Ha\ and weaker Fe~II lines. The pEWs of He~I\lam7065 line shows weak correlations with those of the He~I\lam5876, He~I\lam7065, and the Ca~II~NIR lines. Compared with SNe Ib, the intensities of He~I lines are more correlated with that of O~I line in SNe IIb.
For SNe Ic and Ic-BL, no correlation is found between the pEWs of the C~II/O~I lines and those of other lines.
The pEWs of the Na~I and Ca~II lines exhibit strong/moderate correlations in SNe Ic/Ic-BL spectra.
However, the correlation between the pEWs of Na~I and Fe~II is relatively weaker in SN Ic spectra, while that between Na~I and Si~II is stronger. 
The distributions of the pEW relationship between Fe~II and Ca~II of the four subtypes appear to belong to a single population. Nevertheless, the correlation appears to be strongest in SNe Ic, followed by SNe Ic-BL. 

\begin{figure}
	\centering
	\includegraphics[width=0.9\linewidth]{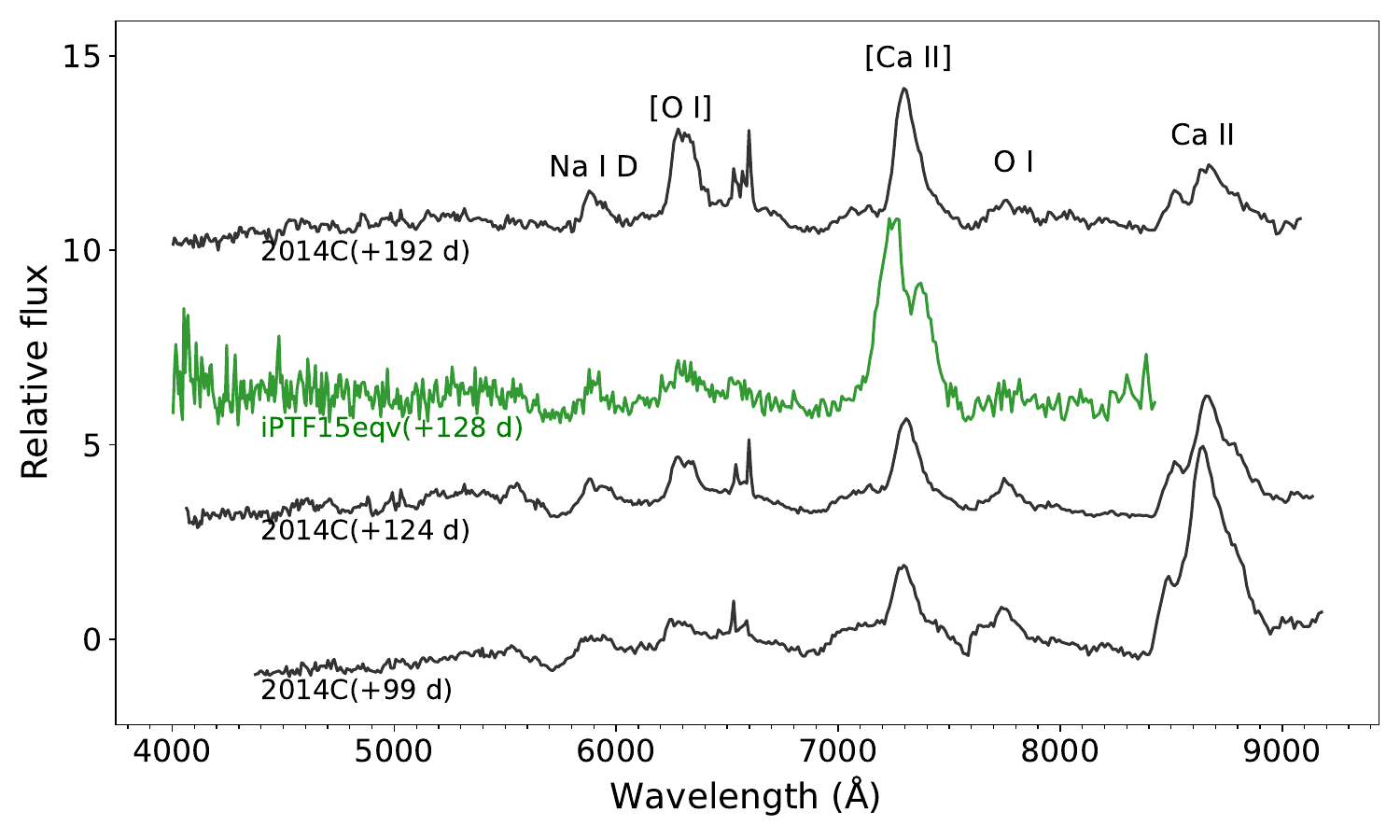}
	\includegraphics[width=0.9\linewidth]{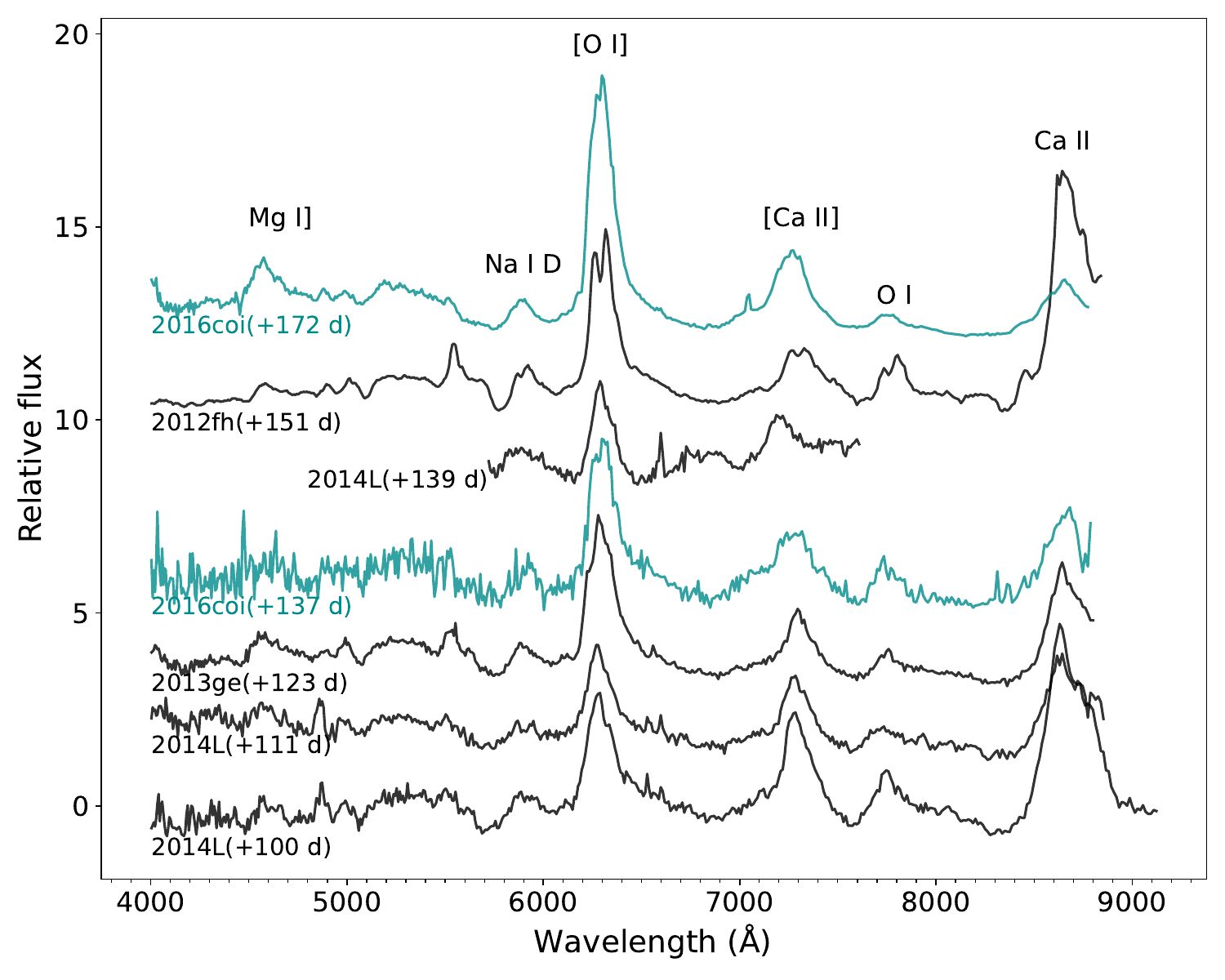}
	\caption{\textit{Top}: Nebular spectra of SNe Ib (black) and one type IIb (green). \textit{Bottom}: Nebular spectra of SNe Ic (black), and the peculiar He-rich broad-lined SN~2016coi (cyan). SN names and the corresponding phases are displayed below each spectrum.}
	\label{fig:neb-spec}
\end{figure}

\begin{figure}
	\centering
	\includegraphics[width=1.0\linewidth]{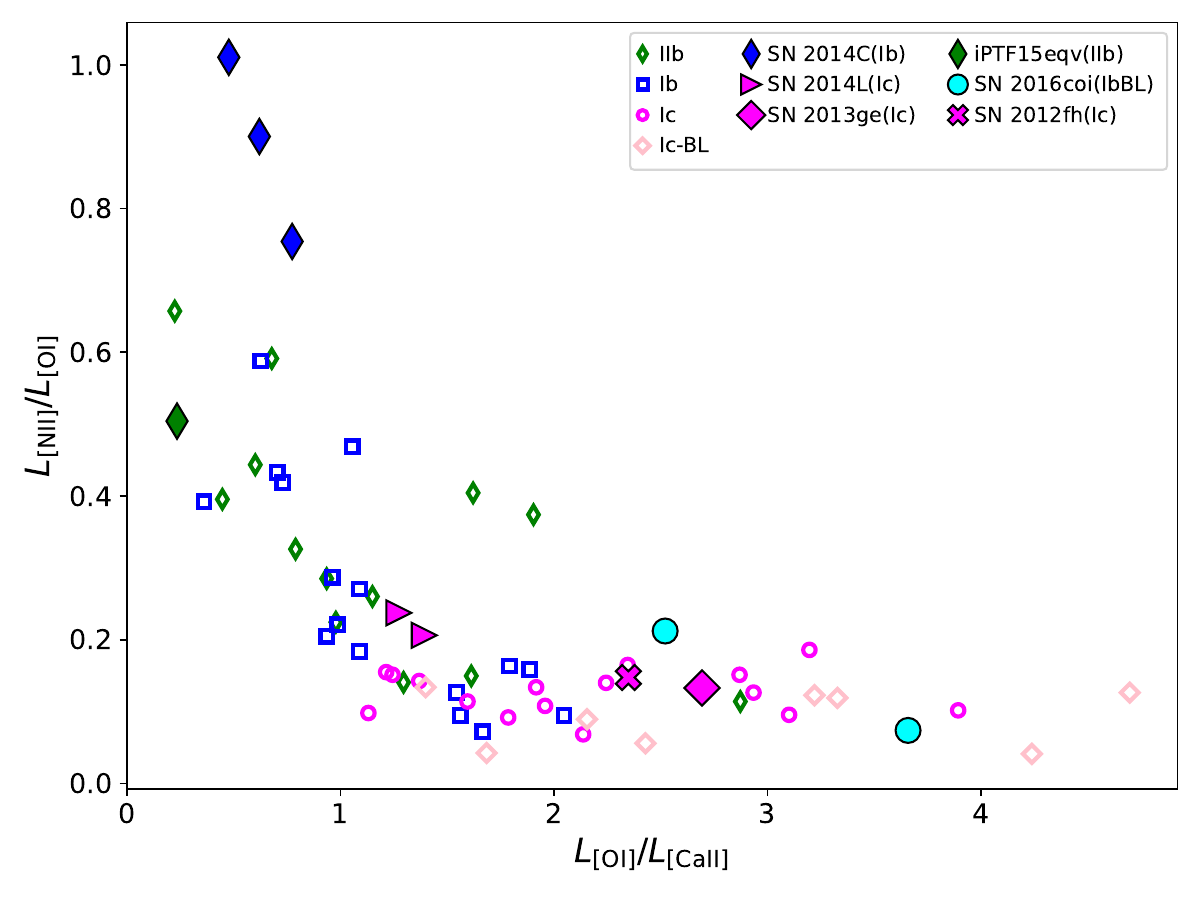}
	\caption{The distribution of $L_{\mathrm{[NII]}}/L_{\mathrm{[OI]}}$ and $L_{\mathrm{[OI]}}/L_{\mathrm{[CaII]}}$ measured from the nebular phase spectra of our sample. Data from literature are plotted as empty symbols \citep{Fang2019NatAs}. }
	\label{fig:neb-OCa}
\end{figure}

\subsection{Nebular phase spectra}\label{sec:neb-phase}
Nebular-phase spectra are available for a limited number of SNe in our sample. To ensure a complete transition to the nebular phase, we included only spectra observed at phases >90 days post-maximum in this analysis.
The nebular phase spectra of each subtype is shown in Fig.~\ref{fig:neb-spec}. Note that in SN~2014C, the narrow emission near 6560~\AA\ is H$\alpha$ formed due to interaction with the circumstellar medium \citep{2018MNRAS.478.5050M,2025ApJ...978..163Z}. The most dominant emission lines common to all subtypes of SESNe are O~I$\lambda$7774, [O~I]$\mathrm{\lambda\lambda}$6300,6364, Ca~II NIR triplet, [Ca~II]$\lambda\lambda$7291,7324 and Na~I~D. Notably,  Mg~I]$\lambda$4571 line is rather weak in SNe IIb and Ib. 
Luminosities of [O~I]$\lambda\lambda$6300,6364 and [Ca~II]7291,7323 are related to the CO core mass, hence the initial mass of the progenitor stars. We measured $L_{\mathrm{[OI]}}$, $L_{\mathrm{[NII]}}$ and $L_{\mathrm{[CaII]}}$ following the method described in \cite{Fang2019NatAs} and \cite{2022ApJ...928..151F}. In Fig.~\ref{fig:neb-OCa} we show the distribution of $L_{\mathrm{[NII]}}/L_{\mathrm{[OI]}}$ and $L_{\mathrm{[OI]}}/L_{\mathrm{[CaII]}}$. Measurements from \cite{Fang2019NatAs} are also plotted as empty symbols for comparison. 
As shown in Fig.~\ref{fig:neb-spec}, the line strength ratio of [O I] to [Ca II] in SNe IIb/Ib is markedly lower than that in SNe Ic, suggesting that SNe Ic have more massive progenitors. SN 2014C exhibits the highest  $L_{\mathrm{[OI]}}$, $L_{\mathrm{[NII]}}$ in the whole dataset, probably due to extra hydrogen emission from late-time interaction with hydrogen-rich circumstellar material.
Although the presence of He was spectroscopically confirmed in the photospheric phase spectra of SN~2016coi, its nebular-phase spectrum, however, shows close resemblance to that of a typical SN Ic or SN Ic-BL, exhibiting remarkably stronger [O~I] emission relative to [Ca~II]. 

\section{Summary}
In this paper, we present the spectra dataset of stripped-envelope supernovae (SESNe) collected by the Tsinghua University-Yunnan Observatory supernova group. Observations were conducted during the years from 2010 to 2020 using the 2.16-m telescope at Beijing Xinglong Observatory and the 2.4-m telescope at Lijiang Observatory. The dataset contains 249 spectra for 62 SESNe, including 20 spectra of 12 SNe IIb, 76 spectra of 20 SNe Ib, 85 spectra of 19 SNe Ic, and 68 spectra of 11 SNe Ic-BL. The SNe in our sample have an average redshift of 0.015, with the farthest one at $z=0.06$. The phases of the dataset cover a range from $-$16 to over 190 days since the maximum light. 

For each SN, we performed detailed spectroscopic classification to resolve discrepancies or misclassifications in public databases. The time of maximum light was determined using our follow-up photometric observations or published photometric data. For SNe lacking reliable light curves, the epoch of each spectrum was estimated via spectral cross-correlation using SNID. 

For each spectrum, we identified the dominant spectral lines according to previous empirical identification results, and measured the pseudo-equivalent widths (pEWs) and blueshift velocities of each line. 
A key focus was the absorption feature near 6200\AA\ in early-phase ($\lesssim$+10 days) SNe Ib spectra, which has been subject to divergent interpretations in the literature. We analysed this feature by comparing SNe Ib to SNe IIb and Ic, testing both \Ha\ and Si~II\lam6355 identifications. Our results show that its properties in SNe Ib align more closely with \Ha\ in SNe IIb, leading us to identify it as \Ha. This indicates that residual hydrogen is common in SNe Ib, with some early spectra showing \Ha\ strengths comparable to those in SNe IIb. This supports a continuous stripping sequence from type IIb to Ib progenitors. The main spectroscopic differences are that in SNe IIb, \Ha\ continues to strengthen after maximum light, whereas in SNe Ib, it weakens rapidly before maximum. Additionally, \Ha velocities are systematically higher in SNe Ib.

Expanding on the analytical methodology employed in prior studies of SESNe spectral samples \citep{2016ApJ...827...90L,2018A&A...618A..37F,2023A&A...675A..83H}, we examined the correlations in both velocity and strength among the common spectral features shared by the four subtypes. Velocities of nearly all common spectral lines exhibit overlapping distributions across different subtypes and follow similar correlation patterns, suggesting they represent variations within a continuous population. An exception is the \Ha-O~I\lam7774 pair between type IIb and type Ib, where SNe Ib show systematically higher \Ha\ velocitiesthan SNe IIb, marking a distinct deviation from the trend defined by the latter.
A velocity gradient is observed across subtypes of SESNe, with SNe Ic exhibiting the highest line velocities, followed by normal SNe Ic, SNe Ib, and SNe IIb, consistent with prior statistical studies.
In SNe Ic and Ic-BL, the velocities of different ions are more tightly correlated than in SNe IIb and Ib, implying more compact progenitor cores in the He-poor subtypes. 
Furthermore, the distribution of Fe~II line strength inferred for SNe Ib aligns more closely with that of SNe Ic/Ic-BL, being systematically stronger than in SNe IIb. 

In the correlation between spectral line intensities (pEWs), SNe IIb and Ib are most clearly distinguished by their early-phase ($\lesssim$+10 d) \Ha-Fe~II relation: SNe IIb exhibit stronger \Ha\ lines alongside weaker Fe~II lines. However, at later epochs, as the strength of \Ha\ lines in SNe IIb declines to levels comparable to those in SNe Ib, the two subclasses become indistinguishable in the \Ha-Fe~II diagnostic diagram. At this stage, distinguishing between them may require additional spectral diagnostics, such as the strength of He~I lines. Besides, He~I line intensities are more correlated with that of O~I line in SNe IIb than in SNe Ib.
Oxygen lines in SNe IIb/Ib are weaker than in SNe Ic/Ic-BL. The distributions of spectral line intensities show no significant differences between type Ic and type Ic-BL.

Our dataset includes 11 nebular-phase spectra. In spectra of SNe IIb/Ib, [Ca~II] emission dominates over [O I], whereas SNe Ic, including the helium-rich broad-lined SN 2016coi, display stronger [O~I] lines. We measured the luminosities of the [O~I], [N~II], and [Ca~II] lines and calculated their ratios. SNe Ic exhibit significantly higher [O~I]/[Ca~II] luminosity ratios than SNe IIb/Ib, suggesting that their progenitors had more massive CO cores and, consequently, higher initial masses.

The spectroscopic data presented here expands the sample of SESNe, providing additional material for future studies on the distinctions, connections, and progenitor properties of such supernovae. The analytical methodology adopted in this work, which incorporates strengths from multiple existing approaches, can serve as a reference for more in-depth statistical studies upon the establishment of a more comprehensive spectroscopic database in the future.

\section*{Data avalability}
All spectral data used in this work are available via the WiseREP. 
Full Table~\ref{tab:spec-list} and data behind Figs.~\ref{fig:Spearman-V}, \ref{fig:Spearman-EW} are only available in electronic form at the CDS via anonymous ftp to cdsarc.u-strasbg.fr (130.79.128.5) or via http://cdsweb.u-strasbg.fr/xxxx.

\begin{acknowledgements}
	We acknowledge the support of the staff of the Lijiang 2.4-m and Xinglong 2.16-m telescopes. This work is supported by the National Natural Science Foundation of China (NSFC, grants 12288102 and 12033003), and the Tencent Xplorer Prize. DFX is supported by the National Natural Science Foundation of China (grant 12503047). JZ is supported by the B-type Strategic Priority Program of the Chinese Academy of Sciences (Grant No. XDB1160202), the National Key R\&D Program of China with grant 2021YFA1600404, the National Natural Science Foundation of China (NSFC grants 12173082 and 12333008), the Yunnan Fundamental Research Projects (YFRP; grants 202501AV070012 and 202401BC070007), the Top-notch Young Talents Program of Yunnan Province, the Light of West China Program provided by the Chinese Academy of Sciences, and the International Centre of Supernovae, Yunnan Key Laboratory (grant 202302AN360001). Funding for the LJT has been provided by the CAS and the People's Government of Yunnan Province. The LJT is jointly operated and administrated by YNAO and Center for Astronomical Mega-Science, CAS. H.L. was supported by the National Natural Science Foundation of China (NSFC grants No. 12403061) and the innovative project of ‘Caiyun Postdoctoral Project’ of Yunnan Province. Y.-Z. Cai is supported by the National Natural Science Foundation of China (NSFC, Grant No. 12303054), the National Key Research and Development Program of China (Grant No. 2024YFA1611603), and the Yunnan Fundamental Research Projects (Grant Nos. 202401AU070063, 202501AS070078).
	Chengyuan Wu is supported by the National Natural Science Foundation of China (No. 12473032), the Yunnan Revitalization Talent Support Program-Young Talent project, and the Yunnan Fundamental Research Project (No. 202501AW070001). JNF acknowledges the support from the National Natural Science Foundation of China (NSFC) through the grants 12090040, 12090042 and 12427804.
	This work is partly supported by the China Manned Space Program with grant no. CMS-CSST-2025-A13, the Tianchi Talent Introduction Plan, and the Central Guidance for Local Science and Technology Development Fund under No. ZYYD2025QY27.
	
\end{acknowledgements}

\bibliographystyle{aa}
\bibliography{ref}
\begin{appendix}

\onecolumn
\section{Additional Tables and Figures}
\setlength{\tabcolsep}{2pt}
\begin{longtable}{ccccccccc}
	\caption{Information of SESNe observed in the THU sample.\label{tab:sne-list}}\\
	\hline\hline
	SN&R.A.&Dec.&Type    &Redshift\tablefootmark{a}&$E(\mathrm{B-V})_{\mathrm{MW}}$\tablefootmark{b}&Max. $B$&Max. $V$&Peak date refs.\tablefootmark{c} \\
	&(J2000)&(J2000)&    &&&(MJD)&(MJD)&\\
	\hline
	\endfirsthead
	\caption{\textit{continued.} Information of SESNe observed in the THU sample.}\\
	\hline\hline
	SN    &R.A.    &Dec.    &Type    &Redshift\tablefootmark{a}    &$E(\mathrm{B-V})_{\mathrm{MW}}$\tablefootmark{b}&Max. $B$    &Max. $V$    &Peak date refs.\tablefootmark{c} \\
	&(J2000)&(J2000)&    &&&(MJD)&(MJD)&\\
	\hline
	\endhead
	\hline
	\endfoot
	\hline
	css141005 & 02:23:15.64 &$-$07:05:20.80 &   IIb &    0.045 & 0.023 &   ... & 56944 &G21 \\
	iPTF13bvn & 15:00:00.18 &$+$01:52:53.5 &    Ib &  0.00449 & 0.045 & 56474 & 56476 &C13,F16 \\
	iPTF15eqv & 10:52:08.33 &$+$32:56:39.4 &   Ib(Ca-rich) &  0.00529 & 0.021 & 57243 &   ... &TNT,M17 \\
	PSNJ0110 & 01:10:11.91 &$+$33:13:53.6 &   IIb &    0.018 & 0.052 &   ... & 57249 &SNID \\
	PSNJ0123 & 01:23:24.38 &$+$09:25:54.5 &    Ib &  0.00758 & 0.041 &   ... & 56193 &SNID \\
	SN 2010ln & 03:20:53.62 &$+$38:15:11.9 &    Ib &   0.0168 & 0.219 &   ... & 55521 &SNID \\
	SN 2011bl\tablefootmark{e} & 13:34:14.69 &$+$37:12:34.7 &    Ic &   0.0184 & 0.007 &   ... & 55658 &SNID \\
	SN 2011dh & 13:30:05.12 &$+$47:10:11.3 &   IIb &    0.002 & 0.032 & 55730 & 55733 &TNT \\
	SN 2011fu & 02:08:21.41 &$+$41:29:12.30 &   IIb & 0.018489 & 0.068 & 55845 & 55847 &M15 \\
	SN 2011jf\tablefootmark{e} & 02:38:54.61 &$+$27:50:48.7 &    Ic &   0.0153 & 0.152 &   ... & 55922 &SNID \\
	SN 2012ap\tablefootmark{f} & 05:00:13.72 &$-$03:20:51.2 & Ic-BL &   0.0121 & 0.045 & 55974 & 55976 &L15 \\
	SN 2012au & 12:54:52.18 &$-$10:14:50.2 &    Ib &   0.0045 & 0.042 & 56004 & 56006 &T13 \\
	SN 2012C & 09:37:30.48 &$+$32:50:31.5 &    Ic &   0.0145 & 0.015 &   ... & 55949 &SNID \\
	SN 2012cw & 10:13:47.95 &$+$03:26:02.6 &    Ic & 0.004486 & 0.028 &   ... & 56094 &SNID \\
	SN 2012ej & 06:26:51.01 &$+$36:07:17.4 &    Ic &   0.0089 & 0.089 &   ... & 56174 &SNID \\
	SN 2012fh\tablefootmark{e} & 10:43:34.05 &$+$24:53:29.0 &    Ic & 0.001935 & 0.029 &   ... & 56070\tablefootmark{d} &SNID,Z22 \\
	SN 2013ge\tablefootmark{e} & 10:34:48.46 &$+$21:39:41.9 &    Ic & 0.004356 & 0.020 & 56616 & 56620 &TNT \\
	SN 2014ad & 11:57:44.44 &$-$10:10:15.7 & Ic-BL &   0.0057 & 0.039 & 56735 & 56740 &TNT,S18 \\
	SN 2014as & 14:00:54.49 &$+$40:58:59.6 & Ic-BL & 0.012469 & 0.013 &   ... & 56772 &TNT \\
	SN 2014bl & 13:25:38.81 &$+$25:57:33.9 &    Ic &   0.0377 & 0.013 &   ... & 56812 &TNT \\
	SN 2014C\tablefootmark{f} & 22:37:05.60 &$+$34:24:31.9 &    Ib & 0.002722 & 0.081 & 56668 & 56670 &B14,Z25 \\
	SN 2014dj & 00:57:40.18 &$+$43:47:34.2 &    Ic &    0.018 & 0.071 &   ... & 56914 &SNID \\
	SN 2014ds & 08:11:16.45 &$+$25:10:47.4 &   IIb &   0.0137 & 0.038 &   ... & 56966 &Z22 \\
	SN 2014eh & 20:25:03.86 &$-$24:49:13.3 &    Ic & 0.010614 & 0.056 & 56973 & 56976 &Z22 \\
	SN 2014L\tablefootmark{f} & 12:18:48.68 &$+$14:24:43.5 &    Ic & 0.008029 & 0.034 & 56693 & 56694 &Z18 \\
	SN 2015ah & 23:00:24.63 &$+$01:37:36.8 &    Ib &    0.016 & 0.071 &   ... & 57251 &SNID \\
	SN 2015bi & 14:32:15.31 &$+$26:19:32.02 &   IIb & 0.016014 & 0.017 &   ... & 57411 &SNID \\
	SN 2016adj\tablefootmark{e} & 13:25:24.12 &$-$43:00:57.9 &    Ic & 0.001825 & 0.102 & 57431 & 57431 &S16,S24,S22,B18 \\
	SN 2016ajo & 18:44:12.49 &$+$24:09:29.7 &    Ib &    0.016 & 0.107 &   ... & 57437 &SNID \\
	SN 2016bau & 11:20:59.02 &$+$53:10:25.6 &    Ib & 0.003856 & 0.015 & 57474 & 57478 &Z22 \\
	SN 2016bll & 08:33:18.30 &$+$19:20:44.8 &    Ib &    0.019 & 0.028 &   ... & 57453 &SNID \\
	SN 2016cce & 13:33:27.20 &$+$05:28:57.8 &    Ic &    0.022 & 0.029 &   ... & 57515 &SNID \\
	SN 2016coi\tablefootmark{f} & 21:59:04.08 &$+$18:11:10.46 &    Ib(BL) &   0.0036 & 0.075 & 57548 & 57552 &P18 \\
	SN 2016frp & 00:21:32.54 &$-$05:57:24.29 &    Ib &    0.027 & 0.031 &   ... & 57640 &SNID \\
	SN 2016G & 03:03:57.74 &$+$43:24:03.50 & Ic-BL &   0.0091 & 0.139 & 57409 & 57411 &TNT \\
	SN 2016hkn & 02:08:34.23 &$+$29:14:11.10 &   IIb &    0.022 & 0.053 &   ... & 57693 &SNID \\
	SN 2016iae & 04:12:05.53 &$-$32:51:44.75 &    Ic & 0.003468 & 0.013 & 57708 & 57712 &P19 \\
	SN 2016jdw & 13:16:19.62 &$+$30:40:32.67 &    Ib &   0.0189 & 0.011 & 57760 & 57764 &P19 \\
	SN 2016M &07:16:37.750 &$+$67:53:32.30 &   IIb &    0.036 & 0.037 &   ... & 57404 &SNID \\
	SN 2016P & 13:57:31.13 &$+$06:05:51.6 & Ic-BL &  0.01462 & 0.024 & 57413 & 57417 &G20 \\
	SN 2017bgu & 16:55:59.47 &$+$42:33:36.01 &    Ib &   0.0085 & 0.019 & 57816 & 57818 &TNT \\
	SN 2017cxz & 17:19:19.78 &$+$57:53:55.80 &    Ib &   0.0289 & 0.028 &   ... & 57860 &SNID \\
	SN 2017ein\tablefootmark{f} & 11:52:53.25 &$+$44:07:26.2 &    Ic &   0.0027 & 0.019 & 57910 & 57913 &X19 \\
	SN 2017giq & 23:57:54.73 &$+$28:30:12.0 &    Ic &     0.03 & 0.075 &   ... & 57993 &SNID \\
	SN 2017gpn\tablefootmark{f} &  3:37:44.97 &$+$72:31:59.00 &   IIb &   0.0073 & 0.303 & 58001 &   ... &B21 \\
	SN 2017ifh & 06:35:03.56 &$+$50:26:27.90 & Ic-BL &    0.039 & 0.117 & 58074 & 58076 &TNT \\
	SN 2017iro & 14:06:23.11 &$+$50:43:20.20 &    Ib & 0.006191 & 0.016 & 58095 & 58098 &TNT \\
	SN 2017iuk & 11:09:39.52 &$-$12:35:18.34 & Ic-BL &   0.0368 & 0.045 & 58104 & 58105 &D18 \\
	SN 2017jdn & 10:23:45.51 &$+$53:06:20.50 &   IIb &   0.0317 & 0.020 &   ... & 58099 &SNID \\
	SN 2018beh & 09:31:23.03 &$+$17:48:27.90 &    Ib &     0.06 & 0.034 & 58243 & 58246 &G22 \\
	SN 2018gep & 16:43:48.22 &$+$41:02:43.37 & Ic-BL &    0.032 & 0.009 & 58374 & 58375 &P21 \\
	SN 2018gsk & 04:09:11.62 &$+$08:38:51.83 &    Ic &   0.0116 & 0.227 &   ... & 58387 &SNID \\
	SN 2018ie &10:54:01.040 &$-$16:01:21.72 & Ic-BL & 0.014233 & 0.060 & 58138 &   ... &P19 \\
	SN 2018if &09:14:23.840 &$+$49:35:32.90 & Ic-BL &    0.031 & 0.012 & 58141 & 58142 &TNT \\
	SN 2019ehk\tablefootmark{f} &12:22:56.130 &$+$15:49:33.60 &    Ib &    0.005 & 0.023 & 58615 & 58616 &J20 \\
	SN 2019yvr &12:45:08.134 &$-$00:27:32.73 &    Ib &    0.005 & 0.022 & 58854 & 58855 &K21 \\
	SN 2019yz &15:41:57.301  &$+$00:42:39.45 &    Ic & 0.006388 & 0.101 &   ... & 58516 &ZTF \\
	SN 2020aaxf &08:20:41.967 &$-$01:24:52.37 &   IIb &    0.014 & 0.069 & 59197 & 59198 &ZTF \\
	SN 2020adow &08:33:42.262 &$+$27:42:43.56 & Ic-BL & 0.007505 & 0.032 & 59217 & 59219 &TNT \\
	SN 2020hvp &16:21:45.390 &$-$02:17:21.37 &    Ib & 0.005247 & 0.140 & 58973 & 58975 &TNT \\
	SN 2020oi\tablefootmark{f} &12:22:54.925 &$+$15:49:25.05 &    Ic &  0.00524 & 0.023 & 58863 & 58866 &R21 \\
	SN 2020zgl\tablefootmark{e} &23:28:01.150 &$-$02:09:53.64 &    Ib &   0.0065 & 0.051 & 59178 & 59180 &ZTF \\
\end{longtable}
\tablefoot{
	\tablefoottext{a}{Redshifts are spectroscopic redshifts sourced primarily from TNS and WiseRep.}\\
	\tablefoottext{b}{Foreground Galactic extinctions are from \cite{2011ApJ...737..103S}, as accessed via the NASA/IPAC Extragalactic Database.}\\
	\tablefoottext{c}{Method or references for the maximum time of SNe. TNT--TNT photometry, ZTF--ZTF photometry, SNID--in SNID release version 5.0 via templates-2.0 \citep{2007ApJ...666.1024B} and SESNe template by \cite{2016ApJ...827...90L}, B14--\cite{2014Ap&SS.354...89B}, B18--\cite{2018MNRAS.481..806B}, B21--\cite{2021MNRAS.501.5797B}, C13--\cite{2013ApJ...775L...7C}, D18--\cite{2018A&A...619A..66D}, F16--\cite{2016A&A...593A..68F}, G20--\cite{2020MNRAS.497.3770G}, G21--\cite{2021MNRAS.505.3950G}, G22--\cite{2022ApJ...941..107G}, J20--\cite{2020ApJ...898..166J}, K21--\cite{2021MNRAS.504.2073K}, L15--\cite{2015RAA....15..225L}, M15--\cite{2015MNRAS.454...95M}, M17--\cite{2017ApJ...846...50M}, P18--\cite{2018MNRAS.478.4162P}, P19--\cite{2019MNRAS.485.1559P}, P19--\cite{2019MNRAS.485.1559P}, P21--\cite{2021ApJ...915..121P}, R21--\cite{2021ApJ...908..232R}, S16--\cite{2016TNSCR1117....1S}, S18--\cite{2018MNRAS.475.2591S}, S22--\cite{2022ApJ...939L...8S}, S24--\cite{2024A&A...686A..79S}, T13--\cite{2013ApJ...772L..17T}, X19--\cite{2019ApJ...871..176X}, Z18--\cite{2018ApJ...863..109Z}, Z22--\cite{2022MNRAS.512.3195Z}, Z25--\cite{2025ApJ...978..163Z}.}\\
	\tablefoottext{d}{SN~2012fh was discovered more than 100 days after peak and photometric observations started long after peak. The peak dates obtained from light curve fitting by \cite{2022MNRAS.512.3195Z} is quite close to the first data point. So peak date is determined by SNID fitting result.}\\
	\tablefoottext{e}{These SNe are reclassified in this work. \\ 
		SN 2011bl: Original classification of SN 2011bl is SN Ib/c (Latest Supernova) but no public spectrum is available. Our SNID analysis shows that its spectrum matches well with that of the normal SN Ic 2004aw at approximately $-$3 days. We therefore classify it as a normal SN Ic.\\
		SN 2011jf: This object has conflicting public classifications (SN Ib-pec on WiseRep; SN Ib/c on TNS). Both our spectra and a spectrum from WiseRep are consistent with the normal SN Ic PTF12gzk near maximum light, though they also show some similarity to broad-lined SNe Ic at phases around +14 days. Given that our spectra were obtained very soon after discovery, we favour a normal SN Ic classification for SN 2011jf.\\
		SN 2012fh: Public classifications for this object are discrepant (SN Ib/c on TNS; SN Ib on WiseRep). Only nebular-phase spectra are available. As shown in Fig.~\ref{fig:neb-OCa}, its nebular spectral parameters are markedly offset from the distribution of SNe Ib and are instead consistent with the median of the SNe Ic distribution. We therefore reclassify it as a type Ic.\\
		SN 2013ge: Public classifications for this object are inconsistent (SN Ic on TNS; SN Ib/c on WiseRep). \cite{2016ApJ...821...57D} noted the possible presence of weak He I lines and classified it as SN Ib/c, although they acknowledged this identification was uncertain. On the other hand, Some other works found the spectra of SN 2013ge very similar to those of SNe Ic such as SNe 2007gr, 2017ein. We decided to follow the classification in other literatures \citep{2019MNRAS.482.1545S,2022ApJ...925..175S,2022ApJ...928..151F} to classify it as a normal type Ic.\\
		SN 2016adj: This object has discrepant public classifications (SN Ib on TNS; SN IIb on WiseRep). However, a recent study by the same group responsible for the original classification report has reclassified it as a type Ic, noting that weak hydrogen lines in its spectra may originate from interaction with hydrogen-rich circumstellar material \citep{2024A&A...686A..79S}. We therefore adopt the type Ic classification.\\
		SN 2020zgl: It is publicly classified as SN Ib-pec (TNS and WiseRep). Our SNID analysis shows a good match to normal SNe Ib, such as SN 2004gv, at comparable phases. Consequently, we classify it as a normal type Ib. }\\
	\tablefoottext{f}{The spectroscopic data for these supernovae have been published in the following references: 
		SN 2012ap--\cite{2015RAA....15..225L}, 
		SN 2014C--\cite{2025ApJ...978..163Z},
		SN 2014L--\cite{2018ApJ...863..109Z},
		SN 2016coi--\cite{2018MNRAS.478.4162P},
		SN 2017ein--\cite{2019ApJ...871..176X},
		SN 2017gpn--\cite{2021MNRAS.501.5797B},
		SN 2019ehk--\cite{2020ApJ...898..166J},
		SN 2020oi--\cite{2021ApJ...908..232R}.}
}

{\centering
	\setlength{\tabcolsep}{6pt}
	\begin{longtable}{cccccc}
		\caption{Journal of spectroscopic observations of our SESNe sample.\label{tab:spec-list}}\\
		\hline\hline
		SN    &UT Date (yy-mm-dd)    &JD    &Phase (d)    &Telescope    &Wavelength Range (\AA)\\
		\hline
		\endfirsthead
		\caption{\textit{continued.} Journal of spectroscopic observations of our SESNe sample.}\\
		\hline\hline
		SN\tablefootmark{a}    &UT Date (yy-mm-dd)    &JD    &Phase (d)    &Telescope    &Wavelength Range (\AA)\\
		\hline
		\endhead
		\hline
		\endfoot
		\hline
		css141005 &2014-10-13 &  2456945.20 &  $+$0.3 &         LJT &          3899--8264\\
		iPTF13bvn &2013-07-05 &  2456479.07 &  $+$2.3 &         XLT &          3461--8823\\
		iPTF15eqv &2015-12-17 &  2457374.38 &$+$128.0 &         XLT &          3747--8478\\
		PSNJ0110 &2015-08-20 &  2457254.22 &  $+$4.9 &         XLT &          4208--8850\\
		PSNJ0123 &2012-10-12 &  2456213.15 & $+$19.5 &         XLT &          3513--8871\\
		SN 2010ln &2011-01-07 &  2455569.08 & $+$46.8 &         XLT &          4504--8671\\
		SN 2011bl &2011-04-08 &  2455655.23 &  $-$3.2 &         XLT &          4001--8868\\
		SN 2011dh &2011-06-03 &  2455716.12 & $-$16.9 &         XLT &          3646--8844\\
		SN 2011dh &2011-06-18 &  2455731.06 &  $-$2.0 &         LJT &          3181--9213\\
		SN 2011dh &2011-06-20 &  2455733.10 &  $-$0.0 &         XLT &          3482--8841\\
		SN 2011dh &2011-06-20 &  2455733.14 &  $+$0.0 &         LJT &          5032--9533\\
		SN 2011dh &2011-07-03 &  2455746.07 & $+$12.9 &         XLT &          3392--8699\\
		SN 2011fu &2011-09-30 &  2455835.13 & $-$11.8 &         XLT &          3887--8497\\
		SN 2011fu &2011-10-21 &  2455856.21 &  $+$8.8 &         XLT &          4004--8867\\
		SN 2011jf &2011-12-23 &  2455919.01 & $-$3.4 &         XLT &          3801--8897\\
		SN 2011jf &2011-12-24 &  2455920.19 & $-$2.3 &         LJT &          3481--9144\\
	\end{longtable}
	\tablefoot{
		This table is available in its entirety in machine-readable form at the CDS.
	}
}

\begin{figure*} 
	\centering
	\includegraphics[width=0.45\columnwidth]{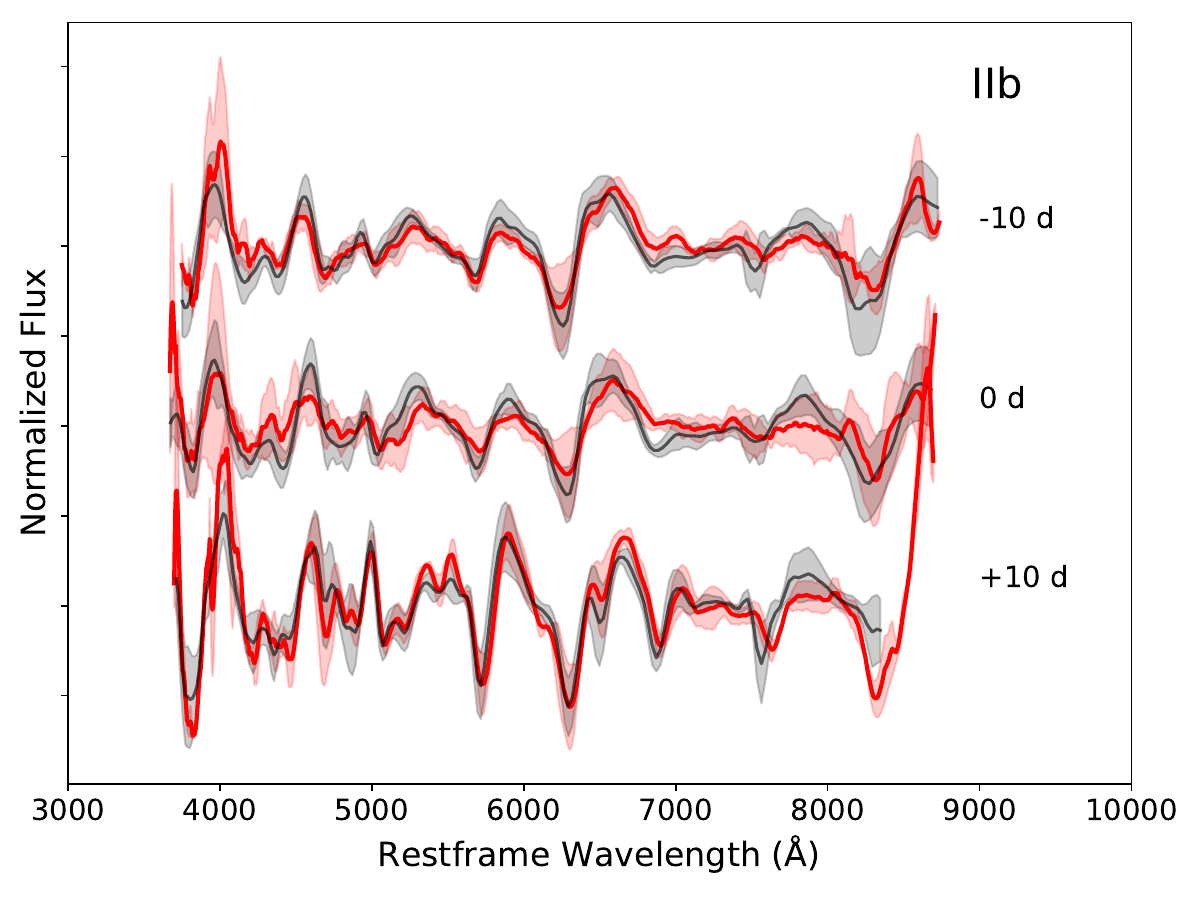}
	\includegraphics[width=0.45\columnwidth]{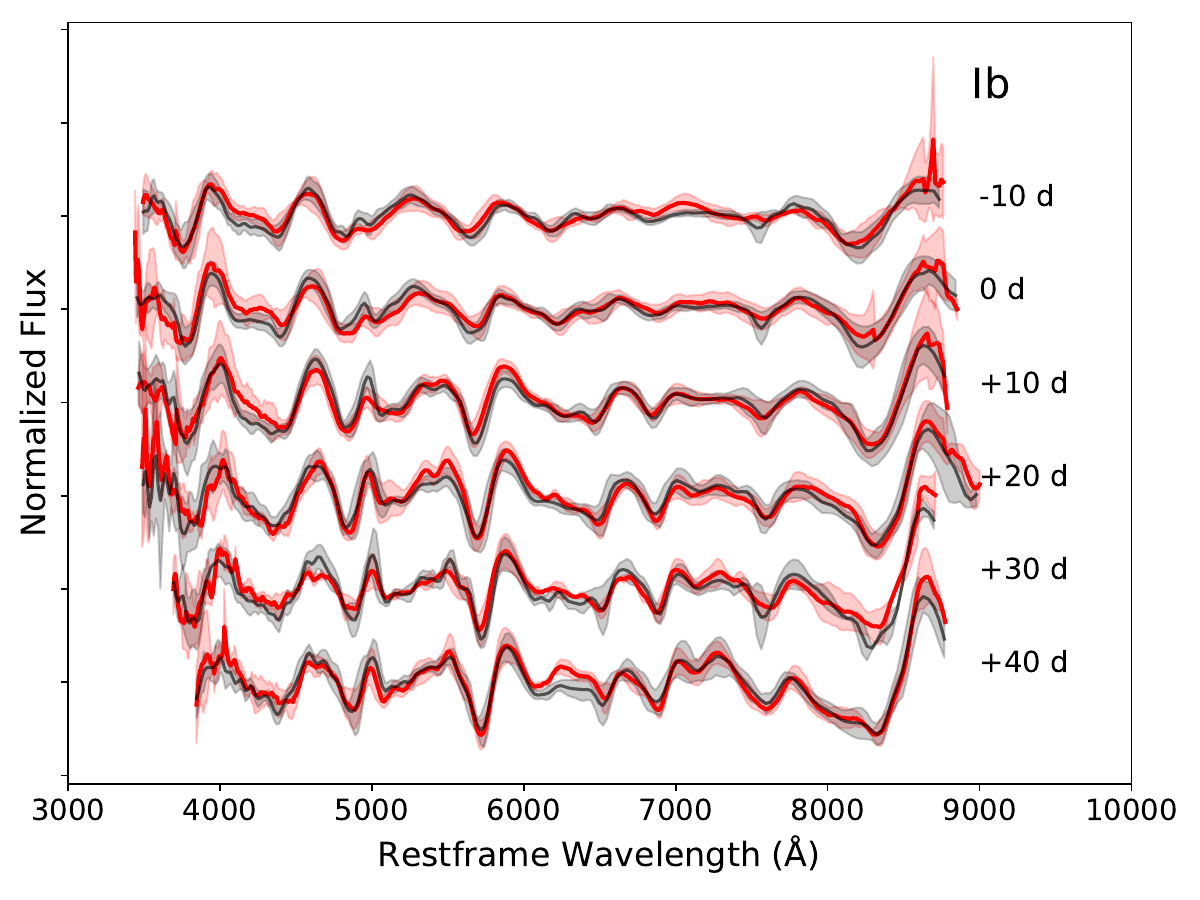}
	\includegraphics[width=0.45\columnwidth]{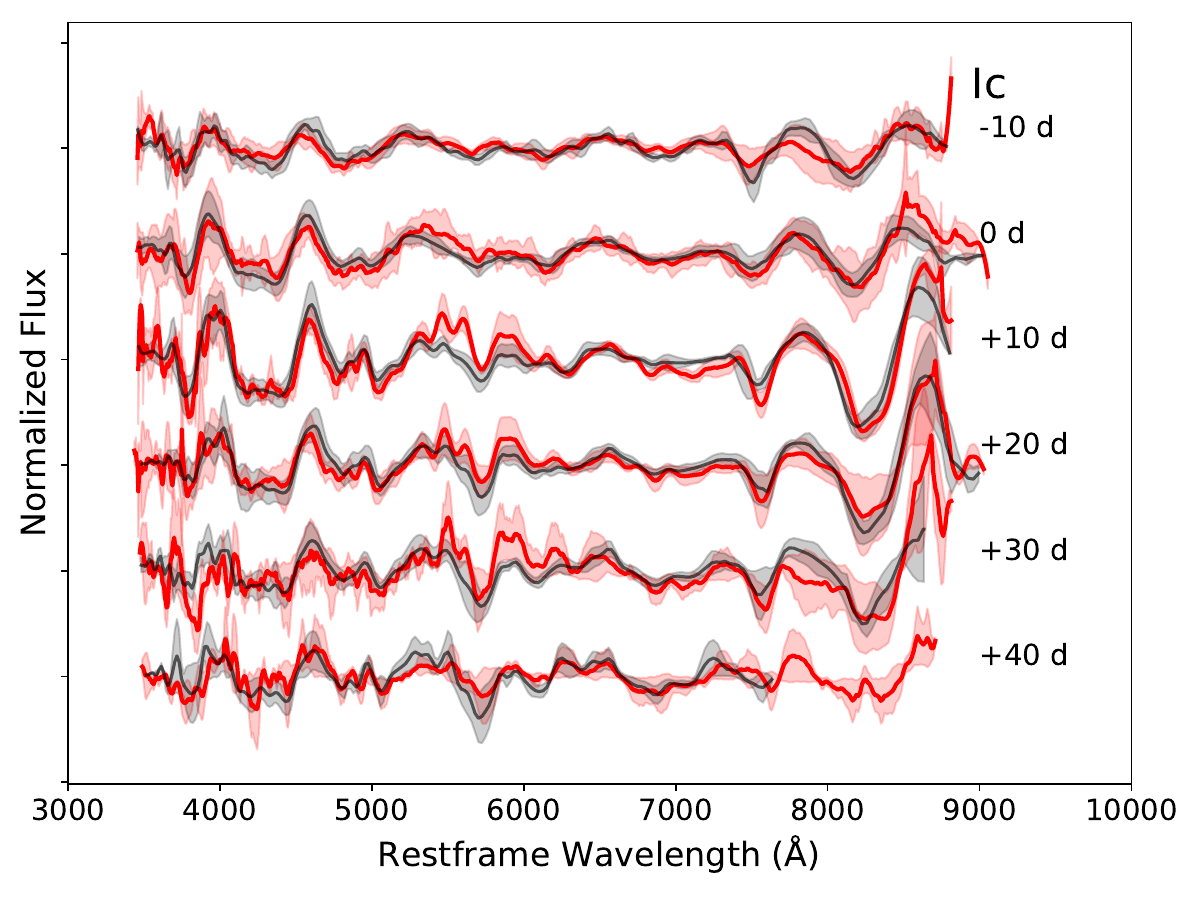}
	\includegraphics[width=0.45\columnwidth]{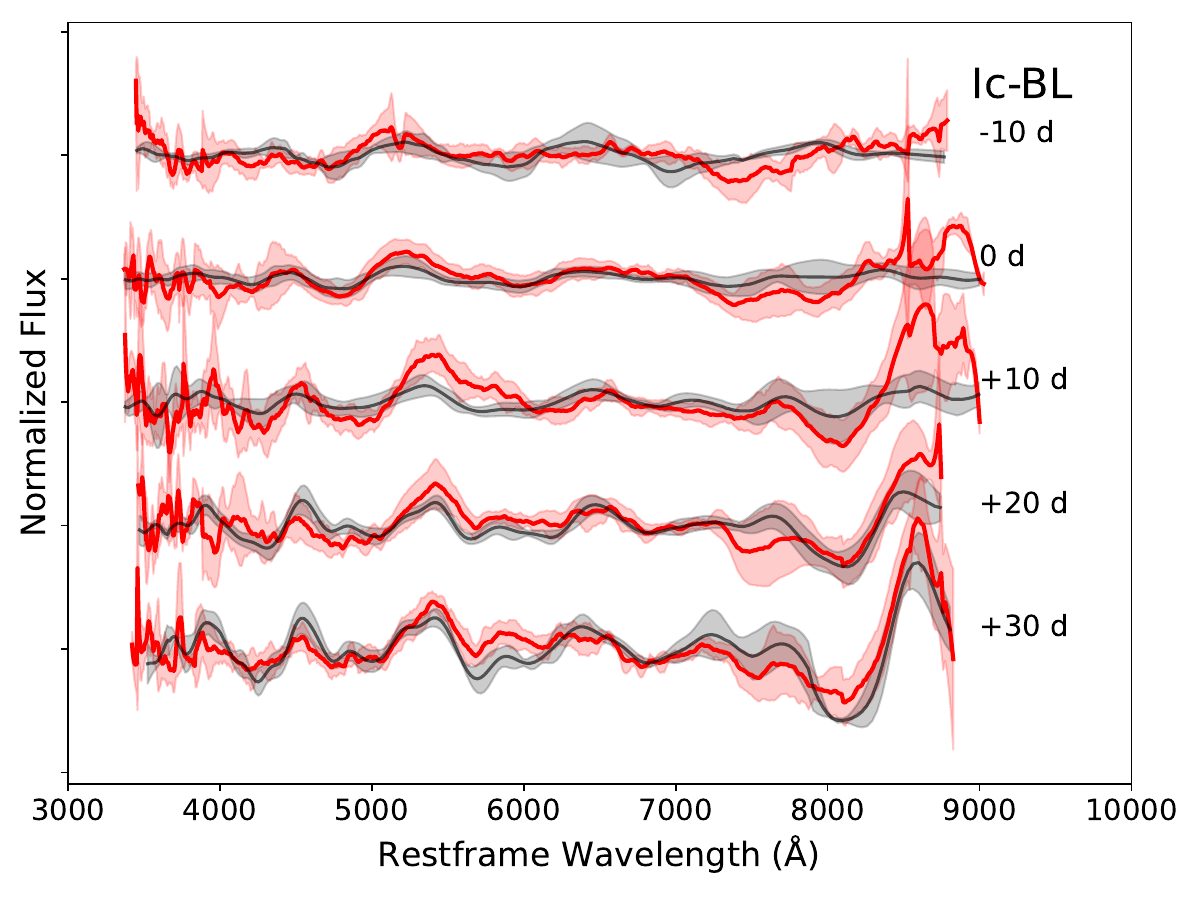}
	\caption{Comparision of our mean spectra and standard deviations (red) to those of \cite{2016ApJ...827...90L} (black).}
	\label{fig:mean-spec-comp}
\end{figure*}

\clearpage

\begin{figure*}
	\centering
	\includegraphics[width=0.155\linewidth]{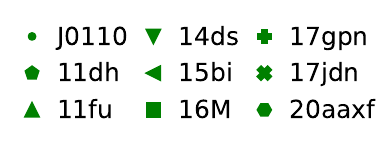}
	\includegraphics[width=0.313\linewidth]{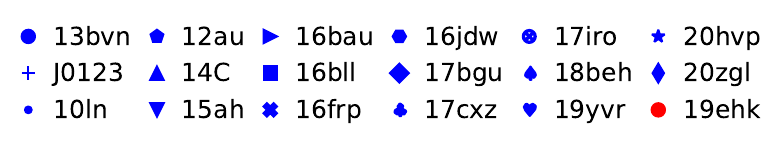}
	\includegraphics[width=0.313\linewidth]{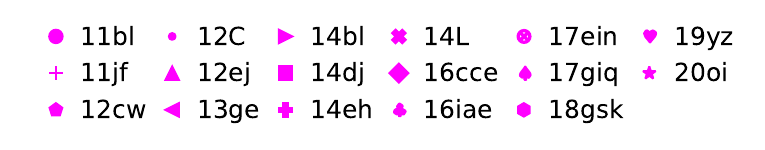}
	\includegraphics[width=0.204\linewidth]{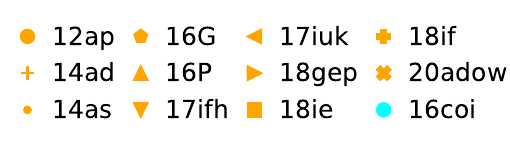}
	
	\includegraphics[width=0.217\linewidth]{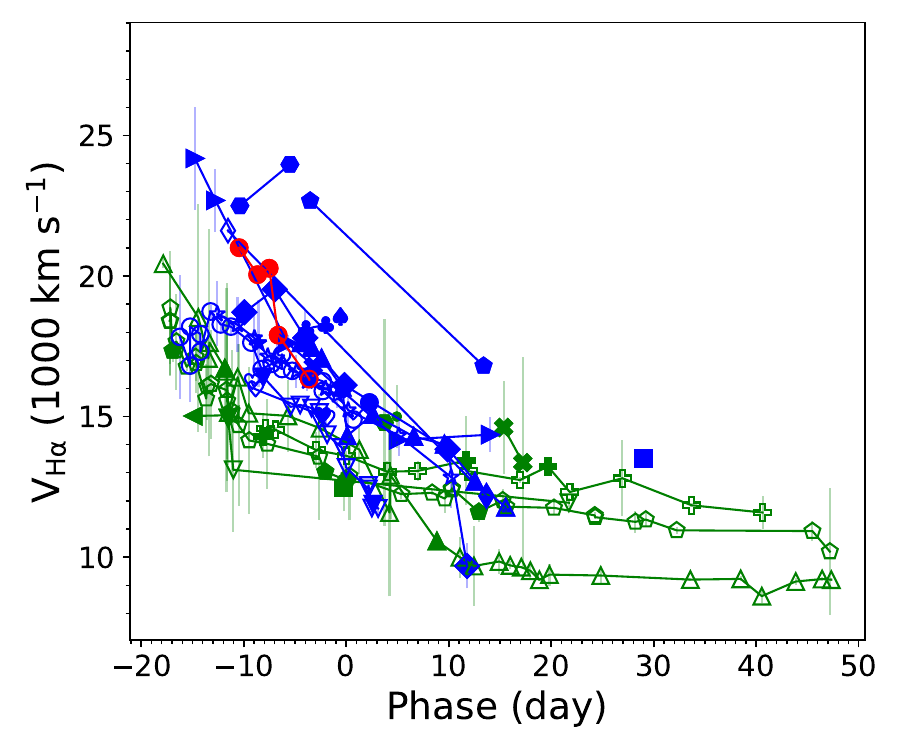}
	\includegraphics[width=0.217\linewidth]{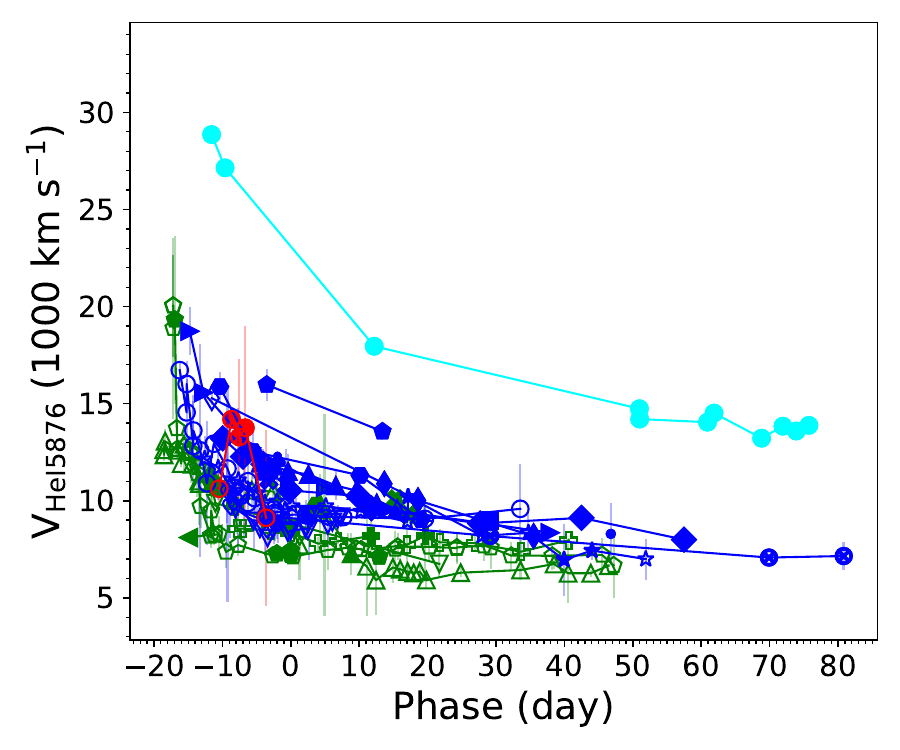}
	\includegraphics[width=0.217\linewidth]{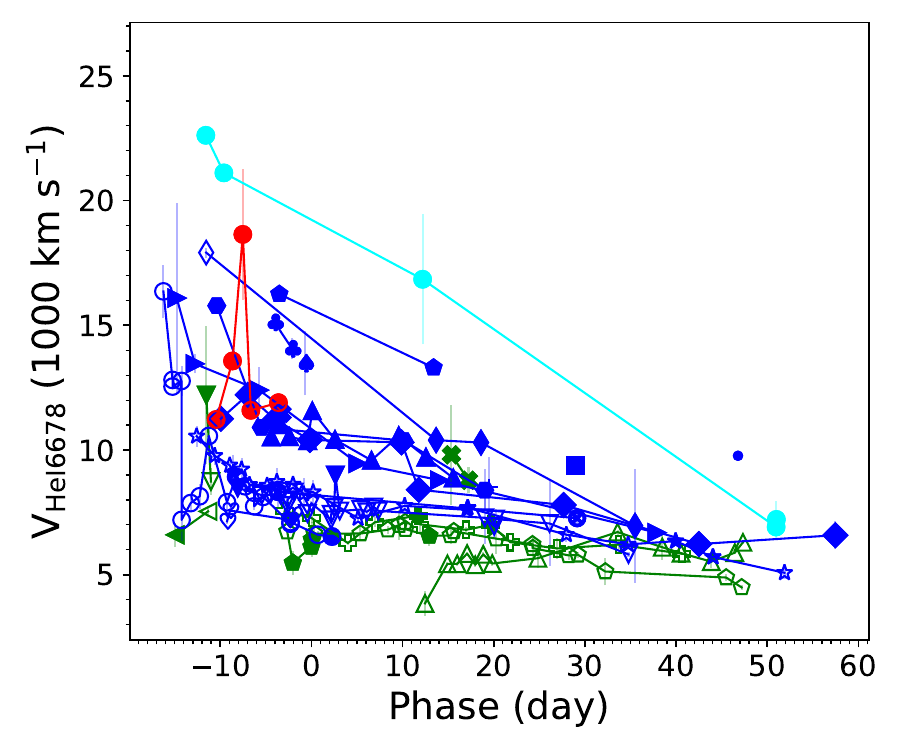}
	\includegraphics[width=0.217\linewidth]{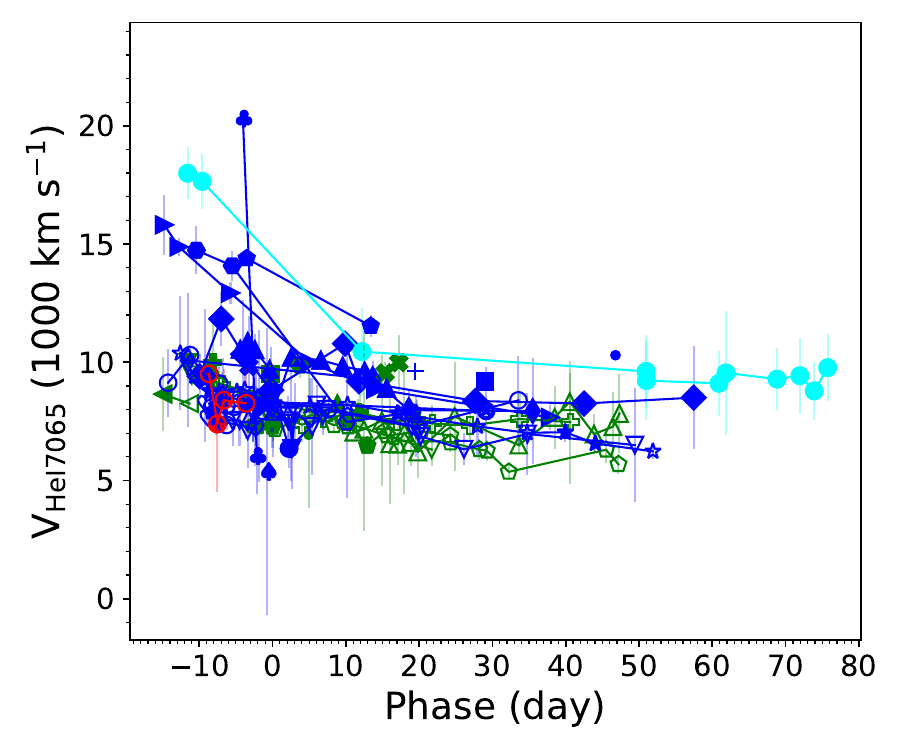}
	\includegraphics[width=0.217\linewidth]{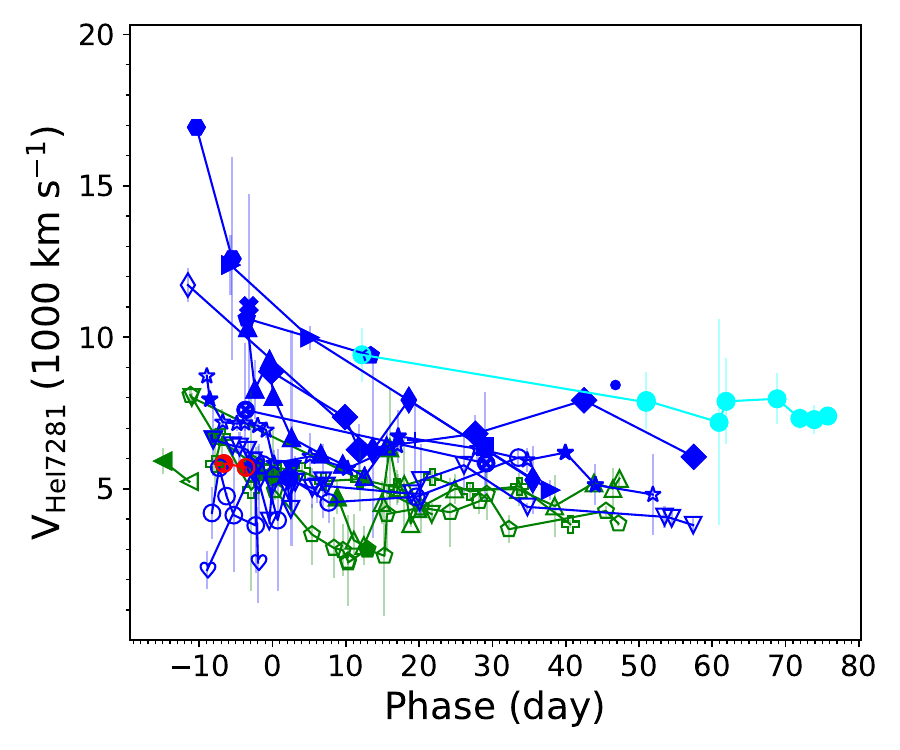}
	\includegraphics[width=0.217\linewidth]{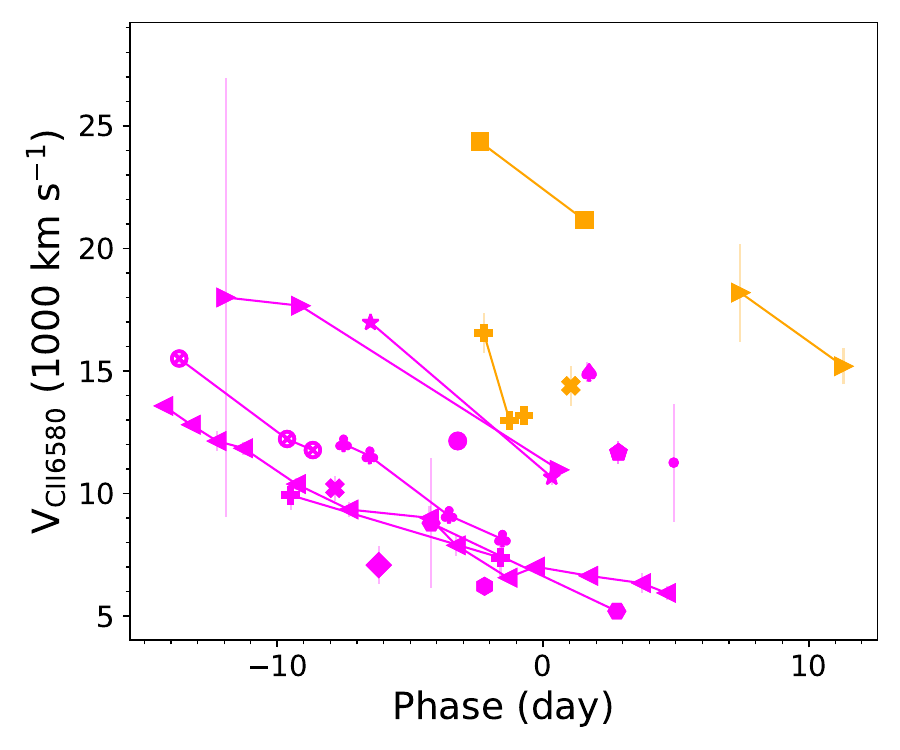}
	\includegraphics[width=0.217\linewidth]{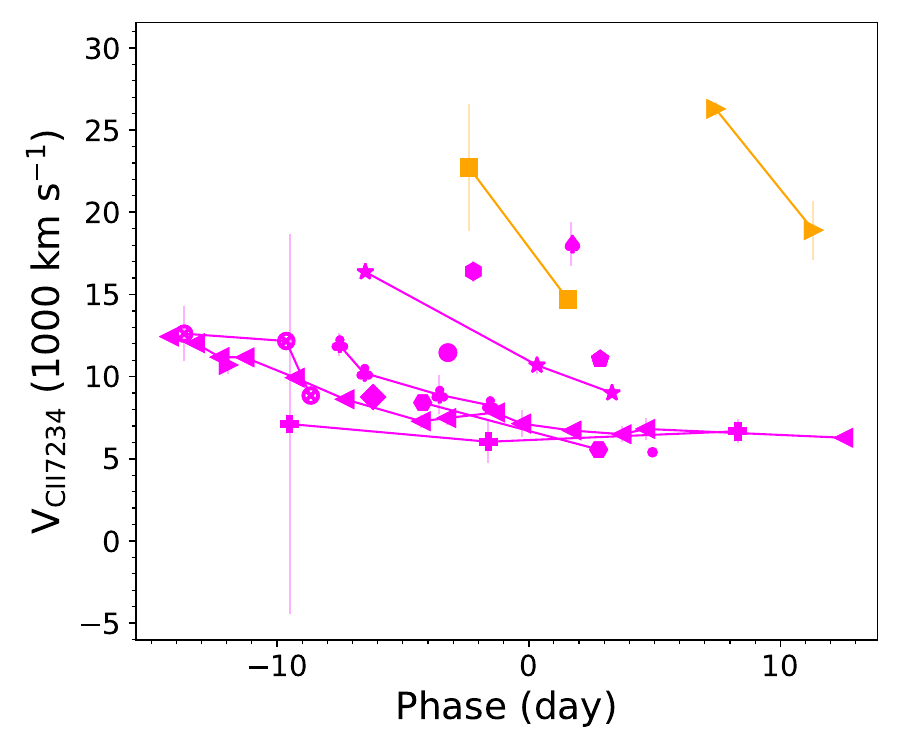}
	\includegraphics[width=0.217\linewidth]{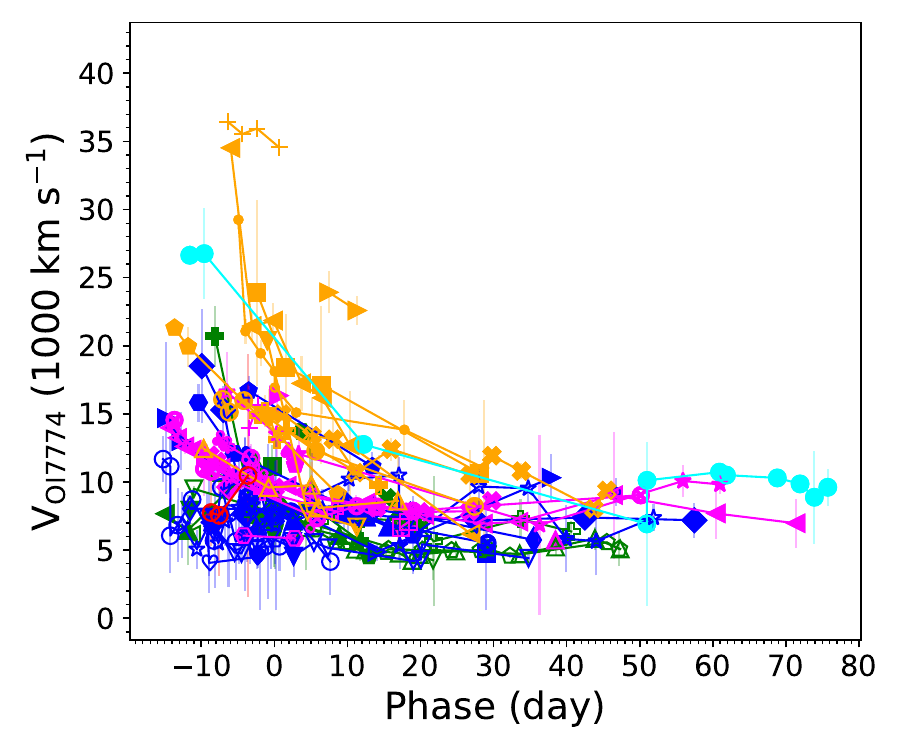}
	\includegraphics[width=0.217\linewidth]{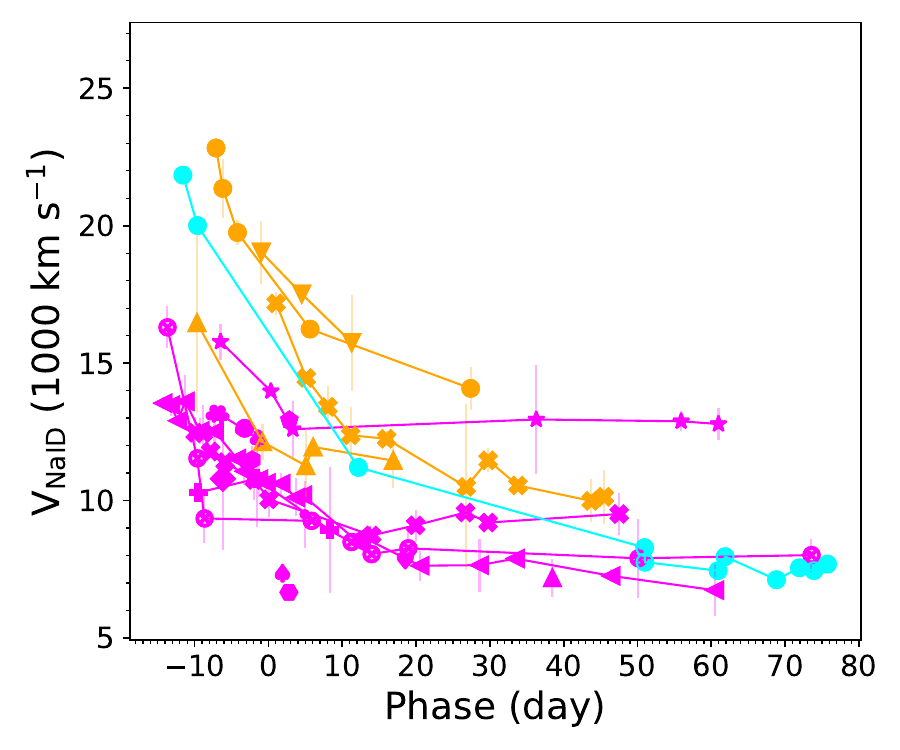}
	\includegraphics[width=0.217\linewidth]{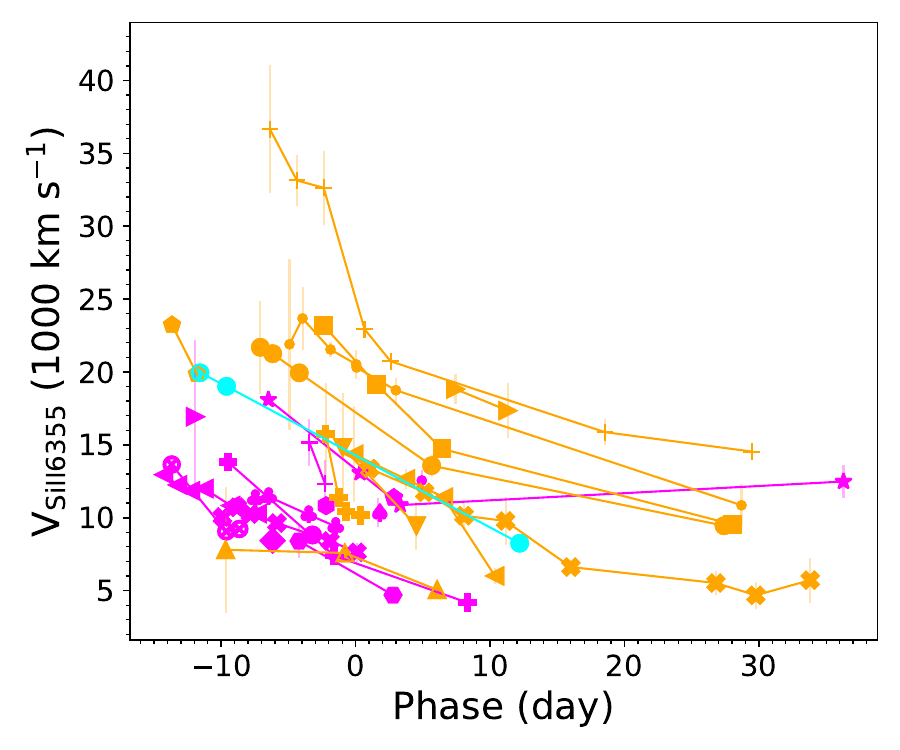}
	\includegraphics[width=0.217\linewidth]{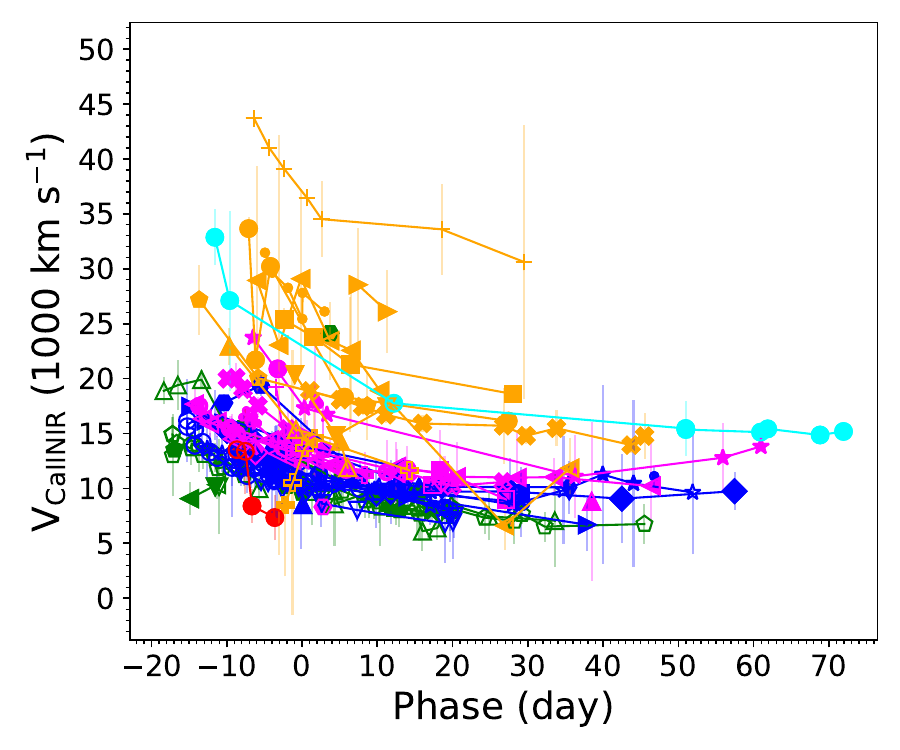}
	\includegraphics[width=0.217\linewidth]{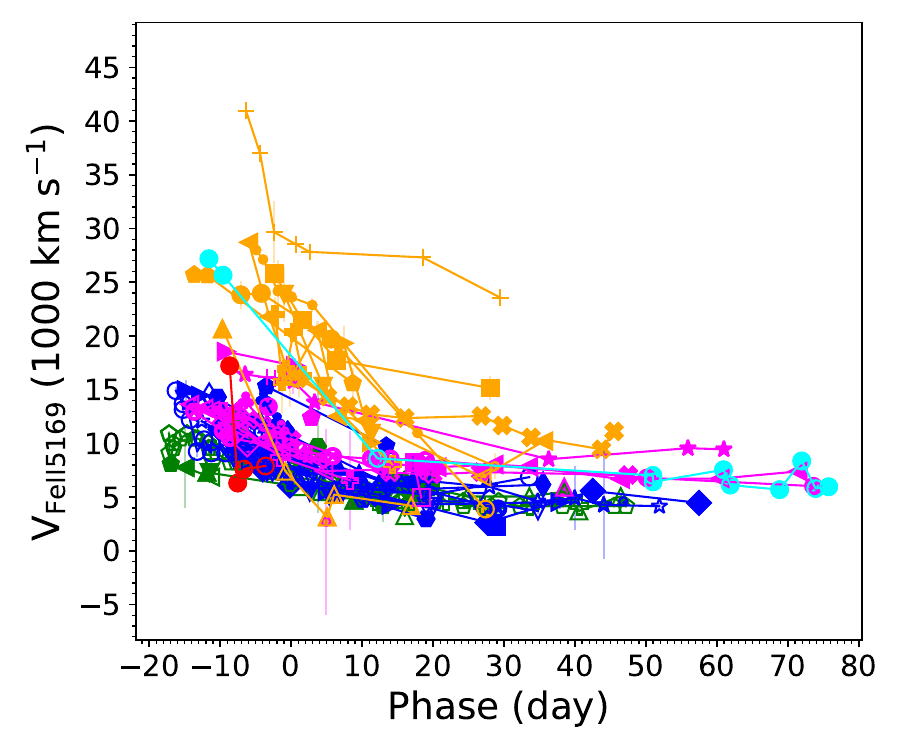}
	
	\caption{Evolution of line velocities of in the four subtypes of SESNe. In each figure, different subtypes are plotted in distinct colors: green--IIb, blue--Ib, magenta--Ic, orange--Ic-BL. The same supernovae are plotted with symbols shown in the top legend and connected by lines. SN 2016coi and SN 2019ehk are represented in red and cyan, respectively, as their spectral phenotypes differ from the four subclasses.}
	\label{fig:evol-V-all}
\end{figure*}

\begin{figure*}
	\centering

	\includegraphics[width=0.217\linewidth]{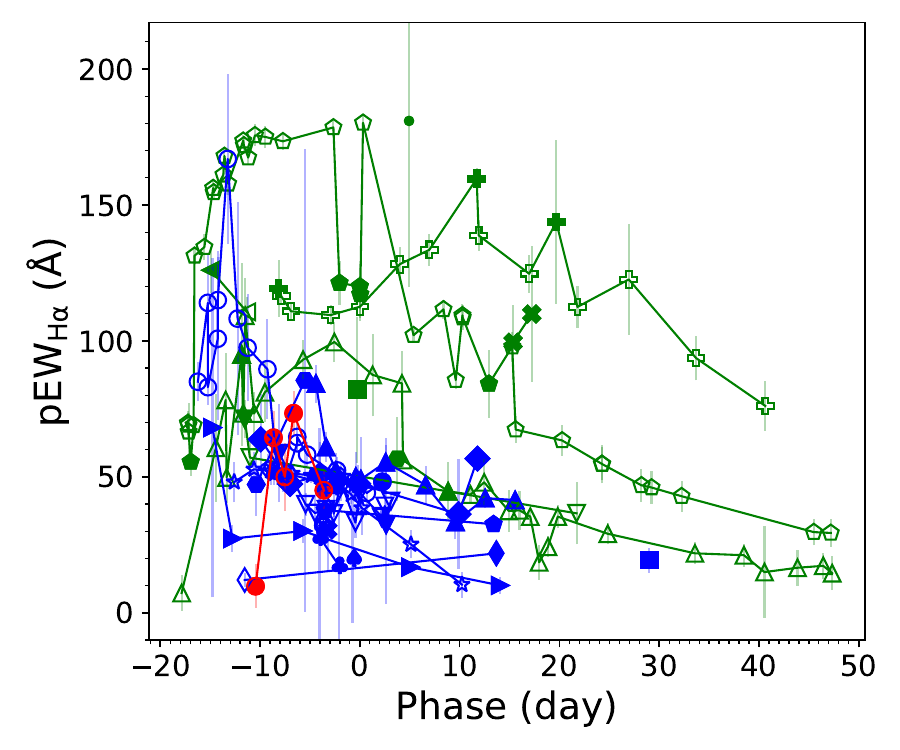}
	\includegraphics[width=0.217\linewidth]{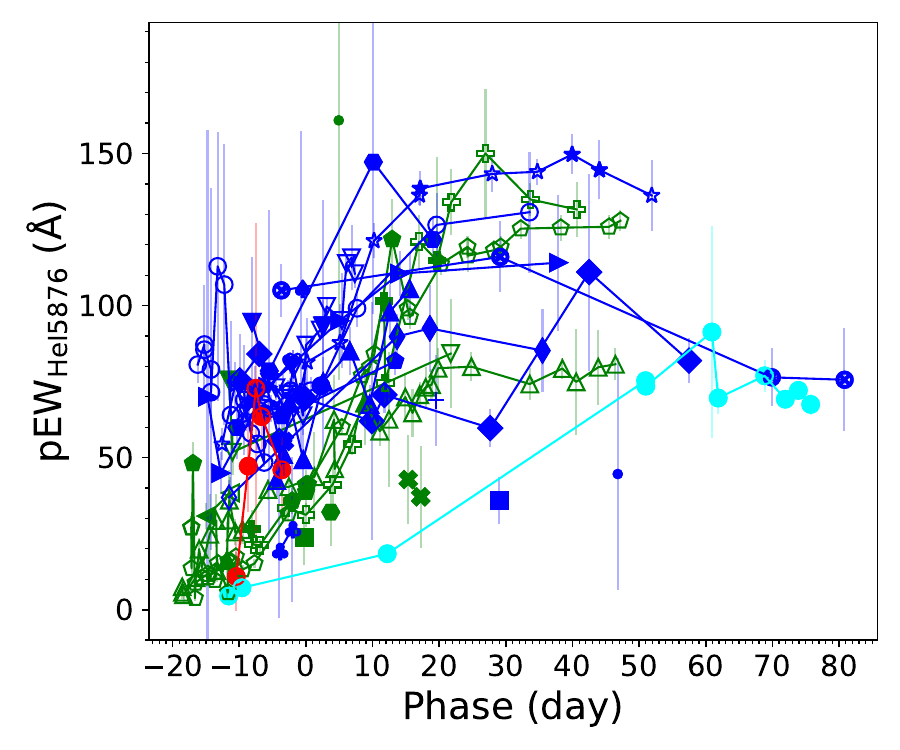}
	\includegraphics[width=0.217\linewidth]{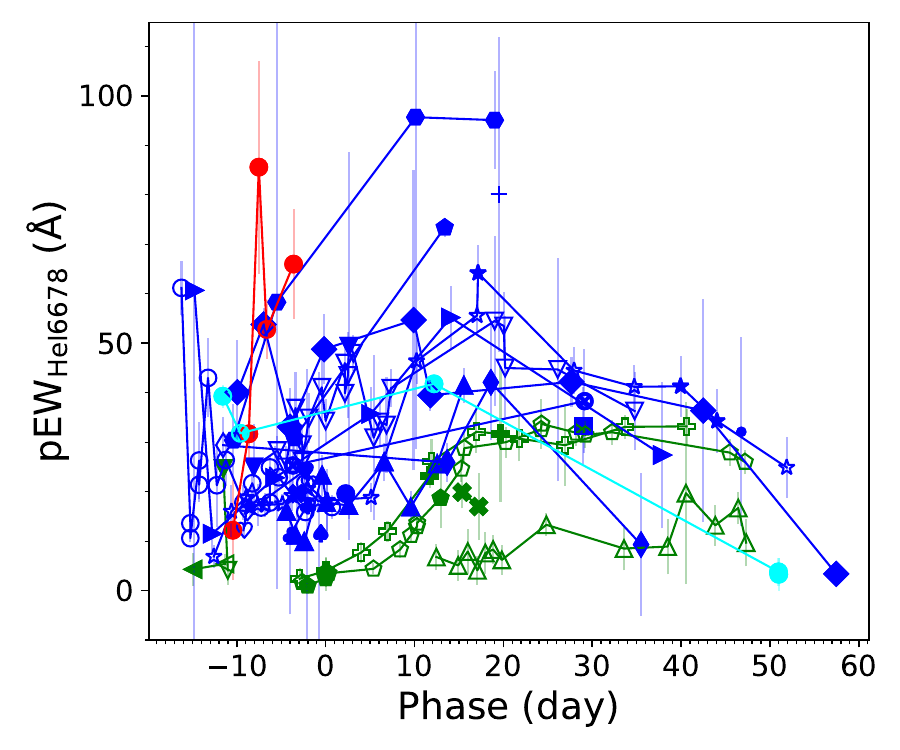}
	\includegraphics[width=0.217\linewidth]{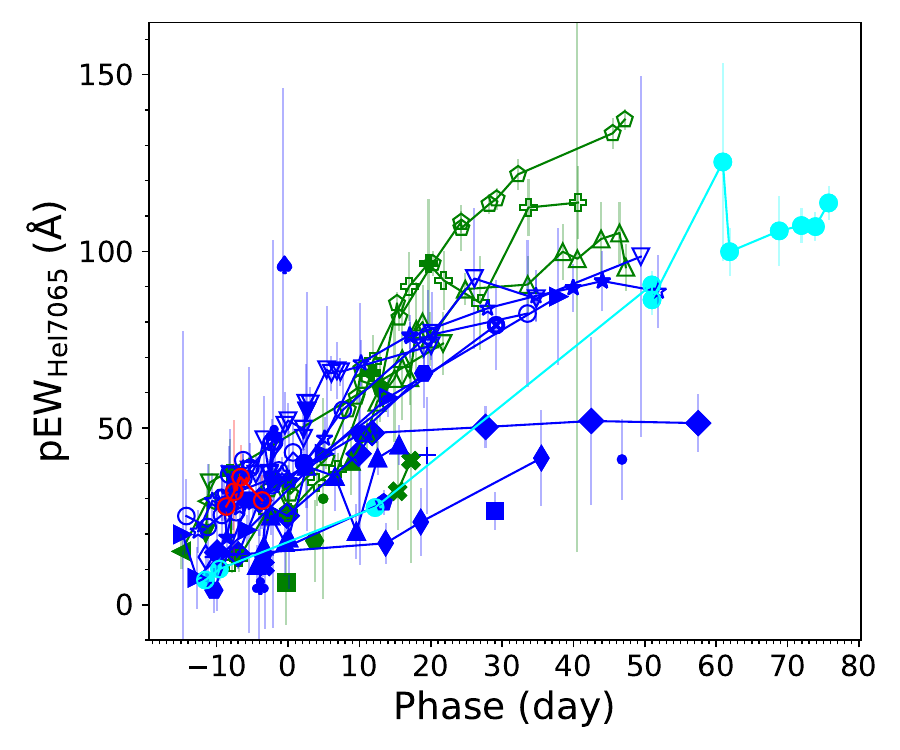}
	\includegraphics[width=0.217\linewidth]{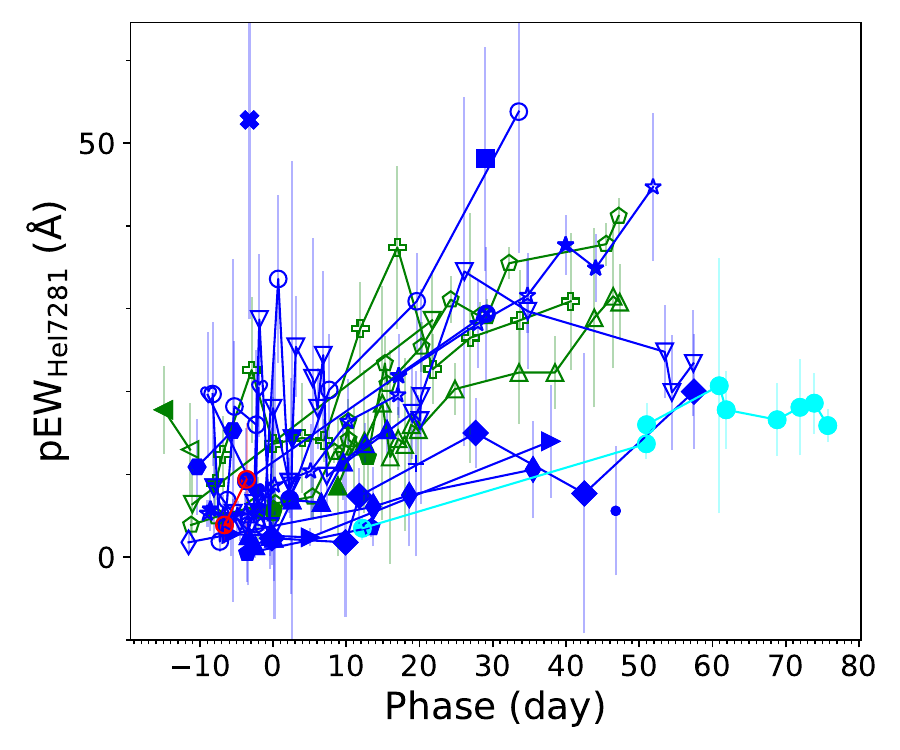}
	\includegraphics[width=0.217\linewidth]{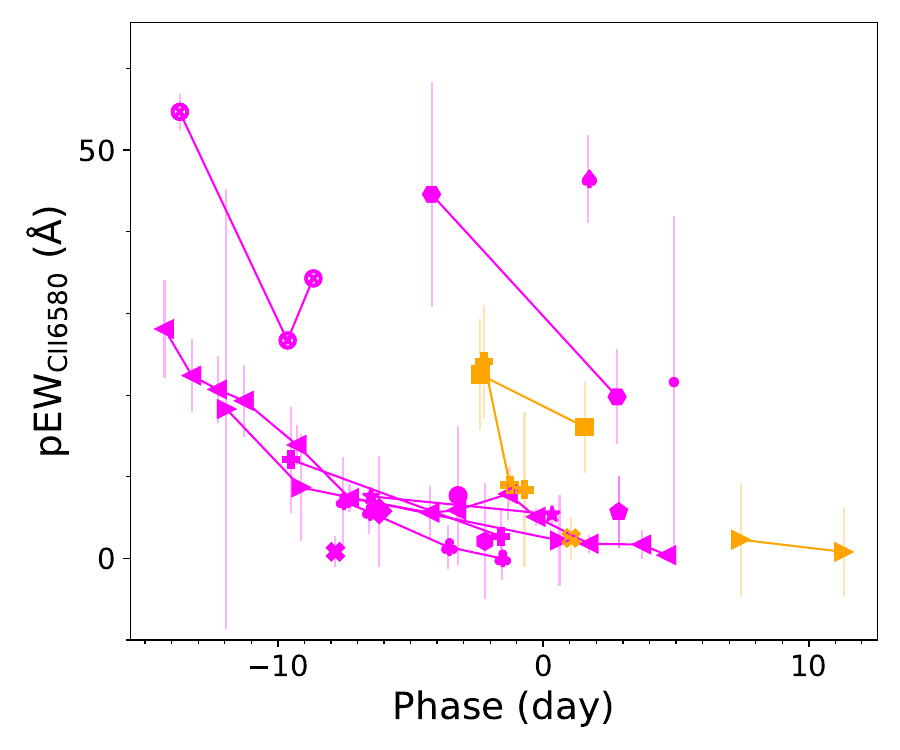}
	\includegraphics[width=0.217\linewidth]{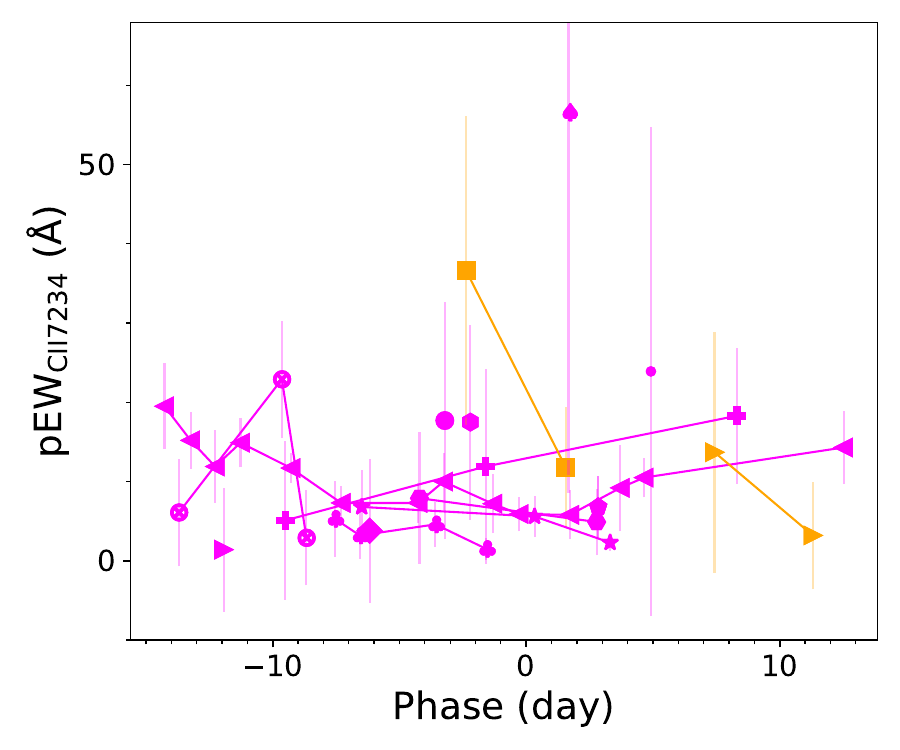}
	\includegraphics[width=0.217\linewidth]{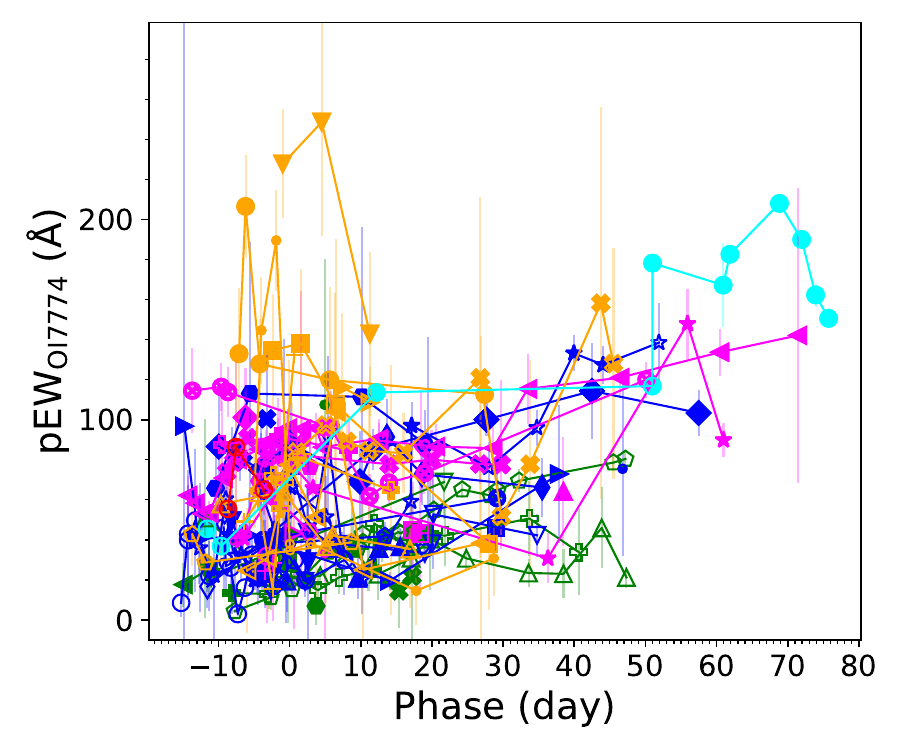}
	\includegraphics[width=0.217\linewidth]{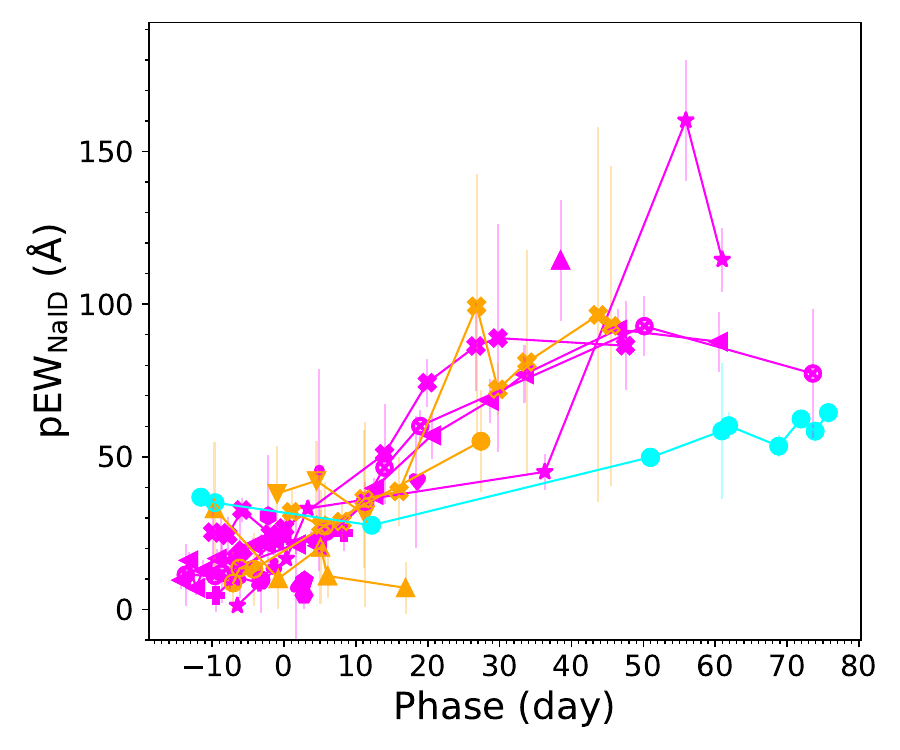}
	\includegraphics[width=0.217\linewidth]{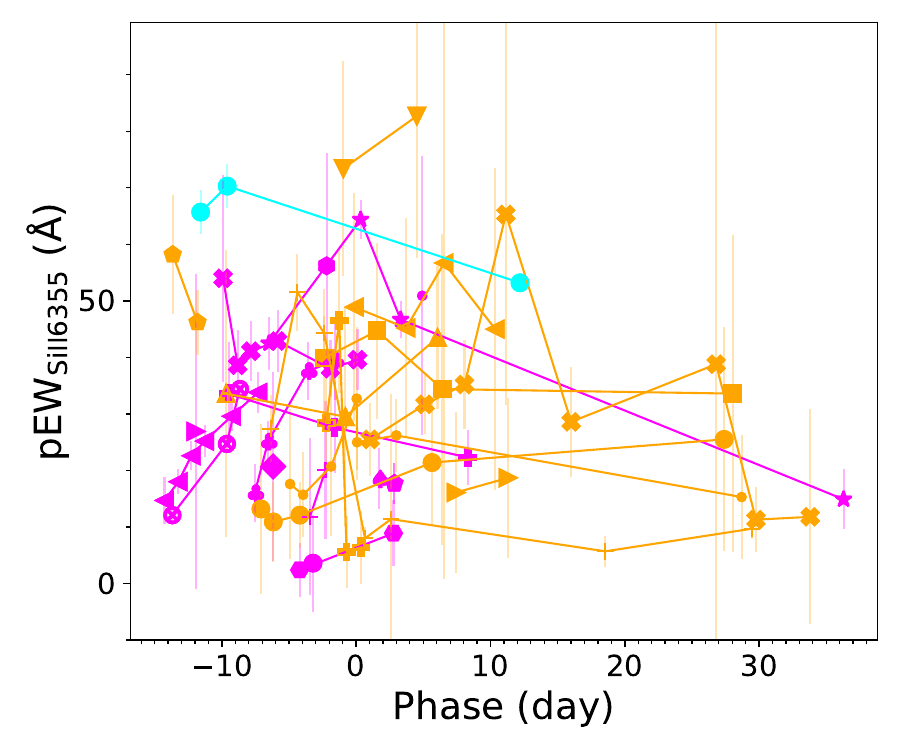}
	\includegraphics[width=0.217\linewidth]{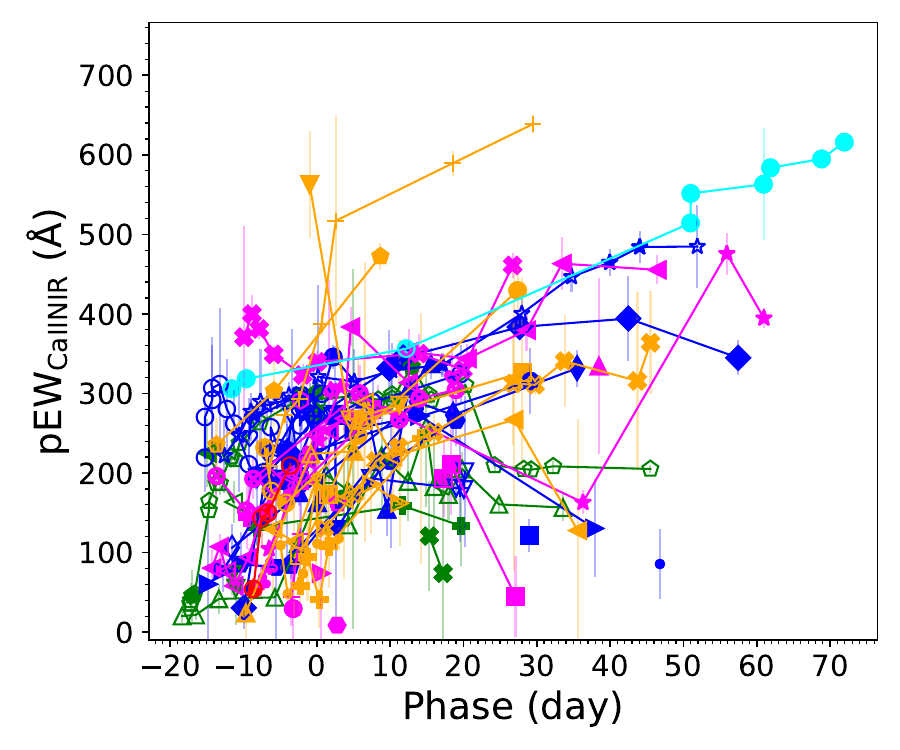}
	\includegraphics[width=0.217\linewidth]{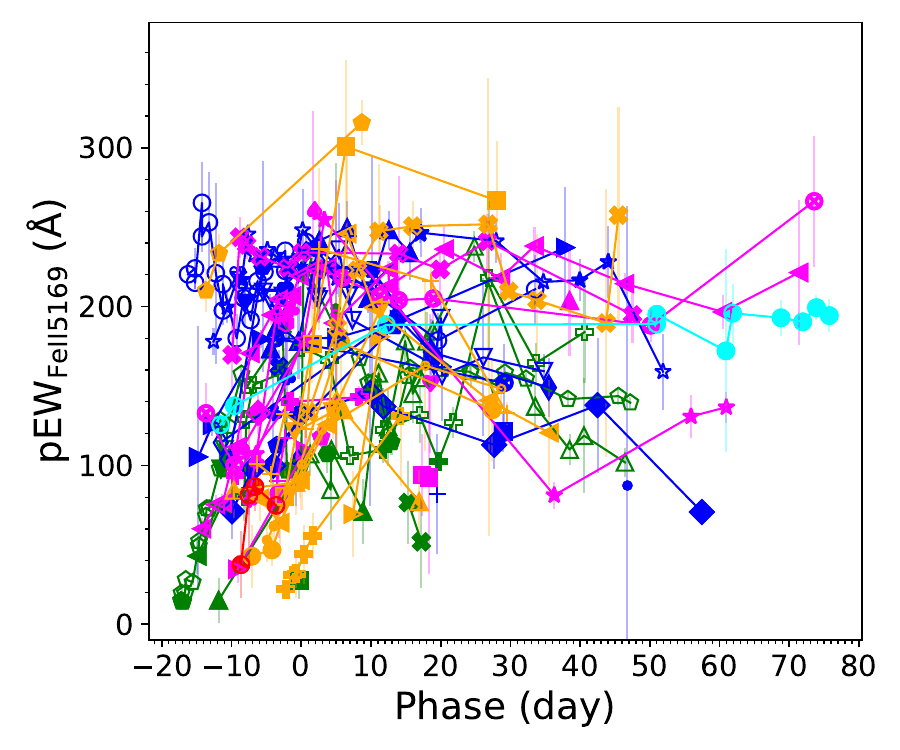}
	
	\caption{Evolution of line intensities (pEW) of in the four subtypes of SESNe. Data points follow the same color/symbol scheme as in Fig.~\ref{fig:evol-V-all}.}
	\label{fig:evol-EW-all}
\end{figure*}

\begin{figure*}
	\centering
	\includegraphics[width=0.162\linewidth]{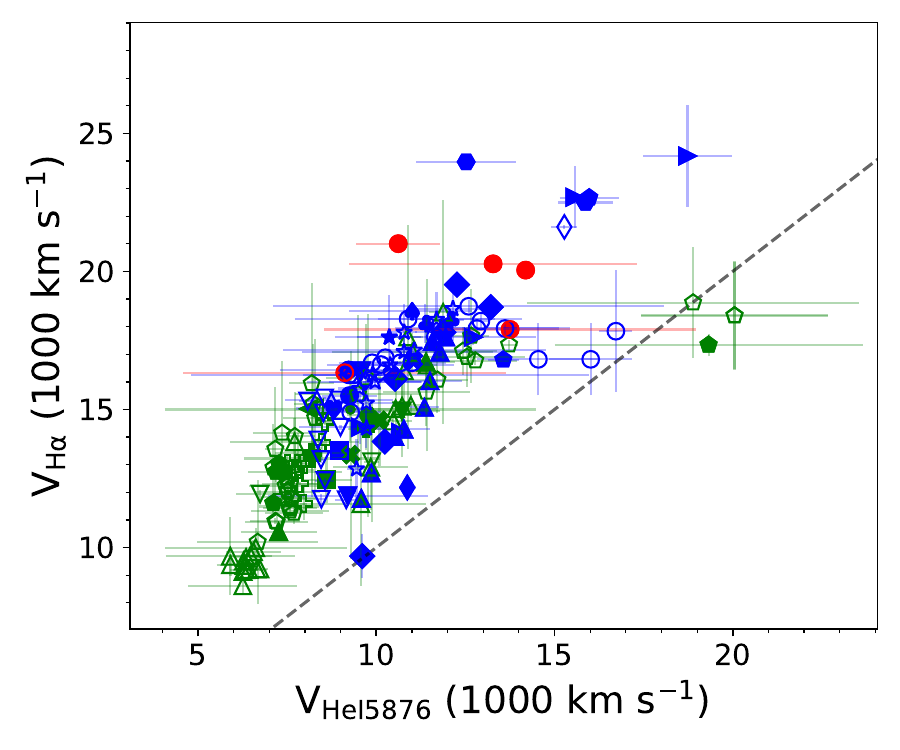}
	\includegraphics[width=0.162\linewidth]{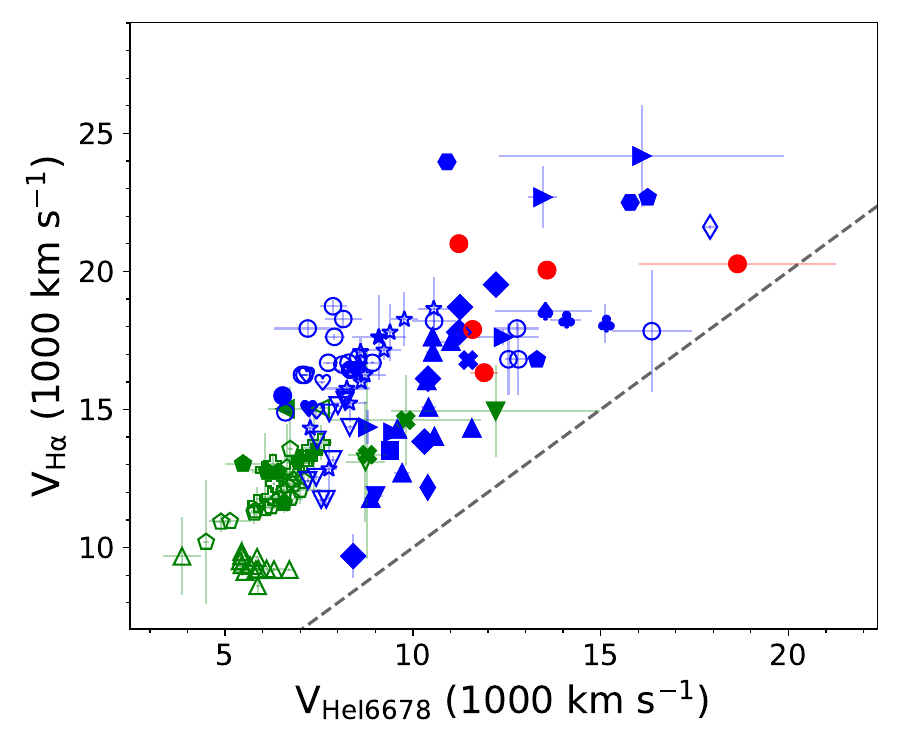}
	\includegraphics[width=0.162\linewidth]{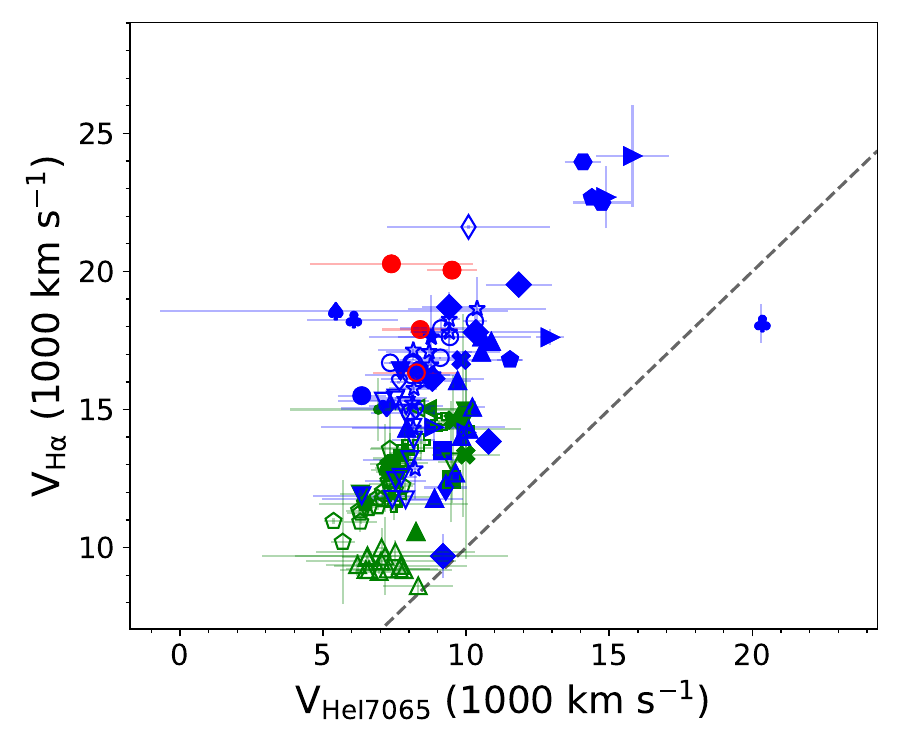}
	\includegraphics[width=0.162\linewidth]{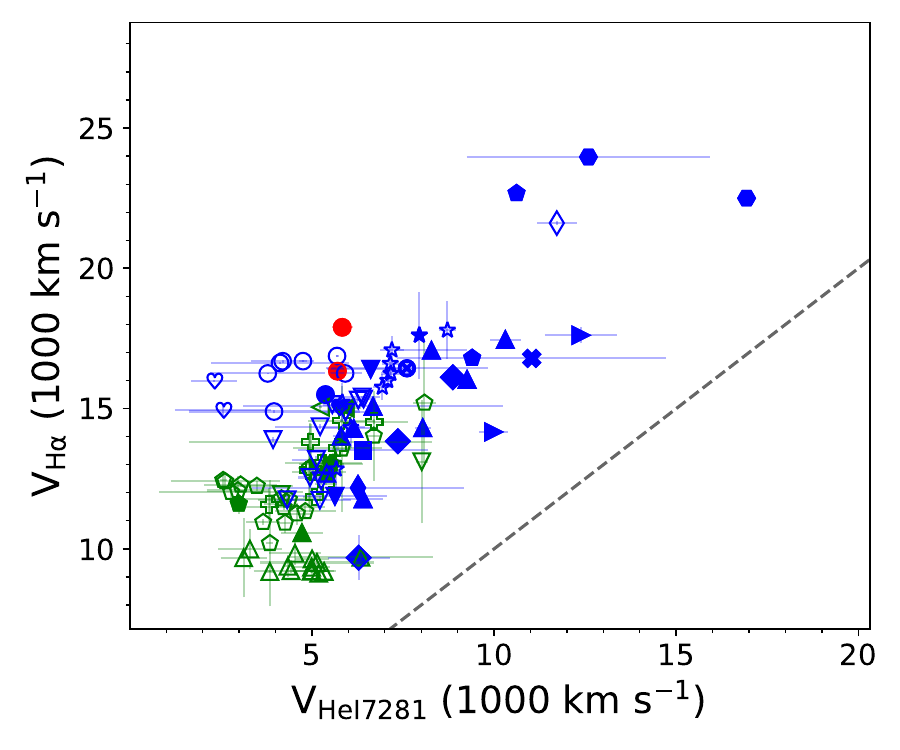}
	\includegraphics[width=0.162\linewidth]{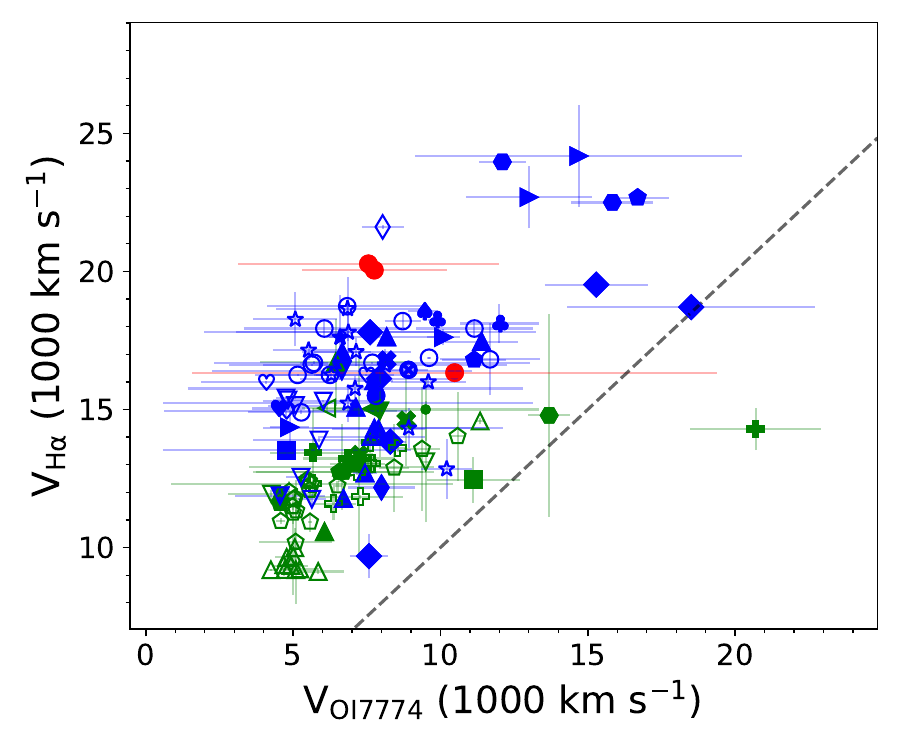}
	\includegraphics[width=0.162\linewidth]{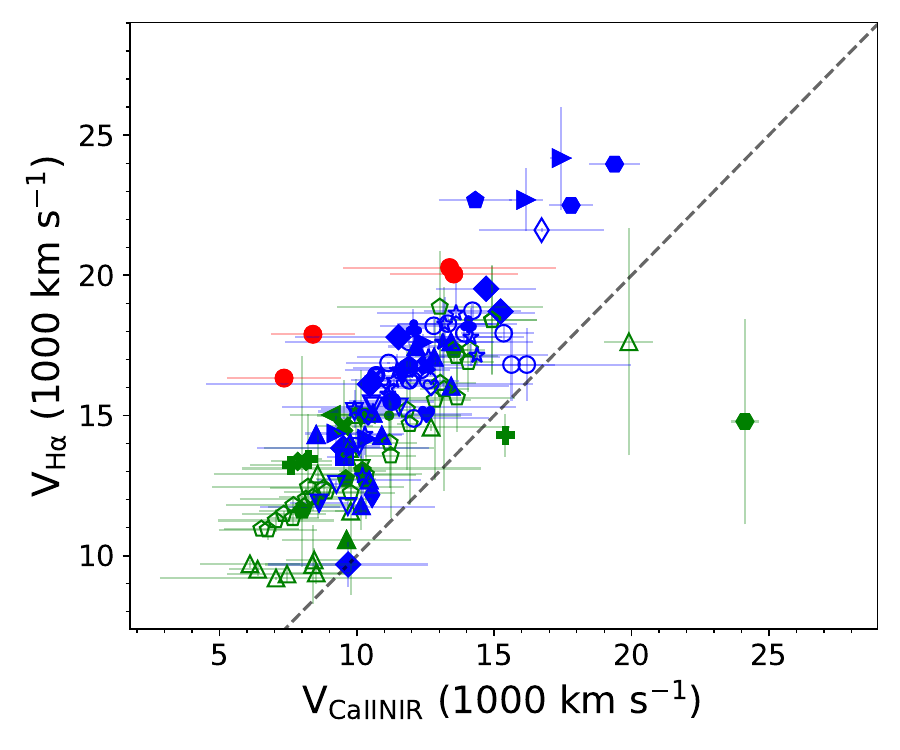}
	\includegraphics[width=0.162\linewidth]{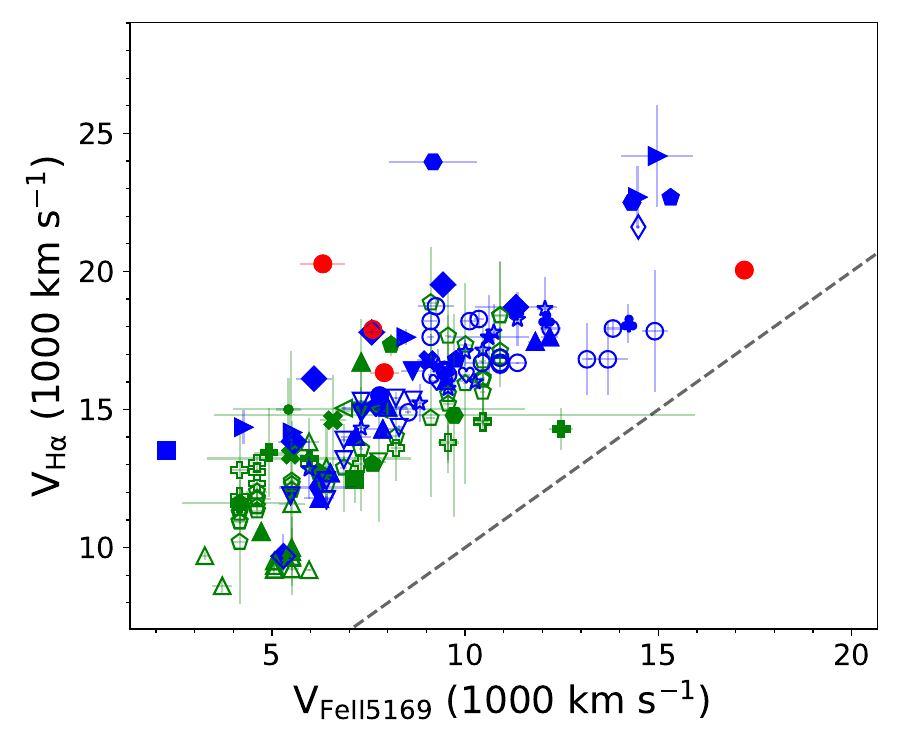}
	\includegraphics[width=0.162\linewidth]{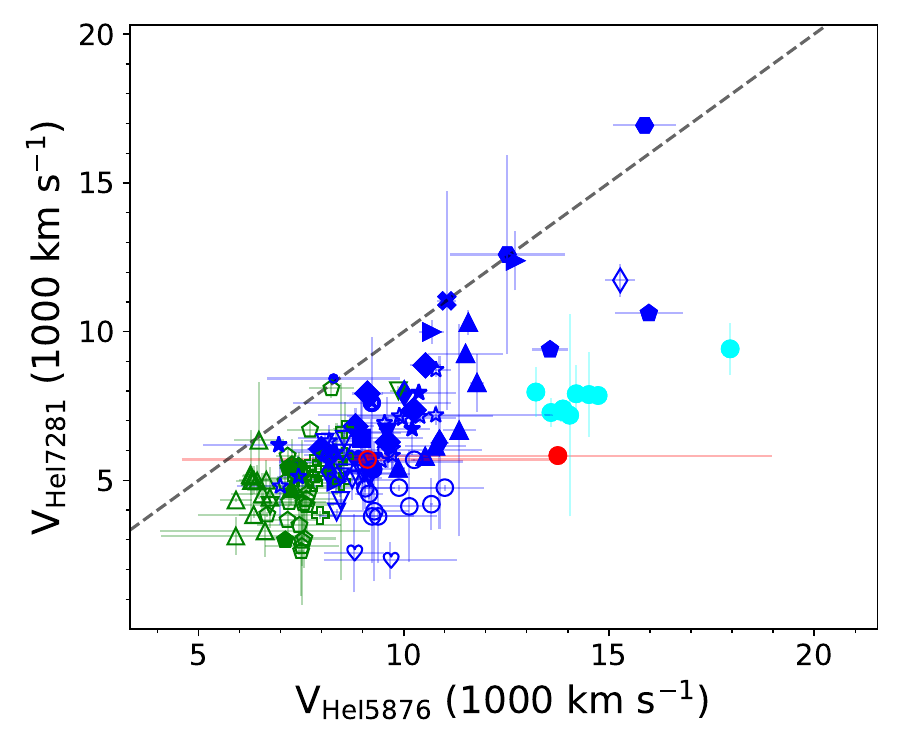}
	\includegraphics[width=0.162\linewidth]{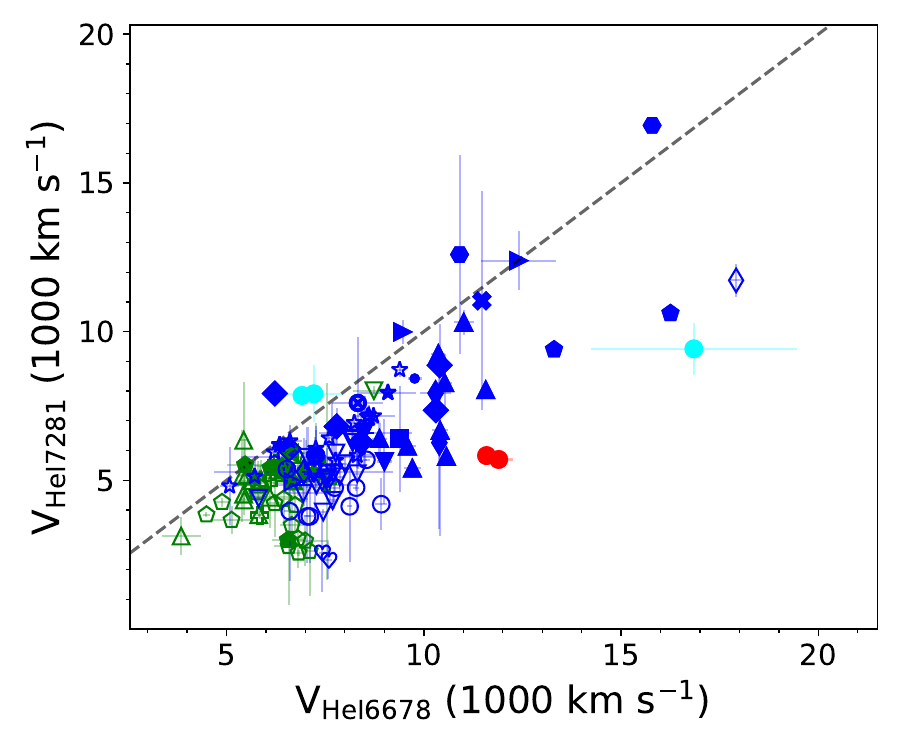}
	\includegraphics[width=0.162\linewidth]{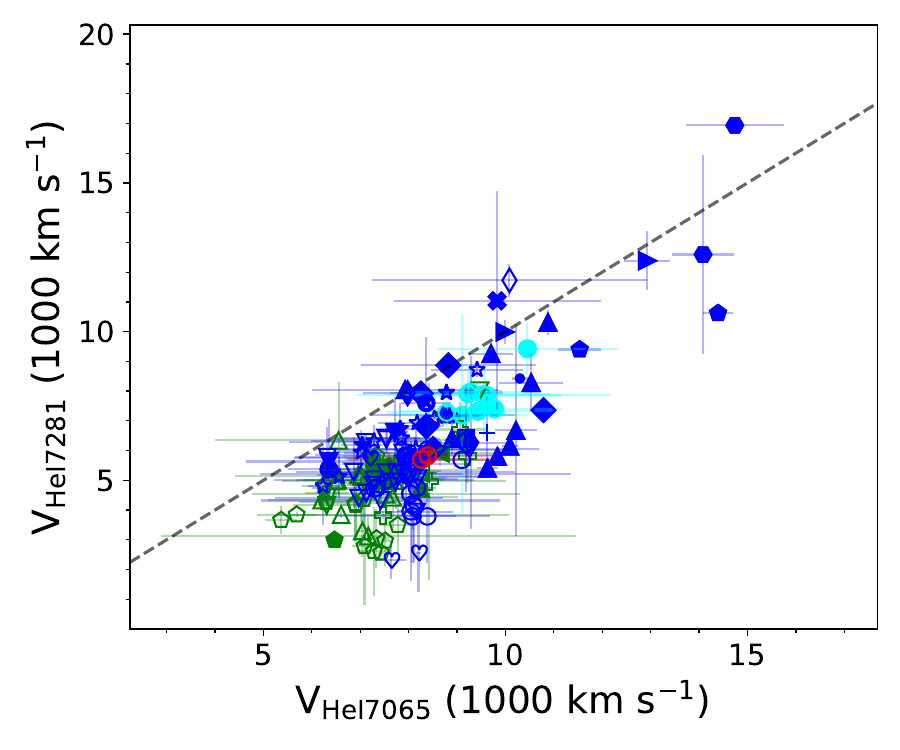}
	\includegraphics[width=0.162\linewidth]{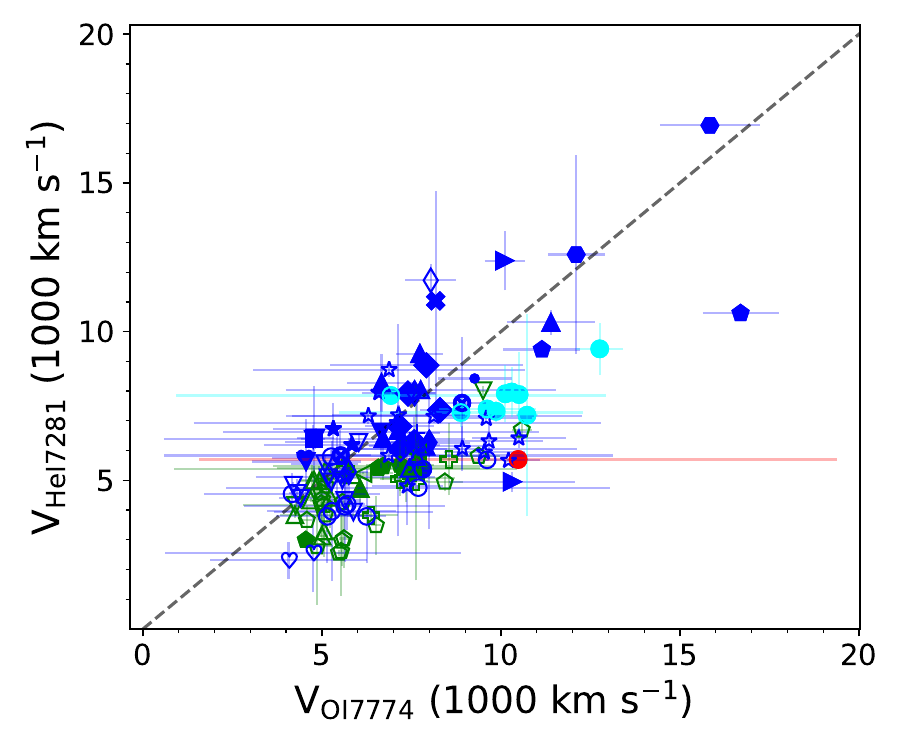}
	\includegraphics[width=0.162\linewidth]{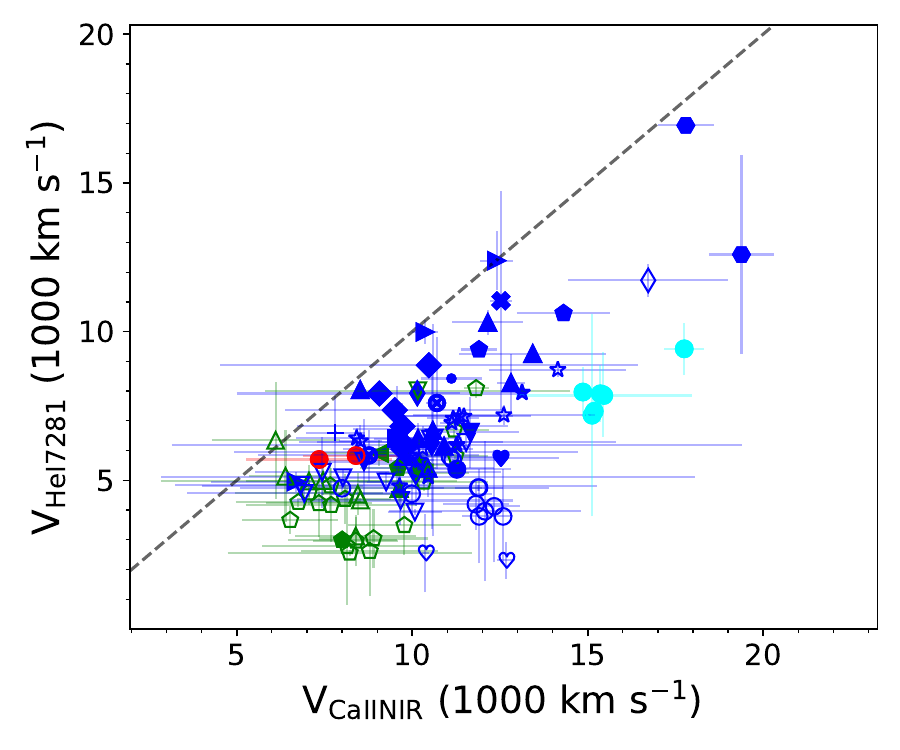}
	\includegraphics[width=0.162\linewidth]{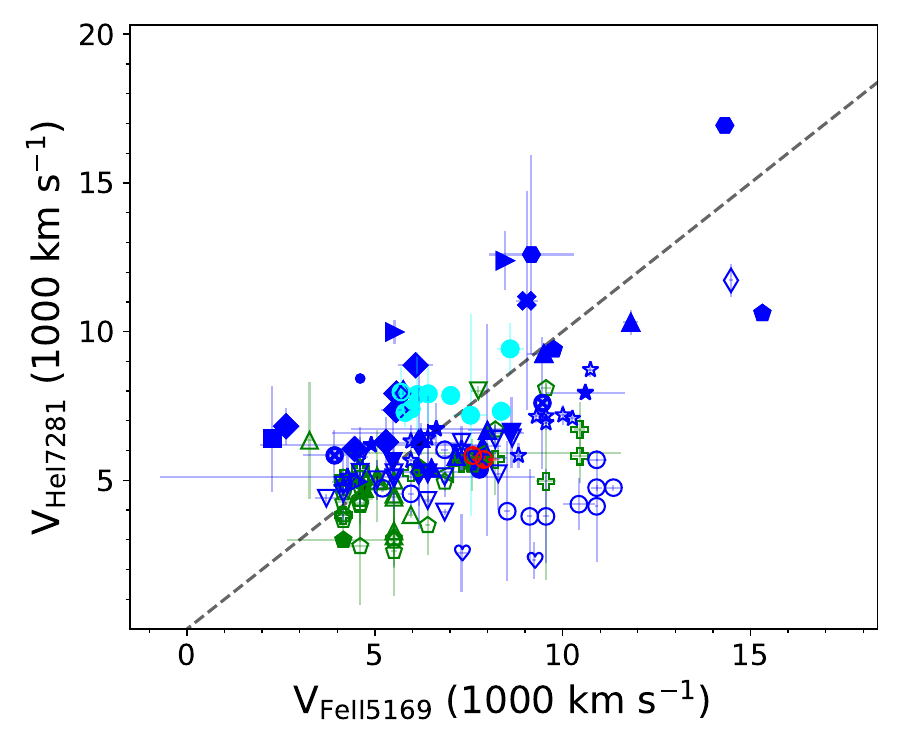}
	\includegraphics[width=0.162\linewidth]{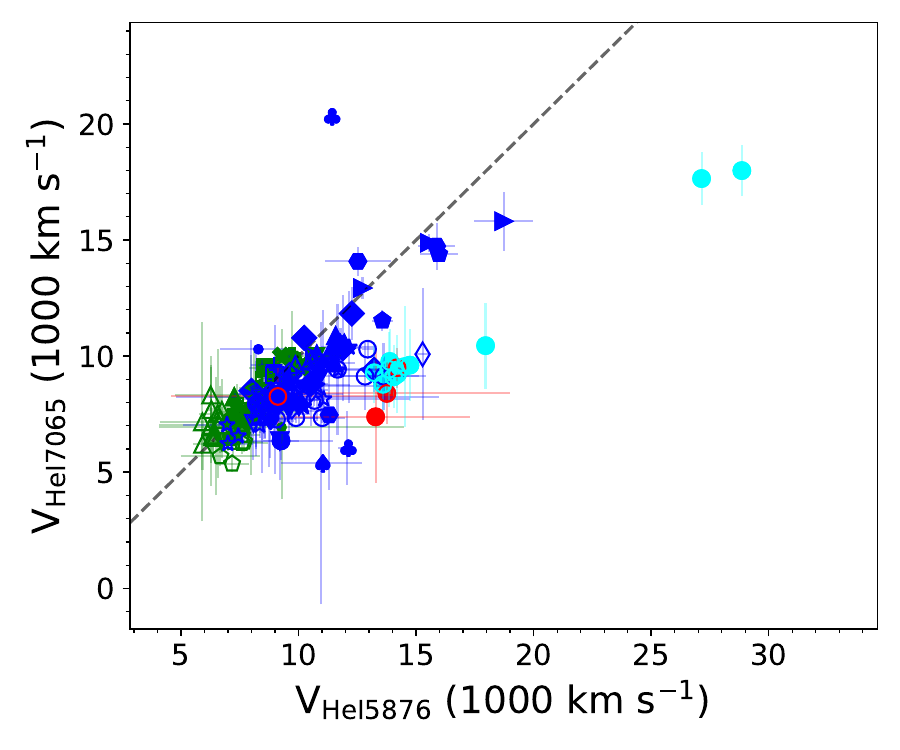}
	\includegraphics[width=0.162\linewidth]{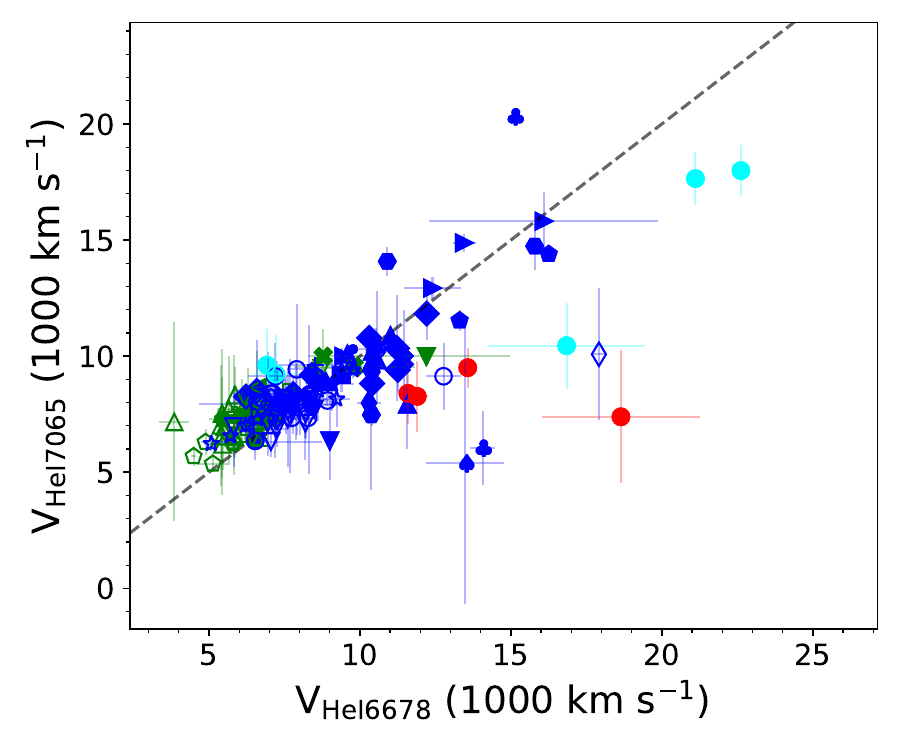}
	\includegraphics[width=0.162\linewidth]{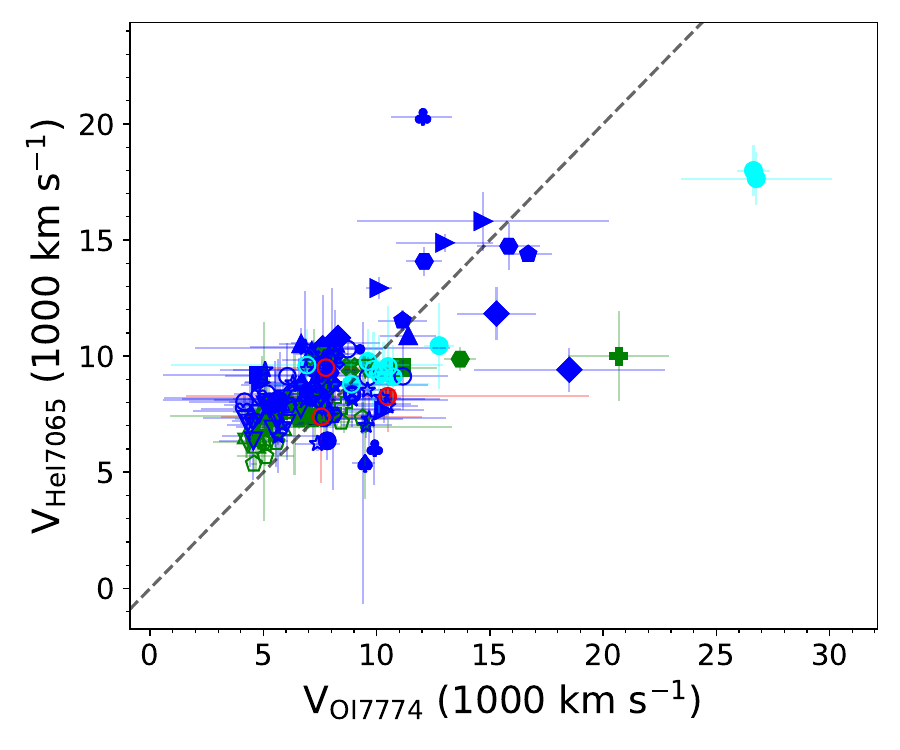}
	\includegraphics[width=0.162\linewidth]{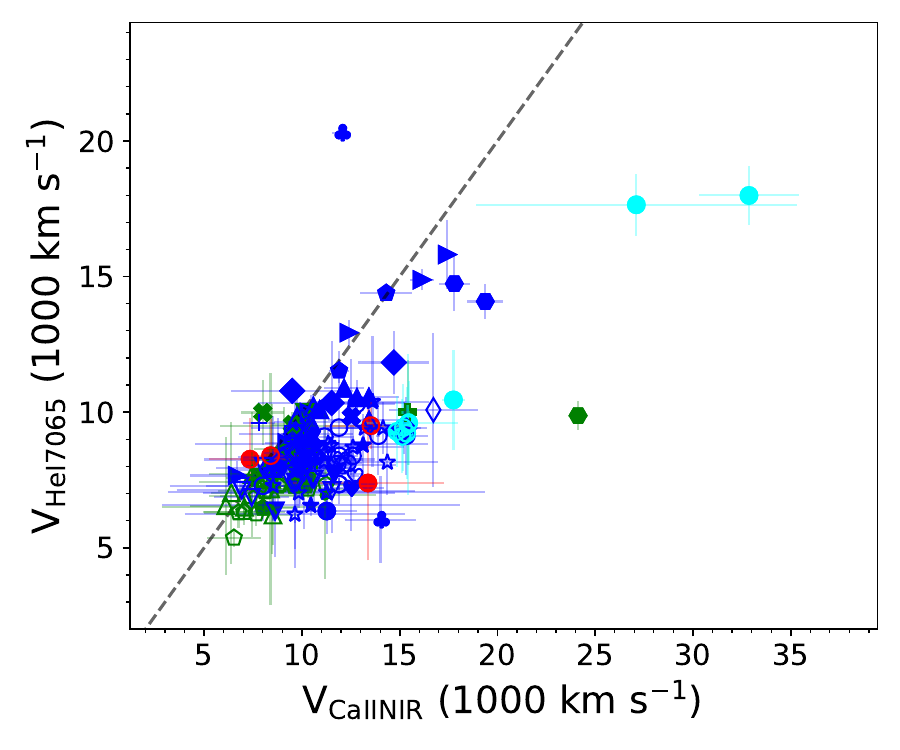}
	\includegraphics[width=0.162\linewidth]{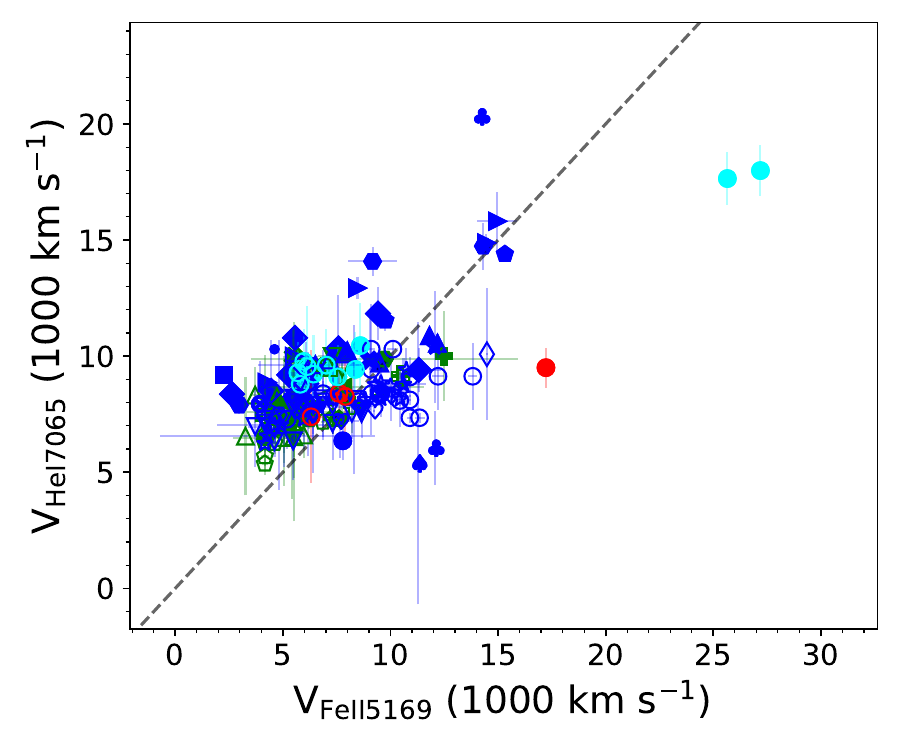}
	\includegraphics[width=0.162\linewidth]{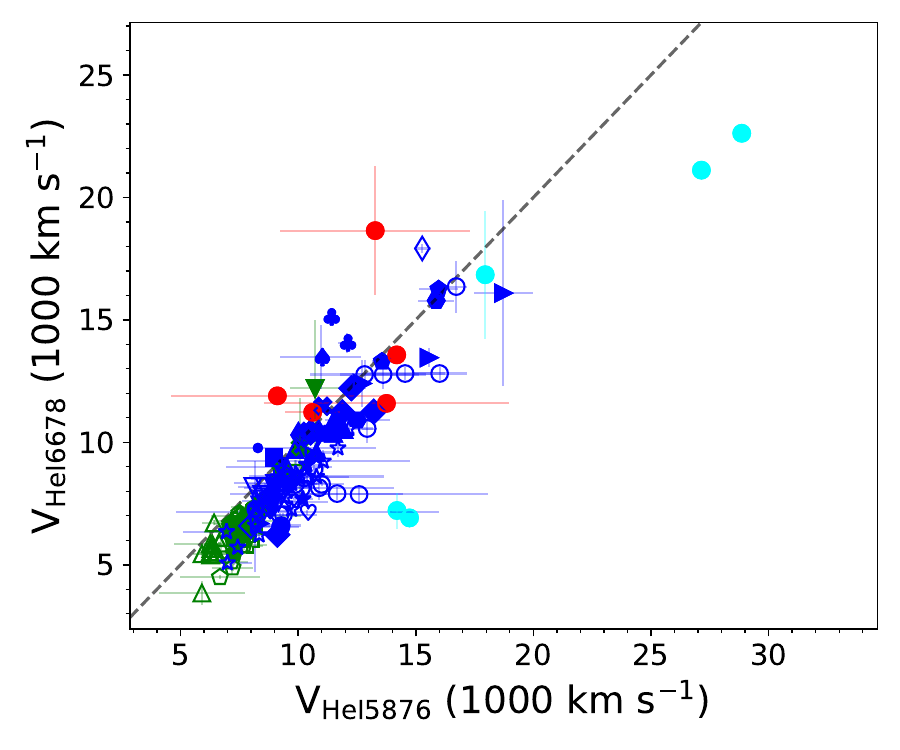}
	\includegraphics[width=0.162\linewidth]{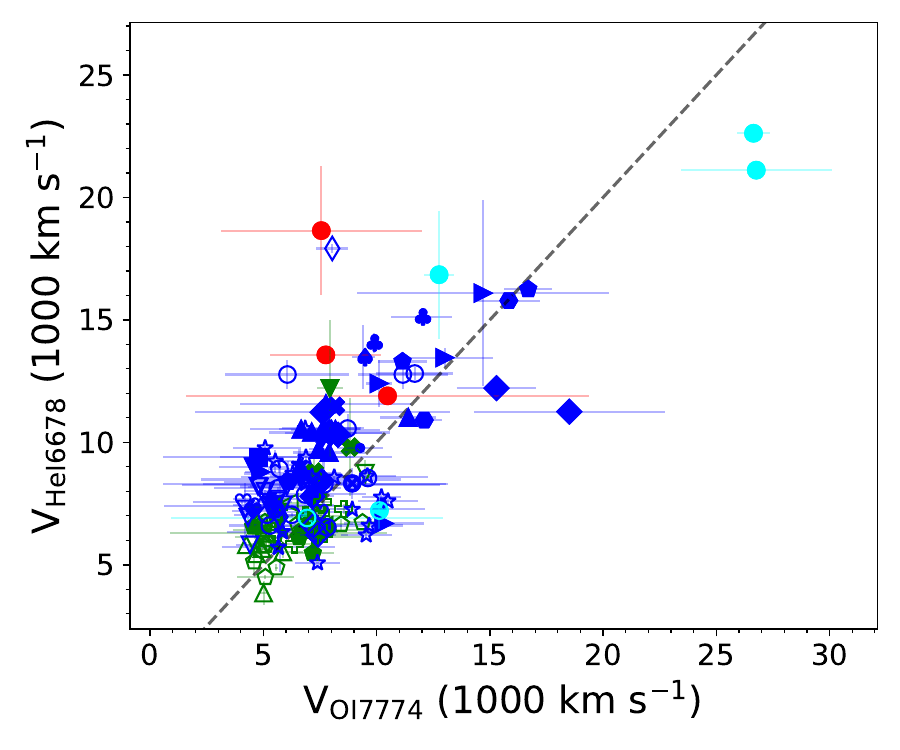}
	\includegraphics[width=0.162\linewidth]{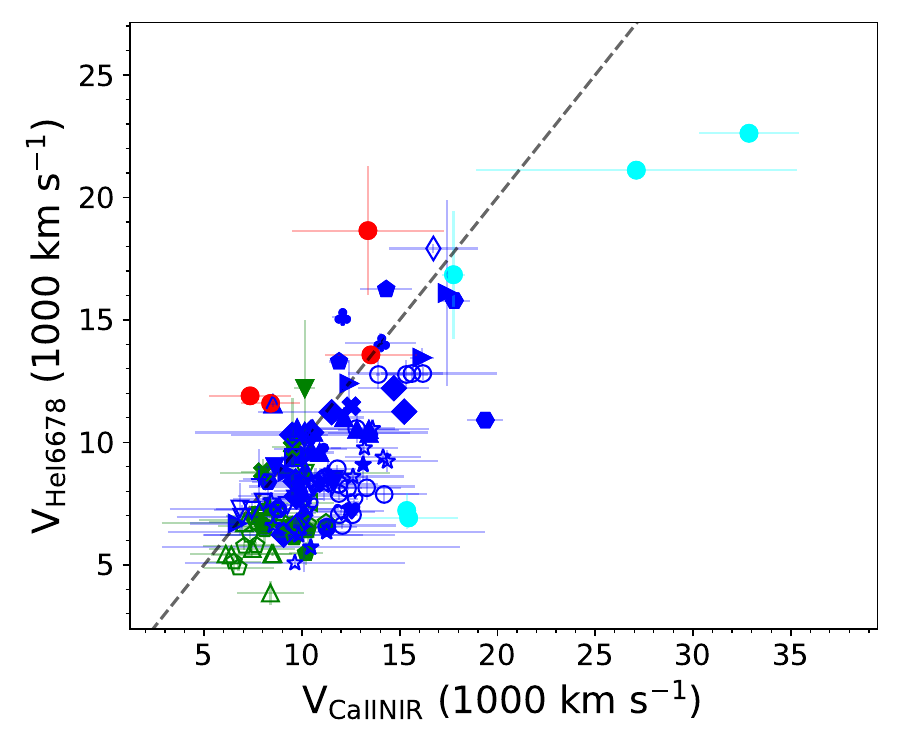}
	\includegraphics[width=0.162\linewidth]{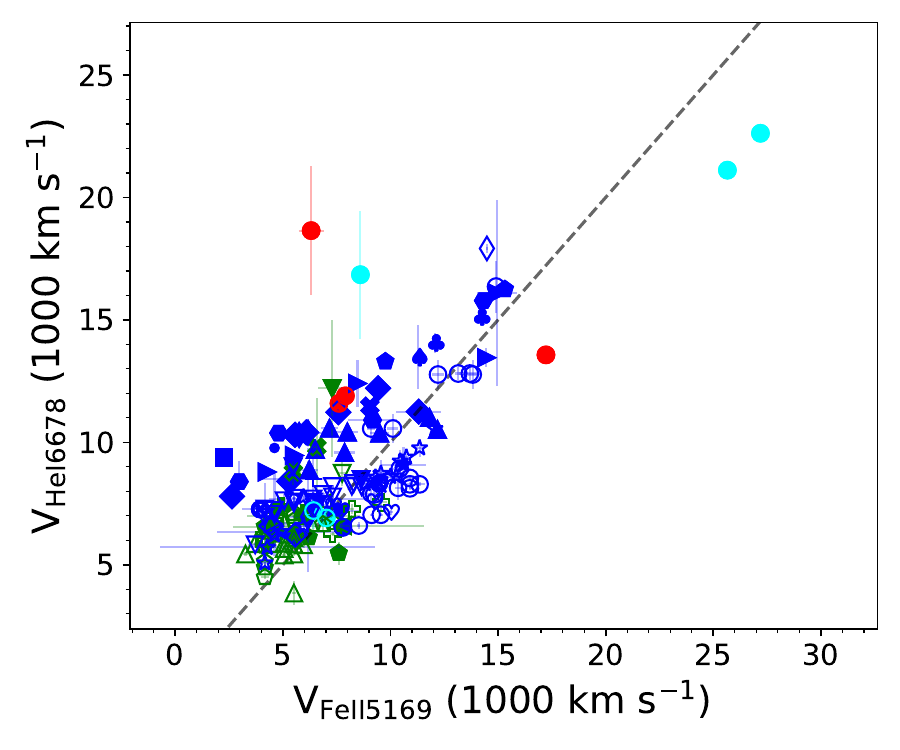}
	\includegraphics[width=0.162\linewidth]{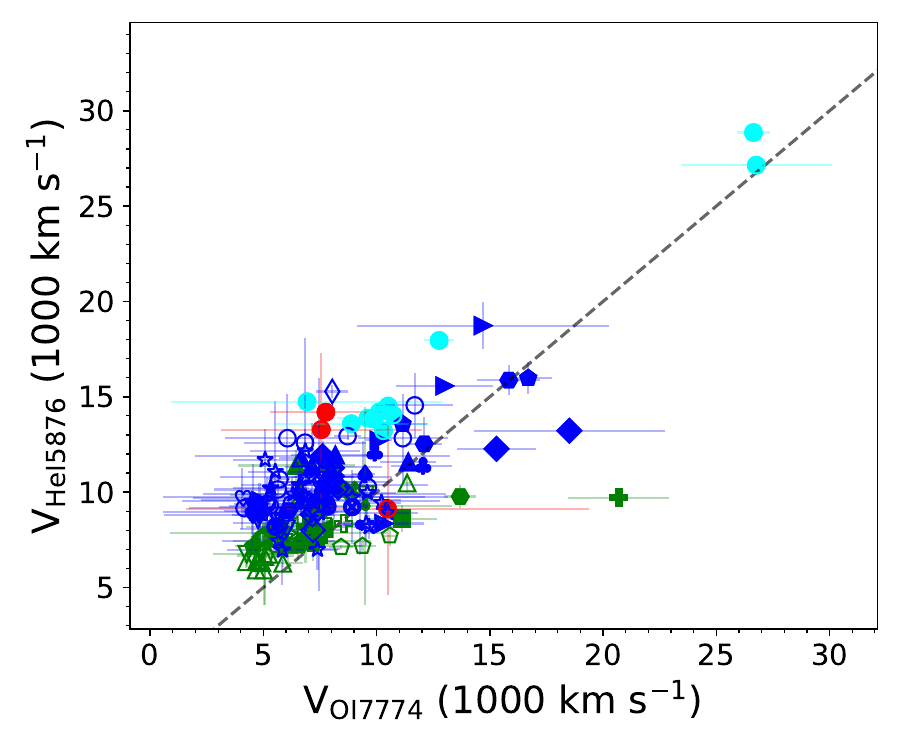}
	\includegraphics[width=0.162\linewidth]{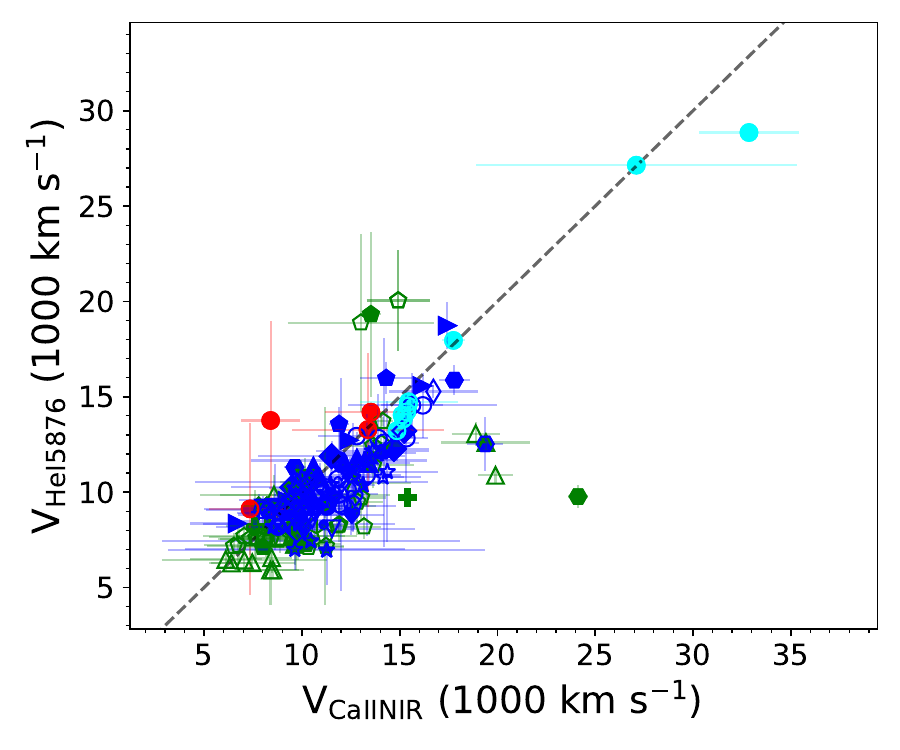}
	\includegraphics[width=0.162\linewidth]{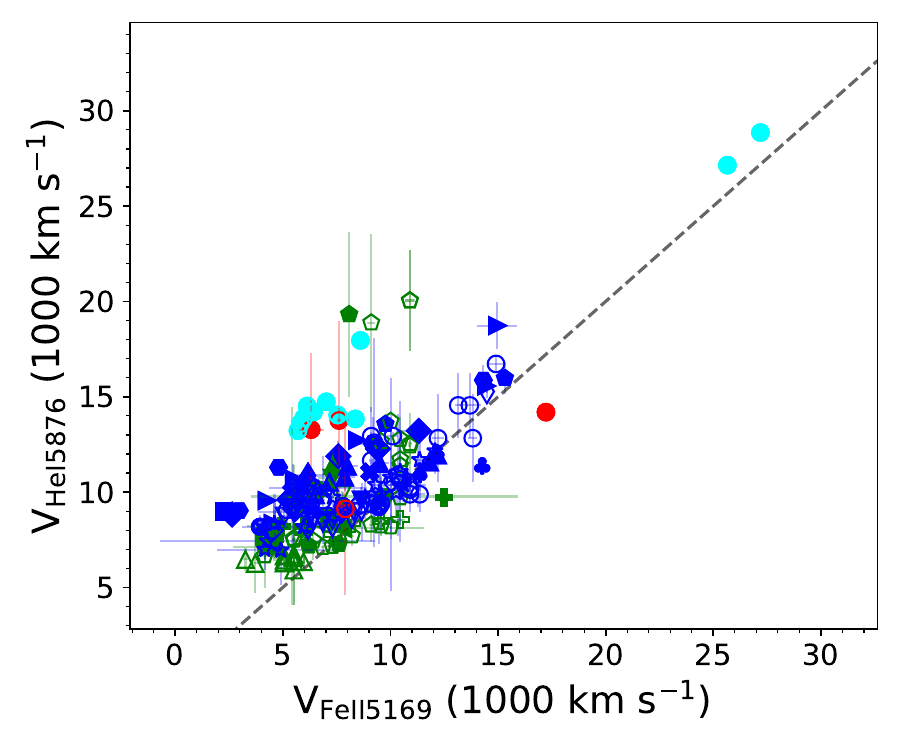}
	\includegraphics[width=0.162\linewidth]{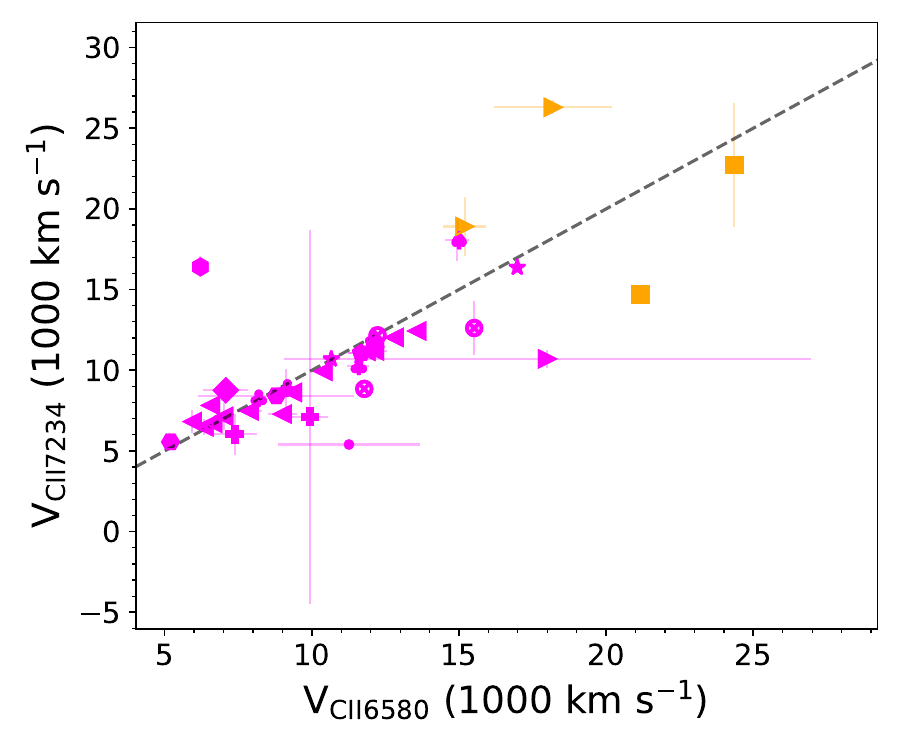}
	\includegraphics[width=0.162\linewidth]{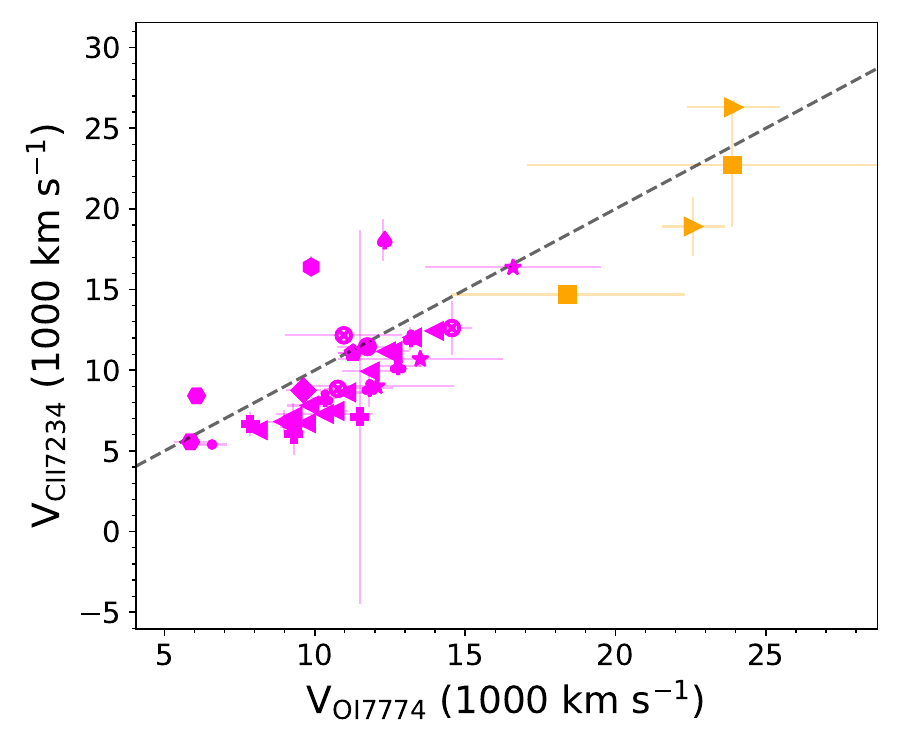}
	\includegraphics[width=0.162\linewidth]{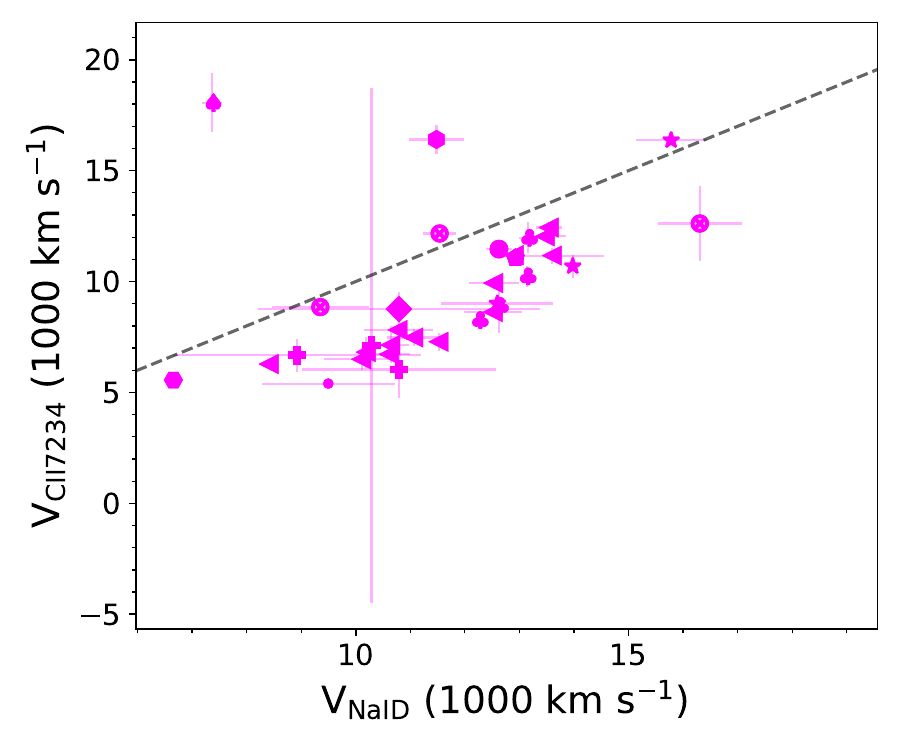}
	\includegraphics[width=0.162\linewidth]{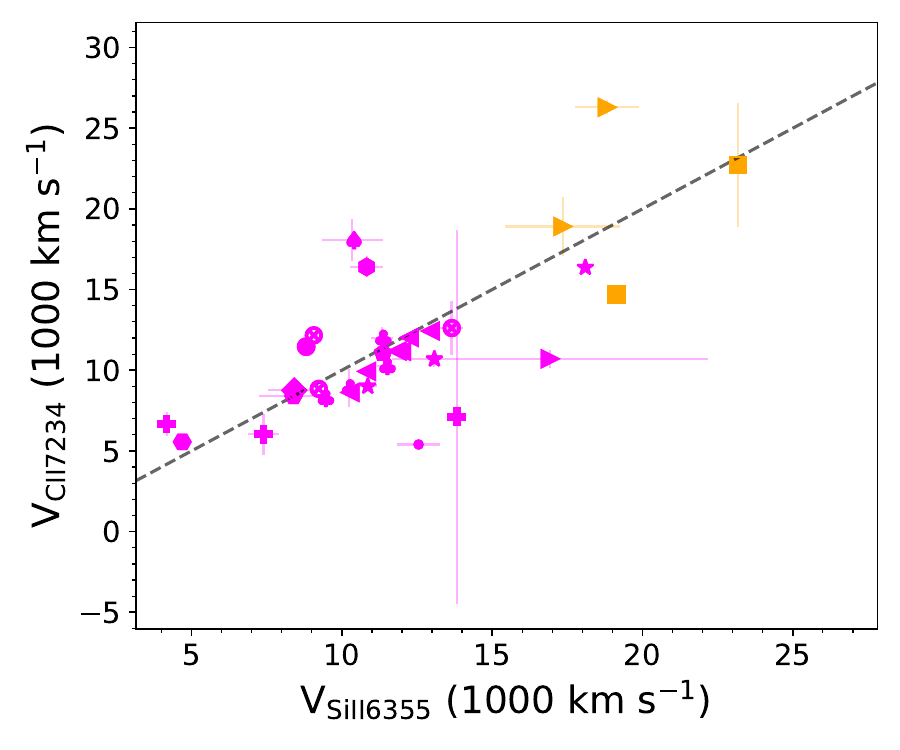}
	\includegraphics[width=0.162\linewidth]{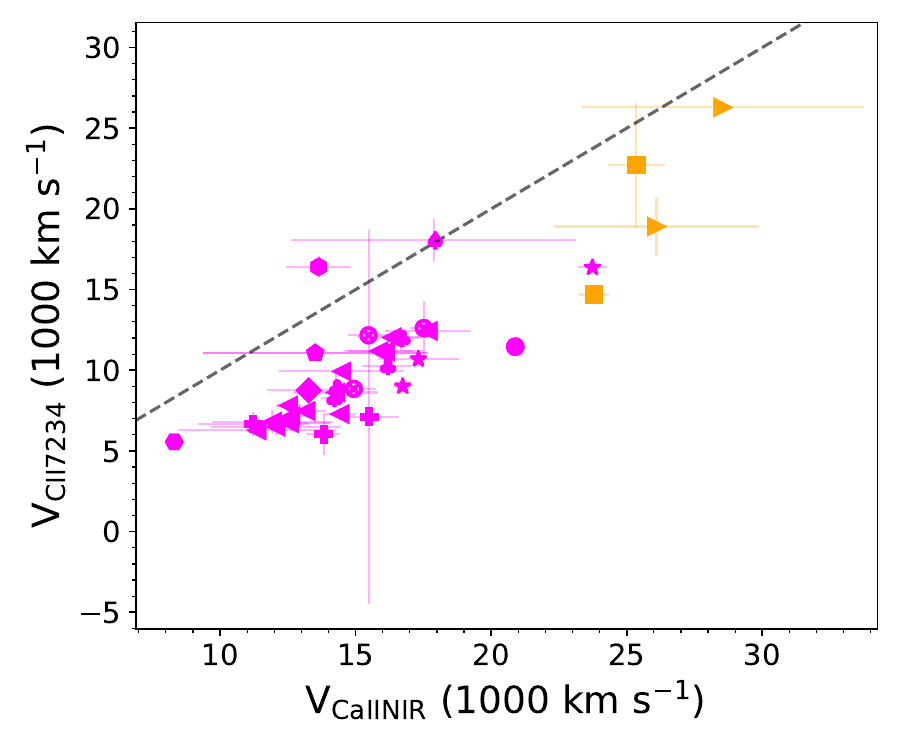}
	\includegraphics[width=0.162\linewidth]{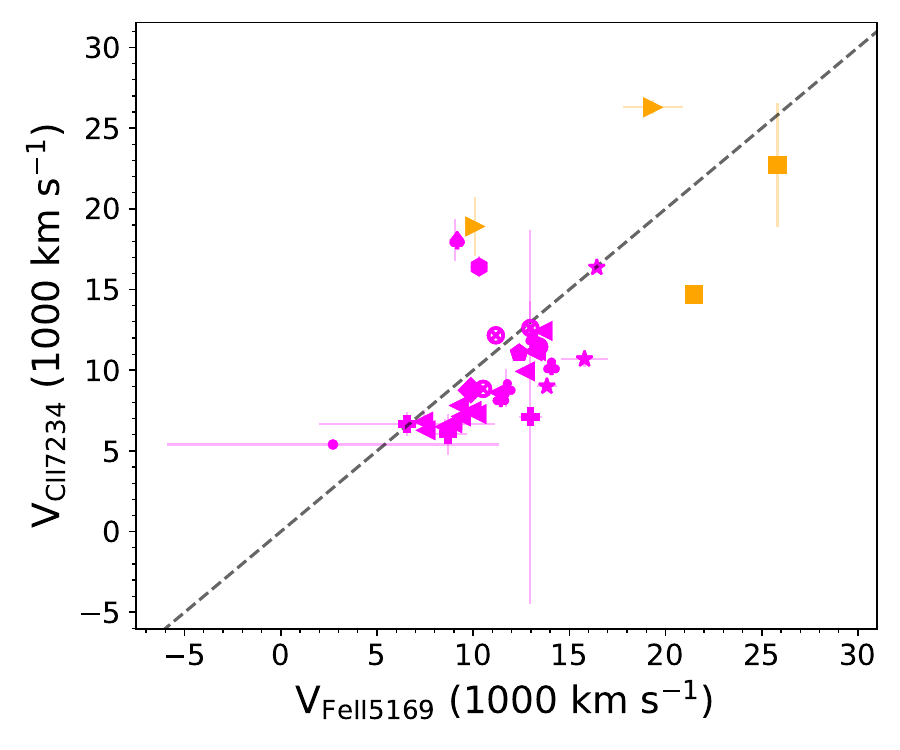}
	\includegraphics[width=0.162\linewidth]{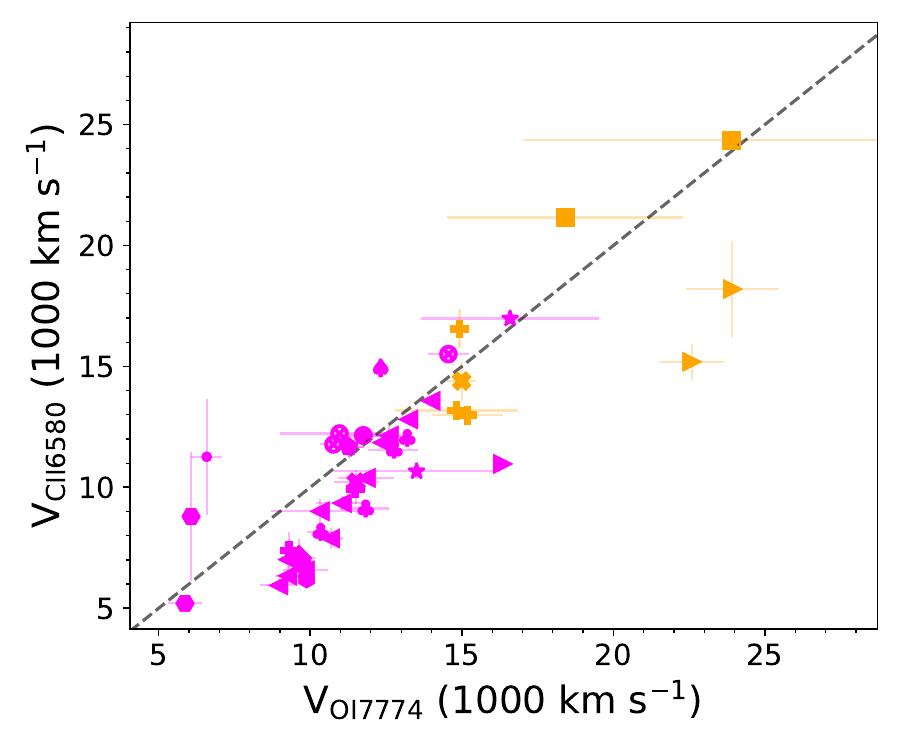}
	\includegraphics[width=0.162\linewidth]{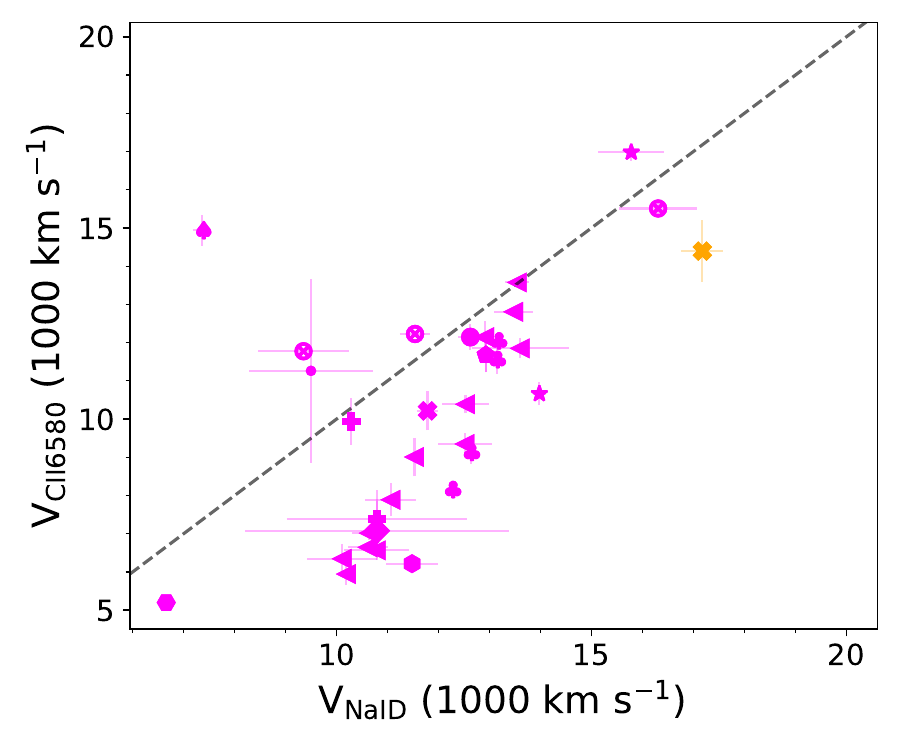} 
	\includegraphics[width=0.162\linewidth]{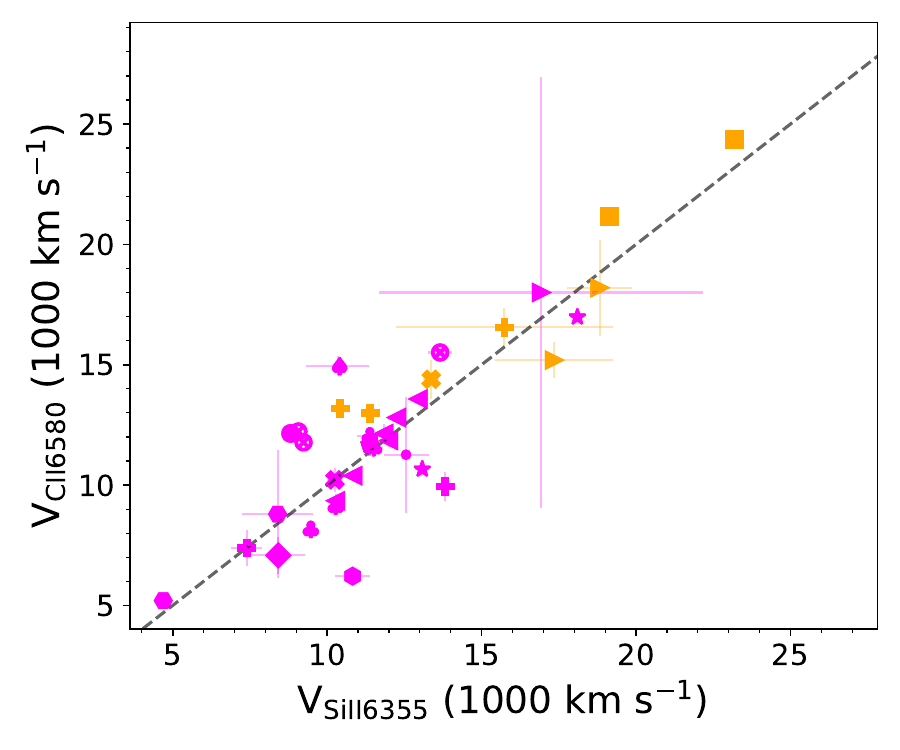}
	\includegraphics[width=0.162\linewidth]{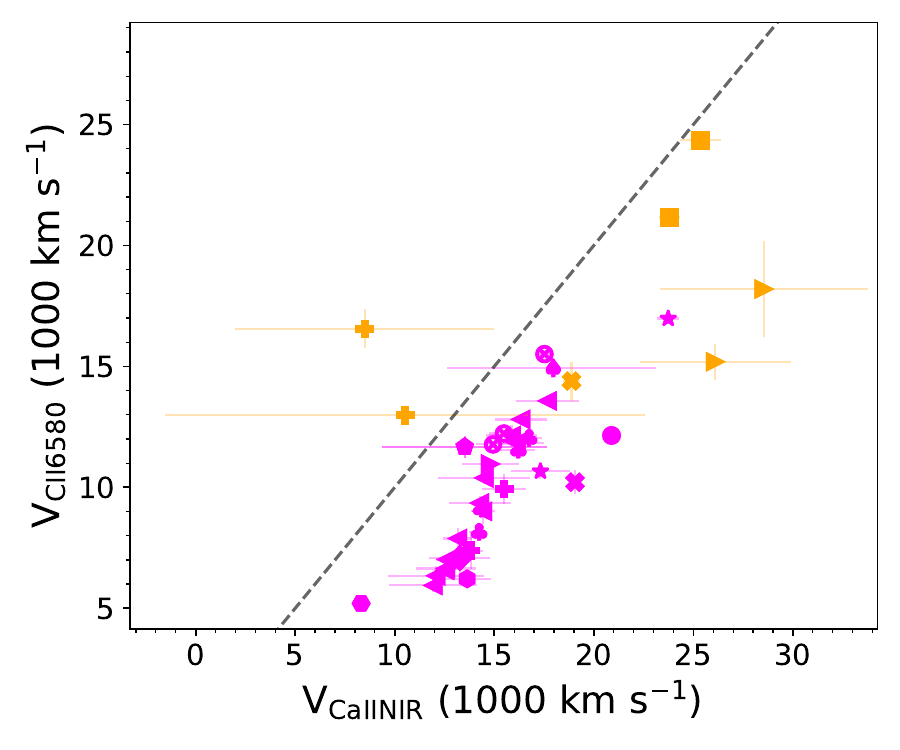}
	\includegraphics[width=0.162\linewidth]{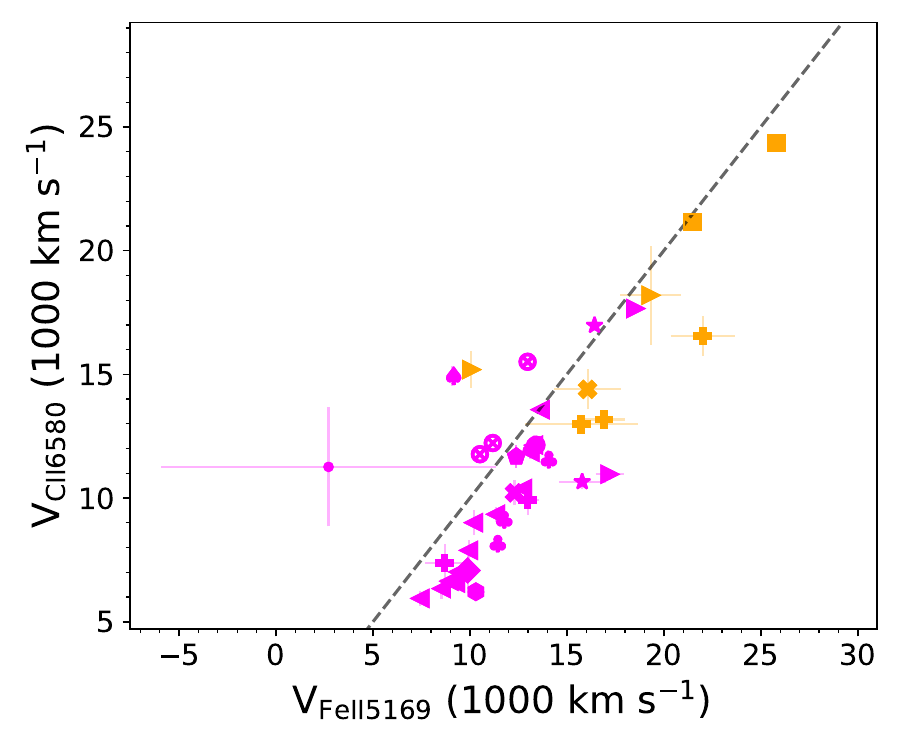}
	\includegraphics[width=0.162\linewidth]{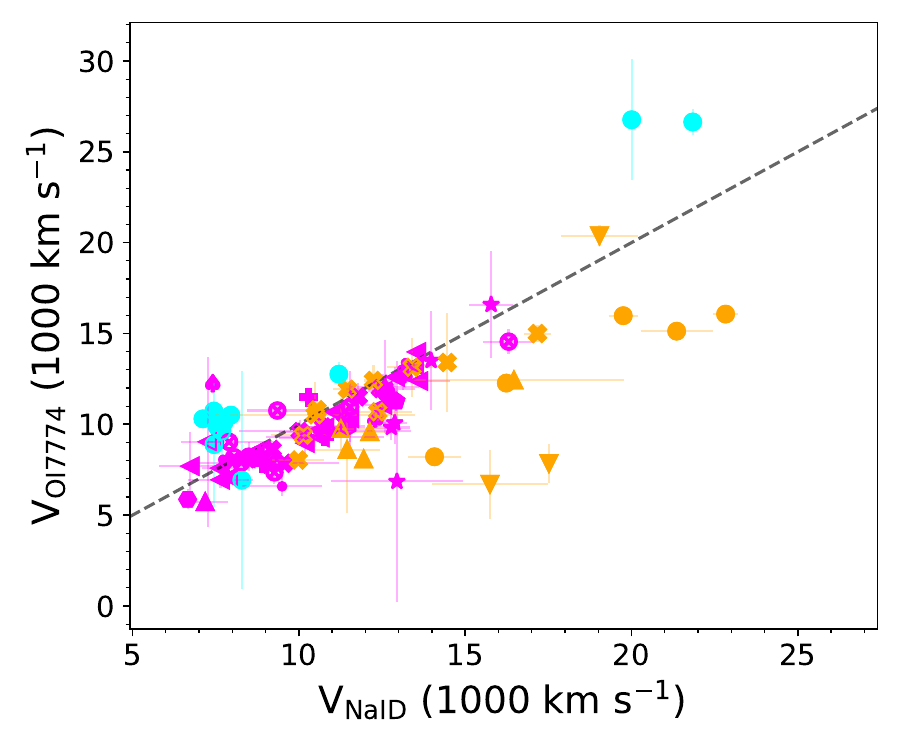}
	\includegraphics[width=0.162\linewidth]{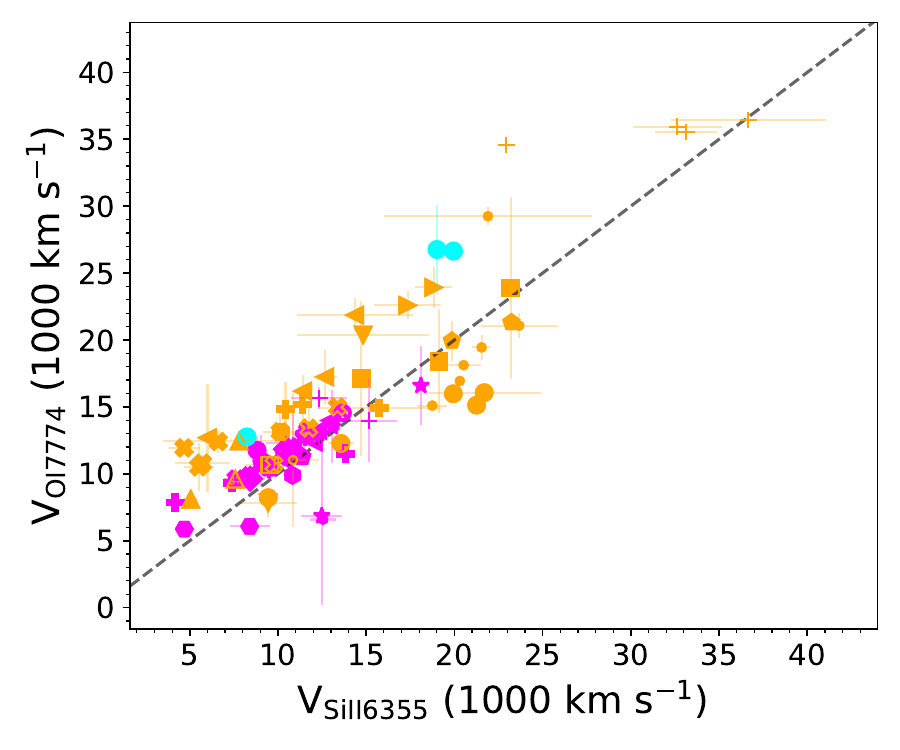}
	\includegraphics[width=0.162\linewidth]{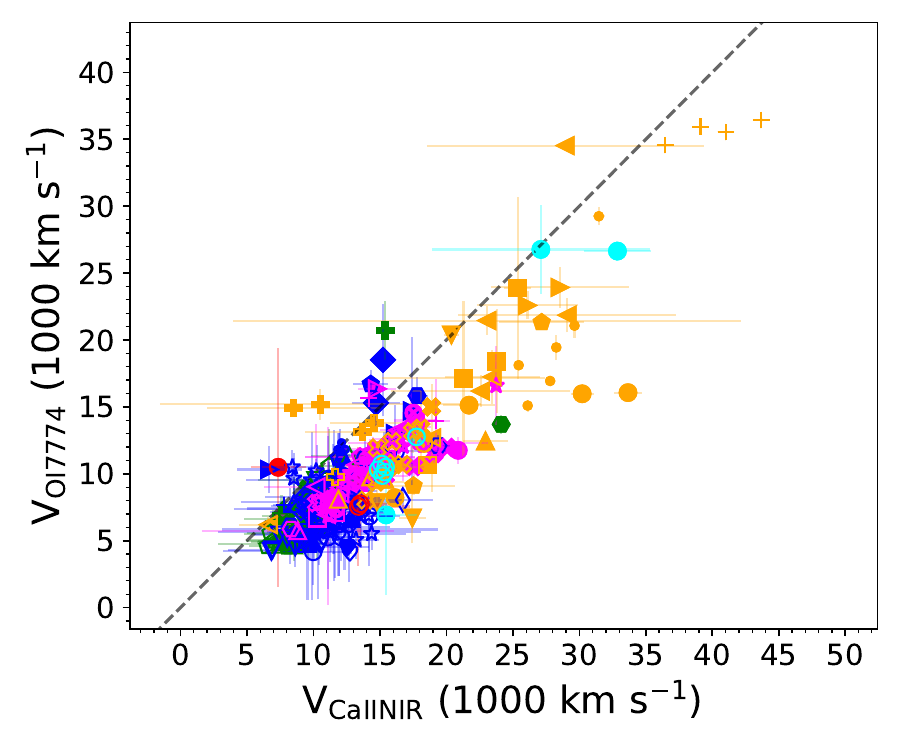}
	\includegraphics[width=0.162\linewidth]{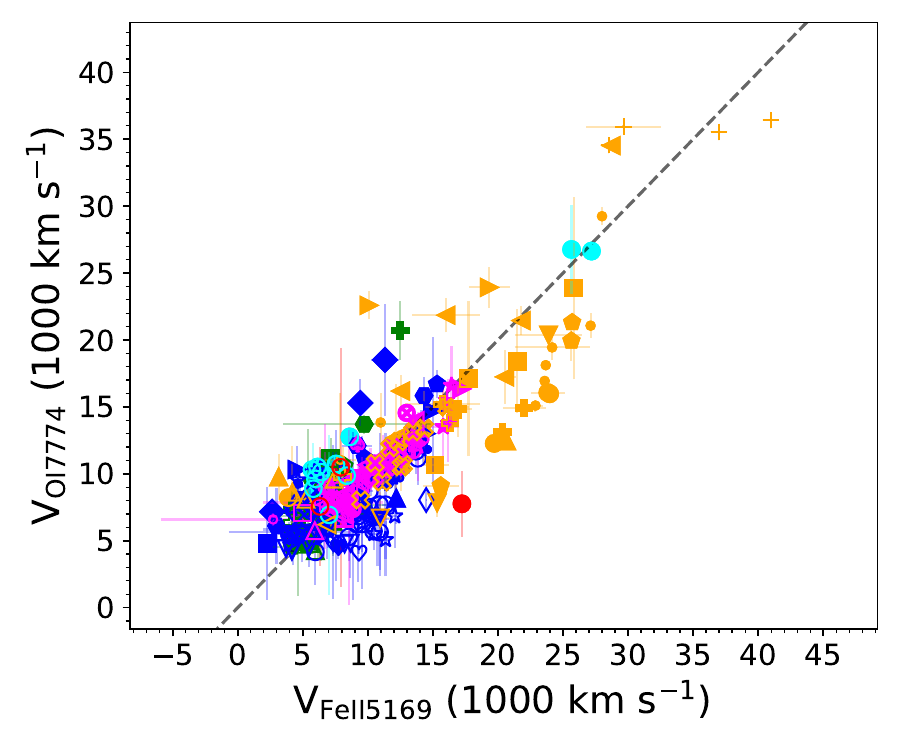}
	\includegraphics[width=0.162\linewidth]{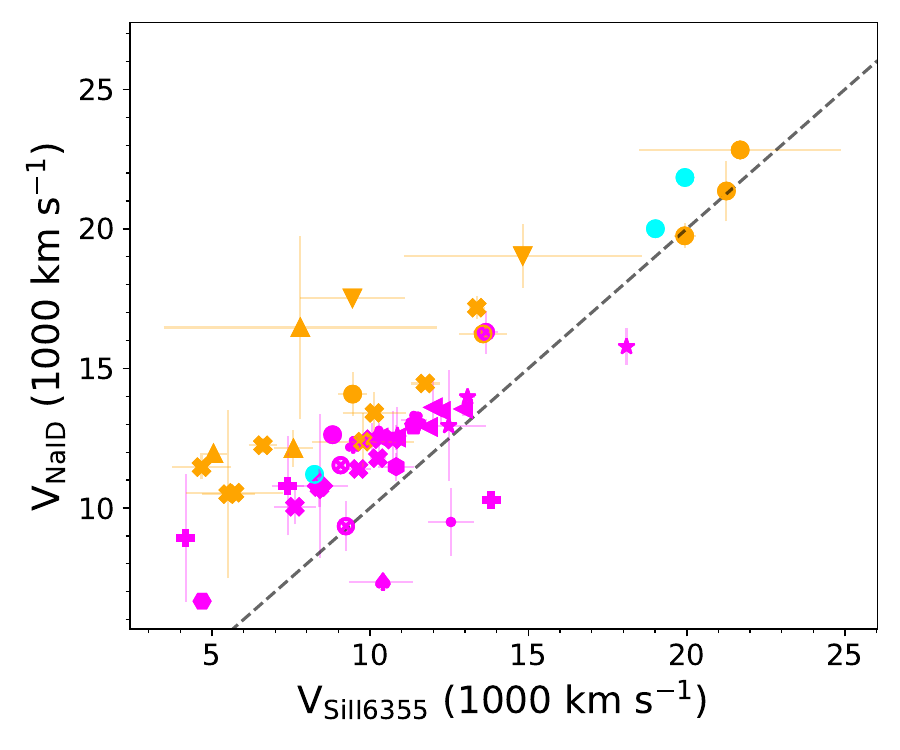}
	\includegraphics[width=0.162\linewidth]{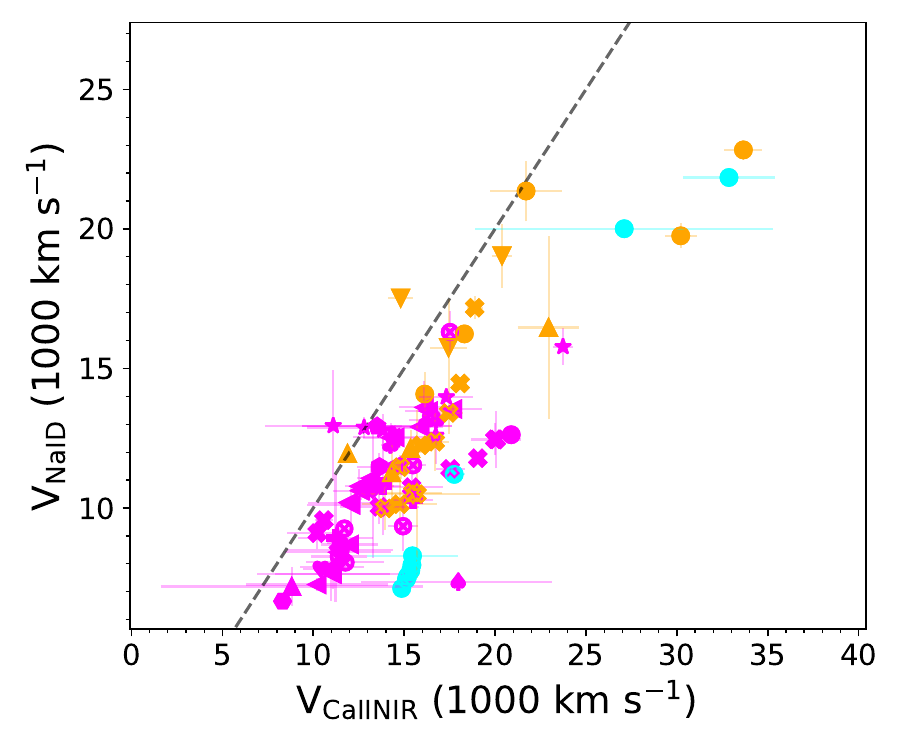}
	\includegraphics[width=0.162\linewidth]{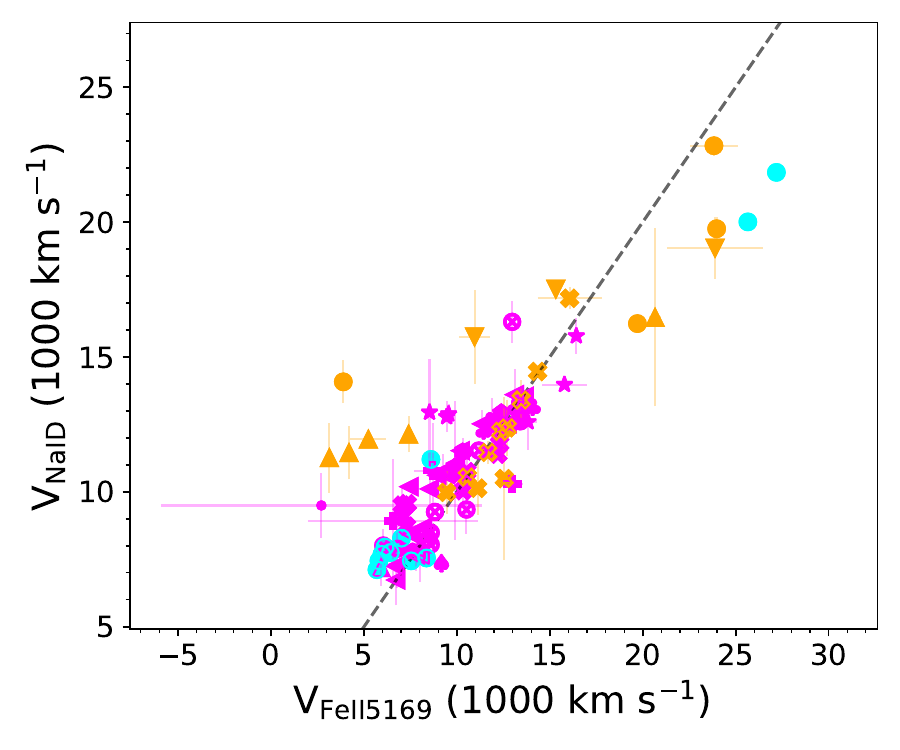}
	\includegraphics[width=0.162\linewidth]{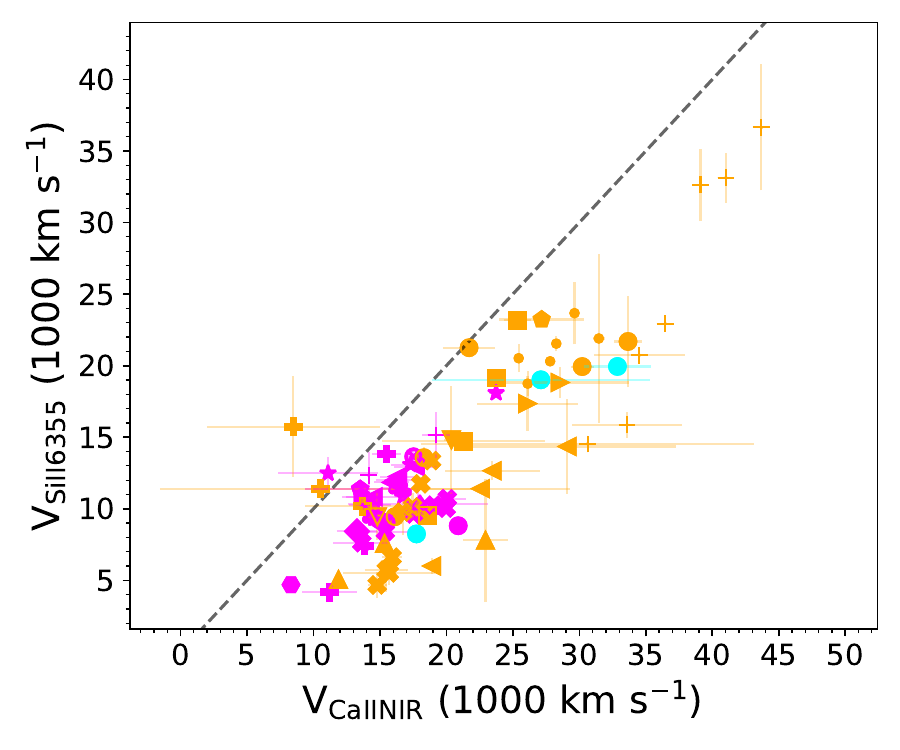}
	\includegraphics[width=0.162\linewidth]{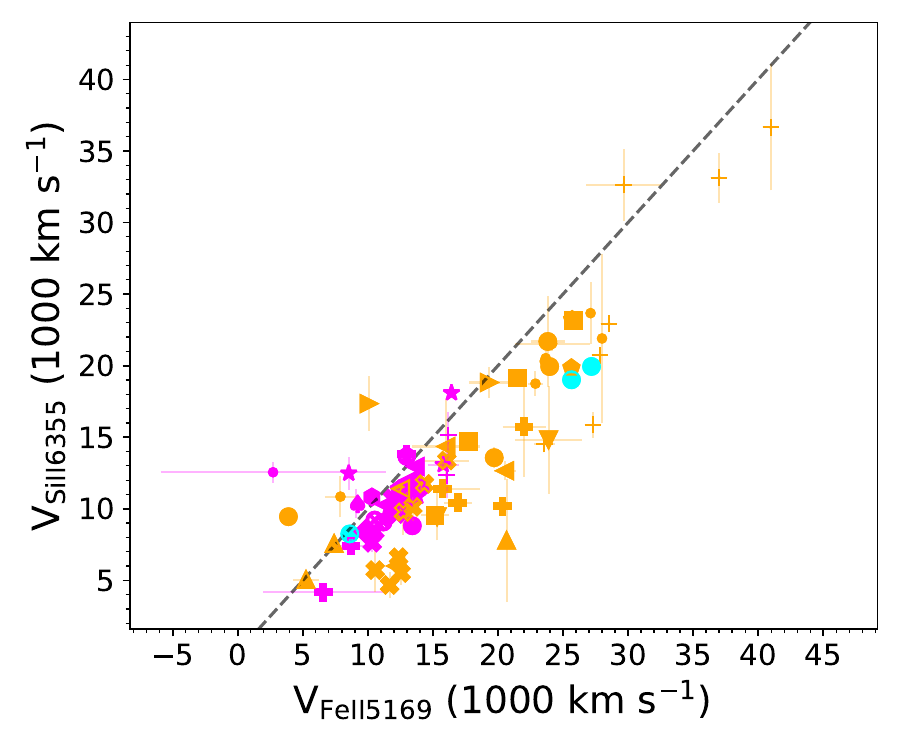}
	\includegraphics[width=0.162\linewidth]{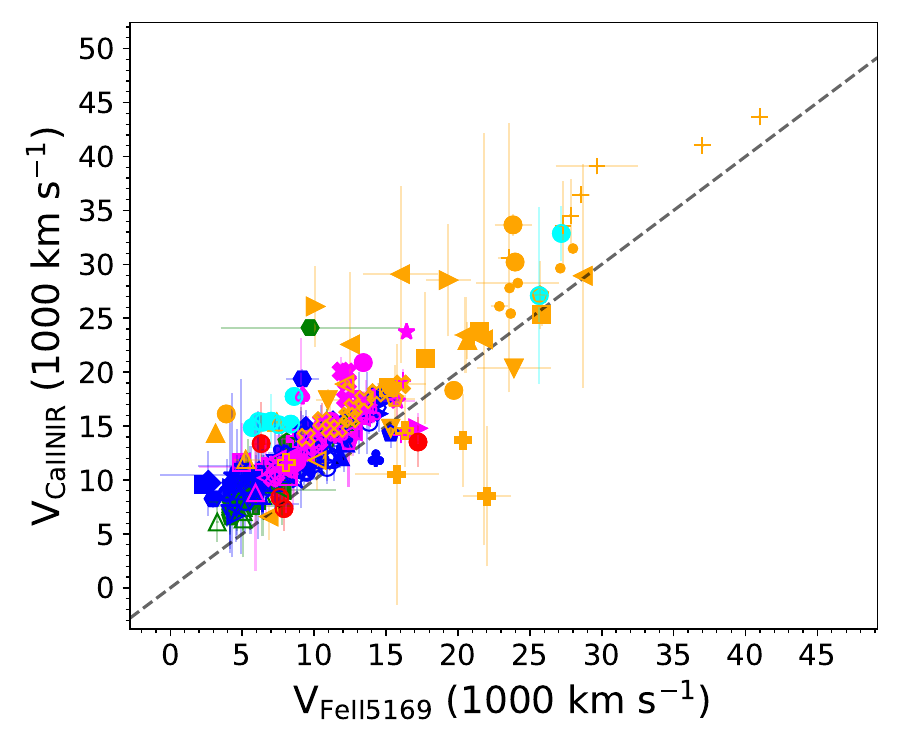}
	
	\caption{Correlation of line velocities of SESNe. The blask dashed line in each panel denotes equal velocities of the two lines. Data points follow the same color/symbol scheme as in Fig.~\ref{fig:evol-V-all}.\label{fig:corr-V-all}}
\end{figure*}

\begin{figure*}
	\centering
	\label{fig:corr-EW-all}
	\includegraphics[width=0.162\linewidth]{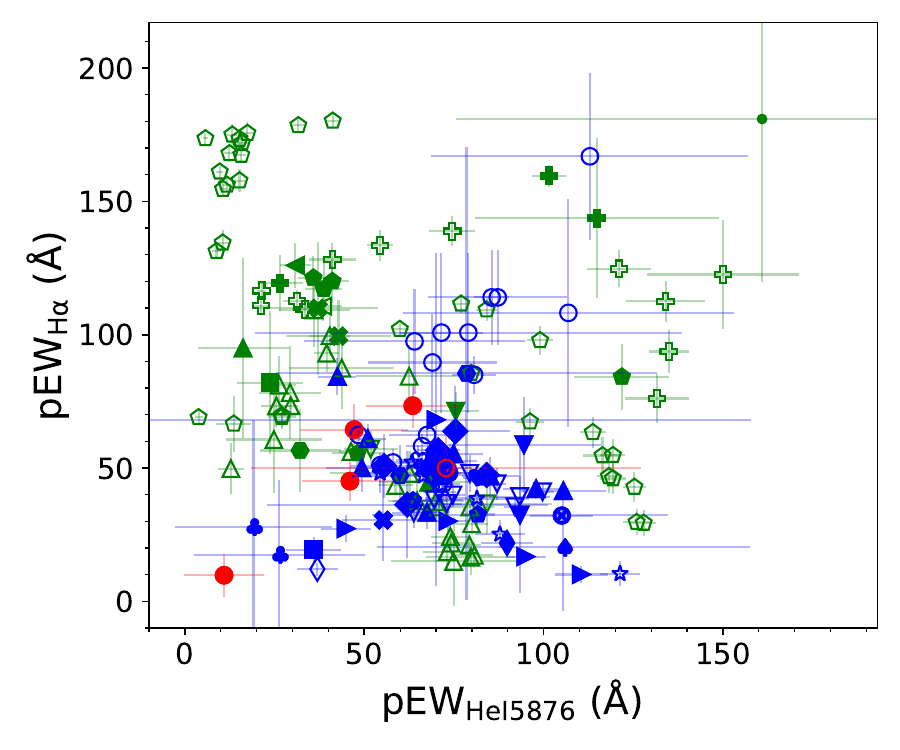}
	\includegraphics[width=0.162\linewidth]{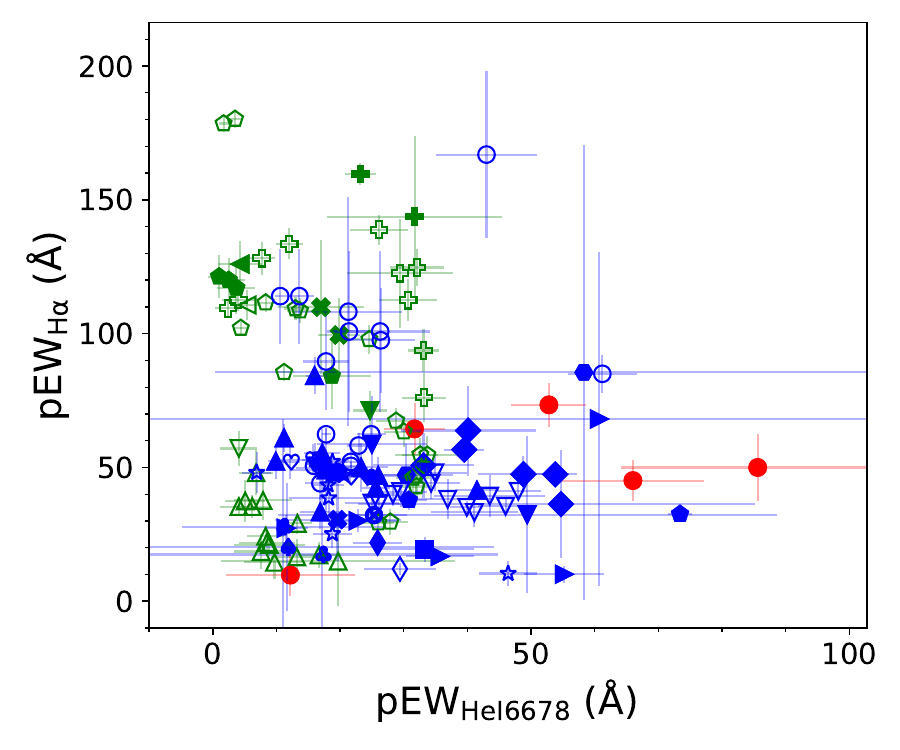}
	\includegraphics[width=0.162\linewidth]{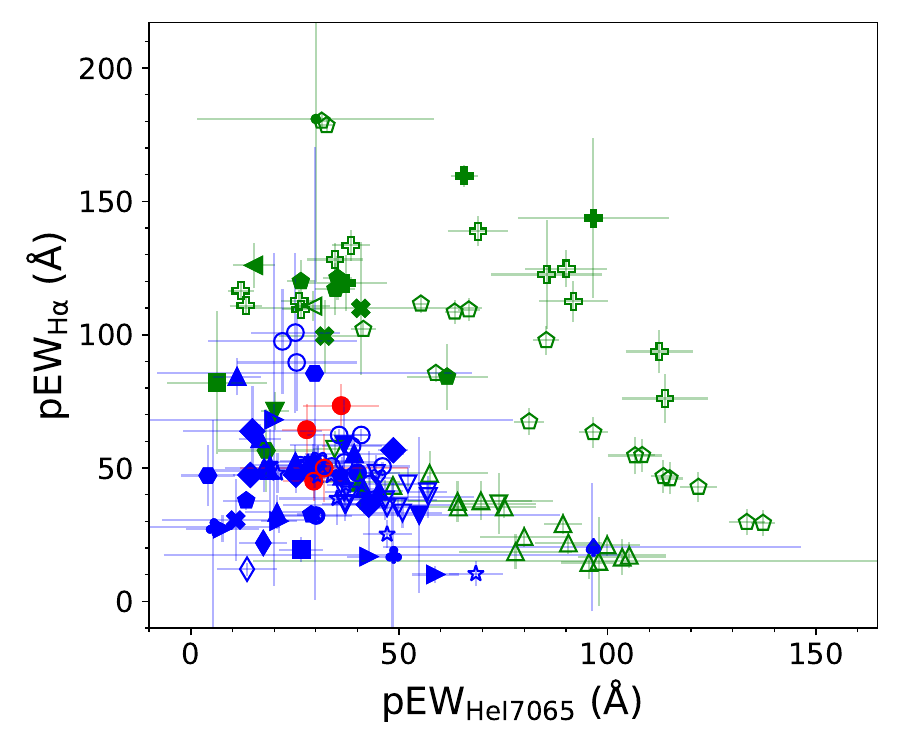}
	\includegraphics[width=0.162\linewidth]{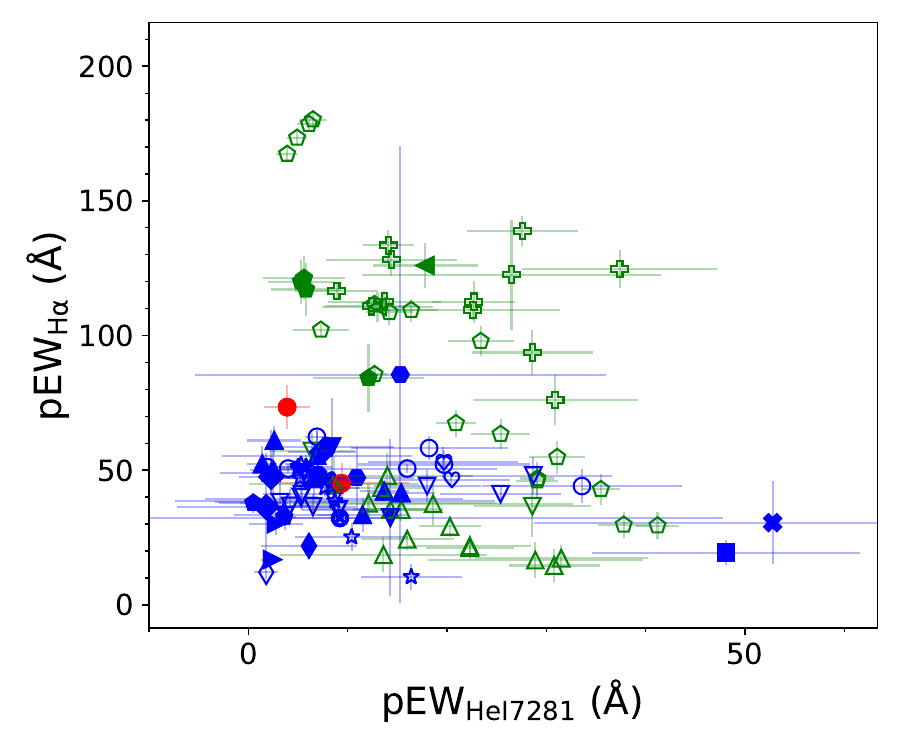}
	\includegraphics[width=0.162\linewidth]{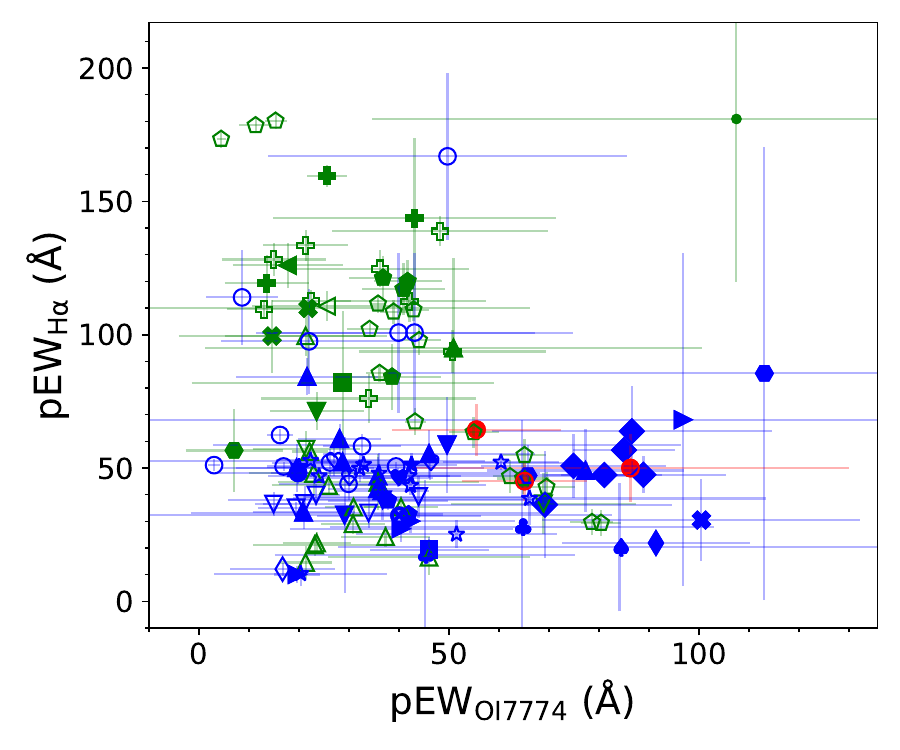}
	\includegraphics[width=0.162\linewidth]{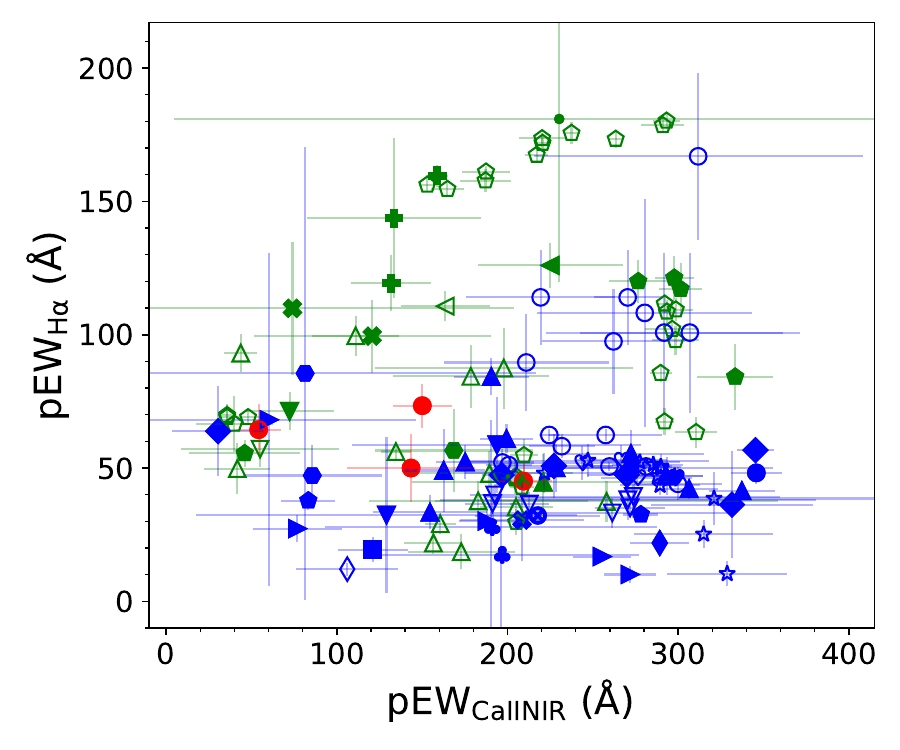}
	\includegraphics[width=0.162\linewidth]{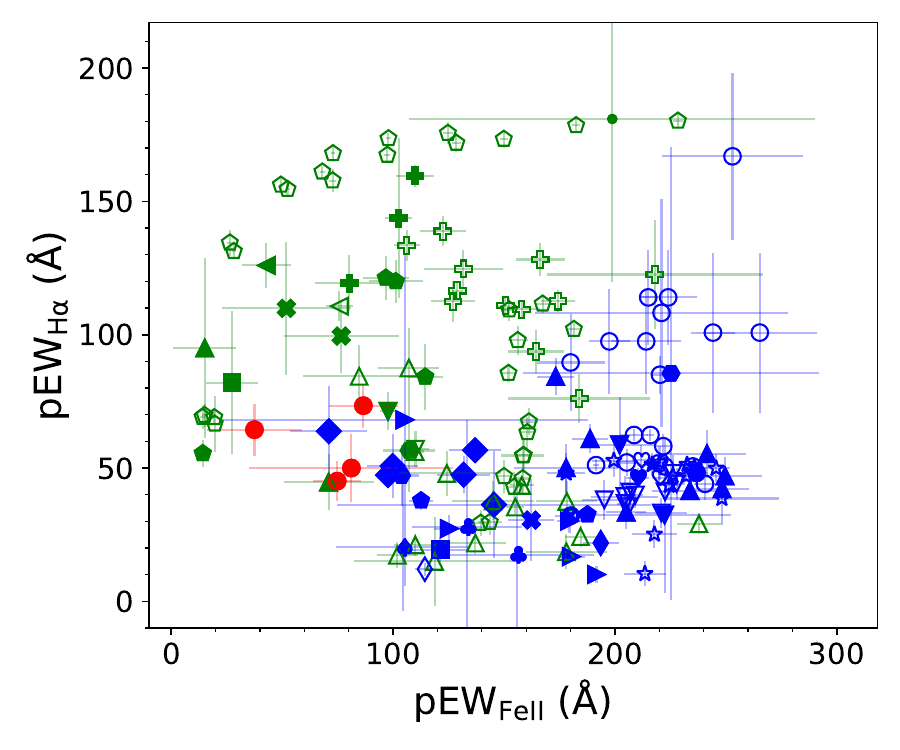}
	\includegraphics[width=0.162\linewidth]{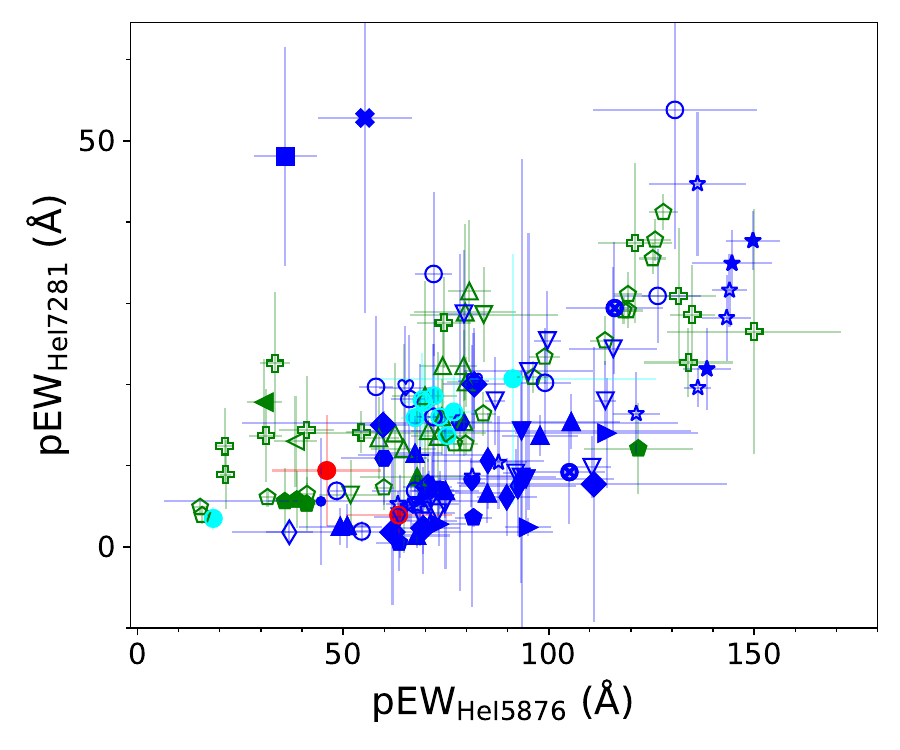}
	\includegraphics[width=0.162\linewidth]{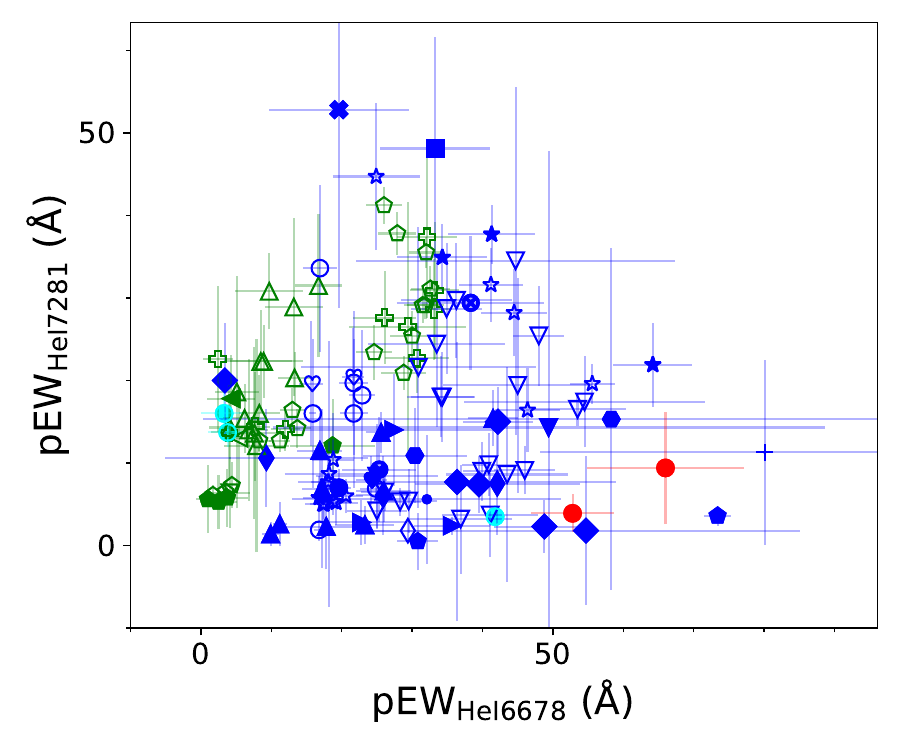}
	\includegraphics[width=0.162\linewidth]{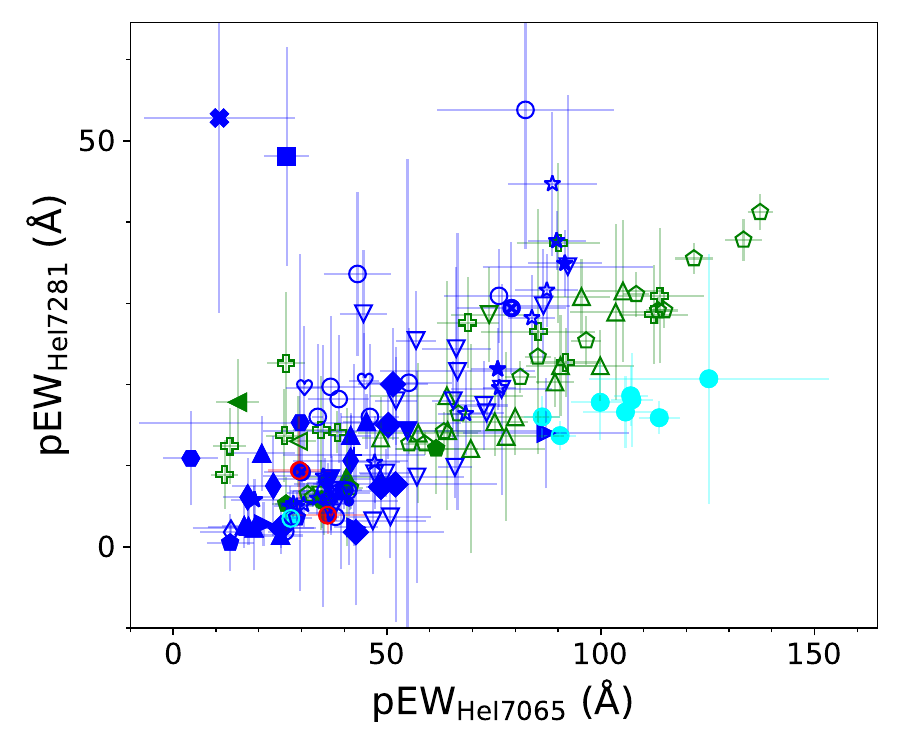}
	\includegraphics[width=0.162\linewidth]{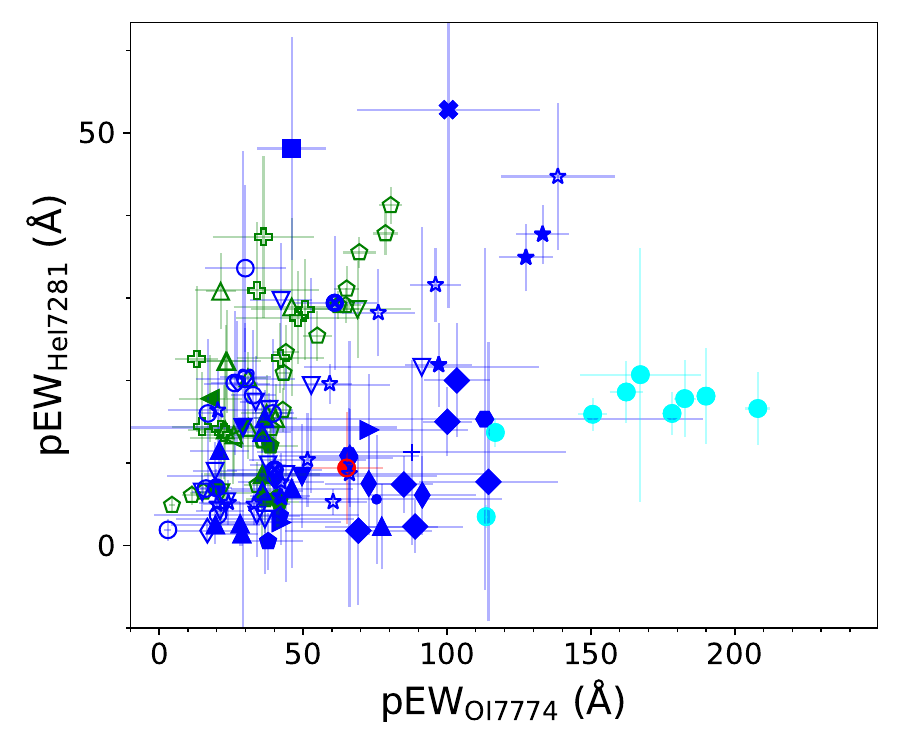}
	\includegraphics[width=0.162\linewidth]{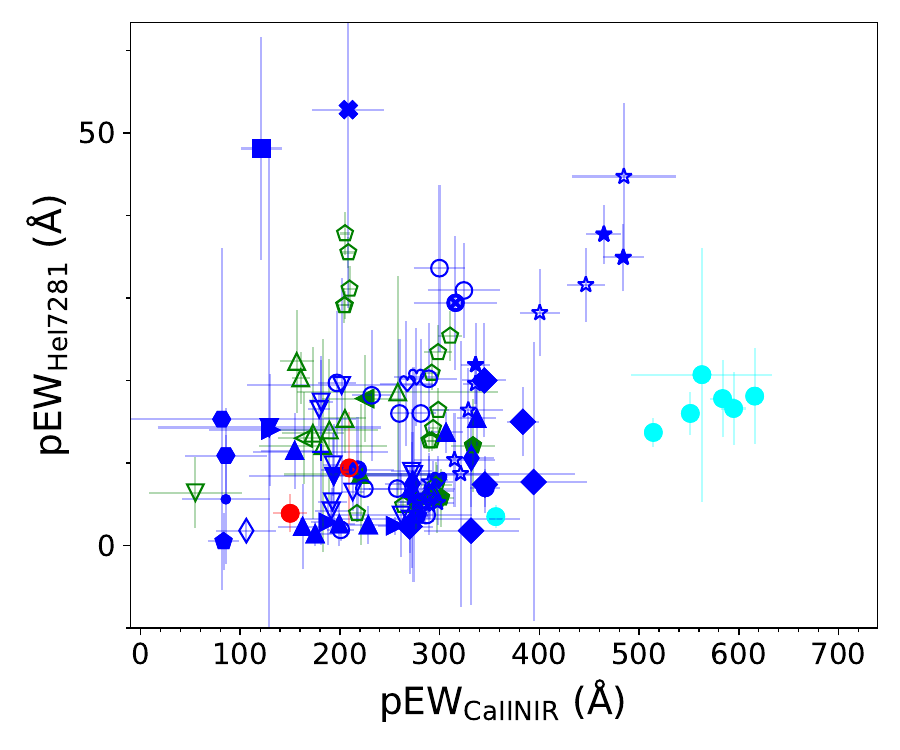}
	\includegraphics[width=0.162\linewidth]{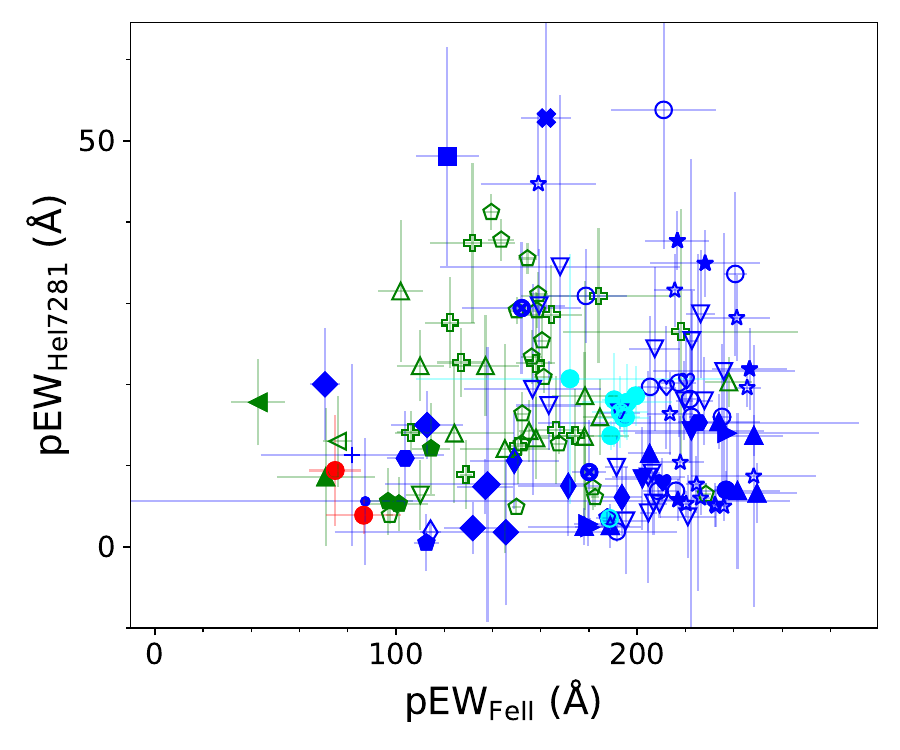}
	\includegraphics[width=0.162\linewidth]{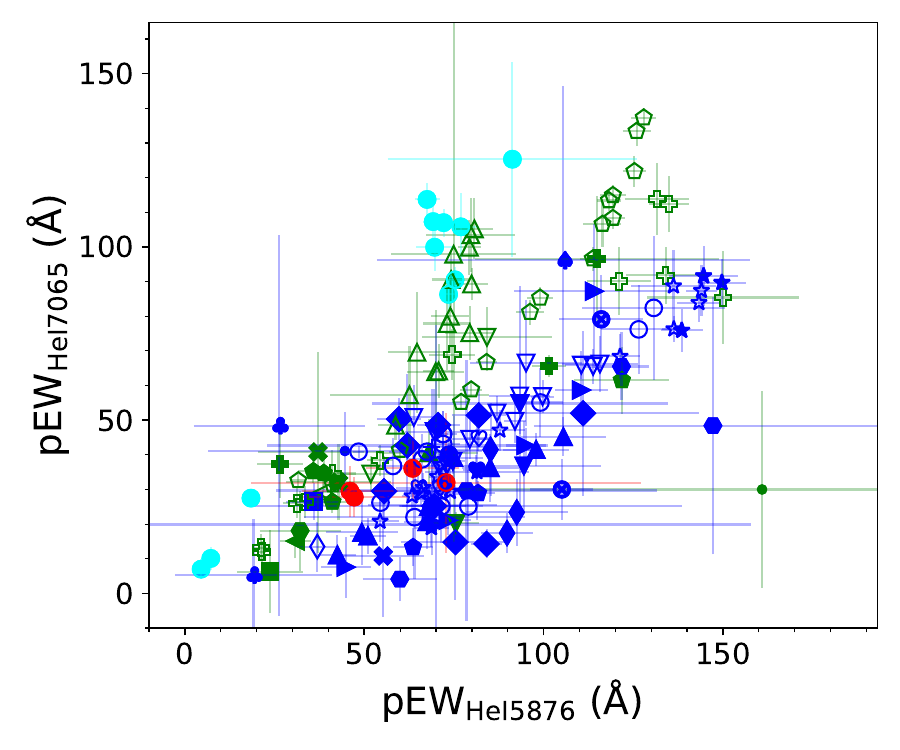}
	\includegraphics[width=0.162\linewidth]{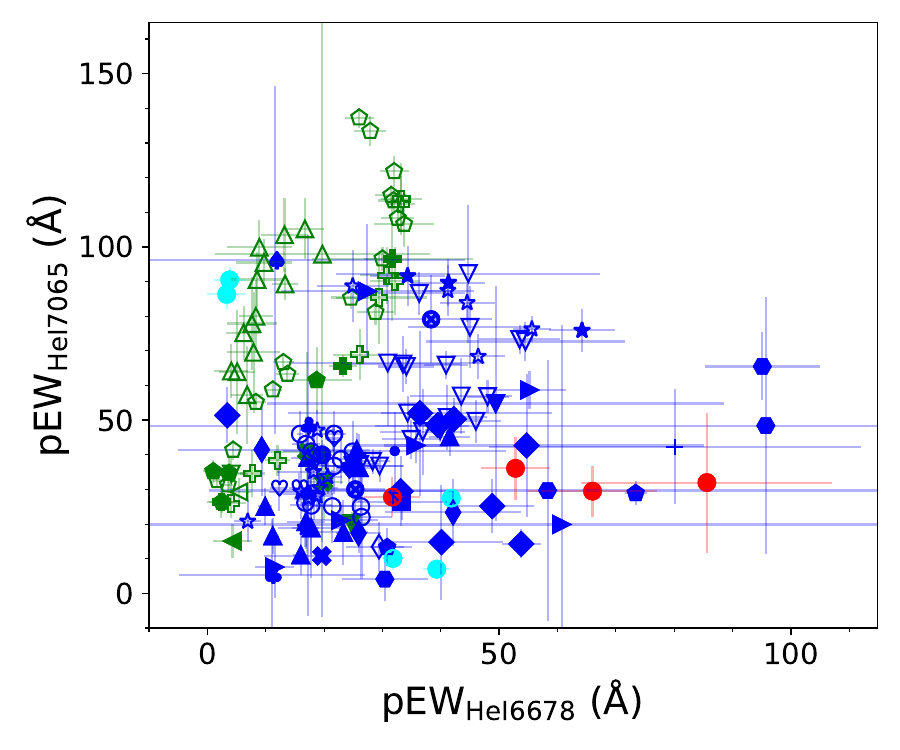}
	\includegraphics[width=0.162\linewidth]{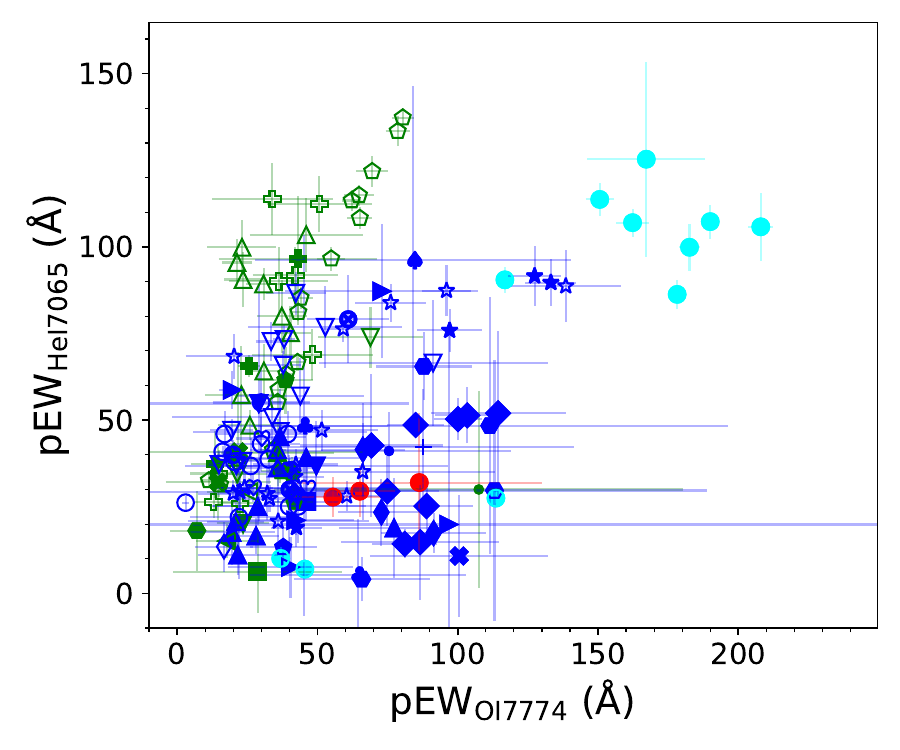}
	\includegraphics[width=0.162\linewidth]{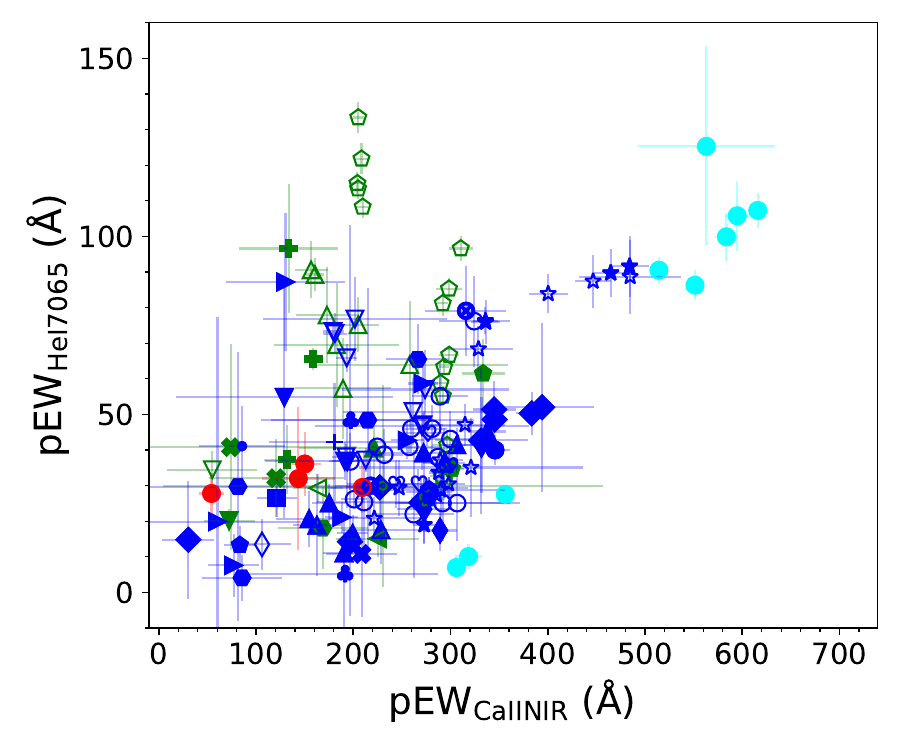}
	\includegraphics[width=0.162\linewidth]{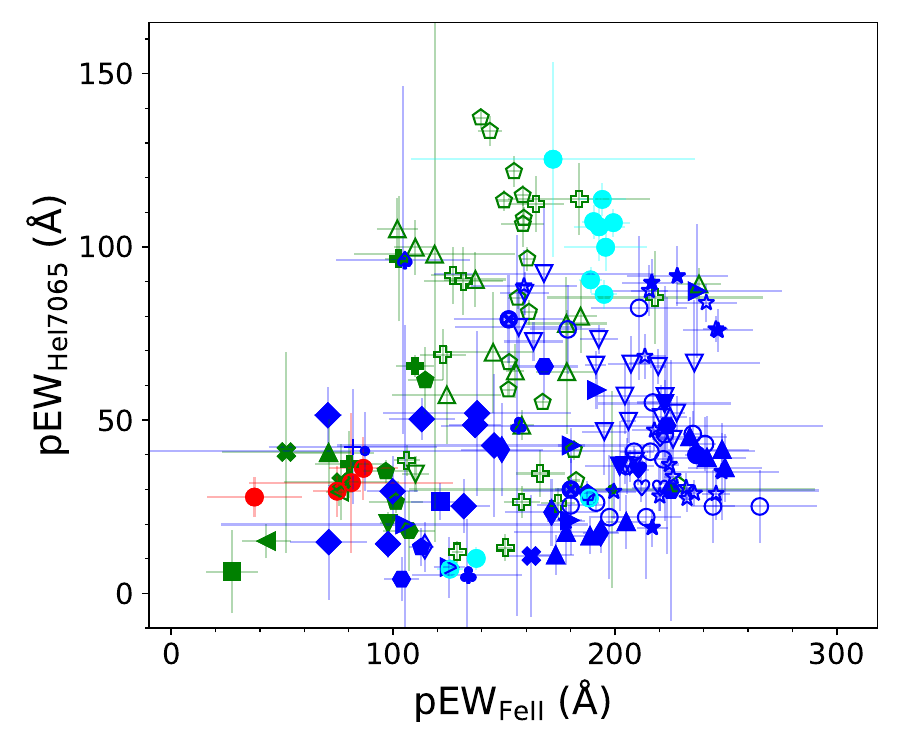}
	\includegraphics[width=0.162\linewidth]{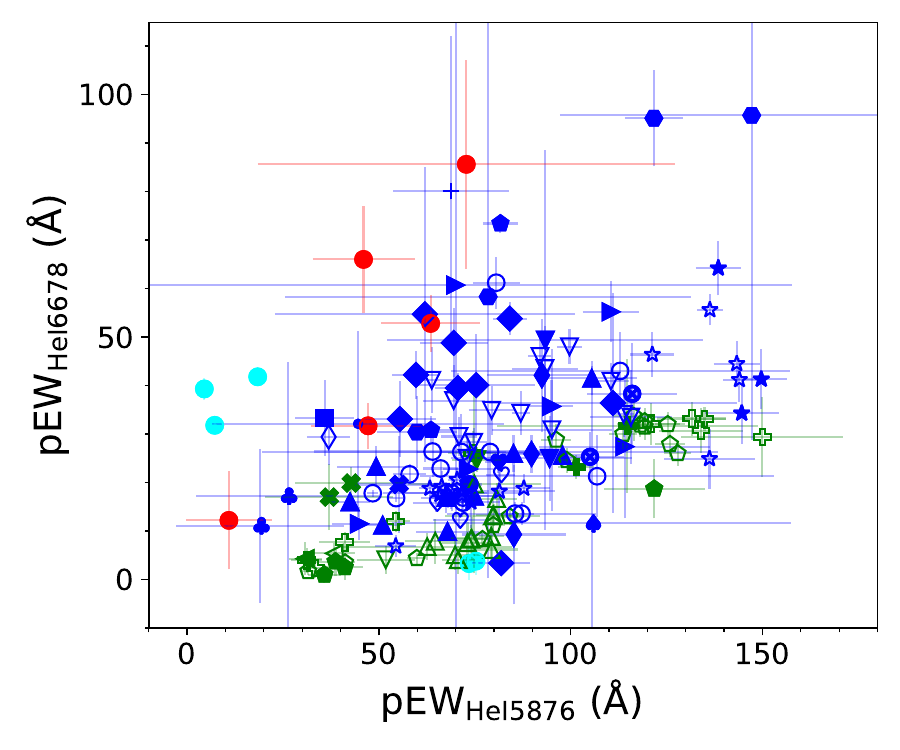}
	\includegraphics[width=0.162\linewidth]{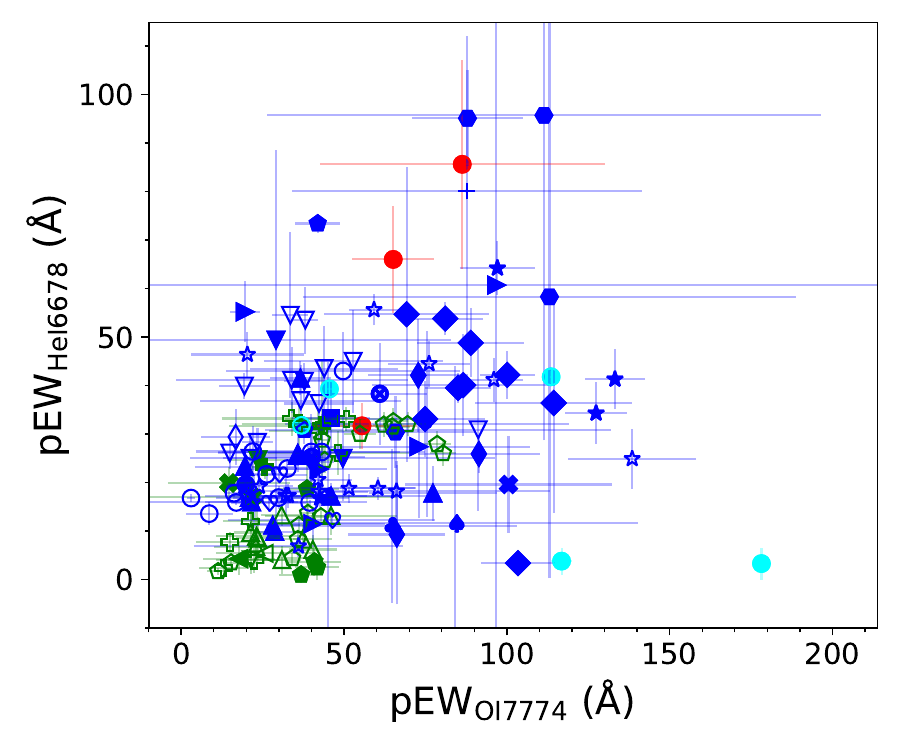}
	\includegraphics[width=0.162\linewidth]{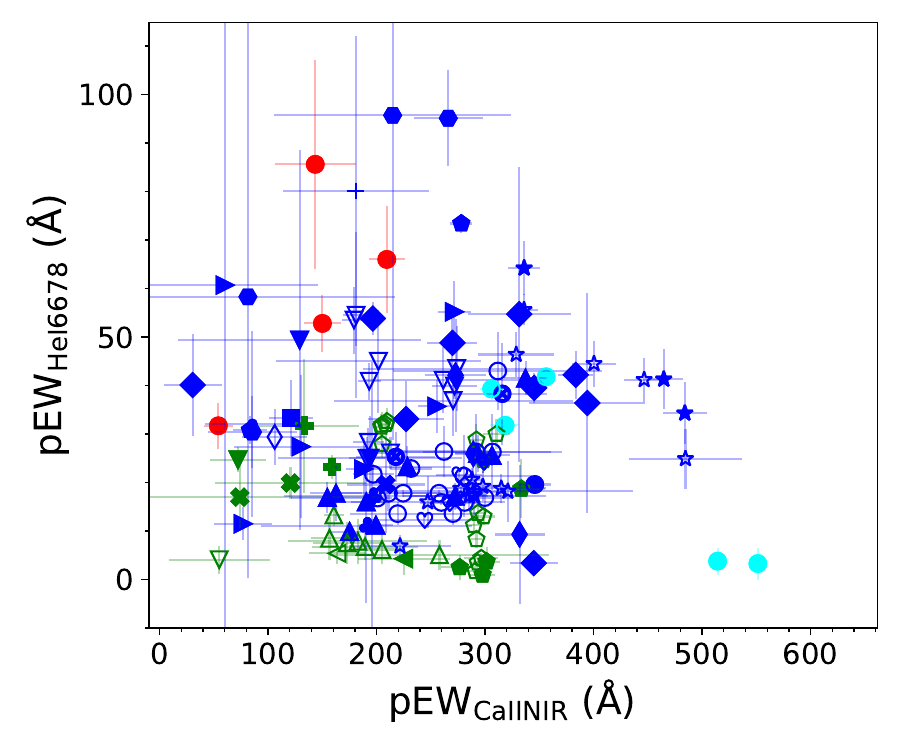}
	\includegraphics[width=0.162\linewidth]{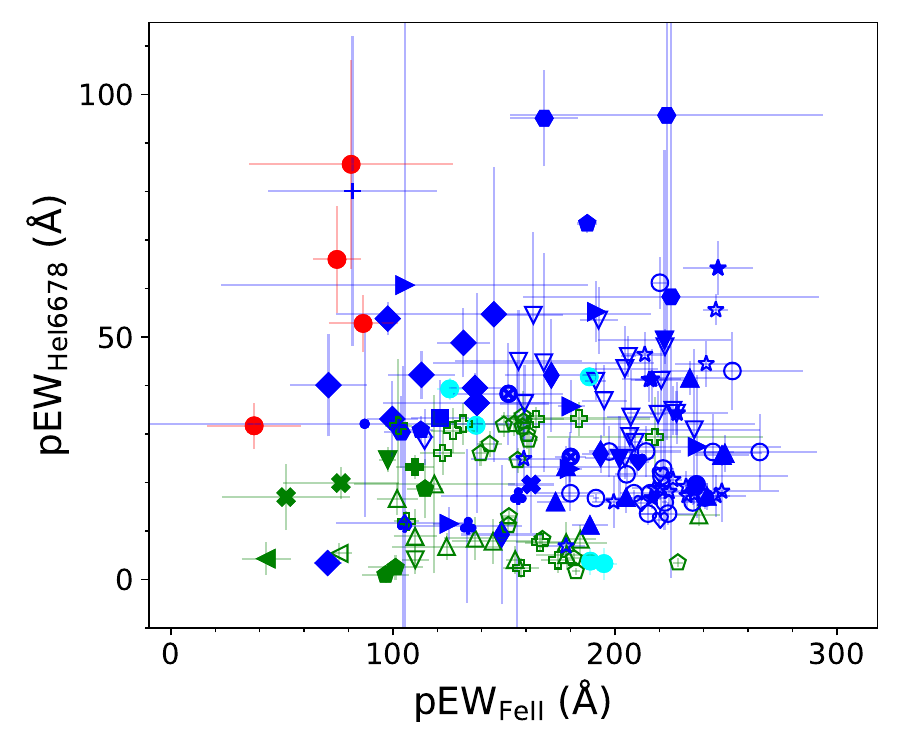}
	\includegraphics[width=0.162\linewidth]{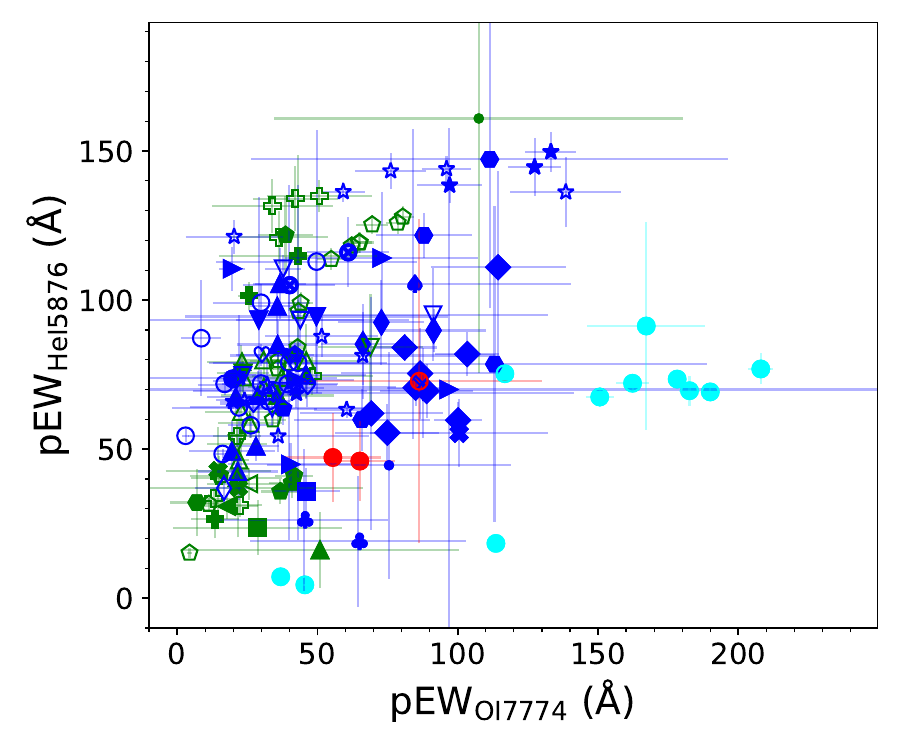}
	\includegraphics[width=0.162\linewidth]{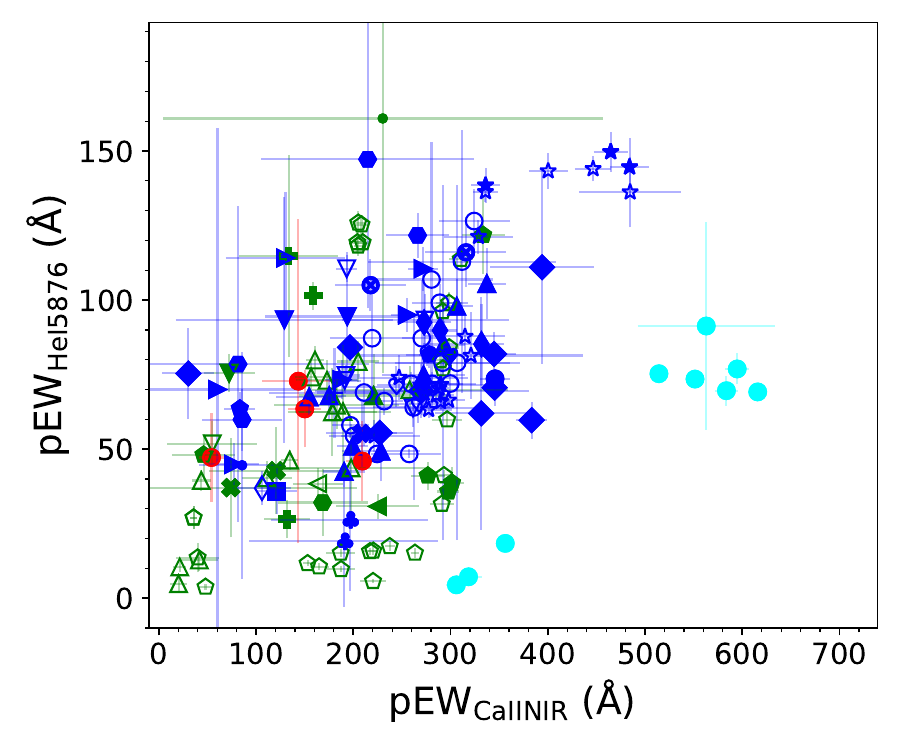}
	\includegraphics[width=0.162\linewidth]{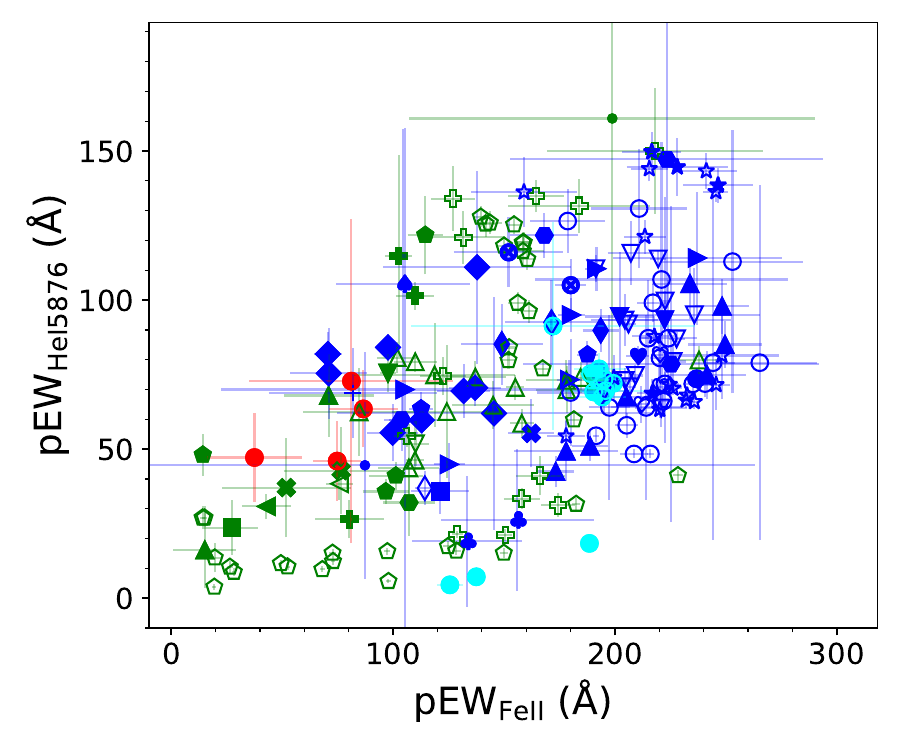}
	\includegraphics[width=0.162\linewidth]{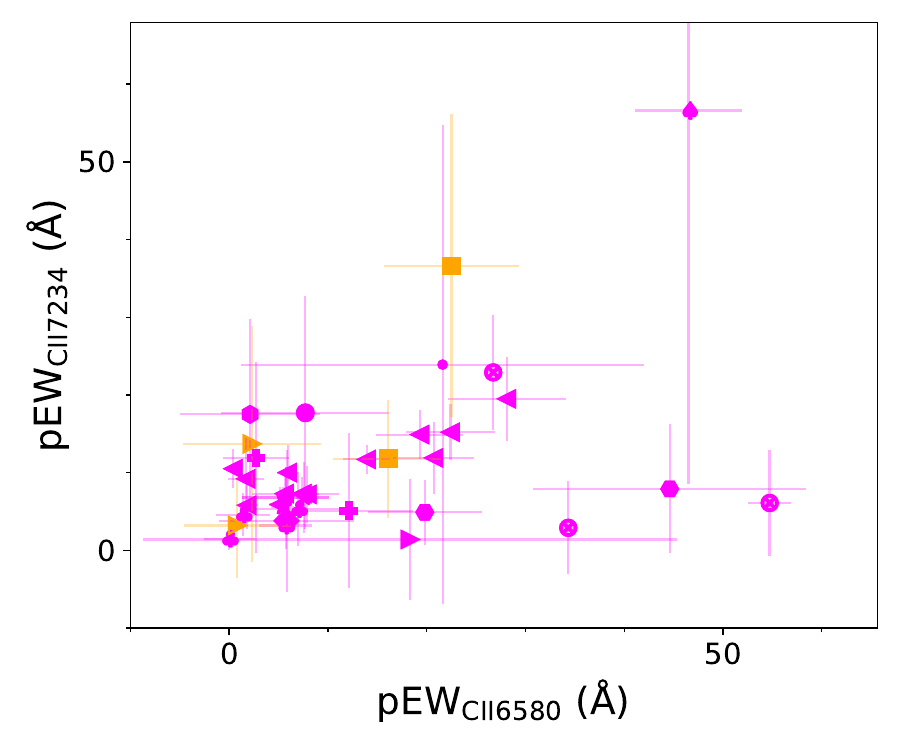}
	\includegraphics[width=0.162\linewidth]{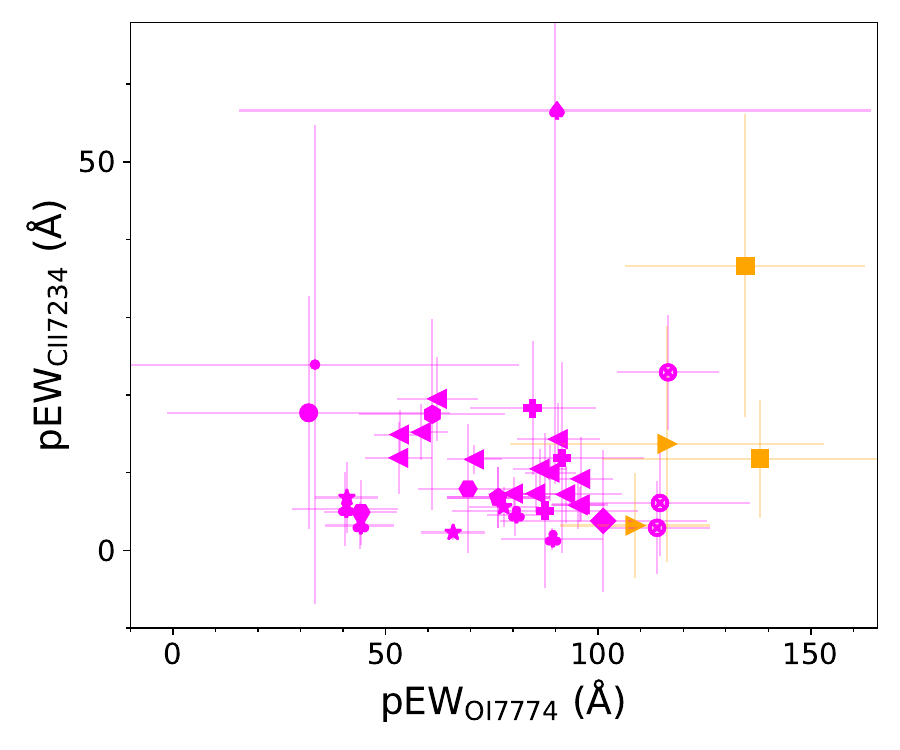}
	\includegraphics[width=0.162\linewidth]{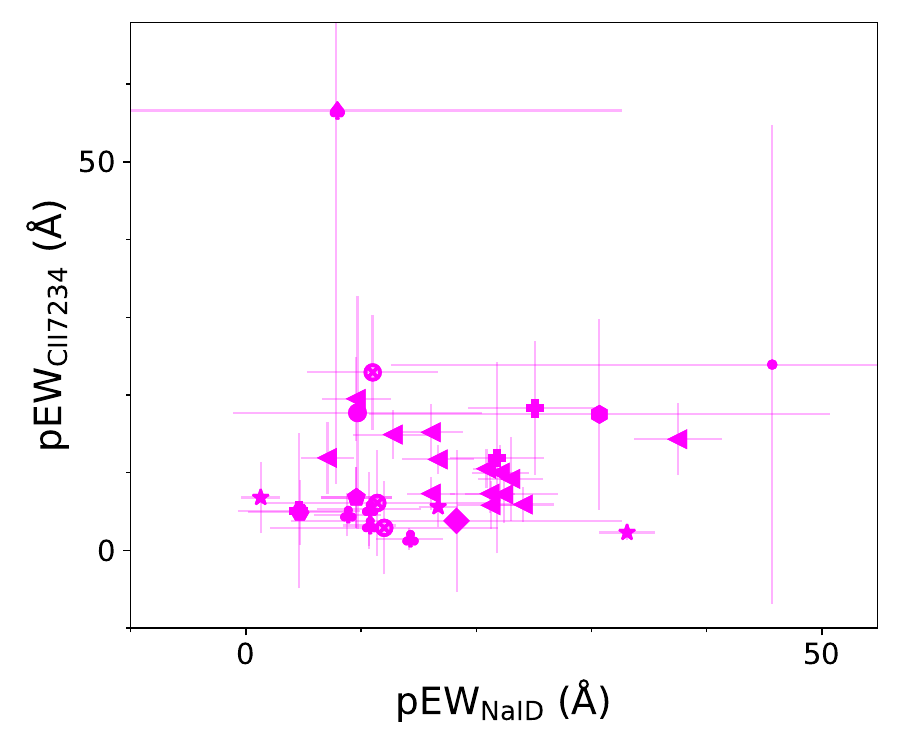}
	\includegraphics[width=0.162\linewidth]{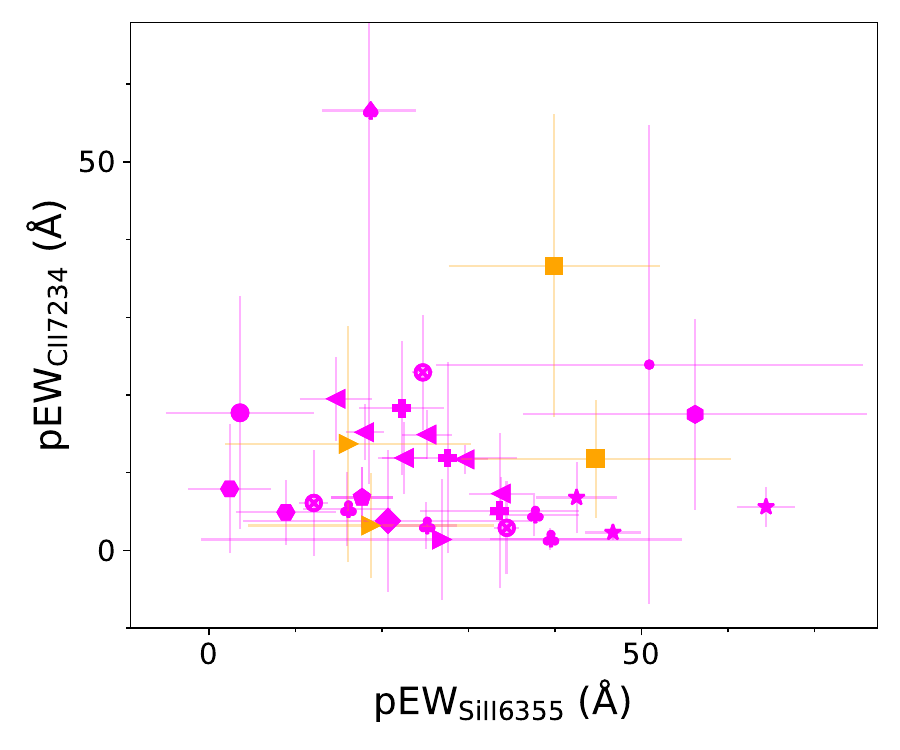}
	\includegraphics[width=0.162\linewidth]{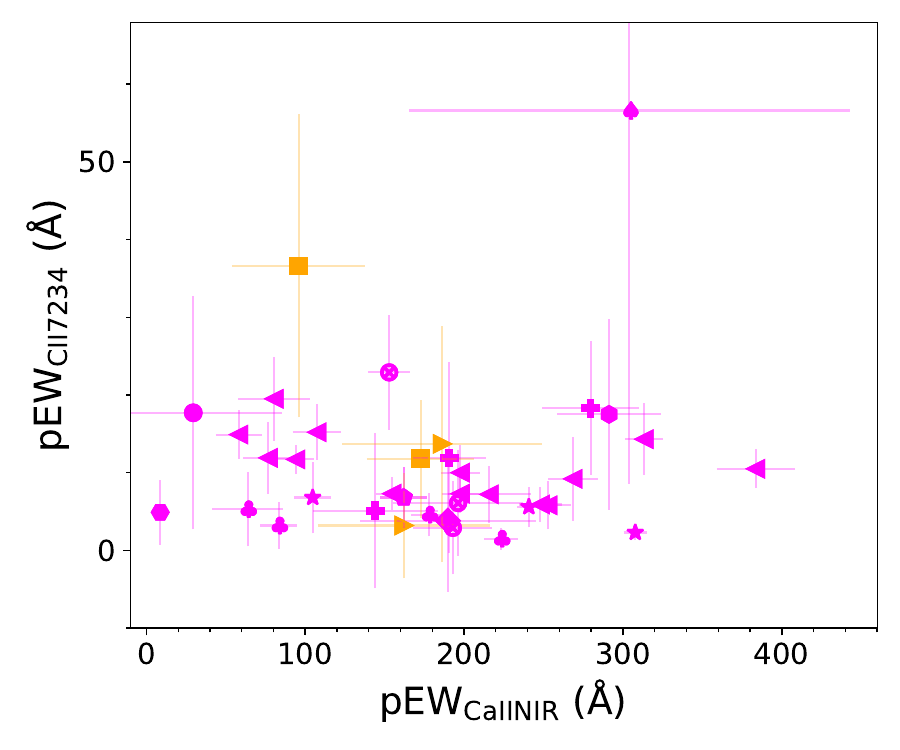}
	\includegraphics[width=0.162\linewidth]{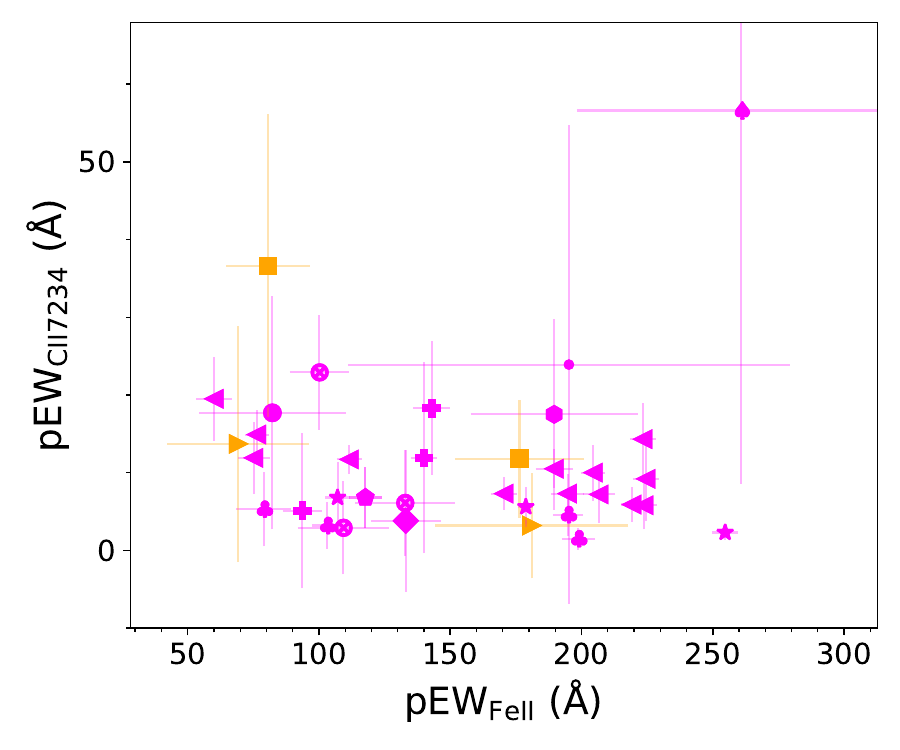}
	\includegraphics[width=0.162\linewidth]{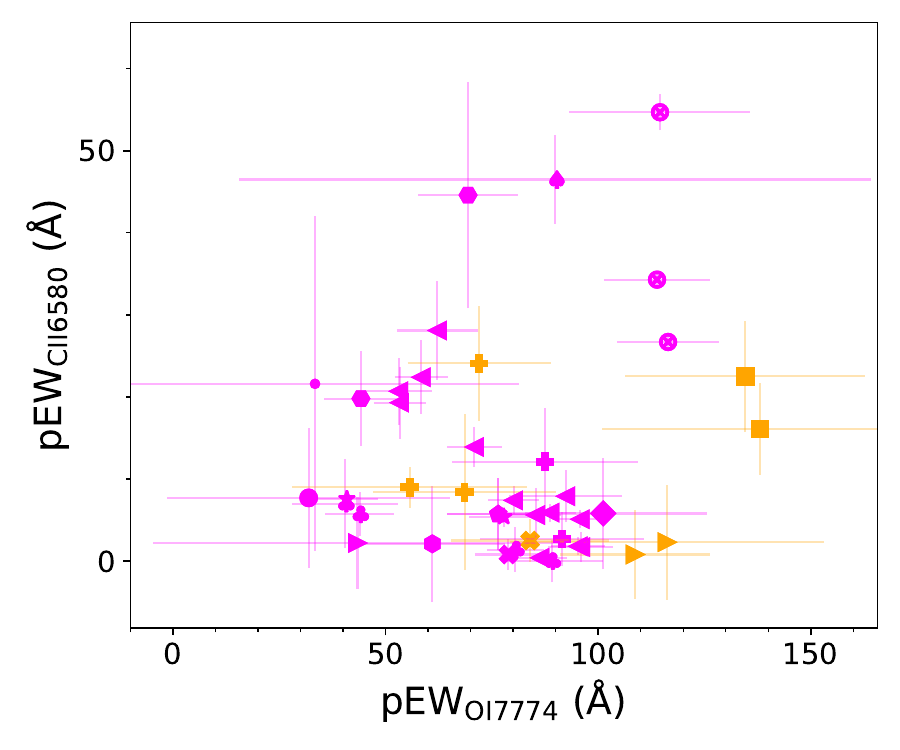}
	\includegraphics[width=0.162\linewidth]{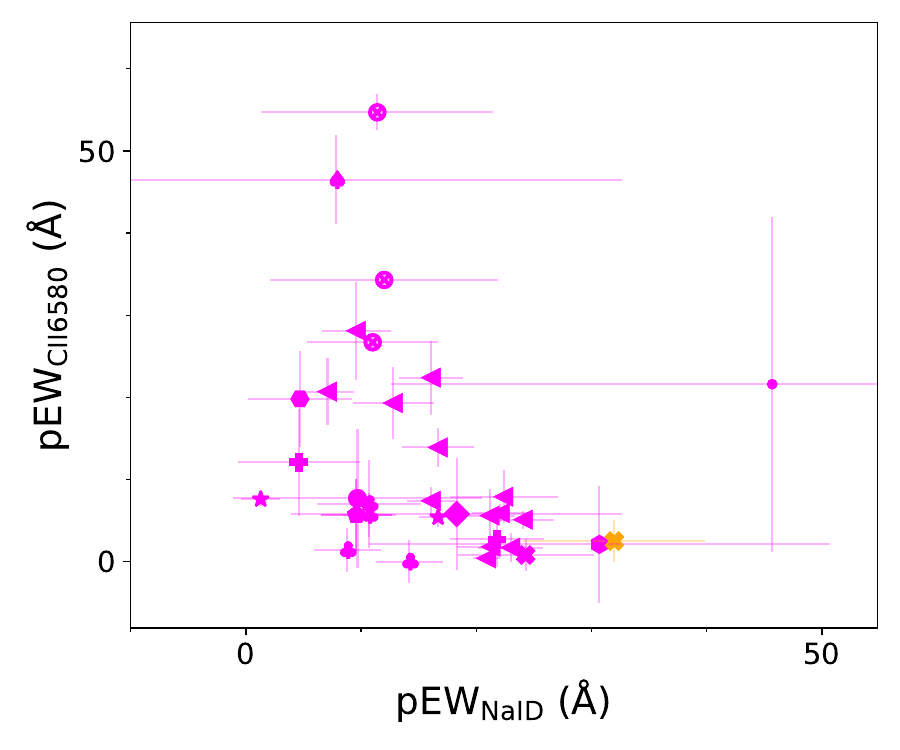} 
	\includegraphics[width=0.162\linewidth]{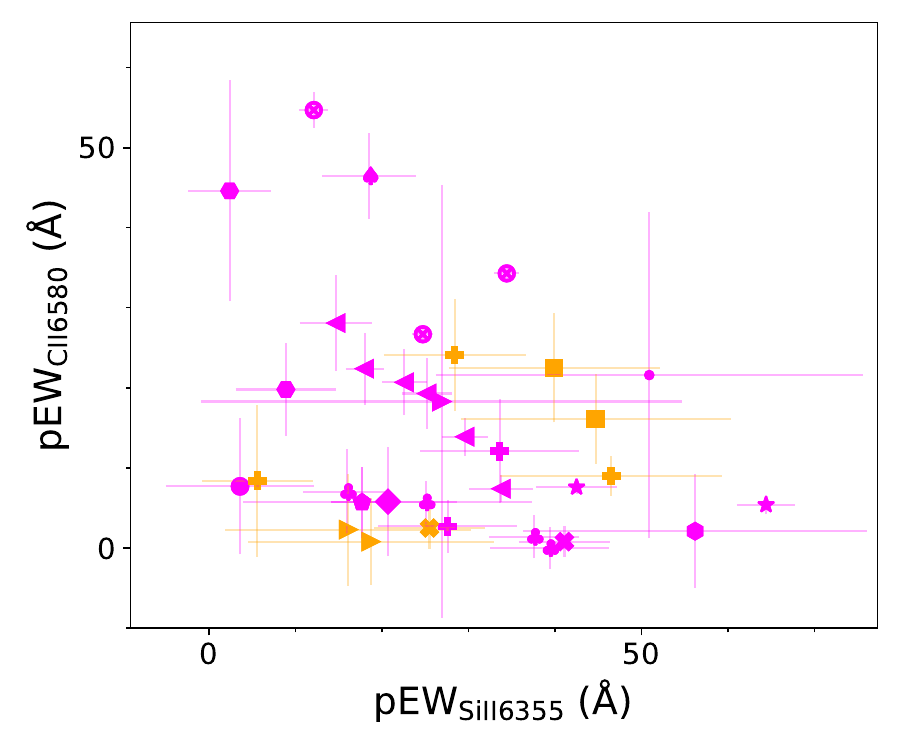}
	\includegraphics[width=0.162\linewidth]{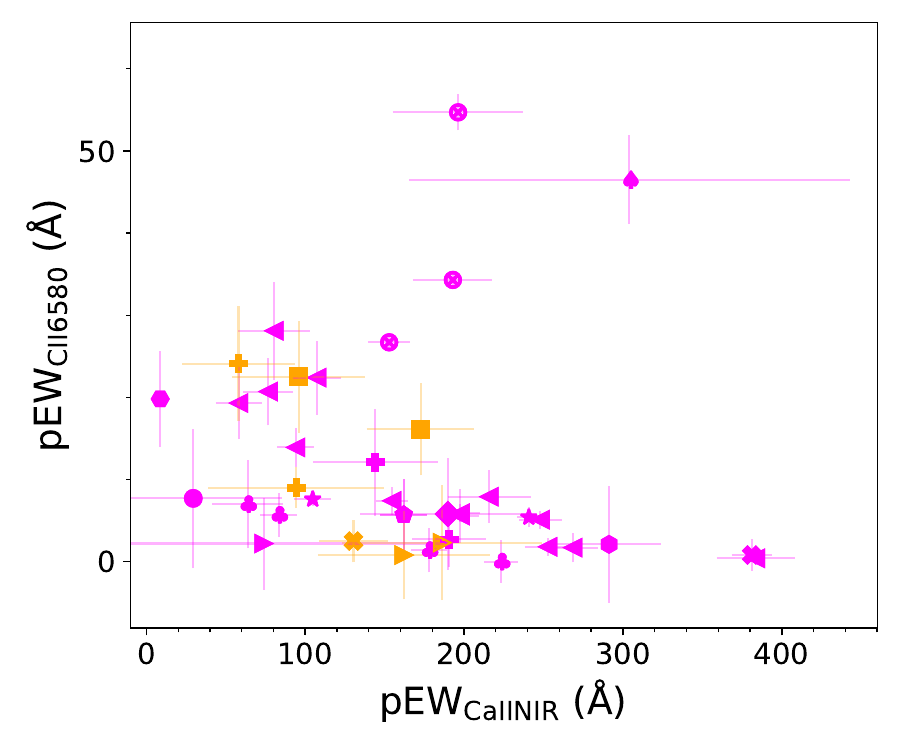}
	\includegraphics[width=0.162\linewidth]{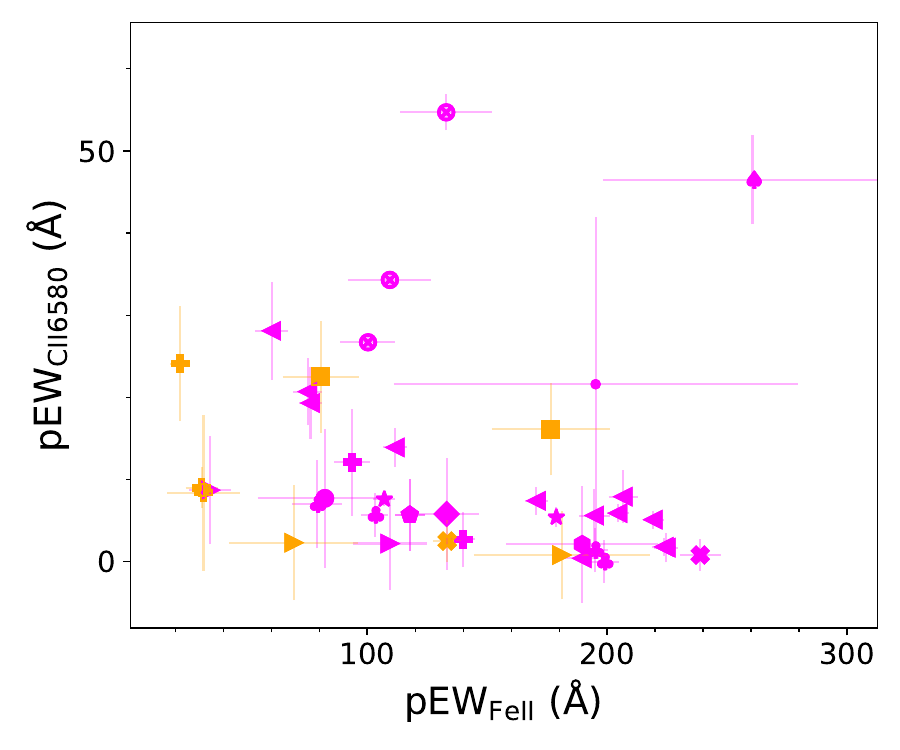}
	\includegraphics[width=0.162\linewidth]{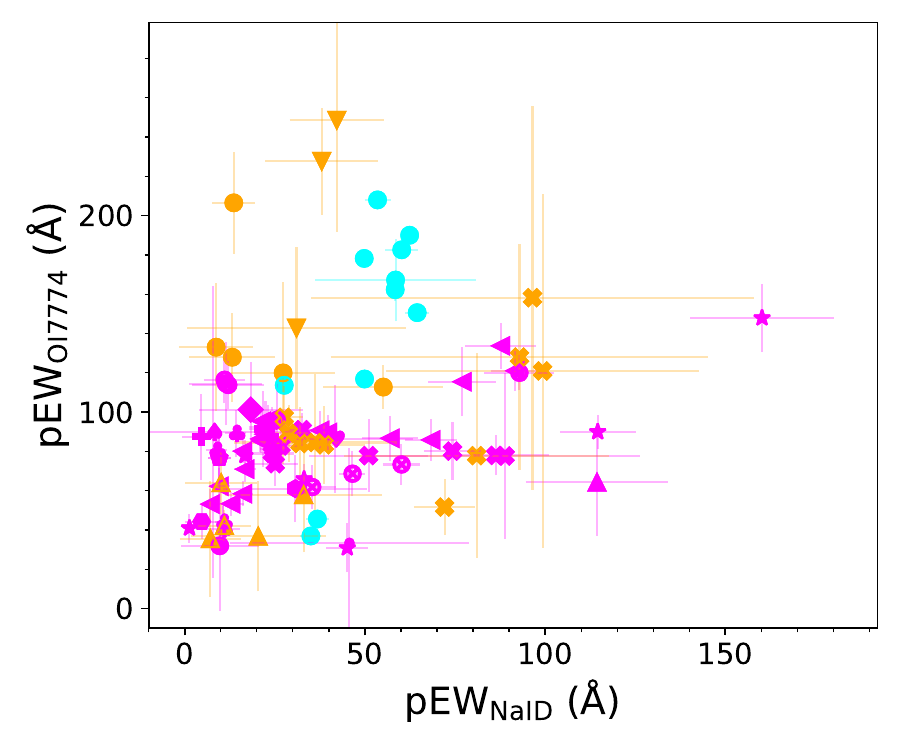}
	\includegraphics[width=0.162\linewidth]{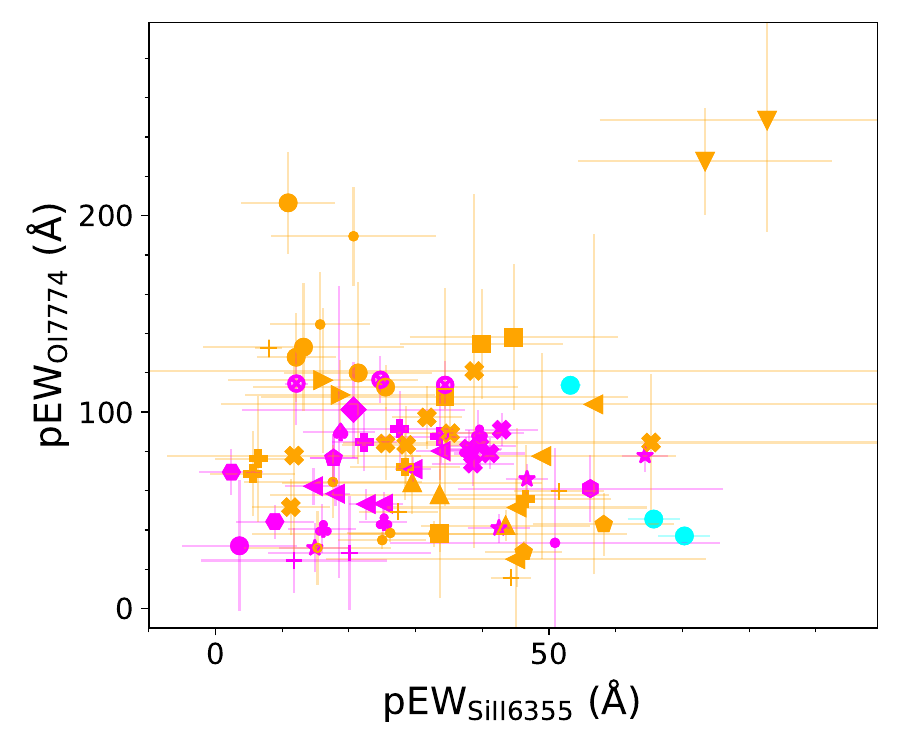}
	\includegraphics[width=0.162\linewidth]{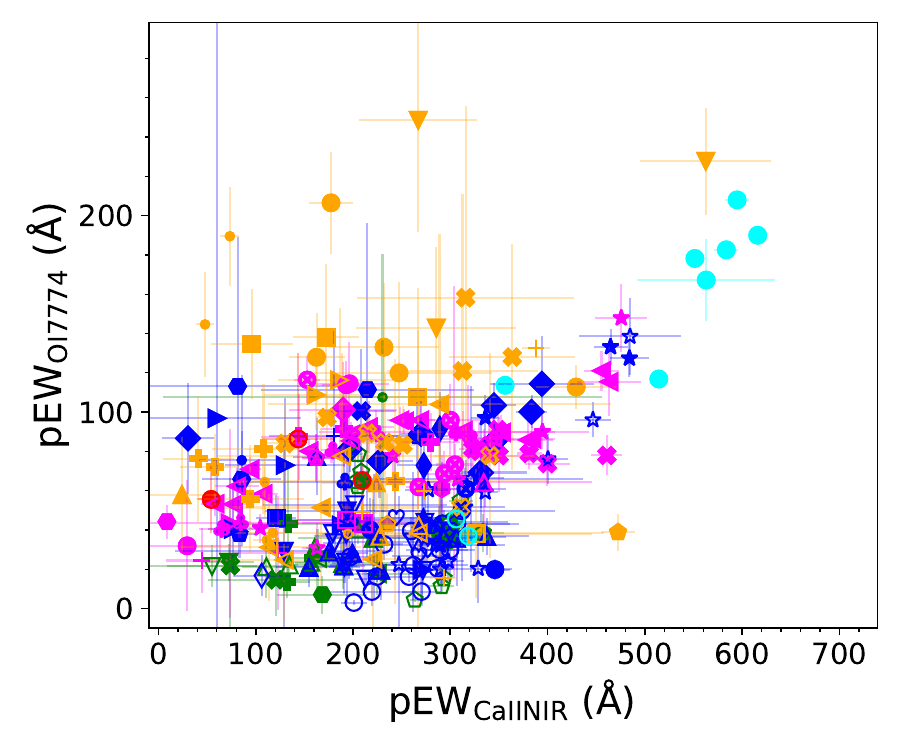}
	\includegraphics[width=0.162\linewidth]{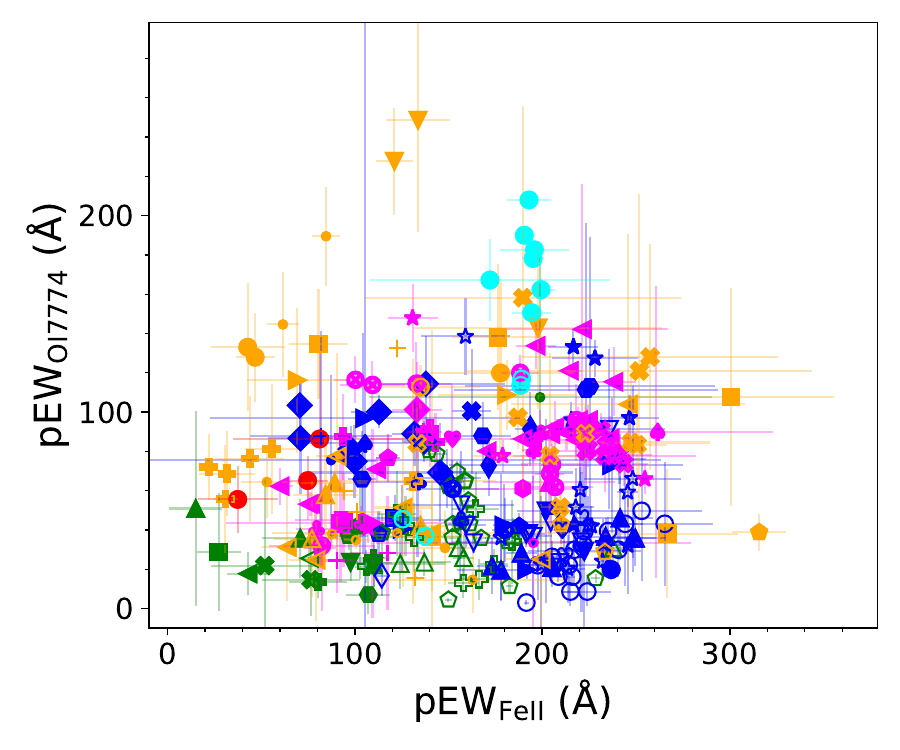}
	\includegraphics[width=0.162\linewidth]{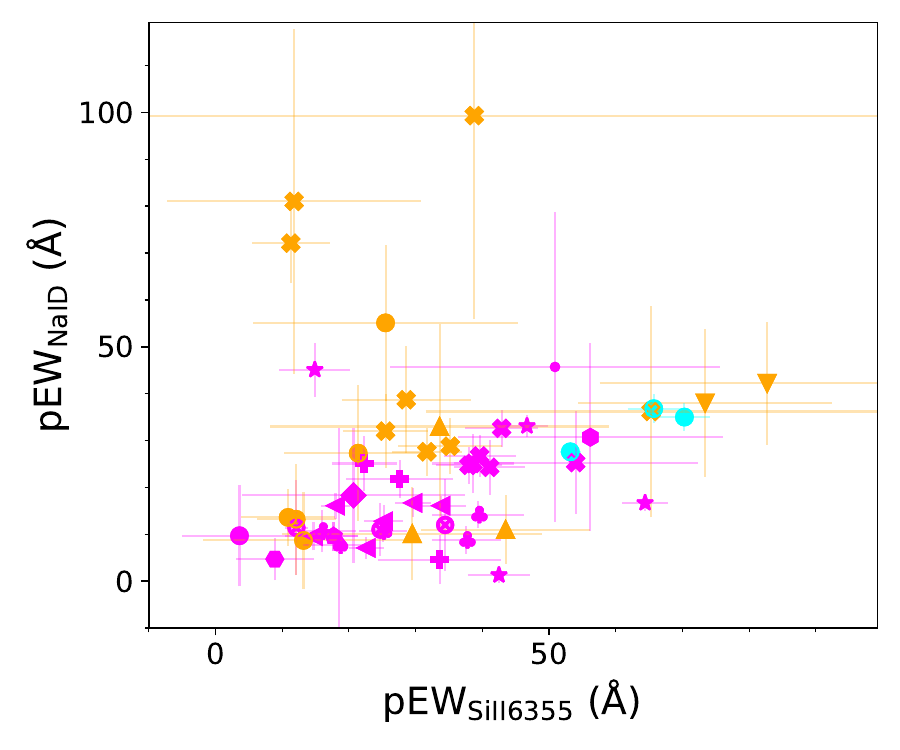}
	\includegraphics[width=0.162\linewidth]{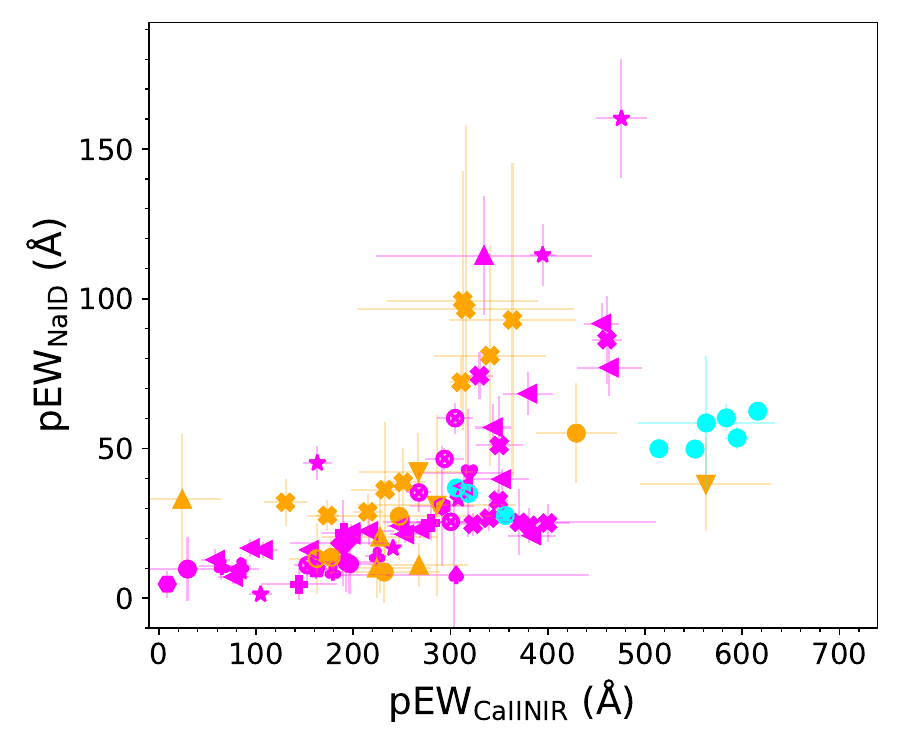}
	\includegraphics[width=0.162\linewidth]{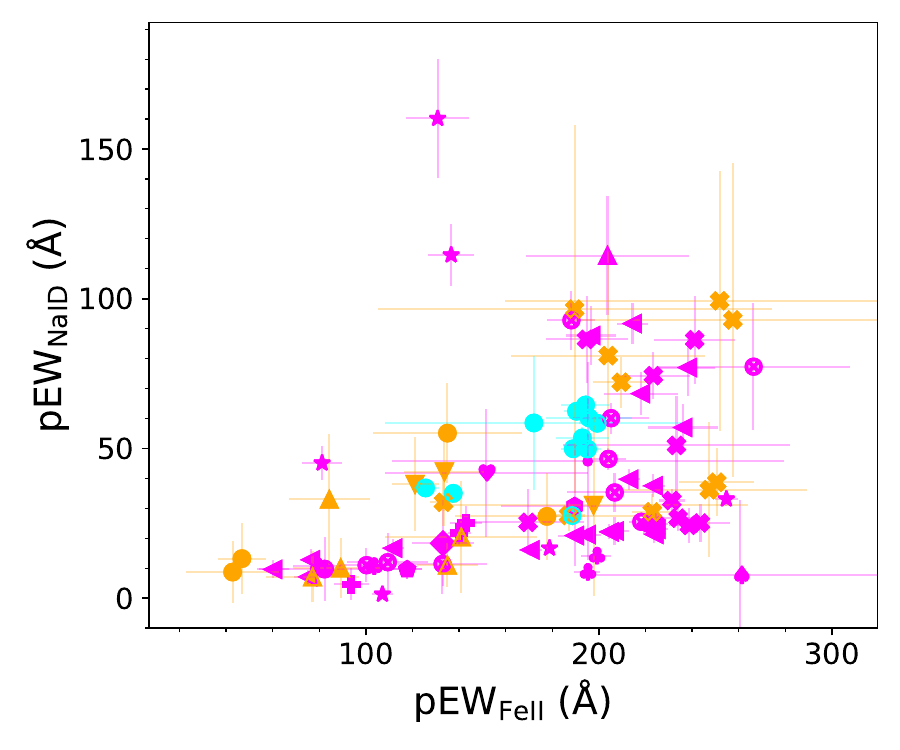}
	\includegraphics[width=0.162\linewidth]{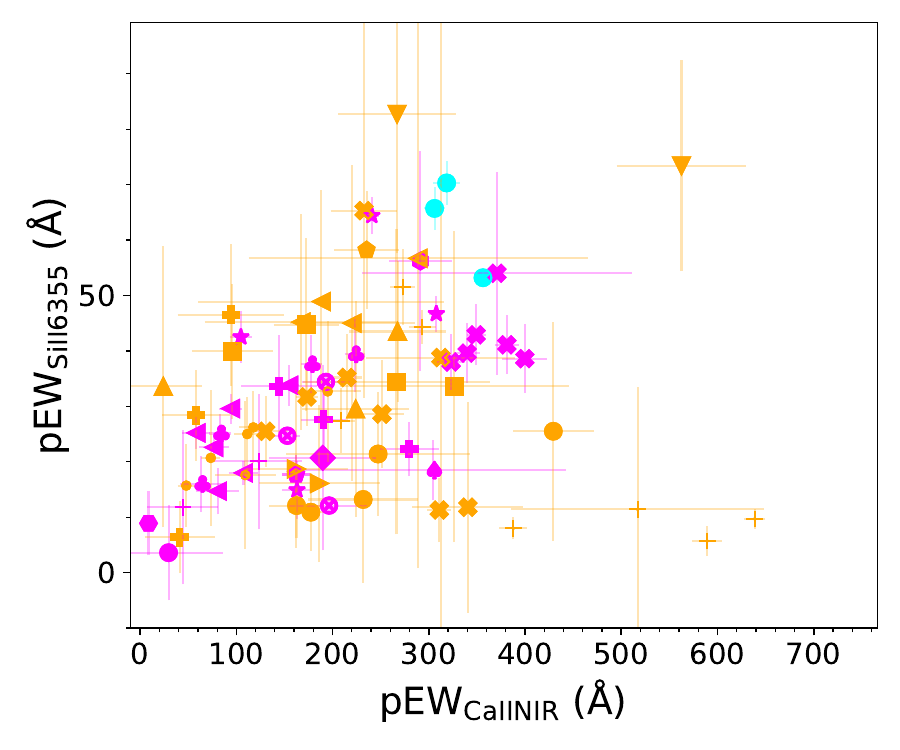}
	\includegraphics[width=0.162\linewidth]{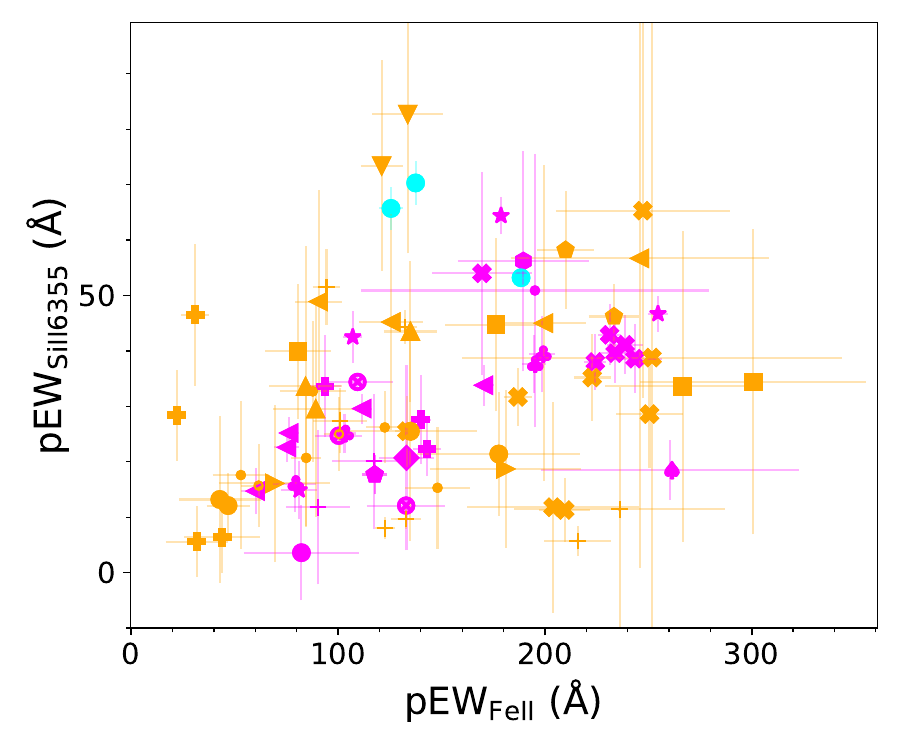}
	\includegraphics[width=0.162\linewidth]{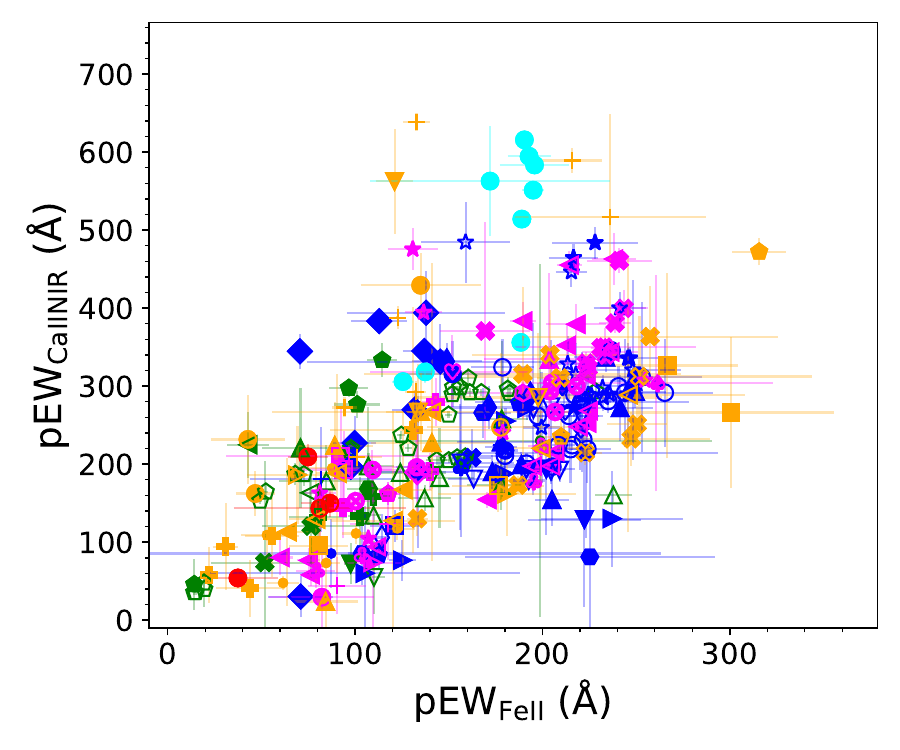}
	
	\caption{Correlation of line intensities of SESNe. Data points follow the same color/symbol scheme as in Fig.~\ref{fig:evol-V-all}.}
\end{figure*}
\clearpage

\FloatBarrier
\clearpage

\end{appendix}
\end{document}